**PUBLICATION PRODUCTIVITY AND CITATION ANALYSIS OF
THE MEDICAL JOURNAL OF MALAYSIA:
2004 to 2008**

**SANNI, SHAMSUDEEN ADEMOLA**

**DISSERTATION SUBMITTED AS A PARTIAL FULFILLMENT OF
THE REQUIREMENTS FOR THE DEGREE OF MASTER IN
LIBRARY AND INFORMATION SCIENCE**

**FACULTY OF COMPUTER SCIENCE
AND INFORMATION TECHNOLOGY
UNIVERSITY OF MALAYA
KUALA LUMPUR**

**JANUARY 2011.**

# Abstract

The very few studies that address the publication and citation practices among researchers in the medical fields in Malaysia motivated this study. This study adopted bibliometric methods to examine article productivity and citation analysis of the oldest medical journal in Malaysia, *Medical Journal of Malaysia* (*MJM*) from years 2004 to 2008 using data extracted from the *Malaysian Abstracting and Indexing System (MyAis)* database. A total of 580 papers were analyzed. The result indicates that *MJM* sustained consistent publication productivity with at least 100 to an average of 116 papers per year. A total of 2177 authors singly or jointly contributed articles and author's productivity pattern conforms slightly to *Lokta's law* of author productivity. Most productive authors are senior medical researchers affiliated to higher institutions and foreign contribution was very small (6.20%), as most of the papers were contributed by Malaysians (88.96%). More than 90% of the papers were joint-authored with 3 authorship being the highest pattern of collaborative work. High degree of collaboration index of 0.9 was observed using *Subramayam's formula*. Most collaboration were between institutions within Malaysia. Analysis of keywords indicated active contributions in topics such as diabetes, cancer, endoscopy, hypertension, and tuberculosis. Only 61 original research articles acknowledge funding, mainly from the Ministry of Science, Technology and Innovations as well as universities own research funds. The average number of references used per year is 1391 citations, while the average number of citation per article is 12. Most cited references are from scholarly journals (87.67%). The age of citations used were mainly about 1 to 11 years old (63.87%) with publications of over 20 years, still being referenced suggesting the longevity of the usefulness of works published in the medical field. Through *Bradford's law*, analysis of journals cited indicated the proportion of number of journals in the three productive zones is 43: 210: 1270, in the ratio 1: 5: 25. The core journals revealed 22 titles. Citation to *MJM* articles were obtained from *Google Scholar* which revealed that 76.8% of all the articles have been cited at least once both by main stream as well as Malaysian journals and the citing authors came from seventy six different countries. *MJM* recorded its highest Impact Factor (*IF*) in year 2007 (0.616) followed by 2006 (0.456), 2009 (0.378) and 2008 (0.367), while the result of a five years *IF* is 0.577.

# Acknowledgements

All praises and thanks are due to Allah, The Owner of knowledge, Most Gracious, Most Merciful, who has guided me through the journey of life up to this moment.

Worthy of thanks are my dear parents, who granted me this academic opportunity by sponsoring my masters program. I wish to categorically thank my father, Alhaj Ishaq Kunle Sanni for his financial support, patience, prayers, and love all along and my mother, Alhaja Rafiah Idowu Sanni, for her loving support, regular prayers, advice, songs and melodies.

I am very grateful to my supervisor Professor Dr. Zainab Awang Ngah who never stopped believing in me from the first day. Her love, encouragement, guidance, assistance and goodwill have helped me a lot through my life as a student and research-assistance under her tutelage in Malaysia.

To all my teachers and lecturers, Professor Dr. Zainab Awang Ngah who taught me how to organize information; Ass. Professor. Dr. Dhiljit Sing, who guided me through the methods of research; Ass. Professor Dr. Abrizah Abdullah, from whose class I fell in love with literatures and scholarly communications; Ass. Professor. Dr. Edzan Nasir, from whom I learnt the intricacies of reference materials, Mrs. Leuvelin Cacha, who taught me how to manage internet resources, to mention just a few; and to all characters that have transferred knowledge to me directly and indirectly. To all my family members, brothers and sisters, far and near, friends and other wonderful colleagues who have shown concern and support, I am forever indebted to you all.

# Table of Content

**CHAPTER 1**

INTRODUCTION

**CHAPTER 2**

LITERATURE REVIEW

**CHAPTER 3**

METHODOLOGY

**CHAPTER 4**

DATA ANALYSIS

# List of Figures

# List of Tables

# CHAPTER 1

# INTRODUCTION

## 1.1. Introduction

The fundamental process of research is communication, and scholarly communication is central to the growth and development of the scientific community. By scholarly communication, we mean the study of how scholars in any field use and disseminate information through formal and informal channels (Borgman, 1990). Scientists are expected to promote their disciplines by highlighting their contributions to human well being (Nwagwu, 2007). One way of achieving this is by studying the literature of a discipline (Al-Qallaf, 2003) through varieties of medium: books, journals articles, conference proceedings etc. Among these varieties, journal articles appear to be the principal channel used in disseminating original research and commentary on current developments. They are the primary formal channels for communicating theories, methods, and empirical results (Rice, 1990), the most appropriate source for the study of information dissemination, knowledge transfer and the diffusion of innovation within a profession (Hashimah, 1997; Schubert, 2002), which make them the most important vehicle of scientific communication (Sen and Zainab, 1996).

Identifying high-quality science is necessary for science to progress (Bornmann and Daniel, 2009), and one approach to facilitating identification of sound medical evidence is to identify high-quality journals that are likely to publish high-quality research (Lee et al, 2002), a major issue for publishers, researchers and authors (Omotayo, 2004), extremely relevant to promotion committees, funding agencies, national academies and politicians, all of whom need a means by which to recognize and reward good research and good researchers (Bornmann and Daniel, 2009; Durieux and Gevenois, 2010). It follows by saying that, the quality of a scientific article of today would depend on the prestige of the periodical in which it is published and on the number of times the article is cited in the literature (Cunha-Melo et al 2006; Zainab, 2006; Nwagwu, 2007; Lee et al, 2010). This is based on the premise that frequently cited articles are most likely to have maintained a sufficiently larger influence in the field or research front than those less cited and therefore are regarded to be very significant to a field.

Tiew (1998) explained that a journal is also a social institution that confers prestige and rewards to those associated with it. Authors who had contributed to journals would be measured by their productivity and hence the contributions will add to their professional credentials among their peer. This would show the impact of and contribution of authors to their respective fields (Chiu and Ho, 2007). Publications also facilitate the establishment of priority and ownership of new inventions and ideas in case of any dispute relating to claim. Besides, editors and referees of journals are also given due recognition and prestige as to editorial and refereeing performed (Tiew, 1998). Thus, this Journal level of analysis reflects the international nature of science (Zitt and Bassecoulard, 1998) and due to their frequency of publication, are suitable elements in analysis of statistical significance (Hashimah, 1997).

Similarly, Tijssen et al (2002) discussed that, part of the important step in establishing a solid reputation of scientific excellence is to track the intellectual influence of tangible research outputs on the external world in terms of their contributions to scientific progress and pushing back international research frontiers. In this tracking process, scholars make use of bibliometrics methods, defined as the "application of mathematics and statistical methods to books and other media of communication" (Pritchard 1969 quoted in Borgman, 1990; Durieux and Gevenois, 2010) or "The quantitative analysis of bodies of literature (such as articles, books and patents) and their references: citations and co-citations". This provides vital information that enables us to understand the structure and patterns in literature, and other medium of scholarly communications as they change over time.

Bibliometrics can provide some useful perspectives on the scholarly communication channel, development, and a measure of research performance of a research field (Moed and de Bruin, 1990; Schubert, 2002; Hu and Rousseau, 2009), the state of science in developing countries (Arvanitis et al, 2000), or an overall picture of the scientific output of a country (Gauthie, 1998), which have relatively high reliability because of the well-maintained databases from which data are drawn (Paisley, 1990), and another advantage of the use of bibliometric data are availability and flexibility, which has proven their usefulness in recent years (Gauthier, 1998).

Hence, since scientific research is a global enterprise that should be universal for the benefit of mankind, a very important task is to examine journals in terms of their publication quality, authors productivity, international visibility, scientific influence and penetration in international literature, in order to determine the impact of these journals in their respective discipline and how the research focus might have changed over the years.

*Medical Journal of Malaysia (MJM)* is the oldest medical journal in the country and few studies have been dedicated to its examination. The only known study that examines *MJM* was by Hashimah in 1997, who conducted a citation analysis of the journal between 1990 -1995. This was more than 10 years ago and it was observed that it is required to draw yet more bibliometric research attention to this medical journal.

In sum, this study would apply bibliometric analysis to examine the structure, content and patterns in the publication of *Medical Journal of Malaysia* between the periods of 2004 to 2008.

## 1.2. Background of the Study

Research activities and publications in Malaysian medical field has a long history, and have witnessed consistent progress in recent years. The medical field continues to grow and expand as an important scientific discipline in the country. Development in this field has great positive impact on clinical services, healthcare, well being, and economies of the nation, as scientist find cures and new treatment methods for diseases, develop strategies to control epidemics, and the country recording positive population growth every year. There has been progress all along, and this progressive trend is not only dependent on the performance of the academia but largely on social and political interest towards medical and health science research.

There is funding allocated through government and private agencies to stimulate the purpose of research and there has been emphasis on the importance of national and international collaboration of scientist from different regions of the world to support Malaysian research community. In addition, the availability of electronic databases and open access database like *MyAis (Malaysian abstracting and indexing system http://Myais.fsktm.um.edu.my/)* which index journals published in the country has reduced major obstacles and bottlenecks, which slows down or even prevent information dissemination and retrieval.

The consequence of all these was the increase in awareness and accessibility to information. Accessibility here refers to the degree of ease with which the journals could be accessed, or the degree of ease with which scholarly journals and articles can be obtained (Zainab, 2006). Besides, it has also helped scientist to publish their work in time and to build on works of others, as a result, new knowledge is exchanged for recognition providing a discussion forum for research in the scientific community.

Borgman and Furner (2002), stated that, "the cycle of scholarly activities is blending in a continuous, looping flow, as people discuss, write, share, and seek information through networked information systems", assisting in the dissemination, sharing and analyses of scientific knowledge regardless of frontiers. These recent development and enhancement of online databases have open more opportunities for students and researchers alike to be aware of available useful information resources and to access published journal articles for studies, decision making and to explore its impact in a particular field of study through bibliometric analysis, which would in effect lead to a more balanced flow of scientific knowledge.

With the 9th Malaysia Plan (2006 - 2010) (Ninth Malaysian Plan, 2006) and the government's aim to place itself in the nation of developed countries by 2020, there is need to continuously assess and evaluate the level of research performance in the country. To achieve its objectives, the Malaysian government rely more on the awareness, conduct and increase use of research evidence to continually improve its performance to meet local and global demands and to support all levels of decision making (Ministry of Health Malaysia, 2008). The government believes that improved research and development is a major step towards realizing Malaysia as a developed nation of healthy individuals and communities, to correspond with vision 2020. In light of this, the government has earmark a sum of RM2.3 billion for setting up National Institute of Cancer, National Forensic Institute and National Institute for Oral Health and RM10.28 billion for disease prevention (Ninth Malaysian Plan, 2006).

In addition, it is evident that scholarly journal was the main channel used by Malaysian biomedical and health science researchers to publish their research outputs. This channel constitutes more than seventy percent of the total amount of publications (Hashimah, 1997; Hazmir, 2008). Such information from bibliometric studies of journal publications would be key to better strategy development and implementation.

Bibliometric analysis would be used to examine the *Medical Journal of Malaysia (MJM)*. Following the work of Tiew (1997) and Utap (2008), citation, content analysis and journal impact factors which are the key performance indicators and the commonest metrics used to report research performance against targets for researchers, journals and institutions (Wiles and Williams, 2010) would be used to examine quantitative investigations of *MJM* articles and to establish relationships between authors and their work, the publication pattern, authorship pattern, productive authors, types of collaboration, institutional affiliations of authors, keywords in articles, research funding, types of reference materials, age of reference materials, list of core journals, citations received by *MJM* and journal impact factor.

### 1.2.1. *Medical Journal of Malaysia*

*Medical Journal of Malaysia* is the oldest scientific journal in Malaysia published since 1890. It originated as the *Journal of the Straits Medical Association (JSMA) (1892 - 1897)*. The Straits Medical Association was established by a group of medical officers who saw a need to form a professional body for medical practitioners in Singapore to discuss and research on local diseases and other medical subjects. The association noted that cooperative research and learning was important for the medical community and that the tropical climate of the region presented unique perspectives to the study of medicine (Chen, 1982; Lim, 1995; Chia and Yeong, 2006). The association published its journal (*JSMA*) in March 1890 under the editorship of Max F. Simon (Dr) who was the Principal Civil Medical Officer of the Straits Settlements. Eighteen months after its establishment, the association's membership grew to 18 ordinary, 7 corresponding and 4 honorary members. The pioneering batch of office-bearers included D.J. Galloway (Dr) (President), W. Gilmore Ellis (Dr) (Vice President), Leask (Dr) (Committee Member), Tripp (Dr) (Committee Member), Von Tunzelmann (Dr) (Committee Member) and E. W. Von Tunzelmann (Dr) (Honorary Secretary and Treasurer).

The association held its preliminary meeting on 11 March 1890 and its attendees included Mugliston (Dr), Leask, Galloway and Von Tunzelmann. During the meeting, it was proposed that the society be called the Straits Medical Society and a committee was appointed to draft rules for the society. On 22 April 1890, the society held its third meeting, and the Straits Medical Association was formally adopted as the society's name. The first office-bearers were also elected and sixteen association rules were accepted.

On 17 December 1892, the association decided to initiate talks with the British Medical Association with regards to its intention to be registered as a branch of the British Medical Association. On 1 January 1894, the association was admitted as a branch of the British Medical Association and became known as the Malaya Branch of the British Medical Association. The assets and liabilities of the Straits Medical Association, including the library and museum, were transferred to the newly formed Malaya Branch of the British Medical Association (Chia and Yeong, 2006).

In light of this new development, the publication of the *Journal of the Straits Medical Association* (*JSMA*) was discontinued. However, it was later revived in 1904 as the *Journal of the Malaya Branch of the British Medical Association (1904 - 1907)*. Due to lack of contributions, the journal could not sustain its publication and suffered from a series of false starts. Some of the names under which the journal was published include *the Malaya Medical Journal (1911 - 1912), the Transactions of the Malaya Branch of the British Medical Association (1922 - 1923), the Malayan Medical Journal (1926 – 1937), The Journal of the Malayan Branch of the British Medical Association (1937 - 1941), Medical Journal of Malaya (1946 - 1971) and Medical Journal of Malaysia (1971 - present)* (Chen, 1982; Lim, 1995; Chia and Yeong, 2006). Besides the presentation and publication of papers, the association was also instrumental in the drafting of three ordinances, namely, the Medical Registration Act, the Pharmacy Act and the Poisons Act.

Reflecting the rapid political developments that were taking place during the post-war era, the association went through sweeping changes during the 1950s. On 24 October 1959, the Malayan Medical Association was formed through an amalgamation of the Malaya Branch of the British Medical Association and the Alumni Association of Malaya. The Malaya Branch of the British Medical Association represented doctors who were recruited from overseas while the Alumni Association, established in 1923, comprised locally recruited doctors who graduated from the King Edward VII College of Medicine in Singapore (Chen, 1982; Lim, 1995; Chia and Yeong, 2006).

With the formation of the Malayan Medical Association (later known as Malaysian Medical Association) (MMA), the Journal became the official organ, supervised by an editorial board. Some of the early Hon. Editors were Mr. H.M. McGladdery (1960 - 1964),

Dr. A.A. Sandosham (1965 - 1977), Prof. Paul C.Y. Chen (1977 - 1987) (Malaysian Medical Association, 2010 http://www.mma.org.my/).

The journal is the oldest medical journal in the country and one of the oldest in the region. It is published quarterly and can be found in medical libraries in many parts of the world. The Journal also enjoys the status of being listed in *Index Medicus*, the internationally accepted reference index of medical journals. The editorial columns often reflect the association's views and attitudes towards medical problems in the country, and the articles published deals with all aspects of medicine (Malaysian Medical Association, 2010, http://www.mma.org.my/). Being the oldest journal, it is felt that a thorough study of the journal is appropriate to understand the trend and growth of medical article publication activity in Malaysia.

### 1.2.2. Publication Practices of *Medical Journal of Malaysia*

The *Medical Journal of Malaysia (MJM)* is published quarterly and accepts articles of interest on all aspects of medicine, especially if it relates to Malaysia, in the form of original papers, continuing medical education (CME) articles, case reports, short communications and correspondence. The journal also publishes brief abstracts, of original papers published elsewhere, concerning medicine in Malaysia. *MJM* allow authors to submit the names of two possible reviewers whom they feel are qualified and suitable to review their paper. The reason is to hasten the process of peer review. Reviewers normally are not involved in the work presented and are always from another institution that would be approached by *MJM*.

### 1.2.3. Types of Papers *MJM* Currently Hold

This information was taken from the *Malaysian Medical Association (MMA http://www.mma.org.my/)* website. The types of materials accepted for publication are as follows.

**(a) Manuscripts**

Manuscripts are submitted online through *MJM* Editorial Manager and instructions for registration and submission are found on the webpage. Authors are able to monitor the progress of their manuscript at all times via the *MJM* Editorial Manager. For authors and reviewers encountering problems with the system, an online Users' Guide and FAQs can be accessed via the "Help" option on the taskbar of the login screen. Manuscripts may also be submitted in triplicate to the office of the editor.

**(b) Editorials**

These are commissioned authoritative commentaries on topics of current interest or articles in the *Medical Journal of Malaysia*. The members of the current editorial board are:

*Editor-in-Chief* -- Azhar Md Zain

*Editorial Advisor* -- John T Arokiasamy

*Ex-officio* -- Khoo Kah Lin

*Editor* -- Lim Kean Ghee

*Members* -- Abdul Hamid Abdul Kadir, Gurdeep Singh Mann, Siva Achanna, Sivalingam Nalliah, Lim Thiam Aun, Kelvin Lim Lye Hock, Sim Kui Hian, H Krishna Kumar, Teoh Siang Chin, S Fadilah Abdul Wahid, Harvinder Singh, Andrew Tan Khian Khoon, Zabidi Azhar Mohd Hussin, Tan Si Yen, and John George

*Office Manager* -- Matilda Cruz

**(c) Review Articles**

Review Articles are usually solicited articles written by Malaysian experts providing a clear, up to date account of a topic of interest to a local audience. The review normally includes an update of recent developments (from the past 1 to 2 years) including research done in Malaysia. Alternatively, Malaysians of international standing residing abroad are also invited to contribute Review Articles on topics related to their area of expertise.

**(d) Continuing Medical Education (CME) Articles**

A CME article is a critical analysis of a topic of current medical interest. The article always includes the clinical question or issue and its importance for general medical practice, specialty practice, or public health.

**(e) Case Reports**

Papers on case reports are written by authors who have unique lesson in the diagnosis, pathology or management of the case they are reporting, and are able to report the outcome and length of survival of a rare problem. Another type of paper accepted is called "Short Communications" which consist of a Summary and the Main Text.

## 1.2.4. Publication Requirements

Articles accepted are expected to adhere o the following requirements.

**(a) Written Language**

The common formats accepted are in English. However, papers may be submitted in Bahasa Malaysia, which is accompanied by a short summary in English. Scientific names, foreign words and Greek symbols are clearly indicated and underlined.

**(b) Acknowledgments**

Authors are expected to acknowledge grants awarded in aid of the study, which state the number of the grant, name and location of the institution or organisation, as well as persons who have contributed significantly to the study.

**(c) References**

References are numbered consecutively in the order in which they are first mentioned in the text. References are identified in text, tables and legends by Arabic numerals (in parenthesis).

**(d) Form of Reference**

*MJM* adopt the form of references adopted by the US National Library of Medicine that is used in the *Index Medicus* and which has been approved by the National Library of Medicine. The titles of journals are abbreviated according to the style used in the ***Index Medicus***.

## 1.2.5. Health Research and Development (R&D) In Malaysia

The issue of healthcare is a subject that concerns virtually all living beings, and part of the prerequisite of a well-developed country is a sound and healthy Research and Development (R&D) program for improved health care and disease prevention. Medical and Health Science research in Malaysia is sponsored by the public and private sectors, and non-governmental organizations. The major provider and financier of Medical and Health Science research is the Ministry of Health, which gives high priority to research under the auspices of The National Institute of Health (NIH) with strategies to enhancing research and development to support "Evidence-based decision-making". The National Institute of Health (NIH) is a network of seven research institutes: Institute of Medical Research (IMR), Clinical Research Center (CRC), Institute of Public Health (IPH), Institute of Health System Research (IHSR), Institute of Health Management (IHM), Institute of Health Promotion (IHP), and National Institute of Natural Product, Vaccines and Biological (NINPVB).

The oldest amongst these institutes is the Institute of Medical Research (IMR), which was established in 1900. In 1967, it was declared Malaysia's National Centre for Tropical Medicine under the SEAMEO (Southeast Asian Ministers of Education Organization). Research programmes of the Institute are geared towards the various priority areas of research, which is in line with the health problems in the country. Research findings and results are used by managers and administrators, in the various ministries and other government agencies, in forming, implementing and evaluating programmes and activities for the diagnosis, prevention and control of diseases in the country (National Institute of Health Malaysia, 2008).

Efforts to develop Malaysia's human capital have strengthened education and research capacity in the health sector. The World Health Organization (WHO) also partners with Malaysia on the issue of health information and health research, provide support for critical health systems research, epidemiological and clinical research that would impact on policy and decision-making and improve the effectiveness and efficiency of the health system (Ministry of Health Malaysia, 2008).

## 1.3. Statement of Problem

Scientist and researchers publish their work in scholarly journals as an avenue to present new research and or critique an existing idea. In so doing, they make their work available for people to read, use, and reproduce in another form for the benefit of the entire society. When a work is used by an author, the author cites and makes reference to the used work for recognition. This process occurs often in scholarly publishing and studying this pattern of citations and the structure is a common culture and practice in developed countries. This has helped scientist discover new knowledge especially with the popularity of electronic scholarly communication, which is growing without bounds. However, the problem is that, there are very few systematic studies that cover the problem of citation practices among people in developing countries or the diffusion of information across national boundaries (Hashimah, 1997). There are also very small number of bibliometric studies carried out on Malaysian scholarly journals (Tiew, 1998; Utap, 2008). Hence, the study of the state of the oldest medical journal in Malaysia will help to understand the trend of contributions, who are contributing, subjects covered by the contributors and the referencing patterns indicated in the contributions.

Another major factor affecting scientific journals generally from developing countries is their accessibility and visibility (Tiew, 1998; Omotayo, 2004; Zainab, 2006; Utap, 2008), which obscure them from analysis and evaluation in related field, since the more visible and accessible a research work is, the easier measurement becomes. Therefore, evaluation of journals has become inevitable if journals published in the country are to be in the hierarchy of highly rated international journals. In respect to this, it was observed that very few studies have been devoted to the study of the medical journals in Malaysia. The only known study was that of Hashimah (1997), who conducted a citation analysis of the *Medical Journal of Malaysia*. This was more than ten years ago and the scholarly community is interested in understanding how a discipline or field changes over time. This underscores the need for a study of this nature.

### 1.4. Objectives of the Study

The main purpose of the present study is to apply bibliometric analysis, such as citation count and impact factor, in examining journal articles published in the *Medical Journal of Malaysia* between years 2004 to 2008. The 5-year data was harvested from an open access database *(MyAis http://Myais.fsktm.um.edu.my),* which provided the necessary data to support a bibliometric study related to Malaysian medical related authors and their work. Hence, the objectives of this study are to determine the following:

(1) To find out the articles productivity of *Medical Journal of Malaysia* from year 2004 to 2008.

(2) To find out the productive authors, who published in *Medical Journal of Malaysia* by

    (a) Finding the number of contributing authors per publication year.

    (b) Testing the productivity pattern using *Lotka's law*.

    (c) Identifying the productive authors.

(3) To find out the co-authorship pattern in terms of:

    (a) The degree of collaboration among the contributing authors by using *Subramanyam formula.*

    (b) The country affiliation of contributing authors.

    (c) The country collaboration per articles contributed.

    (d) The institutional affiliation of contributing authors.

    (e) The institutional collaboration per articles contributed.

(4) To determine the content of *MJM* in term of:

    (a) The keyword distribution of articles.

    (b) The number of words in article title.

(5) To identify the agencies and organizations funding research as expressed by acknowledgement at the end of articles published.

(6) To study the pattern of citations referenced by articles published by *Medical Journal of Malaysia* in term of :

    (a) The distribution of references per year.

    (b) The distribution of citations per article.

    (c) Distribution of citations according to bibliographic forms.

    (d) Age of the referenced literature.

    (e) The half-life of references used.

    (f) The referenced core journals using *Bradford's Law of journal distribution*.

    (g) Language of reference sources.

(7) To find out the pattern of citations received by articles published in *Medical Journal of Malaysia* through:

    (a) Total citations received for articles published between year 2004 to 2008 through *Google Scholar (Harzing Publish and Perish)*.

    (b) Journal self-citations.

    (c) Format types of documents citing *MJM*.

    (d) Scholarly journals citing *MJM* between years 2004 to 2008.

    (e) Country Affiliations of authors citing *MJM* between years 2004 to 2008.

    (f) Journal impact factor of *MJM* articles published from year 2004 to 2008.

## 1.5. Research Questions

The research question follows the objectives of the study:

(1) What is the article productivity of *Medical Journal of Malaysia* between the years 2004 and 2008?

(2) Who are the productive authors publishing in *Medical Journal of Malaysia* and does author productivity follows the *Lotka's law* of author productivity?

(3) What is the co-authorship pattern of articles published in *Medical Journal of Malaysia* and what is the degree of collaboration indicated by published articles?

(4) What is the nature of content of *MJM* articles in respect with the keyword distribution in articles and number of words in titles?

(5) What are the names of the agencies and other organizations funding research as expressed by acknowledgement at the end of *MJM* articles.

(6) What are the patterns, age and format of referenced materials cited by authors publishing in *Medical Journal of Malaysia?*

(7) What is the pattern of citation received by articles published in *Medical Journal of Malaysia* and what is the calculated impact factor (*IF*) for *MJM*?

## 1.6. Significance of the Study

The process of sharing, preserving, and communicating research findings has always been recognized and acknowledged because of the vital role it plays in the development of the community. This process of information and knowledge sharing continues to grow among the teaching, learning and research community, and the medium mostly employed are scholarly journals. Sustaining this growth requires scientist to keep track of patterns, structure, and approaches applied to journal publications. In this way, research communities and other interested groups would be able to identify individuals, institutions, research area, and affiliations based upon their role and contributions to the journal and the medical profession in general.

In order to retain credibility amongst the stakeholders, public and private agencies that support research, and to make a strong case for more funding, it is ideal to devote a study that provide information on the quality and status of research, how productive and consistent the authors have been in the creation, dissemination and diffusion of knowledge and likewise how this research impacts the society.

Bibliometric study on the *Medical Journal of Malaysia* would also provide useful information on factors that have contributed to the progress of research. For example, research strategy, and international collaborations of authors. Examining the acknowledgement pages of articles in this journal would allow identifying institutions and agencies that have been providing financial aids and support towards the success of research. Furthermore, findings from this evaluation can be used as an instrument in decision-making by the government and funding agencies, and in developing research structure and likewise improving the quality of medical research in the country. It would also help to improve the public support to, and at the same time assess the future prospects of medical and health science in the country. In addition, it would also help to assess the adequacy and allocation processes of funding.

Like all other bibliometric studies on journals, it is hoped that this study would encourage further studies on scholarly journals in Malaysia.

## 1.7. Scope and Limitation of the Study

The study examines a single journal, which is *Medical Journal of Malaysia* through a bibliometric approach. It confines its scope to articles published within a 5-year period between 2004 and 2008.  The sample was retrieved from *MyAis (Malaysian Abstracting and Indexing System http://Myais.fsktm.um.edu.my)*, which is the only known open access database that contains bibliometric information about the journal under study. For accuracy, the articles retrieved from *MyAis* were crosschecked with those made available in the website of the *Malaysian Medical Association (MMA http://www.mma.org.my/)* which publishes *Medical Journal of Malaysia.* Thus, like most bibliometric researches, this study is limited and affected by the shortcomings of the databases (*MyAis and Google Scholar*) from which data were retrieved. Some of which may be:

1. Duplicate records.

2. Misspelling of author's names or names spelled the same way.

3. Incorrect, incomplete addresses or citations.

4. Changes in journal titles etc.

*MyAis* provide details about article titles, author's names and affiliations, acknowledgement, keywords, and references, while *Google Scholar* provides details on the citations received by *MJM*. Recordings were made of all the data and information obtained.

## 1.8. Summary of Contents

The content of this study is divided into 5 chapters. Chapter 1 deals with the background, use and applications of bibliometric methodologies in scientific discipline. The chapter went down memory lane to discuss the history and publication practices of *MJM* from inception, in order to justify the significance of the study. The aim and significance of the study were highlighted to understand the approach in answering the research questions. In Chapter 2, the study focuses on reviewing literatures that explain the meaning and also presents results of previous bibliometric studies. In Chapter 3, the study described the research methods employed and how publication data were tracked down to generate tables and figures. Chapter 4 present and describe the results obtained through the analysis of bibliometric data used in the study, while Chapter 5 further explain the findings obtained from data analysis which leads to conclusions and summary as  recommendations were made for future research.

# CHAPTER 2

# LITERATURE REVIEW

## 2.1. Introduction

The focus of this chapter is to describe and explain the meaning of bibliometrics through review of literatures relevant to the areas of knowledge in which bibliometric research have been applied and areas of knowledge that explain the use of bibliometric methods. Literatures employed for this task were retrieved from several online databases which include: Lisa, Library Literature, Springer Link, Emerald Fulltext, Science Direct, Proquest, and Google Scholar. Some of the search terms used were: "Journal Scientometrics" "Bibliometrics and medicine" "Scientometrics and medicine" "webometrics and medic*; informetrics and medic*" "communication pattern in medicine" "publication productivity and medic*, etc. The findings of the studies used in this review would allow us to understand, explain, describe and evaluate the communication behavior of researchers in the medical field.

## 2.2. Bibliometric Studies

Bibliometric measurements, though controversial, are useful in providing measures of research performance in a climate of research competition and marketisation (Wiles and Williams, 2010). Borgman (1990) wrote that the most widely accepted definitions of Bibliometrics are:

(1) To shed light on the processes of written communication and of the nature and course of development of a discipline (in so far as this is displayed through written communication), by means of counting and analyzing the various facets of written communication (Pritchard, 1968).

(2) The assembling and interpretation of statistics relating to books and periodicals…to demonstrate historical movements, to determine the national or universal research use of books and journals, and to ascertain in many local institutions the general use of books and journals (Raisig, 1962).

Glänzel (2003) further the above definition to explain that, books, monographs, reports, theses and papers in serials and periodicals are units of bibliometric analyses. Since certain standards are postulated for such units, the scientific paper published in refereed scientific journals has proven to be the unit most suitable for bibliometric studies. This is as a result of factors such as: the reviewing system, the criterion of originality of research results, the availability of literature and the more or less transparent rules that controls their publication.

Hence, the scientific paper has become the basic unit of bibliometric research. Webster (2005) noted that bibliometric analyses can provide reliable tools in mapping the development of scholarly disciplines which can be of use in research policy, as well as in domain analysis in information science, library collection development or publishing. Moreover, it was explained by Hendrix (2008) that bibliometric statistics are used by institutions of higher education to evaluate the research quality and productivity of their faculty. On an individual level, tenure, promotion, and reappointment decisions are considerably influenced by bibliometric indicators, such as gross totals of publications and citations and journal impact factors. Bibliometric analyses can be conducted on the bases of any sufficiently large publication list compiled and issued, for instance, by a scientific institution. Nevertheless, most reliable sources are the big specialized or multidisciplinary databases that have been first provided in printed form but later on also in electronic form (magnetic tape, CD-ROM, on-line) (Glänzel, 2003).

Furthermore, bibliometric methodology comprises components from mathematics, social sciences, natural sciences, engineering and even life sciences, hence, bibliometrics is one of the rare truly interdisciplinary research fields today which extend to almost all scientific fields (Glänzel, 2003).

## 2.3. Bibliometric Measures

Bibliometrics studies offer a powerful set of methods and measures for studying the structure and process of scholarly communication (Borgman and Furner, 2002). These measures are referred to as bibliometric indicators; defined as: specific concrete examples that demonstrate research impact as a result of a research finding or output (Hendrix, 2008).

Gauthier (1998) explained that bibliometric indicators provide the only overall picture of the scientific output of a country. He then presented two subdivisions of bibliometric indicators: descriptive indicators and relational indicators. Descriptive indicators measure the volume and impact of research at various levels, providing a means of identifying trends in papers, patents and citations and the citations they contain. Relational indicators on the other hand, helped identify links and interactions between the actors of national and international systems of science and technology. Thus, these indicators emphasize the relationships between researchers, institutions and research fields and such interactions constitute the flow of knowledge.

In bibliometrics, as in the other fields, the derived measures or metrics are typically counts of the frequencies with which events of specified types are observed to occur, which (once expressed as ratios of the total number of observed events) may be considered as probabilities of occurrence (Borgman and Furner, 2002). The probability distributions that are thus formed are known as the bibliometric distributions, and these form the basis of certain bibliometric "laws" such as the *Bradford's laws* for journal distribution of productivity in a discipline and *Lotka's law* for the distribution of author's productivity (Borgman and Furner, 2002).

*Lotka's law* describes the frequency of publication by authors in a given field. It states that "the number of persons making 2 contributions is about one-fourth of those making one; the number making 3 contributions is about one-ninth, etc.; the number making n contributions is about $1/n^2$ of those making one; and of all contributors, the proportion that make a single contribution, is about 60 percent (Chung and Cox 1990). Thus, Lotka's empirical finding can be summarized by the equation:

$$a_n = a_1/n^c, \ n = 1, 2, 3,...,$$

Where $\qquad$ $a_n$= the number of authors publishing n papers and
$\qquad\qquad$ $a_1$ = the number of authors publishing one paper

*Bradford's Law* on the other hand, describes how the literature on a particular subject is scattered or distributed in the journals. It serves as a general guideline to librarians in determining the number of core journals in any given field. It states that journals in a single field can be divided into three parts, each containing the same number of articles: 1) core journals on a subject are relatively few in number, that produces approximately one-third of all the articles, 2) a second zone (marginal) contains the same number of articles as the first, but in a greater number of journals, and 3) a third zone (peripheral) contains the same number of articles as the second, but published in a still greater number of journals. The mathematical relationship of the number of journals in the core to the first zone is a constant n and to the second zone the relationship is n². Bradford expressed this relationship as 1: n: n² (Potter, 1988).

Chung and Cox (1990) examined the patterns of productivity in finance literature and the results found bibliometric regularity confirming *Lokta's law*, which state that the number of authors publishing n papers is about $1/n^c$ of those publishing one paper. The results showed that finance literature conforms very well to the inverse square law (c = 2). When data are taken from a large collection of journals, and the law is applied to individual finance journals, it was found that values of c range from 1.95 to 3.26, and also found that top-rated journals have higher concentrations among their contributors. The finding from the study implies that the phenomenon "success breeds success" is more common in higher quality publications.

Delwiche (2003) applied Bradford's Law of Scattering to the list of journals cited in the literature of clinical laboratory science. Three zones were created, each producing approximately one third of the cited references. Thirteen journals were in the first zone, eighty-one in the second, and 849 in the third. The most valuable pieces of information to come out of this study is the identification of the core journals, both for the field of clinical laboratory science overall and for the four specialty areas. The number of journals contained in Zone 1 is larger than for most of the other allied health fields studied in this project, perhaps because of the wide scope of the field.

Nonetheless, some of the most adopted bibliometric indicators are: citation analysis, co-citation analysis, bibliographic coupling and impact factor. Citation analysis is the most celebrated scientometric technique (Haiqi, 1995), defined by (Larsen, 2008) as the quantitative investigations of the reference lists of the documents in a collection and establishes relationships between authors or their work, between journals, between fields, between institutions, and between countries, etc. Co-citation analysis establishes subject similarity between two documents. Bibliographic coupling links papers that cite the same articles, Co-word analysis analyse the co-occurrence of keywords, while the impact factor is considered to be a measure of the resonance of a journal in the scientific community (Larsen, 2008).

Hence, with all its importance as a significant tool, the fundamental demand upon bibliometric indicators is their validity, that is, we have to make sure that we really measure what we are intending and assuming to measure (Glänzel, 2003). This study will employ some bibliometric measures which include: (a) productivity counts, (b) Lokta's law to identify authorship patterns, (c) Bradford's law to identify core journals used by *MJM* authors (d) Subramanyam's formula to measure the collaboration index.

## 2.4. Studies on Citation Analysis

Major issue for scholars in any field is to know where to find up-to-date information on the latest international developments. Few people have time to read everything on library shelves, and with so many different journals being published, it is difficult to know which ones a researcher should actually be focusing on (Smith, 2010). These are one of the issues bibliometric research is aimed at resolving. For instance, in order to create a core list of journals most frequently used by veterinary medical researchers, Crawley-Low (2006) applied bibliometric techniques to analyze the citation patterns of researchers publishing in the *American Journal of Veterinary Research (AJVR)*. He examined the material type, date of publication, and frequency of journals cited. The findings shows that majority of items cited were journals (88.8%), followed by books (9.8%) and gray literature (2.1%). He further divided the journals referenced into 3 even zones to reveal that 24 journals produced 7,361 cited articles in the first zone. One hundred thirty-nine journals were responsible for 7,414 cited articles in zone 2, and 1,409 journals produced 7,422 cited articles in zone 3. He concluded by stating that a core collection of veterinary medicine journals would include 49 veterinary medicine journals from zones 1 and 2.

Schubert (2002) believed that an overview of a particular journal would reflect the development of the research field. As regards, he examined the first 50 volumes of the journal *Scientometrics.* It was discovered that *Scientometrics* has reached and maintained a leading role not only in its immediate field but also in the broader field of Library and Information Science. Authors from 60 countries contributed to the first 50 volumes of the journal and when subject categories were classified into 5, two obvious developments were observed. There was an impressing and steady growth of "Case studies and empirical papers" and "Methodological papers including applications" and there is loss of position of articles on "Sociological approach to bibliometrics, sociology of science" and "Science policy, science management and general or technical discussions". He further presented the geographic and thematic maps of its papers, and also highlights a brief outlook to the future prospects and challenges of the journal.

Gomez (2003) presented the results of a bibliometric analysis of the references cited in the articles published by authors of the Institutode Astrofisica de Canarias during the last decade. The results were in par with the result obtained from rough observations of habits and the surveys conducted earlier among their library patrons. It revealed that their patrons publish mainly in the core astronomical journals, with a preference for A&A, and that above all they cite astronomical journals, ApJ being the most frequently cited. Most of the journals cited were less than 10 years old. The cost-effectiveness of astronomical journals is high whereas that of the physics journals is very low. Fewer books and more journals are being cited in the WWW period and one journal (AJ) has become more visible because of the advent of WWW, and more recent literature seems to be cited more quickly because it becomes available earlier through the WWW. The study was able to determine which kinds of publications and which journals are mainly used, and has provided several interesting quantitative parameters about the use and the cost effectiveness of their library collections.

Hashimah (1997) used citation analysis to investigate information sources used in medical profession in Malaysia with the ultimate purpose of obtaining information to improve the collections and services in the University of Malaya's medical library. She reported that journals constituted 82.8% of the publication format used by medical authors and researchers. The top four most frequently cited authors came from local institutions and 81.8% of the citations were 18 years old or less. Publications in English were relied heavily upon by Malaysian medical professionals. A total of 853 journals titles were cited and majority of them were published in the more developed countries. Ranking of journals by number of citations showed that 14.77% titles accounts for 55.48% of total journal citations. The outcome of the study revealed that bibliometrics research techniques are necessary for future planning and selection of journal titles in the library's collection.

Al-Qallaf (2003) examined the citation in the first 12 years of *Medical Principles and Practice* (*MPP*) in order to identify and describe their trends and patterns. Findings revealed that journal articles were most frequently cited; English language publications dominated the literature; there was a trend of multiple authorship; and the pattern of aging was below the norm for medical literature. He thus concluded that, the results of the study could provide a benchmark to measure the user behavior of a particular group of researchers as well as for the provision of collection development and management decisions. This study is also aimed at identifying and describing the trends and patterns in the publication productivity of *MJM*, to examine the extent of journal citations to other bibliometric format, to study the authorship pattern and find out the half-life of cited literatures.

Fan and McGhee (2008) applied citation analysis to identify the most published authors on the topics of 'cataract' and 'LASIK', the journals in which they publish, and the citation patterns of the most-cited articles by these authors over a 5-year publication period. The results provide the list of 30 most productive authors. The USA and Australia together were the source of more than half of the most-productive authors, and the majority of articles published by the 30 most prolific authors were published in only 10 journals. The impact factors of the publication journals preferred by these authors are influenced by the article citation counts, and not vice versa. This present research would also utilize citation analysis to identify the most prolific authors in *MJM,* and the citation patterns of the articles over a 5-year publication period.

## 2.5. Studies on Content Analysis and Publication productivity

Studies that deal with the publication productivity and content analysis offers an opportunity to identify key areas in a particular discipline, and researchers involved in this areas. In addition, it allows the young and up – coming scientist to identify successful researchers with whom they can interact and also gives a glimpse into newly developing areas of research.

Richter et al (2000) investigated the publication productivity of young scientist in Croatia and found that these young scientist published more scientific papers than the young generations of scientists at the beginning of the nineties; differences between a highly-productive minority, which produced on average half of all scientific publications, and a low-productive majority was already apparent among the young scientists; the productivity of young scientists was in accordance to the productivity patterns typical of a particular scientific fields and discipline. Among the factors which contribute significantly to the explanation of the quantity of scientific publications was attending conferences abroad, followed by scientific qualifications.

Macías-Chapula and Mijangos-Nolasco (2002) did a bibliometric analysis of AIDS documents as produced on Sub-Saharan Africa. AIDSLINE 1980-2000. Results indicated a high pattern of collaboration through multiple authorship. The distribution by type of documents was: 610 (57.98%) journal articles; and 442 (42.02%) corresponded to meeting abstracts. Main journals found were in the following descending order: *AIDS* (68 papers); *Med Trop* (Marseille) (34); *Lancet* (31); *AIDS Res Human Retroviruses* (29); and *BullSoc Pathol Exot* (20). Documents were published mainly in English (84.50%) and French (14.73%). Other languages like German (four documents) and Russian (two documents) were less significant. This study would also provide list of core journals in medical and health science research.

Omotayo (2004) applied bibliometric attributes, including formats and recency of citation, to obtain a content analysis of *Ife Psychologia* from its inception in 1993 to 2002. The result showed that *Ife Psychologia* is a multi-disciplinary international journal that has been able to continue publishing in spite of the problems afflicting journal publishing in Africa, and that the subject coverage spanned over 14 fields, including psychology, education, sociology and mental health. Journals constituted the most common (47.7%) format of publication cited by *Ife Psychologia* contributors, followed by books (45.2%). In addition, contributors to *Ife Psychologia* derive their supporting literature from sources other than the journal itself or their own previous works.

Keiser and Utzinger (2005) investigated trends that occurred in the leading literature on tropical medicine over the past 50 years and observed that there was a strong increase in the number of articles published from 250 in 1952 to 726 in 2002; over the same time span, the median number of authors per article increased from 1 (four journals) or 2 (American *Journal of Tropical Medicine and Hygien*e) to 2.5 (*Leprosy Review*) up to 6 (*Acta Tropica and American Journal of Tropical Medicine and Hygiene*); research collaborations between countries of different HDI ranks increased concomitantly. They also observed that, in 2002, 19.4 - 43.7% of all manuscripts were submitted by authors from different HDI countries, indicating that tropical medicine has become a global endeavor. However, in four of the five journals investigated, the overall percentage of researchers affiliated with low HDI countries decreased over the past 50 years and only a slight positive trend can be observed over the last decade.

Bloom et al (2007) explored the positions of five leading general medical journals (*Lancet, British Medical Journal – BMJ, Journal of American Medical Association – JAMA, New England Journal of Medicine – NEJM, and Annals of Internal Medicine – AIM*) toward the issues of collective violence. They calculated the proportion of war-related articles in the total number of articles published in these five high-impact journals, and in the total number of articles indexed in *PubMed* during the last 60 years. The results showed a continuous increase in the proportion of war-related articles. The findings suggest that the leading general medical journals have taken an active editorial stance toward the issues of war and peace. It was then concluded that high-impact medical journals can make an important contribution to efforts aimed at reducing the risks and consequences of war and violence.

Willett (2007) applied bibliometric analysis to review the articles published in volumes 2-24 of the *Journal of Molecular Graphics and Modelling* (formerly the *Journal of Molecular Graphics*), focusing on the changes that have occurred in the subject over the years, and on the most productive and most cited authors and institutions. He reported that the most cited papers are those describing systems of algorithms, and that the proportion of these types of article is decreasing as more applications of molecular graphics and molecular modeling are reported. The journal is international in scope, receiving papers from around the world, but with the USA being by far the largest source of articles. Thus, it is also part of the objective of this present study to determine the internationality of *MJM* through the number of contributing foreign authors and the extent of collaboration with authors from different regions of the world.

Morley and Ferrucci (2008) bibliometrically analyzed the publication productivity in Geriatrics from 1995 – 2006. The findings report that the top 10 geriatricians with the highest total number of citations came from the editorial board of the top journals in the field; the most productive area in geriatrics has been dementia research; the other area with the most productivity has been epidemiology, particularly in the areas of disability, frailty, sarcopenia, hormones, nutrition, and cytokines. Other well-defined islands of high productivity, which were limited to a few scientists, are delirium, cardiovascular disease, falls, and depression. Surprisingly to the authors are areas of research missing from the list which are at the core of geriatric medicine, such as some classical geriatric syndromes, nursing home care, home care, and palliative care. They suggested that, the areas of research with less productivity represent areas of opportunity for funding agencies and for the next generation of geriatricians.

## 2.6. Studies on International Research Collaborations

It is very important to map countries productivity in the field of scientific knowledge with that of global research production. Some of the studies in this direction were that of Stern and Arndt (1999) who investigated the growth of international contributors to dermatologic literature. Data were drawn from the Institute for Scientific Information, Philadelphia, Pa. Results shows that authors from 121 countries were credited in whole or in part with authorship of original articles. Ten countries accounted for 82% of all articles published as original articles and 87% of citations to these articles. From 1981 to 1996, the proportion of citations attributed to most western European except Scandinavian countries grew significantly (p, .05, t test), but the proportion credited to authors from the United States fell significantly (p,.05, t test). The conclusion was that international representation of author cited articles appearing in the dermatology literature were increasing. The growth of scholarly contributions has been especially great for authors from Western Europe except Scandinavia.

Borkenhagen et al (2002) placed five volumes of the *Journal Psychotherapy Research* under bibliometric scrutiny. They did so by comparing North American and European contributions. The result indicated that 56% articles from the U.S. American/Canadian region provided the largest part of the original papers in the journal, whereas 43% of the articles written by European authors and 2% from other countries (Israel) cover the remaining research matter. They concluded that *Journal Psychotherapy*

*Research* has succeeded in providing a balanced ratio between European and North American contributions.

Falagas et al (2006) applied bibliometric analysis to provide estimates of research productivity in the main journal of the field of tropical medicine by different world regions for the period 1995–2003, using the *PubMed* and the Institute for Scientific Information (*ISI*) "*Web of Science*" database. The result shows that the developing areas of the world produce a considerable amount of research in tropical medicine, and the authors suggested that they probably still need help by the developed nations to produce more research in this field, given the specific geographic distribution of tropical diseases.

Matthews et al (2009) did an exploratory bibliometric analysis of Australia's international collaboration in science and technology. It was revealed that most of the growth in Australia's research publications is associated with international collaboration rather than purely domestic efforts. The proportion of Australian publications in the *Science Citation Index* with international co-authorship has increased from almost 21% in 1991 to over 44% of total publications in 2005. The output of internationally collaborative papers is growing at almost double the rate of purely domestic papers. By the end of this study, we would also be able to determine the numbers and percentage of international collaborative papers among articles published in *MJM* during the period under study.

Wang et al (2009) looked at the effect of cooperation between Chinese scientific journals and international publishers on journals' impact factor (IF), and reported that the data do not suggest that cooperation has improved the journals' IF thus far. It appears that cooperation is generally limited to international distribution, and this has a weak influence on the quality of the journal and its IF, even though the papers can be accessed by worldwide users through publishers' international distribution networks. They submitted that, cooperation with international publishers is one step, but actively working on the quality of the journals is another – and more important – step. This is an important recommendation to take note in this present and future bibliometric research.

Davarpana and Behrouzfar (2009) investigated the extent of the visibility of Iranian international journals through the *Institute for Scientific Information (ISI)* databases for the years 2000-2006. To achieve this, they analyzed the impact factor (IF), total citation (TC), citation rates, self-citation, foreign citation, international citation (IC), international authorship, and subject distribution for the collections. The results indicated that: the visibility rate of Iranian journals is low compared to their international counterparts; the international visibility of Iranian journals differs among disciplines; the increasing citation rate is less than the increase in publication rate; and the majority of authors who published in these journals were Iranians. They concluded that mere inclusion in the *ISI* does not necessarily lead to an increase in visibility. More needs to be done to increase this visibility, and recommended that to change the current conditions of Iranian scientific journals and raise their influence, the quality and visibility of articles needs to be improved. Efforts need to be made to increase their international effect by attracting international researchers to publish in Iranian journals, by encouraging the collaboration of Iranian and non-Iranian researchers, and to publish their results in Iranian journals.

Lee (2009) conducted bibliometric analysis to assess the contributions and achievements of the *Korean Journal of Parasitology (KPJ)* based on the citation data retrieved from 4 major databases; *SCI, PubMed, Synapse*, and *Scopus*. It was found that the *KJP* articles were constantly cited by the articles published in major international journals represented in these databases. More than 60% of 1,370 articles published in the *KJP* from 1963 to June 2009 were cited at least once by *SCI* articles. The overall average times cited by *SCI* articles are 2.6. The rate is almost 3 times higher for the articles published in the last 10 years compared to 1.0 for the articles of the 1960s. The *SCI* journal impact factor for *KPJ* in 2008 is calculated as 0.871. The IF is reported to have increased and it is expected to increase further with the introduction of the *KJP* in the SCI database in 2008.

## 2.7. Studies on Countries' Research Production

One of the most tangible production units of science in any country is research publications, and indeed the number of publications and the amount of citations these papers generate are often used as indicators of quantity and quality of research (Levi et al, 2009).

Gomez et al (1995) applied a series of bibliometric and socioeconomic indicators to study the production in Spanish biomedical main-stream science in the years 1986-1989. It was observed that Madrid and Catalufia were in an outstanding position as regards to geographical distribution, institutions involved, and active centers. The average level of the Spanish research output was basic; clinical papers were mostly published in national journals which are scarcely covered by the database used; the results of Spanish research were published in journals of a similar impact to those used by other European Union countries, although the number of citations received was much smaller, as has already been observed for other disciplines in Spain and peripheral countries. The consequences of the results were also discussed.

Gupta et al (2002) investigated Indian collaboration with South-East-Asian countries in Science and Technology during the period 1996 to 2000 using the *Science Citation Index*. Results show that India had S&T collaborations with the seven South East

Asian countries during the period of study; the strongest collaboration was with Malaysia especially in chemistry, the second strongest collaboration was with Singapore. India's S&T collaboration with the five other countries, viz. Philippines, Thailand, Indonesia, Vietnam and Laos was mainly through multilateral agreement mainly in clinical medicine, biology and biomedical research. The weakest collaboration was found with Laos with an output of only one paper during the period. The paper suggested that, perhaps opportunities for further collaboration with this country would be the focus in future and strong cooperation in the fields of clinical medicine and biomedical research, with all the seven countries is expected to grow in strength. Other potential areas of collaboration with South East Asian countries are environment and medicine to protect their interests in the regime of intellectual property rights. With this observations and suggestions by Gupta et al (2002), the findings from this present study would determine the focus of collaboration for Malaysian medical and health science researchers, and would be able to suggest future collaborative focus.

Onyancha and Ocholla (2004) did a comparative study on the literature on HIV/AIDS in Kenya and Uganda. They compared variables such as: publication type and date, institutional affiliation, publishing sources, size of publications, gender, and nature of research collaboration. Results show that research funding plays major role in the creation of relevant research centers in these countries and in financing research projects and research affiliates. Most publications were co-authored and focused on women, and a large proportion of HIV/AIDS documents were published outside Africa. The study recommended the government of Kenya and other stakeholders involved in the formulation of policies on research in HIV/AIDS to vigorously campaign for more research funds and other resources that have made Uganda's case a success, and strengthen collaborative links with foreign researchers as well as the rest of Africa.

Mitha and Leach (2006) carried out a study that assess the patterns of authorship and the publications of academic institutions in the sciences, medical institutions and affiliated organisations in South Africa. The focus was on HIV/AIDS literature for the period 1982-2002. Published literature in the sciences and medicine from two internationally recognized databases were used for the measurement. As expected by the author, the results of the study demonstrated exponential growth in the literature, which was due to the multidisciplinary nature of the disease. Publications were scattered in a variety of discipline-based journals. The results also demonstrated that South African researchers were fast becoming internationally recognized in the field of HIV/AIDS research. Hence, part of the aim of this study is also to determine the rate at which Malaysian researchers are becoming internationally recognized in the medical field.

Chuang et al (2007) studied stroke-related research articles published by Taiwan researchers which were indexed in the *Science Citation Index* from 1991 to 2005. They found that the quantity of publications has increased at a quicker pace than the worldwide trend. Over the years, there has been an increase in international collaboration, mainly with researchers in the United States. Article visibility, measured as the frequency of being cited, also increased during the period. It appears that stroke research in Taiwan has become more globally connected and has also improved in quality. They recommended that continuity in research focuses should be maintained, and funding should be allocated on a long-term basis to institutes with a proven record of success, in order to improve the quality and efficiency of stroke research.

Swedish Science Council (2007) did a report on dental research in Sweden, which includes two bibliometric studies conducted for 1995 and 2004. The study clearly suggests that the output of Swedish scientific articles has declined sharply. The decline, about one fourth of the total number of publications, is attributed mainly to a decrease in the average number of articles published by individual authors rather than a decrease in the number of dental researchers that publish articles. The analysis also shows that Swedish dental research is generally of good quality, but there are indications that it is weakening somewhat. The high-quality research, however, appears to be maintaining and perhaps even strengthening its position. The authors concluded that the findings should be approached with caution.

Levi et al (2009) conducted a bibliometric analysis to study research production on internal medicine in the Netherlands compared to other countries of the world, it was revealed that, the Netherlands also has a relatively high number of publications in the two leading clinical journals (the *New England Journal of Medicine* and the *Lancet*). It also estimates the average number of publications per staff member per five years in the eight academic departments of Internal Medicine in the Netherlands. The findings were very encouraging and the position of research in Internal Medicine in the Netherlands is worthy of praises. The author submitted that it is a challenge for the next decades to maintain this position and to become even more productive while keeping up the high quality.

The main aim of the study by Šember, et al (2010) was to analyze the 2007 citation count of articles published by the *Croatian Medical Journal* in 2005-2006 based on data from the *Web of Science, Scopus, and Google Scholar*. The study demonstrated that the Web of Science databases covered the highest-impact scientific journals as the source of citation for the *Croatian Medical Journal*, but that the coverage of Scopus and especially of *Google Scholar* was broader and included additional local sources. It has been shown that the *Web of Science* is a selective source of publication citations. On the other hand, for a sample of high-profile general medicine articles*, Google Scholar* and *Scopus* may retrieve a greater number of citations than *Web of Science*.

Bala and Gupta (2010) analysed the research output of India in neurosciences during the period 1999-2008, and reported that India's global publications' share in neurosciences during the study period was 0.99% (with 4503 papers) and it ranked 21st among the top 26 countries in neurosciences. India was behind China, Brazil and South Korea in terms of publication output, citation quality and share of international collaborative papers in neurosciences. The study concluded that there was an urgent need to substantially increase the research activities in the field of neurosciences in India, and also investments, both at the institutional level as well as in terms of extramural funding from different scientific agencies. In addition, there should be a substantial increase in international collaboration to increase the output and also to improve the quality of research and there should be more collaboration among the Indian institutions. In this study, the research productivity of medical researchers in Malaysia in terms of their contribution to the medical field would be examined.

## 2.8 Studies on Global Research Production

Haiqi (1995) bibliometrically compared three medical library periodicals published in China, Japan and U.S.A for the period 1990-1992 to find each country's trends of research in medical library and information services. They reported that, citation characteristics in the periodicals provided evidence that differences were likely attributable to the fact that the three periodicals were published in different countries; Citation analysis which is the most celebrated scientometric technique, has been increasingly applied to research evaluation in recent years and has been described in various degrees in the literature; the most frequently used research methods were description or survey and experiment or investigation, which were employed in almost all subjects of the three periodicals. The articles cited more periodical articles than monographs; the age curve of cited publications was similar in the three periodicals and a curve line of age of the cited publications showed that references aged more than 10 years being cited less and less; as is often the case, relatively few periodicals produced most of the references cited in the periodicals and the single periodical that each cited most frequently was itself.

Glover and Bowen (2004) examined the bibliometric profile of tropical medicine and international health using data from *PubMed* and *ISI Science Citation Index*. As expected in a European journal, Europe led with 564 papers, the largest number of authored papers, Africa has the second highest representation with 517 authored papers and this may reflect the high concern for tropical diseases in the region in addition to the Sub-Saharan epidemic incidence of HIV/AIDS and tuberculosis. The USA, the most prodigious producer of scientific papers, had authorship of only 158 articles. The UK and Asia each had a similar numbers of papers published. African-authored papers have been cited most with 3512 citations. Geographical distribution of authors demonstrates that the journal they published in reflects fairly the research and views of scientists from all regions of the world. This global distribution would seldom be replicated in the top journals from other *ISI* categories.

Chiu and Ho (2005) conducted a bibliometric analysis of all homeopathy-related publications in *Science Citation Index (SCI)* during the period of 1991 to 2003. They analyzed parameters, which are also considered in our present studies, which include:

authorship, patterns of international collaboration, journal, language, document type, research address, number of times cited, and reprint author's address. It was observed that; 49% of the articles in homeopathy were single authored. The top 3 ranking countries of publication and citations were the UK, the US, and Germany. English remained the dominant language, it comprised 76%, while German contributed 18%, and the remaining where distributed among 8 European languages. More document types and languages, and fewer pages have appeared in homeopathy research. Small-group collaboration was a popular method of co-authorship. In addition, University of Exeter in the UK had the highest frequency as corresponding institute. The most frequently cited article was published in the *JAMA-Journal of the American Medical Association* which is the second highest impact factor journal in the category of General and Internal Medicine.

Webster (2005) applied bibliometric methods to examine the volume of research published, its impact and sources of funding of biomedical research in the UK. The results showed that the UK achieved position as the second largest producer of biomedical research is under threat from Japan and Germany and other countries with traditionally weaker biomedical research base; the strength in malaria and asthma research and relative weakness in surgery and renal medicine was notable; the profile of UK biomedical research has changed significantly in the period analyzed, with a doubling of the level of international collaboration, a significant increase in basic research papers and an increase in the potential impact of UK publications; a relative decrease of acknowledgement of UK Government funding was noted, as were increased in acknowledgements to UK not-for-profit and international organizations.

Falagas et al (2006) performed a bibliometrics analysis to study research around the world in the fields of preventive medicine, occupational/environmental medicine, epidemiology and public health. The source of data was all articles published in the database of the *Institute for Scientific Information (ISI)* during the period 1995 and 2003. From the results, USA led the research production in all three sub-categories. Canada and Western Europe shared the second position in the first two sub-categories, while Oceania researchers ranked second in the field of public health. USA researchers maintain a leadership position in the production of scientific articles, which supported previous findings suggesting that the majority of research published in scientific journals, is carried out in the developed world. Furthermore, the study found that four regions of the world with more than 50% of the world's population have minimal contribution in the scientific fields studied. The authors encouraged less developed regions to support their researchers in the above fields in order to improve scientific production and advancement of knowledge in their countries.

Swaminathan et al (2007) investigated the diversity in research conducted by anesthesia-based researchers using *Medline* and *Ovid databases*. It was observed that despite the study being limited to English-language publications, non-English speaking nations contributed strongly to this diversity and that about 85% of all publications were represented by 46 journals. Randomized controlled trials constituted 4685 (70%) of publications. Turkey had the highest percentage of randomized controlled trials (88%). The United States led the field in quantity (20% of total) and mean impact factor (3.0) of publications. Finland had the highest productivity when adjusted for population (36 publications per million populations). Publications from the United States declined from 23% in 2000 to 17% in 2005. The study concluded that: clinical research attributable to investigators in their own specialty is diverse, and extends beyond the traditional field of anesthesia and intensive care. The United States produces the most clinical research, but per capita output is higher in the European nations.

Ugolini et al (2007) conducted a bibliometric research that compares the scientific production in the field of cancer molecular epidemiology among countries and evaluate the publication trend between 1995 and 2004 using the *PubMed* database as a source of data. The findings revealed that: Europe and USA have the highest scientific production in the field of cancer molecular epidemiology. Nevertheless, some areas of excellence also emerged, such as Northern Europe and Israel. Among the European countries, the analysis confirms the results observed in other biomedical disciplines, with the UK ranking first both in quantity and quality of scientific production.

Chiu and Ho (2007) employed bibilometric analytical technique to examine all tsunami related publications in the *Science Citation Index (SCI)*. From the results, the US and Japan produced 53% of the total output where the seven major industrial countries accounted for the majority of the total production. English was the dominant language (976; 95%) followed distantly by Russian (43; 4.2%), French (7; 0.68%), and Spanish (1; 0.10%). Journal articles were the most-frequently used document type comprising 88% (903) of the total production, followed by reviews (50; 4.9%). Moreover the tsunami publication patterns in the first 8 months after the Indonesia tsunami occurred on 26 December 2004 indicated a high percentage of non-article publications and more documents being published in journals with higher impact factors.

Hazmir (2008) applied bibliometric method to Malaysian biomedical and health sciences publications in the *ISI* database from 1990 to 2005. He reported that Malaysian biomedical and health sciences are becoming a fast growing research field; the most productive period was during the 8MP (Eight Malaysian Plan) and the trend line indicated a continuing upward trend. Journal articles (73.3%) were the main type of publication produced. More than half (63.7%) of authors (4,178) were one-time contributors. Fifteen authors were identified as the most productive producing an average of 2.7 papers each per year; The majority of publications were multi-authored (89.3%) works; Institutions of higher learning especially from the public sector, dominated the production of research publications with University of Malaya, Universiti Sains Malaysia, Universiti Kebangsaan Malaysia and Universiti Putra Malaysia as the main key players. A total number of 2,413 (63.5%) joint papers were identified and 47.4 percent were the result of international collaboration; Clinical Medicine is the most actively researched area.

In line with the other bibliometric studies reported in this chapter, the present study will explore the publication productivity amongst medical researchers published in the oldest medical journal in Malaysia, the *Medical Journal of Malaysia*. Among the findings that would be demonstrated by this present study are: productivity of *MJM*, the most productive authors, the degree of collaborations, the pattern of collaborations, authors institutional affiliations, authors country affiliations, the subject area of research, the pattern of citations, bibliographic format of citations, half-life of reference materials, the core journals cited, the distribution of citations and impact factor received by *MJM* articles during the period under study.

# CHAPTER 3

# METHODOLOGY

## 3.1. Introduction

This chapter describes the source, and the approach used in the collection, manipulation, compilation of information and measurement of statistical data aimed at analyzing the content and structure of articles published in the *Medical Journal of Malaysia* in line with the objectives of the study. The increasing availability of databases containing publication and citation information has spurred the development of the subject of bibliometrics. This involves the analysis of a set of publications characterized by bibliographic variables such as the author(s), the place of publication, the associated subject keywords, and the citations (Willett, 2007). While information obtained from indexing and abstracting databases such as the *ISI* database are most popularly used to assess the quality of publications, Zainab (2006) considered that a single database that provides information about the productivity of individual or group of academics as well as institutions for a particular country or region is also as good. This idea is also welcomed by (Herther, 2008), who suggested many scholarly databases that offers cited reference data and statistics as a sources of citation information. Therefore, the study relies on data contained in an open access database named *MyAis (Malaysian abstracting and indexing system http://Myais.fsktm.um.edu.my/)* as a source of data, due to the fact that it covers the period of the years under study and has features that provide relevant information needed for a bibliometric study of this nature.

Descriptive statistics, relational methods and bibliometric analysis would be the main methods employed. Gauthier (1998) quoted Polanco (1995) explained that bibliometric methods serve three main functions, and these are, description, evaluation, scientific and technological monitoring. Durieux and Gevenois (2010) also explained that, there are three types of bibliometric indicators: *quantity indicators,* which measure the productivity of a particular researcher; *quality indicators,* which measure the quality (or "performance") of a researcher's output; and *structural indicators*, which measure connections between publications, authors, and areas of research. Thus, as a descriptive tool, it would provide an account of the characteristics, quantities and research trends in publishing activities of the *Medical Journal of Malaysia* and the comparative analyses of the journal's productivity, while as a relational indicator, would serve to identify connections, interactions, and relationships between researchers, fields of research, and affiliations, and such interactions constitute the flow of knowledge.

While explaining the methods of bibliometric studies, both Borgman (1990) and Paisley (1990) explain that "Bibliometric studies of scholarly communities use one more of three theoretical variables: "producers, artifacts, and concepts". Producers are the originators of information at any level of author's research teams, institutions, fields, or countries, other principles and levels of aggregations, women or men, authors of different ages, and other geographical units. Artifacts are the information products themselves; books, journals articles, encyclopedia articles, conference presentations, and so on, while the third is the Communication concept, which comprises words, themes, citations, or presentation details.

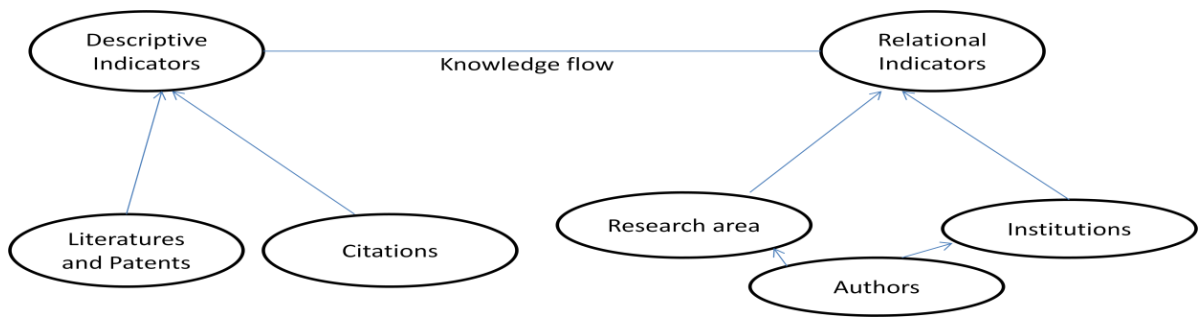

**Figure 3. 1 Relationship of Some Parameters in Bibliometric Studies**

Figure 3.1 indicates some of the parameters that are considered when applying bibliometric research indicators. It is to express relationships between authors, affiliations and research area. Therefore, through bibliometric studies, it would be possible to determine the existing relationship, spread, and variability in the works of authors. These help monitor changes in various scientific disciplines, identify emerging research topics and the relevant contributors. This study would consider some of the parameters discussed to present the article productivity of *MJM*, the most productive authors, the degree of collaborations, the pattern of collaborations, authors institutional affiliations, authors country affiliations, the subject area of research, keywords distribution in articles, the pattern of document citations, bibliographic format of citations, age and half-life of reference materials, the core journals and the distribution of citations received by *MJM* articles during the period under study.

## 3.2. Research Method

Research methods are procedures designed to exploit opportunities for measurement (Paisley, 1990). Bibliometrics is the procedure adopted in this study, which relies on an open access database *MyAis (Malaysian abstracting and indexing system http://Myais.fsktm.um.edu.my/),* with large datasets as a source of data, bounded by its measurement opportunities and capabilities. The use of these large datasets makes possible analyses at a scale that cannot be achieved by traditional methods such as survey and case studies (Borgman, 1990), in addition with its reliability and validity which are the criteria that a research method is being used appropriately (Paisley, 1990). The database is sufficiently comprehensive for this study, and has several features capable to support bibliometric research. The process followed a very carefully planned step by step approach. The first step is the process of extracting data from the bibliographic databases *(MyAis)* and arranging them according to a specified set of field in each of the spreadsheet accommodating data. The study exercise due diligence in taking the second step, which is a thorough data clean up; a very necessary task due to errors resulting from misspelling, incomplete or wrong addresses, publication year and incorrect citations. The next is to code each field according to the significant variables at which content behavior is being observed (papers, institutional affiliations, collaborations, etc) and use these variables to make comparisons in line with the objectives of the study. The file used in the spreadsheet is indicated in figure 3.2.

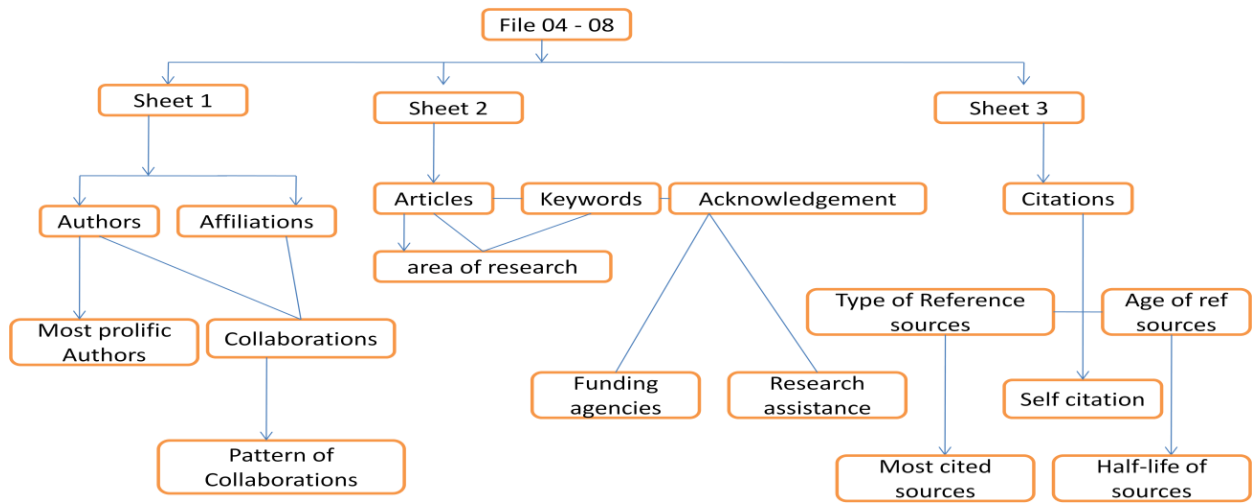

**Figure 3. 2 Chart Representing the Breakdown of Data Elements**

## 3.3. Sample of the Study

All articles published in the *Medical Journal of Malaysia* from year 2004 to 2008 serves as the sample for this study. The study examines a 5 year period in the publication of this journal, which is considered appropriate, looking at the large amount of data been analyzed and the fact that impact factor (*IF*) would be calculated (on citations received) for articles only after 2 years of publication. The data were retrieved from *MyAis (Malaysian abstracting and indexing system http://Myais.fsktm.um.edu.my/)* an open access database that index scholarly journals published in Malaysia. The numbers of articles retrieved for the purpose of this study were 580 articles.

## 3.4. Data Collection

The source of bibliometric data is an important consideration for any study that intends to analyze patterns in scholarly journals. Data for this study was obtained from *MyAis (Malaysian abstracting and indexing system http://Myais.fsktm.um.edu.my/)* which is the only known open access database that index Malaysian journals. The database has all the bibliometric information for the *Medical Journal of Malaysia* from year 2004 to 2008.

Glänzel (2003) provide an example of the databases of the *Institute for Scientific Information (ISI)*, the *Science Citation Index (SCI),* and suggested some unique features that are basic requirements of a database which would serve as a source of bibliometric data. Some of these features pertaining to this study are;

- Full coverage: All papers published in periodicals covered are recorded.

- Completeness of addresses: The addresses of all authors are indicated, allowing analyses of scientific collaboration and the application of full publication counting schemes.

- Bibliographical references: Together with each document their references are processed. Redefining references as sources makes it possible to analyze citation patterns and to construct citation indicators; and.

- Availability: The database is available as printed edition, in electronic form on magnetictapes, as on-line version and as CD-ROM edition.

As regards to the features suggested above, *MyAis (Malaysian abstracting and indexing system)* measured up for this bibliometric study. Information obtained from the database includes: Title of articles, Year of publication, Names of authors, Geographical and Institutional affiliation of authors, Keywords in articles, Works cited in an Article, Number of citation per article, Number of articles, Number of authors, Acknowledgement notes, and other related information.

Miyamoto et al (1990) noted that many scientists in different fields consider that bibliographic databases provide interesting material for analysis, aside from their usefulness as a tool for information retrieval. This is also true for *MyAis (Malaysian abstracting and indexing system http://Myais.fsktm.um.edu.my/)* which provides very useful information for journal analysis. This information were collected, compiled, recorded, tabulated and analyzed in order to reach a bibliometric conclusion through observations and measurement. As *Malaysian abstracting and indexing system (MyAis http://Myais.fsktm.um.edu.my/)* is indexed by *Google Scholar*, citations received by *Medical Journal of Malaysia* is retrievable either through *Google Scholar* itself or other bibliometric analysis tool such as *Harzing's Publish or Perish* as this tools generate citation information from data captured from *Google Scholar*. Therefore, data generated by *Harzing's Publish or Perish* was used to obtain the number of citations received by articles in the *Medical Journal of Malaysia*, and *Microsoft Excel 2007* was used to accommodate and manage the data. During citation data collection process, duplicate data were frequently encountered and were diligently sorted to make sure what is intended to be counted, counts. In addition, the study employed online translation software to determine needed information from non-English sources.

The spreadsheets files were created for each year, which contains the following sheets and fields (Table 3.1).

**Table 3. 1 Microsoft Excel Worksheet Created For Each Year**

| Spreadsheet | Data fields | Task | Information Retrieved |
|---|---|---|---|
| **Sheet 1** | Article Title | Attach unique codes to each title | Unique code |
| | Number of words in Title | Count number of words in article title | Number of words per article title |
| | Acknowledgement | Identify number of articles acknowledging funds or assistance | Names of agencies funding research |
| | Keywords per article | Count number of keywords per article | Number of keywords per article |
| | Keywords | Copy each word from the keywords and arrange in a different field | Unique keywords with frequency of occurrence |
| **Sheet 2** | Authors | Identify Single or multi - author | Local vs Local / Local vs Foreign / foreign vs foreign |
| | Names of authors with their affiliations | Arranged names in sorted order | Unique names with frequency of occurrence |
| | Productivity of authors | Test for Lokta's Law | Productivity of authors in conformity with Lokta's |
| | Affiliation of authors | Obtain Collaborating affiliations | Local vs Local / Local vs Foreign / foreign vs foreign |
| | Number of references | Obtain number of references per each article | Total number of references per year |
| | Number of co-authored papers | Obtain number of co-authored papers (if any) | Number of co-authored papers per year and proportion of co-authored papers by applying Subramanyam formula |
| | Number of collaborating affiliations | Identify two or more collaborating affiliations per article | Names and country of collaborating affiliation |
| **Sheet 3** | Reference Sources with year | Obtain Types of reference sources with year | Unique sources by type of material (journals, books, conference, etc) |
| | Journal Citations | Identify names and frequency of journal citations | Test for Bradford's law and obtain list of core journals |
| | Journal self citation | Identify number of *MJM* articles citing *MJM* | Total number of self citations |
| | Year of publication of | Obtain Age of citation | Half-life of citations |
| | cited reference | | |
| **Sheet 4** | Citations received by *MJM* articles from Google scholar | Obtain citation informations from *Harzing publish and perish* | No of citations received by *MJM* articles from 2004 - 2008 |
| | Authors citing *MJM* | Identify citing authors country affiliations | Country's affiliation of authors citing *MJM* articles |
| | Types of documents citing *MJM* | Identify types of document citing *MJM* articles | documents citing *MJM* and frequency of occurrences |

Each of the fields in each of the spreadsheet for each year in Table 3.1 was further arranged in separate sheet during data analysis in order to sort the data and generate tables and graphs.

### 3.4 .1. *MyAis* (Malaysian abstracting and indexing system http://*Myais*.fsktm.um.edu.my/)

Concerned about the visibility of research publication in the country, a research team from Faculty of Computer Science and Information Technology, University of Malaya (in 2006) propose the development of a central database for abstracting and indexing Malaysian scholarly journals and the project was funded by the Malaysian government. Before this, there was no indexing and abstracting service for Malaysian scholarly journals in existence. Since it was launched, it has gained awareness both locally and internationally and has grown tremendously in content.

*Malaysian Abstracting and Indexing System (MyAis)* is an open access system for abstracts and indexes of articles published in refereed scholarly Malaysian journals. The system provides in some cases abstracts and in some other cases full-text access to scholarly journal articles as well as conference proceedings published in Malaysia or abroad. These articles are voluntary contribution from Malaysian academics, academic and professional publishers. The aim is to serve Malaysian educational and research community with information about what has been published in Malaysian refereed journals for each year and in the various disciplines, thereby supporting the needs of an information rich society. The system has been designed to ease both the task of self-achieving and retrieval processes. Zainab (2006) explained that: "The database would be used to obtain information about the productivity and quality of scholarly publications from Malaysia. Information about publication productivity at both individual and institutional levels that may be used for quality assessment, which is just one of the several measures used to assess the quality or rate or rank academic institutions". The main objectives of *MyAis* were further highlighted by Zainab and Nor Badrul (2008):

(i) To provide a system which can control information about Malaysian publication research outputs, both derived from funded and unfunded projects. This involves indexing and abstracting all articles published in Malaysian scholarly journals, Malaysian conference papers, chapters in a book, books and research reports or theses. Uploads of articles are also supported.

(ii) To provide an open access system accessible to all Malaysian learning community.

(iii) To provide the following information, about publication productivity, specially:

    (a) What articles are published in Malaysian journals

    (b) Total publication productivity of authors publishing in those journals

    (c) Total publications productivity by authors' affiliations

    (d) Total publication productivity by broad subjects

    (e) Total publication productivity by year

    (f) Total publications by type of materials (journals, conference proceedings, theses, research reports, etc.)

    (g) To provide the following statistics about citations, specially:

- Total citations received by authors' names, indicating who are citing an author's work, together with information about total views and downloads.

- Top 20 cited authors

- Top 20 cited journals

- Top 20 cited articles

The database currently indexed 30 medical and health science journals published in Malaysia, among which is the *Medical Journal of Malaysia* (*MJM*). Zainab (2008), hoped that this database could act as catalyst for academic journal publishers to improve the quality of their journal publishing, and since information in *MyAis* is indexed by *Google scholar*, it follows that information about journal articles deposited in the database would be visible to the learning community.

## 3.5 Analysis of Data

Analysis of data was carried out through descriptive and relational methods using a bibliometric approach. The source of bibliometric data was *MyAis (Malaysian abstracting and indexing system http://Myais.fsktm.um.edu.my/).* The data was collected and exported to *Microsoft Excel 2007* and arranged according to various fields to allow for proper sorting. A proper arrangement and organization of data in various fields in the spreadsheet tolerate appropriate utilization of statistical functions such as percentages, means, modes, ranges, standard deviations, logarithms etc, which return values and results that were presented in form of tables and depicted in charts and graphs.

Accordingly,

(a) To present results pertaining to the productivity of articles and authors, descriptive tables were utilized, and data were displayed in charts and graphs. The tabular and graphical descriptions showed rankings and treadlines indicating the productivity of articles, authors, affiliated institutions and countries, as well as collaboration patterns. In order to analyze the frequency of publication by authors, *Lokta's law* (Lotka, 1926) of author's productivity was applied, which states that "the number (of authors) making n contributions is about $1 / n^c$ of those making one contribution, where *c* nearly always equals two (c ≈ 2) ; and the proportion of all contributors, that makes a single contribution, is about 60 per cent." (Lotka1926 cited in Glänzel 2003). Thus, *Lotka's* empirical finding can be summarized by the equation:

$$a_n = a_1/n^c, \quad n = 1, 2, 3,.., (1)$$
$$a_1 = a_n . n^c \quad n = 1, 2, 3,.., (2)$$

Where

$n$ = the number of publications,

$a_n$ = the relative frequency of authors with n publications

$a_1$ = the number of authors publishing one paper,

c = a constant.

In addition, to understand the degree of collaborations amongst authors, Subramanyam's formula (1983) was used, which is given as:

C = Nm / (Nm + Ns)

Where,

C = Degree of Collaboration

Nm = Number of Multi Authored Contribution

Ns = Number of Single Authored Contribution.

(b) To present the result pertaining to cited references, the study make use of descriptive tables and figures to indicate total citations referenced in articles published between year 2004 to 2008, types of reference sources used, age and half-life of references. To identify the core medical and medical related journals referenced by authors, *Bradford's law of scattering* (Bradford, 1948) was appllied which states that "if scientific journals are arranged in order of decreasing productivity on a given subject, they may be divided into a nucleus of journals more particularly devoted to the subject and several groups or zones containing the same number of articles as the nucleus when the numbers of periodicals in the nucleus and the succeeding zones will be as 1: b: b² …" (Glänzel, 2003). The results of the Bradford experiments were illustrated descriptively in tables and graph.

(c) As regards to results concerning the number and nature of citations received by *MJM* articles, useful citation information were retrieved from *Harzing Publish and Perish (http://www.harzing.com/)(Google Scholar)* which indexed journal articles from *MyAis (Malaysian abstracting and indexing system http://Myais.fsktm.um.edu.my/)* . This tool was utilized to obtain the number, type, and nature of citations received by articles in the *Medical Journal of Malaysia.* In addition to that was finding results of the journal impact factor (*IF*) calculated by applying the formula created by Garfield, founder of *ISI*. Thus, A journal's *JIF* for year *n* is defined as the ratio between the number of citations during year *n* of the journal's articles published during years n−1 and n−2 (and only then), and the total number of articles published during these two years (Garfield, 1979 quoted in Merlet et al, ( 2007).

$$JIF_n = \frac{C_{n-1} + C_{n-2}}{P_{n-1} + P_{n-2}} \qquad \text{....... (1)}$$

The simplified form is:

$$IF = A/B \qquad \text{........ (2)}$$

Where,

IF = Impact Factor
A = Citations in year $x$ to articles published in previous two years
B = Number of citable articles published in those previous two years

## 3.6 Summary

This chapter has explained and outlined the research method employed in the course of this study. The source of data, the method and approach were dealt with. The subsequent chapters will focus on data analysis and presents the overall findings resulting from the experiments.

# CHAPTER 4

# DATA ANALYSIS

## 4.1. Introduction

In order to answer the seven research questions, this chapter presents the data analysis, divided in seven sections. Each of these sections will give reports on:

1) The article productivity of *Medical Journal of Malaysia* from year 2004 - 2008.

2) The productivity of the authors, who published in *Medical Journal of Malaysia* by:

> (a) Finding the number of contributing authors per publication year;

> (b) Testing the productivity pattern using *Lotka's law;*

> (c) Determining the productive authors.

3) The co-authorship pattern in term of:

> (a) The degree of collaboration among the contributing authors using *Subramanyam's formula*;

> (b) The country affiliation of contributing authors;

> (c) The country collaboration per articles contributed;

> (d) The institutional affiliation of contributing authors;

> (e) The institutional collaboration per articles contributed.

4) The content of *MJM* in term of:

> (a) The keyword distribution of articles;

> (b) The number of words in article title;

5) The Agencies and Organizations funding research.

6) The pattern of citations referenced by articles published by *Medical Journal of Malaysia* in term of :

> (a) The distribution of references per year;

(b) The distribution of references per article;

(c) Distribution of references according to bibliographic forms;

(d) Age of the referenced literature;

(e) The half-life of references used;

(f) The referenced core journals using *Bradford's Law* of journal distribution;

(g) Language of reference sources.

7) The pattern of citations received by articles published in *Medical Journal of Malaysia* through:

(a) Total citations received for articles published between year 2004 to 2008 through *Google Scholar (using Harzing's Publish and Perish);*

(b) Journal self-citations;

(c) Format types of documents citing *MJM;*

(d) Scholarly journals citing *MJM* between year 2004 to 2008;

(e) Countries affiliations of authors citing *MJM* between years 2004 to 2008;

(f) Journal impact factor of *MJM* articles published from year 2004 to 2008.

## 4.2.   Article Productivity of *Medical Journal of Malaysia*: 2004 - 2008

The number of articles published during the five-year (2004 - 2008) period is five hundred and eighty (580). The highest number in article publication was recorded in 2004 with 138 articles, followed by 135 articles in year 2008 as shown in Table 4.1.  These data is based on those retrieved from *(MyAis, Malaysian abstracting and indexing system http://Myais.fsktm.um.edu.my).*

| Year | Number of Articles | Percentage (%) | Cumulative Percentage (%) |
|---|---|---|---|
| 2004 | 139 | 23.97 | 23.97 |
| 2005 | 102 | 17.59 | 41.55 |
| 2006 | 104 | 17.93 | 59.48 |
| 2007 | 100 | 17.24 | 76.72 |
| 2008 | 135 | 23.28 | 100.00 |
| Total | 580 | 100.00 | |

This shows that article publication have been consistently on the increase during the period under study, with an average of 116 articles per year. There are two peak periods observed which are year 2004 (139 articles) and year 2008 (135 articles). This suggests that the journal would be able to maintain its publication consistency in the future.

## 4.3. Number of Contributing Authors per Publication Year

One thousand four humdred and thirty-five (1435) unique authors contributed to the journal and shared two thousand one hundred and seventy seven (2177) authorships between year 2004 and 2008. Since an author can contribute and also collaborate more than once during a particular period of time, this reflect high rate of authors names recorded yearly.

Table 4. 2 Number and Percentage of Authors Contributing Articles Per year

| Year | Number of Articles | Number of Authors | Percentage (%) |
|---|---|---|---|
| 2004 | 139 | 478 | 21.96 |
| 2005 | 102 | 367 | 16.86 |
| 2006 | 104 | 412 | 18.93 |
| 2007 | 100 | 352 | 16.17 |
| 2008 | 135 | 568 | 26.09 |
| Total | 580 | 2177 | 100.00 |

The number and percentage of contributing authors is given in Table 4.2. 568 authors names (unsorted) were the highest recorded in year 2008 and the least number of authors recorded was 352 authors in year 2007. As reflected by total number of publications the number of authors contributing jointly seems to be predominant as total authors nearly always double compared to the number of articles published.

## 4.4. Author's Productivity Pattern in Accordance with *Lotka's Law*

By applying *Lokta's law*, the study aim to find a bibliometric balance in *MJM* that "the number (of authors) making n contribution is about $1 / n^c$ of those making one contribution, where *c* nearly always equals two (c ≈ 2) ; and the proportion of all contributors, that makes a single contribution, is about 60 per cent." (Lotka1926 cited in Glänzel 2003). Chung and Cox (1990) further explained that in the cases examined, it is found that the number of persons making 2 contributions is about one-fourth of those making one; the number making 3 contributions is about one-ninth, etc. the number making n contributions is about $1/n^2$ of those making one; and of all contributors, the proportion that make a single contribution, is about 60 percent.

First the study compare its own observation as regards to author's productivity with that of *Lokta's* by assuming that (c = 2) (in the equation detailed in chapter 3 pg. 54). The result is shown in Table 4.3.

| No of Publication (n) | Frequency of authors with n publications observed ($a_n$) | Observed Percentage (%) | Frequency of authors with n publications expected when c=2 | Expected Percentage (%) when c=2 |
|---|---|---|---|---|
| 1 | 1084 | 75.54 | 1084 | 63.40 |
| 2 | 204 | 14.22 | 271 | 15.85 |
| 3 | 65 | 4.53 | 120 | 7.04 |
| 4 | 34 | 2.37 | 68 | 3.96 |
| 5 | 19 | 1.32 | 43 | 2.54 |
| 6 | 8 | 0.56 | 30 | 1.76 |
| 7 | 7 | 0.49 | 22 | 1.29 |
| 8 | 4 | 0.28 | 17 | 0.99 |
| 9 | 1 | 0.07 | 13 | 0.78 |
| 10 | 1 | 0.07 | 11 | 0.63 |
| 11 | 1 | 0.07 | 9 | 0.52 |
| 12 | 3 | 0.21 | 8 | 0.44 |
| 14 | 1 | 0.07 | 6 | 0.32 |
| 15 | 2 | 0.14 | 5 | 0.28 |
| 19 | 1 | 0.07 | 3 | 0.18 |

The result from Table 4. 4 above indicates that when (c=2) the proportion of all contributors, that makes a single contribution, is (63.40%), and few authors contributed more than one paper. This result confirmed the findings by Chung and Cox (1990) for "Finance Literatures". Hence, the study aim to determine whether *Lotka's Law* is also applicable to the publication productivity of authors in *MJM*, by testing the equations indicated in chapter 3 (pg. 52.) . First, to find our c value we apply equation (2)

$a_1 = a_n . n^c$  n= 1, 2, 3,.., From Table 4.3

Let n = 1, and $a_n$ = 1084

$a_1 = 1084. 1^c$

$a_1 = 1084$

Let n = 2, and $a_n$ = 204, and $a_1$ = 1084

$1084 = 2^c . 204$

$2^c = 1084 / 204$

$2^c = 5.31$

clog2 = log 5.31

c (0.301) = 0.725

c = 0.725 / 0.301

c = 2.4

Calculated value of c = 2.4, and by imputing this value in equation (2)

$a_1 = a_n . n^c$  n= 1, 2, 3,..,

The result generated is presented in Table 4.4.

**Table 4. 5 Author's Productivity Pattern Observed Compared with Expected (c=2.4)**

| No of Publication (n) | Frequency of authors with n publications observed ($a_n$) | Observed Percentage (%) | Frequency of authors with n publications expected when c=2.4 | Expected Percentage (%) when c=2.4 |
|---|---|---|---|---|
| 1 | 1084 | 75.54 | 1084 | 73.17 |
| 2 | 204 | 14.22 | 205 | 13.86 |
| 3 | 65 | 4.53 | 78 | 5.24 |
| 4 | 34 | 2.37 | 39 | 2.63 |
| 5 | 19 | 1.32 | 23 | 1.54 |
| 6 | 8 | 0.56 | 15 | 0.99 |
| 7 | 7 | 0.49 | 10 | 0.69 |
| 8 | 4 | 0.28 | 7 | 0.50 |
| 9 | 1 | 0.07 | 6 | 0.38 |
| 10 | 1 | 0.07 | 4 | 0.29 |
| 11 | 1 | 0.07 | 3 | 0.23 |
| 12 | 3 | 0.21 | 3 | 0.19 |
| 14 | 1 | 0.07 | 2 | 0.13 |
| 15 | 2 | 0.14 | 2 | 0.11 |
| 19 | 1 | 0.07 | 1 | 0.06 |
|  | 1435 | 100.00 | 1481 | 100.00 |

It was found that the frequency of authors publications observed and the frequency of authors expected when c = 2.4 is very close (Table 4.4). This implies that the author's productivity pattern in *MJM* from year 2004 - 2008 conforms slightly to *Lokta's law* with little marginal c value.

## 4.5. Core Authors

In Table 4.5 is the list of names and affiliations of most productive authors, out of the 1435 unique authors. Ruszymah B.H.I, who is affiliated to Universiti Kebangsaan Malaysia is the most productive of all, with 19 articles to her name, followed by Aminuddin B.S (15), Gendeh B.S (15), Chua, K.H (14), Chua, K.B (12), Philip, R (12), Prepageran, N (12), Halim A.S (11), Kwan, M.K. (10). This shows that, Ruszymah B.H.I produced an average of (3.8 articles) per year during the 5-year period, while Aminuddin B.S and Gendeh B.S have an average of 3 articles each. It also follows that Chua, K.H, Chua, K.B, Philip, R, Prepageran, N, Halim A.S, and Kwan, M.K, have produced an average of 2 articles per year during the same period.

**Table 4. 6 List of Most Productive Authors and their Affiliations**

| Group | Cohort | Name of Authors | No of Articles | Affiliations |
|---|---|---|---|---|
| 1 | Cohort: 1 | Ruszymah B.H.I. | 19 | University Kebangsaan Malaysia |
| 2 | Cohort: 2 | Aminuddin B.S. | 15 | University Kebangsaan Malaysia |
| | | Gendeh, B.S. | 15 | University Kerbangsan Malaysia |
| 3 | Cohort: 1 | Chua, K.H. | 14 | Universiti Kebangsaan Malaysia |
| 4 | Cohort: 3 | Chua, K.B. | 12 | Ministry of Health, Malaysia |
| | | Philip, R | 12 | Hospital Ipoh |
| | | Prepageran, N. | 12 | University of Malaya |
| 5 | Cohort: 1 | Halim A.S. | 11 | Universiti Sains Malaysia |
| 6 | Cohort: 1 | Kwan, M.K. | 10 | University of Malaya |
| 7 | Cohort: 1 | Sukumar, N. | 9 | Universiti Kebangsaan Malaysia |
| 8 | Cohort: 4 | Abdullah J.M | 8 | Universiti Sains Malaysia |
| | | Sherina M.S | 8 | Universiti Putra Malaysia |
| | | Sopyan I. | 8 | International Islamic University Malaysia |
| | | Zulmi W. | 8 | Universiti Sains Malaysia. |
| 9 | Cohort: 7 | Gurdeep, S. | 7 | Hospital Ipoh |
| | | Kumarasamy,V | 7 | Ministry of Health, Malaysia |
| | | Loh, K.Y., | 7 | International Medical University |
| | | Rampal, L. | 7 | Universiti Putra Malaysia |
| | | Raymond, A.A. | 7 | Universiti Kebangsaan Malaysia |
| | | Tan, G.C. | 7 | Universiti Kebangsaan Malaysia |
| | | Teng, C.L. | 7 | International Medical University |
| 10 | Cohort: 8 | Chan, K.Y. | 6 | Universiti Kebangsaan Malaysia |
| | | Gopala, K.G. | 6 | University of Malaya |
| | | Hamidon B.B. | 6 | Universiti Kebangsaan Malaysia |
| | | Harvinder, S. | 6 | Hospital Ipoh |
| | | Saim L | 6 | Universiti Kebangsaan Malaysia |
| | | Shashinder, S | 6 | University of Malaya |
| | | Sivalingam, N | 6 | International Medical University |
| | | Zulfiqar M.A | 6 | Universiti Kebangsaan Malaysia |
| 11 | Cohort: 19 | Biswal, B.M. | 5 | Universiti Sains Malaysia |
| | | Chan, S.C | 5 | Royal College of Medicine Perak |
| | | Choon, S.K. | 5 | University of Malaya |
| | | Faisham W.I | 5 | Universiti Sains Malaysia. |
| | | Hamzaini A.H. | 5 | Universiti Kebangsaan Malaysia. |
| | | Khalid B.A.K | 5 | Universiti Kebangsaan Malaysia |
| | | Kuljit, S | 5 | University Malaya |
| | | Leong, C.F. | 5 | Universiti Kebangsaan Malaysia |
| | | Loh, L.C. | 5 | International Medical University |
| | | Mafauzy M | 5 | Universiti Sains Malaysia |
| | | Mallina, S | 5 | Hospital Ipoh |
| | | Naing L | 5 | Universiti Sains Malaysia |
| | | Norizah I | 5 | Kementerian Kesihatan |
| | | Reddy, S.C | 5 | Universiti Putra Malaysia |
| | | Rosalind, S | 5 | Hospital Ipoh |
| | | Saw, A | 5 | University of Malaya |
| | | Subha, S.T. | 5 | Universiti Putra Malaysia |
| | | Teoh, C.M | 5 | University Kebangsaan Malaysia |
| | | Yeap, J.S | 5 | International Medical University |
| 12 | Cohort: 34 | | 4 | |
| 13 | Cohort: 65 | | 3 | |
| 14 | Cohort: 204 | | 2 | |
| 15 | Cohort: 1084 | | 1 | |

## 4.6. Co-Authorship Pattern

Table 4.6 represents the co-authorship pattern per year. A total number of 580 articles were produced during the period under study and only 56 articles (9.65%) were single authored. Joint authorship by 3 authors have the highest frequency of 135 articles (23.28%) followed by 4 authors with 114 articles (19.66%). The highest number of collaboration was 15 authors with frequency of 1 (0.17%), followed by 12 authors, then 11 authors, both with frequencies of 2 (0.34%).

**Table 4. 7 Co-Authorship Pattern per Year**

| Co-Authorship Pattern | 2004 | 2005 | 2006 | 2007 | 2008 | Total | Percentage (%) |
|---|---|---|---|---|---|---|---|
| single author | 17 | 13 | 9 | 7 | 10 | 56 | 9.66 |
| multiple authors by 2 | 23 | 21 | 18 | 19 | 15 | 96 | 16.55 |
| multiple authors by 3 | 39 | 26 | 20 | 25 | 25 | 135 | 23.28 |
| multiple authors by 4 | 25 | 12 | 23 | 25 | 29 | 114 | 19.66 |
| multiple authors by 5 | 20 | 14 | 15 | 15 | 24 | 88 | 15.17 |
| multiple authors by 6 | 9 | 7 | 12 | 8 | 18 | 54 | 9.31 |
| multiple authors by 7 | 3 | 3 | 3 | 0 | 7 | 16 | 2.76 |
| multiple authors by 8 | 2 | 3 | 2 | 0 | 4 | 11 | 1.90 |
| multiple authors by 9 | 1 | 0 | 0 | 1 | 2 | 4 | 0.69 |
| multiple authors by 10 | 0 | 0 | 1 | 0 | 0 | 1 | 0.17 |
| multiple authors by 11 | 0 | 2 | 0 | 0 | 0 | 2 | 0.34 |
| multiple authors by 12 | 0 | 1 | 0 | 0 | 1 | 2 | 0.34 |
| multiple authors by 13 | 0 | 0 | 0 | 0 | 0 | 0 | 0.00 |
| multiple authors by 14 | 0 | 0 | 0 | 0 | 0 | 0 | 0.00 |
| multiple authors by 15 | 0 | 0 | 1 | 0 | 0 | 1 | 0.17 |
| | 139 | 102 | 104 | 100 | 135 | 580 | 100.00 |

In the biomedical field, multiple authorship is extensively practiced (Nwagwu, 2007), which is important, as they provide a platform for exchange of expertise and experience. Table 4.6 suggests that these interchange of expertise is very common in the medical field. The reason as explained by Gauthier (1998) is that characteristically in biomedical science, problems are commonly interdisciplinary in nature and thus it is especially crucial to foster collaborative behavior. Wiles et al, (2010) supported this idea and wrote that as companies have amalgamated into transnational conglomerates in order to compete, it is likely that there is greater collaboration within and across research institutions. On the average, more than 90% of all the articles were collaboratively produced.

## 4.7. Degree of Collaboration

The degree of collaboration between authors is described in Table 4.7, showing the number of single and multi-authored articles published per year.

Table 4. 8 Number of Single and Multi Authored Articles per Year

| Year | No of Single authored Titles | No of Multi-author Titles | Total | Percentage (%) of collaboration |
|------|------------------------------|---------------------------|-------|---------------------------------|
| 2004 | 17 | 122 | 139 | 87.77 |
| 2005 | 13 | 89 | 102 | 87.25 |
| 2006 | 9 | 95 | 104 | 91.35 |
| 2007 | 7 | 93 | 100 | 93.00 |
| 2008 | 10 | 125 | 135 | 92.59 |
| **Total** | **56** | **524** | **580** | |

Utap (2008) applied the formula of Subramanyam (1983), to calculate the proportion of co-authored publications in a single journal. Applying Subramanyam formula (detailed in chapter 3, pg. 53) the result shows that the degree of collaboration in *MJM* between years 2004 and 2008 is (0.9). It ranges from 0.87 to 0.93. This indicates that collaboration between authors in the field of medical sciences is a common practice.

## 4.8. Malaysian and Foreign Contribution based on Affiliation

The distribution of authoring and citing countries can also be used to measure the impact of a journal (Zitt and Bassecoulard, 1998). Illustrated in Table 4.8 based on affiliation is the number of Malaysian and foreign contributors per year. In all, 91.04% (1982) authors are affiliated to Malaysia, while 8.96% (195) are foreign.

| Year | Malaysian | Foreign | Total |
|------|-----------|---------|-------|
| 2004 | 438 | 40 | **478** |
| 2005 | 334 | 33 | **367** |
| 2006 | 394 | 18 | **412** |
| 2007 | 324 | 28 | **352** |
| 2008 | 492 | 76 | **568** |
| **Total** | **1982** | **195** | **2177** |
| **Percentage (%)** | **91.04** | **8.96** | |

Table 4.9 shows the characteristics of collaboration among Malaysia contributors' only, foreign contributors only and also joint contribution by either Malaysia with foreign or foreign with foreign authors.

Table 4. 10 Number of Malaysian and Foreign Collaboration per Year

| Year | Purely Malaysian | Collaboration with foreign | Purely Foreign | All Foreign |
|------|------------------|----------------------------|----------------|-------------|
| 2004 | 126 | 3 | 10 | 13 |
| 2005 | 90 | 7 | 5 | 12 |
| 2006 | 95 | 6 | 3 | 9 |
| 2007 | 88 | 5 | 7 | 12 |
| 2008 | 112 | 7 | 16 | 23 |
| **Total** | **511** | **28** | **41** | **69** |

It can be seen that 511 (88.10%) articles were contributed by authors affiliated to Malaysia alone compared to 41 (7.068%) contributed by foreign affiliated authors. Malaysian authors collaborated with their foreign counterpart 28 times, while foreign authors from one country affiliation collaborated with a different country affiliation 5 times. In total, 69 articles were either authored or co-authored by foreign contributors over the five-year period.

## 4.9. Distribution of Foreign Contributing Countries

Sixteen countries from Asia to Europe contributed to *MJM* articles during the period under study as elaborated in Table 4.10. India (12 articles) is the highest foreign contributor with 17.39% of the total foreign contribution, followed by Singapore 13.04% (9), Australia 13.04% (9), Turkey 8.70% (6), United Kingdom 7.25% (5), United States 7.25% (5), and Indonesia 7.25% (5). France, Ireland, Yemen, Canada, and Pakistan contributed just one article each. Again, this result indicate that although *MJM* did receive foreign contributions, the number is however small.

**Table 4. 11 Foreign Contributing Countries and their Shares in Yearly Foreign Contributions**

|    | Countries    | 2004 | 2005 | 2006 | 2007 | 2008 | Total | Percentage (%) |
|----|--------------|------|------|------|------|------|-------|----------------|
| 1  | India        | 2    | 3    | 3    | 2    | 2    | 12    | 17.39          |
| 2  | Singapore    | 2    | 2    | 0    | 3    | 2    | 9     | 13.04          |
| 3  | Turkey       | 3    | 1    | 0    | 0    | 2    | 6     | 8.70           |
| 4  | Iran         | 1    | 0    | 2    | 0    | 1    | 4     | 5.80           |
| 5  | Japan        | 1    | 0    | 0    | 0    | 2    | 3     | 4.35           |
| 6  | Netherlands  | 0    | 0    | 0    | 0    | 3    | 3     | 4.35           |
| 7  | Australia    | 2    | 2    | 1    | 2    | 2    | 9     | 13.04          |
| 8  | Indonesia    | 1    | 0    | 0    | 0    | 4    | 5     | 7.25           |
| 9  | USA          | 0    | 0    | 0    | 2    | 3    | 5     | 7.25           |
| 10 | UK           | 1    | 2    | 0    | 1    | 1    | 5     | 7.25           |
| 11 | Saudi Arabia | 0    | 0    | 1    | 1    | 1    | 3     | 4.35           |
| 12 | France       | 0    | 0    | 1    | 0    | 0    | 1     | 1.45           |
| 13 | Ireland      | 0    | 0    | 0    | 1    | 0    | 1     | 1.45           |
| 14 | Yemen        | 0    | 1    | 0    | 0    | 0    | 1     | 1.45           |
| 15 | Canada       | 0    | 0    | 1    | 0    | 0    | 1     | 1.45           |
| 16 | Pakistan     | 0    | 1    | 0    | 0    | 0    | 1     | 1.45           |
|    | **Total**    | **13** | **12** | **9** | **12** | **23** | **69** | **100.00**     |

The table further revealed that during this period, year 2008 has the highest number of foreign contribution of 23 articles, this suggest that contribution from foreign countries would continue to grow.

### 4.9.1. Collaborations between Different Countries

Collaborations between different countries including Malaysia are depicted in Table 4.11. Australia and Malaysia collaborated more with 7 articles, next is India and Malaysia (5 times), then United Kingdom and Malaysia (4 times), Indonesia and Netherlands (3 times), Singapore and Malaysia (3 times), USA and Malaysia (2 times), Japan and Malaysia (2 times) etc.

**Table 4. 12 Collaborations between Different Countries**

| No | Region of Collaboration | Countries Collaborating | No of Articles |
|----|--------------------------|--------------------------|----------------|
| 1 | Australia and South East Asia | Australia ::: Malaysia | 7 |
| 2 | South Asia and South East Asia | India ::: Malaysia | 5 |
| 3 | Europe and South East Asia | United Kingdom ::: Malaysia | 4 |
| 4 | South East Asia and Europe | Indonesia ::: Netherlands | 3 |
| 5 | South East Asia and South East Asia | Singapore ::: Malaysia | 3 |
| 6 | North America and South East Asia | United States ::: Malaysia | 2 |
| 7 | South Asia and South East Asia | Japan ::: Malaysia | 2 |
| 8 | South Asia and South East Asia | Pakistan ::: Malaysia | 1 |
| 9 | Middle East and South East Asia | Saudi Arabia ::: Malaysia | 1 |
| 10 | North America, Australia and South East Asia | United States ::: Australia ::: Malaysia | 1 |
| 11 | North America and South East Asia | United States ::: India | 1 |
| 12 | North America and South East Asia | Canada ::: Malaysia | 1 |
| 13 | Europe and South East Asia | France ::: Malaysia | 1 |
| | | **Total** | **32** |

The table also revealed collaboration between 3 countries which occurred only once between US, Australia and Malaysia. As seen, during the 5-years period that was examined, there were contributions between different region of the world other than Africa and South-America.

### 4.9.2. Authors from Foreign Countries

*Medical Journal of Malaysia (MJM)* received contribution from 16 different countries, from year 2004 – 2008, making up 8 different regions of the world as in Table 4.12. South East Asia (Singapore and Indonesia) leads with 20.11% (36 authors) of the foreign total, followed by South Asia (India and Pakistan) with 17.88% (32 authors),

Europe (Netherlands, United Kingdom, France, and Ireland) with 16.76% (30 authors) and Middle-Eastern countries (Iran, Saudi-Arabia, and Yemen,) with 10.06% (18 authors). Authors from Turkey alone produced 13.97% (25 authors), Japan 8.38% (15 authors), and Australia 7.82% (14 authors). While North – America (USA and Canada) produced 5.03% (9 authors) of the total foreign contributors.

Table 4. 13 Authors from Foreign Countries and their Percentage Share in Foreign Total

| | Regions | Countries | No of Authors from Foreign Countries | Percentage (%) |
|---|---|---|---|---|
| 1 | South-East Asia | Singapore | 23 | 12.85 |
| | | Indonesia | 13 | 7.26 |
| | | Sub-total | 36 | 20.11 |
| 2 | South - Asia | India | 31 | 17.32 |
| | | Pakistan | 1 | 0.56 |
| | | Sub-total | 32 | 17.88 |
| 3 | Europe | Netherlands | 14 | 7.82 |
| | | United Kingdom | 11 | 6.15 |
| | | France | 3 | 1.68 |
| | | Ireland | 2 | 1.12 |
| | | Sub-total | 30 | 16.76 |
| 4 | Eurasian | Turkey | 25 | 13.97 |
| | | Sub-total | 25 | 13.97 |
| 5 | Middle -East | Iran | 13 | 7.26 |
| | | Saudi-Arabia | 3 | 1.68 |
| | | Yemen | 2 | 1.12 |
| | | Sub-total | 18 | 10.06 |
| 6 | East - Asia | Japan | 15 | 8.38 |
| | | Sub-total | 15 | 8.38 |
| 7 | Australia | Australia | 14 | 7.82 |
| | | Sub-total | 14 | 7.82 |
| 8 | North - America | USA | 8 | 4.47 |
| | | Canada | 1 | 0.56 |
| | | Sub-total | 9 | 5.03 |
| | | **Main Total** | **179** | **100.00** |

At a glance, Figure 4.1 indicates that during these 5 year period of analysis, there were contributions from different region of the world other than Africa and South-America; however countries from Asia have the highest number of contributions. This implies that *MJM* is being accepted by Asian medical researchers as a channel to communicate their research results.

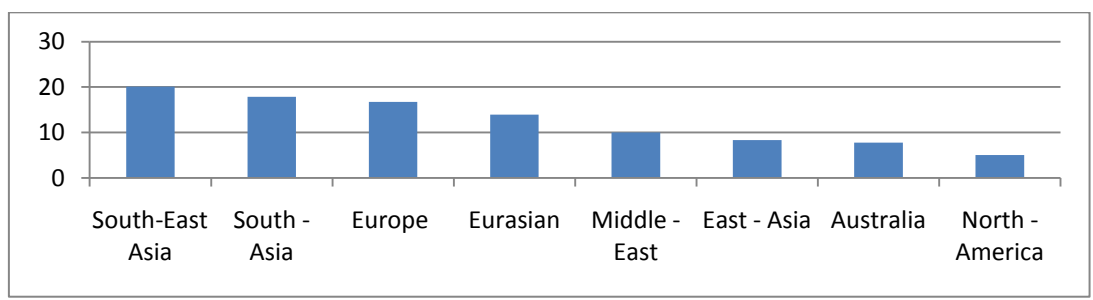

**Figure 4. 1 Percentage of Foreign Authors from Different Regions**

### 4.10. Malaysian Collaboration with Foreign Countries

When Malaysian authors collaborate with foreign authors 28 articles were produced. This is highlighted in Table 4.13 and Figure 4.3. Between year 2004 and 2008 (5 years), Malaysian authors joined Australians to produce 7 articles, with Indians to produce 5 articles, with United Kingdom authors to produce 4 articles, with Singaporeans to produce 3 articles, with the United States authors to produce 2 articles and also Japanese to produce 2 articles. This result supported that of Hazmir (2008), who examined "Malaysian biomedicine and health sciences research in the *ISI* database from 1990 – 2005". He reported that a total of 1,753 (47.4%) international joint papers were contributed by Malaysian institutions with their colleagues from other countries, and the countries that have strong collaboration with Malaysia are the United Kingdom with 278 (7.5%), followed closely by the USA (7.3%), Japan (4.76%), Australia (4.73%), Singapore (3.6%), and China (2.03%). Hence, by these results, it can be said that Malaysia has a strong connection in medical and health science research with Australia, India, UK, Singapore, US, and Japan more than any other country.

**Table 4. 14 Malaysian Collaboration with Foreign Countries**

| | With The Region | With The Country | No of Articles |
|---|---|---|---|
| 1 | Malaysia and Australasia | Malaysia ::: Australia | 7 |
| 2 | Malaysia and South Asia | Malaysia ::: India | 5 |
| | | Malaysia ::: Pakistan | 1 |
| 3 | Malaysia and Europe | Malaysia ::: United Kingdom | 4 |
| | | Malaysia ::: France | 1 |
| 4 | Malaysia and South - East Asia | Malaysia ::: Singapore | 3 |
| 5 | Malaysia and North America | Malaysia ::: United States | 2 |
| | | Malaysia ::: Canada | 1 |
| 6 | Malaysia and Middle - East | Malaysia ::: Saudi Arabia | 1 |
| 7 | Malaysia, Australia and North America | Malaysia ::: Australia ::: United states | 1 |
| 8 | Malaysia and East - Asia | Malaysia ::: Japan | 2 |
| | **Total** | | **28** |

One article each was produced when Malaysian affiliated authors combined separately with authors from Pakistan, Canada, France, and Saudi Arabia. Furthermore, Malaysia collaborated together with authors from Australia and United States in order to produce one article during this period. The results show that international collaborations is still very low, and it is assumed that more international collaboration would lead to more output and more recognitions due to the sharing of ideas and workloads (Chuang et al 2007).

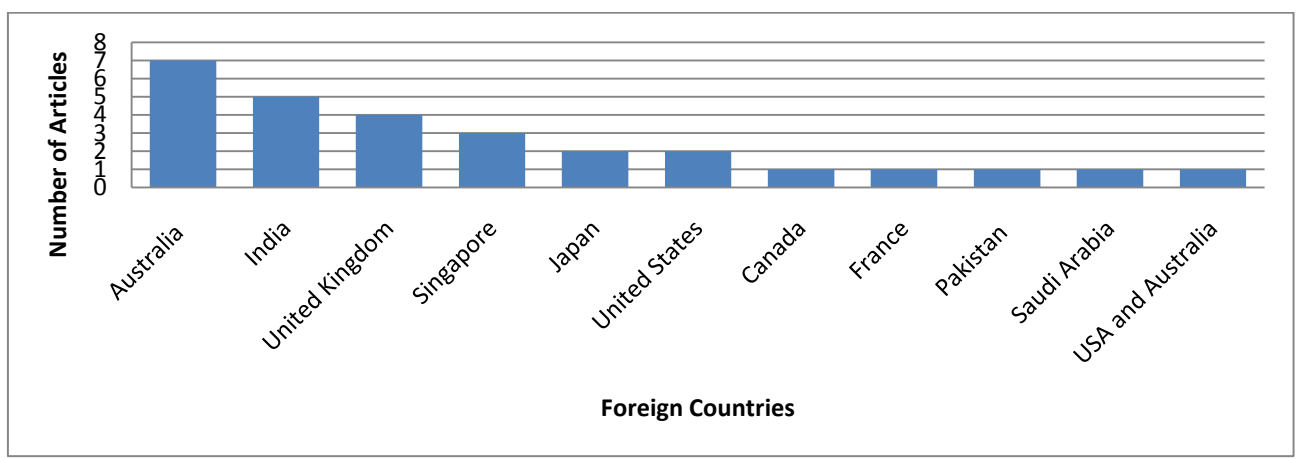

**Figure 4. 2 Malaysian Collaboration with Foreign Countries**

## 4.11. Types of Affiliations

The affiliations encountered were sorted and categorized into 7 different types, and further arranged according to whether they are Malaysian or foreign affiliations as indicated in Table 4.14. One hundred and seventy three (173) unique affiliation types were observed and (35.26 %) of this came from hospitals, followed by higher institutions (34.10%), ministries and government agencies make up (14.45%), medical centers (6.94%), clinics (4.62%), private organizations (4.05%), and international organizations make up (0.58%) of the total.

**Table 4. 15 Publications by Type of Affiliations**

| | Affiliation types | No of Malaysian | No of Foreign | Total | Percentage (%) |
|---|---|---|---|---|---|
| 1 | Hospitals | 48 | 13 | 61 | 35.26 |
| 2 | Institute of Higher Learning | 26 | 33 | 59 | 34.10 |
| 3 | Government Agencies | 19 | 6 | 25 | 14.45 |
| 4 | Medical centers | 9 | 3 | 12 | 6.94 |
| 5 | Clinics | 8 | 0 | 8 | 4.62 |
| 6 | Private organizations | 7 | 0 | 7 | 4.05 |
| 7 | International organizations | 0 | 1 | 1 | 0.58 |
| | **Total** | **117** | **56** | **173** | **100.00** |

As shown in Table 4.14 hospitals (which includes university hospitals) top the list of Malaysian affiliation followed by higher institutions, whereas, higher institution recorded the highest number in foreign affiliation. This indicates that medical professionals affiliated to hospitals are active contributors to *MJM*.

### 4.11.1. Number of Authors per Type of Affiliations

Presented in Table 4.15 is the number of authors per each affiliation type. Most of the authors are affiliated to institution of higher learning (60.91%) followed by hospitals (27.74%). Although hospitals are ranked the highest number in types of unique affiliations (Table 4.14), but there are more authors from higher institutions than Hospitals. This is because authors affiliated to higher institutions are concentrated to few top institutions whereas those affiliated to hospitals spread across many (university, college, state and private hospitals).

**Table 4. 16 Publications by Type of Authors' Affiliations**

| | Affiliation types | No of Authors | Percentage (%) |
|---|---|---|---|
| 1 | Hospitals | 401 | 27.94 |
| 2 | Institutions Of Higher Learning | 874 | 60.91 |
| 3 | Government Agencies | 104 | 7.25 |
| 4 | Medical Centers | 22 | 1.53 |
| 5 | Clinics | 16 | 1.11 |
| 6 | Private Organizations | 15 | 1.05 |
| 7 | International Organizations | 2 | 0.14 |
| 8 | Anonymous | 1 | 0.07 |
| | **Total** | **1435** | **100.00** |

In addition, authors from government ministries make up (7.25%) of the total while those affiliated to private organizations make up only (1.05%).

**4.11.2. Malaysian and Foreign Authors Affiliation**

Table 4.16 – 4.18 compares the number of Malaysian authors and their affiliation type with their foreign counterpart. The 2 authors affiliated to international organizations are foreign authors from World Health Organization (WHO) and there are no foreign authors affiliated to clinics or private organizations.

**Table 4. 17 Malaysian and Foreign Authors Affiliation type**

| | Affiliation types | Malaysian Authors | Foreign Authors | Total |
|---|---|---|---|---|
| 1 | Hospitals | 363 | 38 | 401 |
| 2 | Institute Of Higher Learning | 748 | 126 | 874 |
| 3 | Government Agencies | 95 | 9 | 104 |
| 4 | Medical Centers | 18 | 4 | 22 |
| 5 | Clinics | 16 | 0 | 16 |
| 6 | Private Organizations | 15 | 0 | 15 |
| 7 | International Organizations | 0 | 2 | 2 |
| 8 | Anonymous | 1 | 0 | 1 |
| | **Total** | **1256** | **179** | **1435** |

About 70.39% of the foreign authors are affiliated to higher institutions compared to 59.55% of Malaysian authors (Fig 4.17 and Fig 4.18); also 21.23% of foreign authors are affiliated to Hospitals compared to 28.90% of Malaysian authors.

**Table 4. 18 Malaysian Authors Affiliation Type**

|   | Affiliation types | Malaysian Authors | Percentage (%) |
|---|---|---|---|
| 1 | Hospitals | 363 | 28.90 |
| 2 | Institute Of Higher Learning | 748 | 59.55 |
| 3 | Government Agencies | 95 | 7.56 |
| 4 | Medical Centers | 18 | 1.44 |
| 5 | Clinics | 16 | 1.28 |
| 6 | Private Organizations | 15 | 1.19 |
| 7 | International Organizations | 0 | 0.00 |
| 8 | Anonymous | 1 | 0.08 |
|   | **Total** | **1256** | **100.00** |

There are very few or no foreign authors affiliated to medical centers (2.23 %), clinics (0%), and private organizations (0%). Whereas there are some Malaysian authors affiliated to medical centers (1.44%), clinics (1.28%), and private organizations (1.19%).

**Table 4. 19 Foreign Authors Affiliation type**

|   | Affiliation types | No of Foreign | Percentage (%) |
|---|---|---|---|
| 1 | Hospitals | 38 | 21.23 |
| 2 | Institute Of Higher Learning | 126 | 70.39 |
| 3 | Government Agencies | 9 | 5.03 |
| 4 | Medical Centers | 4 | 2.23 |
| 5 | Clinics | 0 | 0.00 |
| 6 | Private Organizations | 0 | 0.00 |
| 7 | International Organizations | 2 | 1.12 |
| 8 | Anonymous | 0 | 0.00 |
|   | **Total** | **179** | **100.00** |

There are also Malaysian authors affiliated to Ministries and Government agencies (7.56%) compared to Foreign (5.03%). We can thus, observe from the results that researchers and academics from higher institutions and hospitals are active as authors in Malaysia.

### 4.11.3. Number of Authors per Affiliation

Table 4.19 illustrates the names of different types of affiliation, ranked accordingly by numbers of authors. Universiti Kebangsaan Malaysia top the list with (157) authors, followed by Universiti Sains Malaysia (133), Hospital Universiti Kebangsaan Malaysia (102), University of Malaya (101) and University Putra Malaysia (96).

**Table 4. 20 Number of Authors per Affiliation**

| Group | Cohort | Affiliation | No of Authors |
|---|---|---|---|
| 1 | Cohort : 1 | UniversitI Kebangsaan Malaysia | 157 |
| 2 | Cohort : 1 | Universiti Sains Malaysia. | 133 |
| 3 | Cohort : 3 | Hospital Universiti Kebangsaan Malaysia | 102 |
|  |  | University of Malaya | 101 |
|  |  | Universiti Putra Malaysia | 96 |
| 4 | Cohort : 2 | Hospital Kuala Lumpur | 64 |
|  |  | University of Malaya Medical Centre | 59 |
| 5 | Cohort : 2 | Ministry of Health Malaysia | 47 |
|  |  | International Medical University, Seremban | 46 |
| 6 | Cohort : 1 | International Islamic University of Malaysia | 39 |
| 7 | Cohort : 2 | Sarawak General Hospital. | 23 |
|  |  | Penang Hospital. | 22 |
|  |  | Universiti Technology Mara | 20 |
| 8 | Cohort : 2 | Hospital Ipoh.Perak. | 18 |
|  |  | Hospital Sultanah Aminah, Johor Bahru. | 18 |
| 9 | Cohort : 1 | Hospital Selayang | 17 |
| 10 | Cohort : 2 | Hospital Seremban | 16 |
|  |  | Singapore General Hospital | 16 |
| 10 | Cohort : 2 | Hospital Tunku Ampuan Afzan. | 15 |
|  |  | Katurba Medical College | 15 |
| 11 | Cohort : 3 | Asian Institute of Medicine, Science and Technology, Malaysia | 11 |
|  |  | University Medical Centre Groningen | 11 |
|  |  | Queen Elizabeth Hospital, Kota Kinabalu | 11 |
| 12 | Cohort : 3 | Hospital Pakar Sultanah Fatimah | 9 |
|  |  | Institute for Medical Research Malaysia | 9 |
|  |  | Klinik Kesihatan in Negri Sembilan | 9 |
| 13 | Cohort : 4 | Gadjah Mada University, Indonesia | 8 |
|  |  | Maulana Azad Medical College and Associated Lok Nayak Hospital, G.B. Pant Hospital and Guru Nanak Eye Hospital (India). | 8 |
|  |  | Perak Royal College of Medicine | 8 |
|  |  | StemLife Berhad, Malaysia. | 8 |
| 14 | Cohort : 7 | Malaysian Nuclear Agency | 7 |
|  |  | Melaka Manipal Medical College | 7 |
|  |  | Nagoya University (Japan) and University of Tokyo (Japan) | 7 |
|  |  | Shaheed Beheshti University of Medical Sciences, Iran | 7 |
|  |  | Universiti Malaysia Sarawak. | 7 |
|  |  | Yuzuncu Yil University, Turkey | 7 |
|  |  | Putrajaya Hospital. | 7 |
| 15 | Cohort : 1 | Guru Nanak Dev University, India. | 6 |
| 16 | Cohort : 5 | Erciyes University, Turkey. | 5 |
|  |  | Hospital Melaka | 5 |
|  |  | Kyoto Prefectural University of Medicine, Japan. | 5 |
|  |  | Subang Jaya Medical Centre | 5 |
|  |  | University of Abant Izzet Baysal, Turkey | 5 |
| 17 | Cohort : 6 | Harvard University, USA. Brigham and Women's Hospital | 4 |

| | | Institut Jantung Negara | 4 |
|---|---|---|---|
| | | Kuching General Hospital | 4 |
| | | Malaysian Armed Forces | 4 |
| | | National University of Malaysia | 4 |
| | | Universiti Tenaga Nasiona | 4 |
| 18 | Cohort : 19 | Changi General Hospital, Singapore | 3 |
| | | Damansara Specialist Hospital | 3 |
| | | Esramus Medical Center, Netherlands | 3 |
| | | Hospital de la Foundation Rothschild, Paris | 3 |
| | | Hospital Sungai Petani | 3 |
| | | Hospital Universiti Sains Malaysia | 3 |
| | | Hulu Langat District Health Office. Disease Control Unit | 3 |
| | | Inno Bio Diagnostics Sdn Bhd, Malaysia. | 3 |
| | | Institute for Health Systems Research | 3 |
| | | Isfahan University of Medical Sciences, Iran. | 3 |
| | | Kobe University. | 3 |
| | | Mater Children's Hospital, Australia. | 3 |
| | | Mustafa Kemal University | 3 |
| | | Raigmore Hospital (United Kingdom). | 3 |
| | | Royal College of Medicine Perak | 3 |
| | | Sultanah Aminah Hospital | 3 |
| | | Tabriz University of Medical Science, Iran | 3 |
| | | Tengku Ampuan Afzan | 3 |
| | | University College Sedaya International, Kuala Lumpur. School of Pharmacy | 3 |
| 19 | Cohort : 29 | | 2 |
| 20 | Cohort : 76 | | 1 |

## 4.12. Pattern of Collaboration between Authors

Most of the articles published in *Medical Journal of Malaysia (MJM)* from 2004 - 2008 are co-written by two or more authors either from the same or different affiliations. These articles make up to 90% of the total article published during the 5 year period (Table 4.20).

**Table 4. 21 Single and Multi – Author Pattern of Collaboration**

| Year | Single Member Author | Multi-Authors of the same Affiliation | Multi-Authors of different Affiliations in the same Country | Multi-Authors of Different Affiliations from Different Countries | Total |
|---|---|---|---|---|---|
| 2004 | 17 | 68 | 51 | 3 | 139 |
| 2005 | 13 | 49 | 33 | 7 | 102 |
| 2006 | 9 | 64 | 25 | 6 | 104 |
| 2007 | 7 | 63 | 24 | 6 | 100 |
| 2008 | 10 | 68 | 47 | 10 | 135 |
| **Total** | **56** | **312** | **180** | **32** | **580** |

Table 4.20 and 4.21 described the various types of single and multi-author pattern of collaboration. Articles written by single authors are 56 (9.66%), those written by two or more authors from the same affiliations is 312 (53.79%), those written by two or more authors of different affiliation in the same country is 180 (31.03), while those written by authors of different affiliations from different countries is 32 (5.52%). It thus follows that articles written by two or more authors of different affiliations constitute 37.38% (274) of the total.

**Table 4. 22 Single and Multi – Author Pattern of Collaboration**

| Types of Collaboration | No of Articles | Percentage (%) |
|---|---|---|
| Single Member Author | 56 | 9.66 |
| Multi-Members of the same Affiliation | 312 | 53.79 |
| Members of different Affiliations in the same Country | 180 | 31.03 |
| Members of Different Affiliations from Different Countries | 32 | 5.52 |
| **Total** | **580** | **100.00** |

This result indicates that majority of authors contributing to *MJM* mainly collaborate with peers from the same institution or peers from other institutions in Malaysia.

## 4.13. Collaboration between Foreign Affiliations in the same Country

Table 4.22 shows collaborating affiliations belonging to the same foreign country but different Institutions. This category of collaboration produced 10 articles during the period under study. From Turkey (4 articles), Singapore (3 articles), India (1 articles), Indonesia (1 article), and United Kingdom (1 article).

**Table 4. 23 Collaboration between Foreign Affiliations in the Same Country**

| Countries | Collaboration between Affiliations in the same Country (Foreign) Affiliations | No of Articles |
|---|---|---|
| India | Guru Nanak Dev University  ::  District Tuberculosis Centre and Hospital | 1 |
| Indonesia | Gadjah Mada University, Indonesia  ::  Ahmad Dahlan University | 1 |
| Singapore | Alexandra Hospital, Singapore  ::  Singapore Health Services  ::  Changi General Hospital, Singapore | 1 |
| | Tan Tock Seng Hospital  ::  Singapore General Hospital | 1 |
| | Changi General Hospital Singapore  ::  Alexandra Hospital, Singapore | 1 |
| Turkey | Cukurova University, Turkey  ::  Haydarpasa Numune Education :: Research Hospital, Turkey. | 1 |
| | Erciyes University, Turkey  ::  Yuzuncu Yil University, Turkey. | 1 |

| | Mustafa Kemal University :: Dumlupinar University, Turkey :: Mersin University, Turkey | 1 |
| | Mustafa Kemal University, Turkey :: the University, Turkey | 1 |
| UK | Raigmore Hospital (United Kingdom) :: Princess Alexandra Eye Pavilion, UK | 1 |
| | **Total** | **10** |

## 4.14. Collaboration between Malaysian Affiliations

Table 4.23 shows the number of times Malaysian affiliations collaborated during the period under study. The highest number of collaboration between different Malaysian affiliations occurred eight (8) times and that was "Universiti Putra Malaysia with Hospital Kuala Lumpur"and " Universiti Kebangsaan Malaysia Medical Centre with Ampang Puteri Specialist Hospital" This was followed by "International Islamic University Malaysia with Hospital Universiti Kebangsaan Malaysia" and "Universiti Putra Malaysia with Universiti Kebangsaan Malaysia" occurring six (6) times. The results indicate that Universities collaborated actively with hospitals in research activities.

**Table 4. 24 Collaboration between Malaysian Affiliations**

| Group | Collaborating affiliations | No of Times |
|---|---|---|
| 1 | **Cohort : 2** | |
| | Universiti Putra Malaysia and Hospital Kuala Lumpur | 8 |
| | Universiti Kebangsaan Malaysia Medical Centre and Ampang Puteri Spec. Hospital | |
| 2 | **Cohort : 2** | |
| | International Islamic University Malaysia and Hospital Universiti Kebangsaan Malaysia | 6 |
| | Universiti Putra Malaysia and Universiti Kebangsaan Malaysia | 6 |
| 3 | **Cohort : 5** | |
| | International Medical University and Seremban Hospital | 5 |
| | Ministry of health and University Malaya | 5 |
| | Universiti Kebangsaan Malaysia Medical Centre and Subang Jaya Medical Centre | 5 |
| | Universiti Putra Malaysia and University Malaya Medical Centre | 5 |
| | Universiti Sains Malaysia and Universiti Kebangsaan Malaysia. | 5 |
| 4 | **Cohort : 5** | |
| | International Islamic University Malaysia and Hospital Tunku Ampuan Afzan. | 4 |
| | International Islamic University Malaysia and University of Malaya | 4 |
| | International Islamic University Malaysia and University Sains Malaysia | 4 |
| | Universiti Putra Malaysia and Subang Jaya Medical Centre | 4 |
| | Universiti Putra Malaysia and Universiti Kebangsaan Malaysia | 4 |
| 5 | **Cohort : 6** | |
| | Hospital Kuala Lumpur and Penang Hospital | 3 |
| | International Medical University and Klinik Kesihatan Seremban | 3 |
| | International Medical University and Universiti Putra Malaysia | 3 |
| | Universiti Putra Malaysia and Malaysian Armed Forces Health Services Div*ISI*on | 3 |
| | University Malaya Medical Centre and Hospital Queen Elizabeth | 3 |
| | University of Malaya and Tawakal Hospital. | 3 |

| | **Cohort : 28** | |
|---|---|---|
| 6 | Hospital Kuala Lumpur and Subang Jaya Medical Centre | 2 |
| | International Islamic University Malaysia and Tengku Ampuan Afzan Hospital | 2 |
| | International Islamic University Malaysia and Universiti Tenaga Nasional | 2 |
| | International Medical University and Hospital Batu Pahat | 2 |
| | International Medical University and Hospital Kuala Lumpur | 2 |
| | International Medical University and Sarawak General Hospital | 2 |
| | Kudat Health Office  and Beaufort Hospital | 2 |
| | Ministry of health and International Medical University | 2 |
| | Perak Royal College of Medicine and  Hospital Melaka | 2 |
| | Universiti Kebangsaan Malaysia and  Hospital Kuala Lumpur | 2 |
| | Universiti Kebangsaan Malaysia and Hospital Alor Setar | 2 |
| | Universiti Kebangsaan Malaysia and Institute of Medical Research Malaysia | 2 |
| | Universiti Kebangsaan Malaysia and Queen Elizabeth Hospital | 2 |
| | Universiti Kebangsaan Malaysia and Subang Jaya Medical Centre | 2 |
| | Universiti Malaysia Sarawak (UNIMAS) and Sarawak General Hospital, Kuching | 2 |
| | Universiti Putra Malaysia and  Hospital University Kebangsaan Malaysia | 2 |
| | Universiti Putra Malaysia and Ampang Puteri Specialist Hospita | 2 |
| | Universiti Putra Malaysia and Hospital Ipoh | 2 |
| | Universiti Putra Malaysia and Hospital Universiti Kebangsaan Malaysia | 2 |
| | Universiti Putra Malaysia and Island Hospital (Penang) | 2 |
| | Universiti Putra Malaysia and Malaysian Nuclear Agency | 2 |
| | Universiti Putra Malaysia and Tawakal Hospital. | 2 |
| | Universiti Sains Malaysia and Hospital Penang | 2 |
| | University of Malaya and  Universiti Putra Malaysia | 2 |
| | University of Malaya and Institute for Medical Research | 2 |
| | University of Malaya and Universiti Sains Malaysia | 2 |
| | University of Malaya Medical Centre and  Southport District General Hospital | 2 |
| | University of Malaysia Sarawak and Sarawak General Hospital | 2 |
| 7 | **Cohort : 121** | 1 |

## 4.15. Collaboration between Affiliations of Different Countries

Analysis of *MJM* shows that collaboration of affiliations among different countries and different categories of institutional affiliations is a common research practice in the medical field. Table 4.24 indicates that from 2004 – 2008 different affiliations in Malaysia collaborated with their Australian counterparts to produce 7 journal articles. Universiti Kebangsaan Malaysia was involved in 3 of the combinations with Monash University Australia in (2). Other combinations are between Hospitals, Clinics and Higher Institutions of both countries.  University of Wales, and University of Edinburgh, UK combined separately with Universiti Putra Malaysia and University of Malaya respectively to produce (1 article) each. From India, Kasturba Medical College combined with Melaka Manipal Medical College in Malaysia to produce 3 articles.

**Table 4. 25 Collaboration between Affiliations of different Countries**

| | Collaborating Affiliations | No of Articles |
|---|---|---|
| **1** | **Australia and Malaysia** | |
| | Universiti Sains Malaysia and Mater Children's Hospital, Australia. | 1 |
| | Hospital Putrajaya and Royal Melbourne Hospital, Australia | 1 |
| | Universiti Kebangsaan Malaysia and University of Melbourne, Australia | 1 |
| | State Epidemiological Unit, Perak and WHO Collaborating Centre for Reference and Research on Influenza, Melbourne and Monash University Australia and Makmal Kesihatan Awam Kebangsaan | 1 |
| | Kuala Pilah Health Clinic and Kelana Jaya Health Clinic and Monash University, Australia and International Medical University | 1 |
| | Penang Adventist Hospital and University of Queensland | 1 |
| | Universiti Kebangsaan Malaysia and Queen Elizabeth Hospital, South Australia | 1 |
| **2** | **India and Malaysia** | |
| | Madras University, India and AIMST University Malaysia | 1 |
| | Kasturba Medical College and Dr. TMA Pai Hospital, India and Asian Institute of Medicine | 1 |
| | Melaka Manipal Medical College and  Kasturba Medical College, India | 3 |
| **3** | **UK and Malaysia** | |
| | International Medical University and Hospital Putrajaya and Western General Hospital, UK | 1 |
| | University of Malaya Medical Centre and Sheffield Fertility Centre, United Kingdom. | 1 |
| | Universiti Malaysia Sabah and Universiti of Wales and Queen Elizabeth Hospital | 1 |
| | Universiti Putra Malaysia and University of Edinburgh, UK. | 1 |
| **4** | **Singapore and Malaysia** | |
| | Fatimah Hospital, Ipoh and Singapore General Hospital and Mount Elizabeth Medical Centre, Singapore | 1 |
| | Singapore Medical Journal and Melaka Manipal Medical College | 1 |
| | Universiti Sains Malaysia Health Campus and National University Hospital. National University of Singapore | 1 |
| **5** | **Indonesia and Netherlands** | |
| | Gadjah Mada University, Indonesia and Esramus Medical Center, Netherlands | 1 |
| | University of Groningen, Netherlands, and Gadjah Mada University, Indonesia | 1 |
| | Dr. Sardjito General Hospital, Indonesia and University Medical Centre Groningen, The Netherlands and University of Twente (The Netherlands) | 1 |
| **6** | **USA and Malaysia** | |
| | Universiti Kebangsaan Malaysia Medical Centre and Harvard University, USA. Brigham and Women's Hospital | 2 |
| **7** | **Japan and Malaysia** | |
| | Universiti Sains Malaysia and Kobe University | 1 |
| | Nagoya University (Japan) and Universiti Kebangsaan Malaysia and University of Tokyo (Japan) | 1 |
| **8** | **Canada and Malaysia** | |
| | University of Malaya and University of Toronto, Canada | 1 |
| **9** | **France and Malaysia** | |
| | Universiti Sains Malaysia. School of Medical Sciences and Hospital de la Foundation Rothschild, Paris | 1 |
| **10** | **Pakistan and Malaysia** | |
| | Hyderabad Medical Complex, Pakistan and University Malaya Medical Centre | 1 |
| **11** | **Saudi Arabia and Malaysia** | |
| | Hospital Tengku Ampuan Afzan and Hospital Universiti Kebangsaan Malaysia and  King Fahd Medical Centre, Saudi Arabia | 1 |
| **12** | **USA and Australia and Malaysia** | |
| | National Public Health Laboratory and University of Malaya and Center for Diseases Control and Prevention, USA and Australian Animal Health Laboratory ( | 1 |
| **13** | **USA and India** | |
| | Drexel University College of Medicine (USA) and Maulana Azad Medical College and Associated Lok Nayak Hospital india | 1 |
| | **Total** | 32 |

In general, Malaysians and Australians produce 7 articles, with India 5 articles, with

United Kingdom 4 articles, with the United States 2 articles, Japan 2 articles and with Canada, Pakistan, and Saudi Arabia to produce one article each. . Other collaborations are with Indonesia and Netherlands (3), USA and India (1).

## 4.16. Distribution of Keywords in Articles

Table 4.25 lists the keywords (medical related words) that appeared in articles published in *MJM* from 2004 to 2008 according to their frequency of occurrence. The word Diabetes has the highest frequency of (28) followed by Cancer (13), Endoscopic (13), hypertension (13), Tuberculosis (12), etc. The occurrence of the keywords reflected the actively researched areas by medical practitioners and scientists.

**Table 4. 26 Distribution of Keywords in Articles**

| Keywords | Frequency | Keywords | Frequency |
|---|---|---|---|
| Diabetes | 28 | Pituitary | 5 |
| Cancers | 13 | Posterior | 5 |
| Endoscopic | 13 | Prognosis | 5 |
| hypertension | 13 | Survival | 5 |
| Tuberculosis | 12 | Adults | 4 |
| Bone | 10 | Asthma | 4 |
| Children | 10 | cataract | 4 |
| Elderly | 10 | Cerebral | 4 |
| Primary care | 10 | Cervical | 4 |
| Clinical trials | 8 | Community | 4 |
| Medical Students | 8 | Dyslipidaemia | 4 |
| Pregnancy | 8 | Emergency | 4 |
| Risk-factors | 8 | Ethnicity | 4 |
| Urinary | 8 | Foreign body | 4 |
| Blood pressure | 7 | glaucoma | 4 |
| Chronic | 7 | Mortality | 4 |
| HIV/AIDS | 7 | Pediatrics | 4 |
| Knee | 7 | Pancreas | 4 |
| Nasal | 7 | Pulmonary | 4 |
| Neonatal | 7 | Thrombocytopaenia | 4 |
| Outbreaks | 7 | Validity | 4 |
| Prevention | 7 | anesthesia | 3 |
| Stem cells | 7 | Angiotensin | 3 |
| Surgery | 7 | Aspiration | 3 |
| Attitude | 6 | Audit | 3 |
| Congenital | 6 | carcinoma | 3 |
| Hospital | 6 | Chikungunya virus | 3 |
| Knowledge | 6 | Compliance | 3 |
| Abdominal | 5 | Day care | 3 |
| Adolescence | 5 | Epistaxis | 3 |
| Aged | 5 | Fracture | 3 |
| Benign | 5 | General practice | 3 |
| Body Mass Index | 5 | Guidelines | 3 |
| Dengue | 5 | Health care workers | 3 |
| Depression | 5 | Ischaemia | 3 |
| diseases | 5 | Laparoscopic | 3 |
| Epidemiology | 5 | Laryngeal Mask Airway | 3 |
| Lupus | 3 | Pain | 3 |
| Malignant | 3 | Placenta | 3 |
| Management | 3 | Pseudoaneurysm | 3 |
| Maxillary sinus | 3 | Radiation | 3 |
| Nephrotic syndrome | 3 | Rubella | 3 |
| Oral Health | 3 | Selangor | 3 |
| Orthopaedic | 3 | Severe Acute Respiratory Syndrome | 3 |
| Outcome | 3 | Shoulder dislocation | 3 |
| Overweight | 3 | Tracheostomy | 3 |
| Pain | 3 | Treatment | 3 |

Most of the medical researches are aimed towards finding solutions to diseases and combating outbreaks that may be peculiar to a particular country, states, or geographical area. The study found that medical researchers often add the geographical settings to the titles and keywords of their articles (Table 26).

**Table 4. 27 Names of Places and Institution that Appear in the Keywords**

| Keywords | Frequency |
|---|---|
| Malaysia | 47 |
| Kuala Lumpur | 3 |
| Johor | 2 |
| Penang | 2 |
| Sarawak | 2 |
| Saudi Arabia | 2 |

Malaysia appeared as a frequent keyword (47 times), suggesting that there are 49 research articles very peculiar to medical related situations in the country. Kuala Lumpur appeared (3 times), while Johor, Penang, Sarawak, and Saudi Arabia all appeared (2 times).

### 4.16.1. Number of Keywords in Articles

*Medical Journal of Malaysia (MJM)* has 580 articles published from year 2004 to 2008. Among these, 34 articles have no keywords; the highest number of keywords found in an article is 10 keywords (1 article), followed by 9 keywords (1 article). The majority of the articles (188) have 3 keywords each; about 131 articles have 4 keywords each and the least number of keywords is 1, which was identified in 17 articles (Table 4.27).

**Table 4. 28 Number of Keywords in Articles**

| No of keywords | Frequency | Percentage (%) |
|:---:|:---:|:---:|
| 0 | 34 | 5.86 |
| 1 | 17 | 2.93 |
| 2 | 96 | 16.55 |
| 3 | 188 | 32.41 |
| 4 | 131 | 22.59 |
| 5 | 68 | 11.72 |
| 6 | 34 | 5.86 |
| 7 | 4 | 0.69 |
| 8 | 6 | 1.03 |
| 9 | 1 | 0.17 |
| 10 | 1 | 0.17 |
| | **580** | **100.00** |

As shown above, the minimum number of keyword is (1), while the maximum is (10). The mean for the number of keywords is 3.33 keywords. Most of the articles without keywords are editorials.

### 4.16.2. Numbers of Words in Titles

Titles of articles in *MJM* were analyzed to determine the number of words in Titles (Table 4.28). The minimum count is 2 words, identified in 4 articles, while the maximum is 26 words (1 article). The mean for the number of words in article titles is 10.79.

**Table 4. 29 Numbers of Words in Titles**

| No of Words in Titles | Frequency | Percentage (%) |
|:---:|:---:|:---:|
| 2 | 4 | 0.69 |
| 3 | 9 | 1.55 |
| 4 | 14 | 2.41 |
| 5 | 26 | 4.48 |
| 6 | 45 | 7.76 |
| 7 | 50 | 8.62 |
| 8 | 45 | 7.76 |
| 9 | 57 | 9.83 |
| 10 | 43 | 7.41 |
| 11 | 51 | 8.79 |
| 12 | 44 | 7.59 |
| 13 | 44 | 7.59 |
| 14 | 39 | 6.72 |
| 15 | 21 | 3.62 |
| 16 | 24 | 4.14 |
| 17 | 18 | 3.10 |
| 18 | 12 | 2.07 |
| 19 | 15 | 2.59 |
| 20 | 8 | 1.38 |
| 21 | 5 | 0.86 |
| 22 | 3 | 0.52 |
| 23 | 1 | 0.17 |
| 25 | 1 | 0.17 |
| 26 | 1 | 0.17 |
| | **580** | **100.00** |

## 4.17. Agencies and Organizations Funding Medical Research

Financial support is very important to support researchers and scientist with their work. Funding and sponsorship is ascertained by screening the acknowledgments part at the end of each *MJM* articles. This approach also enables the study to determine names of the institutions supporting research. Analysis of papers in the *Medical Journal of Malaysia* published from 2004 to 2008 reveals that only 61 papers (Original research articles) received financial support during the five-year period (Table 4.29).

**Table 4. 30 Organizations Funding Research per Year**

| | Funding | 2004 | 2005 | 2006 | 2007 | 2008 | Total |
|---|---|---|---|---|---|---|---|
| 1 | eSciece Fund grant | 0 | 0 | 0 | 0 | 1 | **1** |
| 3 | International Islamic University Research Center | 0 | 2 | 1 | 0 | 1 | **4** |
| 4 | International Medical University | 0 | 0 | 3 | 3 | 1 | **7** |
| 5 | Iranian budget and programming organization | 1 | 0 | 0 | 0 | 0 | **1** |
| 6 | Johor State Health Dept | 1 | 0 | 0 | 0 | 0 | **1** |
| 7 | Toray Sci. Foundation, Japan for UM | 0 | 0 | 1 | 0 | 0 | **1** |
| 8 | Meditel electronics Sdn Bhd and Agfa ASEAN | 0 | 0 | 1 | 0 | 0 | **1** |
| 9 | Melaka Manipal Medical College | 0 | 0 | 0 | 1 | 0 | **1** |
| 10 | Ministry of Health | 1 | 0 | 0 | 0 | 1 | **2** |
| 11 | Ministry of Sc and Tech (Top down, IRPA) Grant | 5 | 3 | 3 | 0 | 2 | **13** |
| 12 | Ministry of Sport Malaysia, Malaysian Road Safety Council, | 1 | 0 | 0 | 0 | 0 | **1** |
| 13 | National Biotechnology Directorate (NBD) under the 8th MP | 0 | 1 | 0 | 0 | 0 | **1** |
| 14 | National Research Council, Iran | 0 | 0 | 1 | 0 | 0 | **1** |
| 15 | Novartis Malaysia | 0 | 1 | 0 | 0 | 0 | **1** |
| 16 | Novo Nordisk Asia | 0 | 1 | 0 | 0 | 0 | **1** |
| 17 | Renal research fund of Malaysian Society of Nephrology | 1 | 0 | 0 | 0 | 0 | **1** |
| 18 | Selangor State Health Dept. | 0 | 0 | 0 | 1 | 0 | **1** |
| 19 | UNIMAS short term grant | 0 | 1 | 0 | 0 | 0 | **1** |
| 20 | Universiti Tecknology Mara, Institute of Res. Dev. and Comm | 0 | 0 | 0 | 1 | 0 | **1** |
| 21 | Universiti Kebangsaan Malaysia, Fundamental, other grant | 3 | 0 | 0 | 1 | 2 | **6** |
| 22 | Universiti Kebangsaan Malaysia, Faculty of Medicine | 0 | 0 | 0 | 1 | 0 | **1** |
| 23 | University of Malaya short term grant (F Vote) | 0 | 1 | 0 | 2 | 0 | **3** |
| 24 | University of Malaya. Res and Dev. Management Unit | 1 | 0 | 0 | 0 | 0 | **1** |
| 25 | University of Medical Sciences, Tehran, Iran | 0 | 0 | 1 | 0 | 0 | **1** |
| 26 | University Putra Malaysia | 0 | 0 | 1 | 0 | 0 | **1** |
| 27 | USM (FRGS, short term research grant) | 3 | 0 | 2 | 1 | 1 | **7** |
| 28 | Yayasan FELDA, Kuala Lumpur | 1 | 0 | 0 | 0 | 0 | **1** |
| | | **18** | **10** | **13** | **11** | **9** | **61** |

It is worthy to point out here that the categories of papers published by *Medical Journal of Malaysia (MJM)* per issue yearly are: original research papers, continuing medical education (CME) articles, case reports, short communications correspondence and brief abstracts of original papers published elsewhere, concerning medicine in Malaysia. However, for this analysis section (Table 4.30), the numbers of funded papers were considered against the total number of "Original research articles" published yearly. This is because, other papers that belongs to the category of "continuing medical education (CME), case reports, short communications correspondence and brief abstracts are not original research and may not have attracted or in need of sponsorship or financial aid.

Table 4.30 shows the number of "Original research articles" published per year and revealed that just 17.63% of all the papers have been funded.

**Table 4. 31 Research Funding Per Year**

| Year | Original articles | Articles funded | Articles funded (%) |
|------|-------------------|-----------------|---------------------|
| 2004 | 72 | 18 | 5.20 |
| 2005 | 77 | 10 | 2.89 |
| 2006 | 68 | 13 | 3.76 |
| 2007 | 66 | 11 | 3.18 |
| 2008 | 63 | 9 | 2.60 |
|      | **346** | **61** | **17.63** |

In addition, from Table 4.29, it could be noticed that most of the financial aids were by the top public higher institutions in the country, with the Malaysian government playing a key role. There are 2 papers sponsored by Iranian government, 1 co-sponsored by Japan and 2 by private organizations from Malaysia. Others are by government ministries and University research grants. This result is corroborated by Lin and Chiang (2007), who noted that 60% of the papers published in the *American Journal of Chinese Medicine* from 2002 to 2004 were sponsored by various institutions in China with the Chinese government playing a very important role. Hence, the study observed that the Ministry of Science, Technology and Innovations Malaysia is the main funder of research published in *MJM* during the period under examination, through its top down and IRPA grants.

## 4.18. References Used in Articles

Medical researchers give credit to works of other researchers through references when conducting their research, as the norm in all scientific discipline. Table 4.31 revealed that the more the number of articles, the more the number of references. The total number of references recorded during the 5-year period by the 580 published articles is 6958.

**Table 4. 32 Distribution of References in Articles per Year**

| Year | No of Articles | Cited References | Average Citation |
|------|----------------|------------------|------------------|
| 2004 | 139 | 1898 | 14 |
| 2005 | 102 | 1364 | 13 |
| 2006 | 104 | 1200 | 12 |
| 2007 | 100 | 1258 | 13 |
| 2008 | 135 | 1238 | 9 |
| **Total** | **580** | **6958** | **12** |

The average number of references per year is 1391 citations, while the average number of references per article is approximately 12. The range of average references per article is from 9 to 14 citations. Table 4.32 shows that from the 580 articles produced, most references are within the range of 0 - 10 (325 articles) (56.0%), followed by 11 – 20 (156 articles) (26.90%). The largest range is between 80 – 90 references (1 article) (0.17%) and 71 – 80 references (2 articles) (0.34%).

**Table 4. 33 Range of References per Article**

| Range of References | No of Articles | Percentage (%) |
|---------------------|----------------|----------------|
| 80 - 90 | 1 | 0.17 |
| 71 - 80 | 2 | 0.34 |
| 61 - 70 | 2 | 0.34 |
| 51 - 60 | 4 | 0.69 |
| 41 - 50 | 4 | 0.69 |
| 31 - 40 | 19 | 3.28 |
| 21 - 30 | 67 | 11.55 |
| 11--20 | 156 | 26.90 |
| 0 - 10 | 325 | 56.03 |
|  | **580** | **100.00** |

### 4.18.1. Distribution of References According to Bibliographic Format

Table 4.33 outlined the types and kinds of sources commonly refer to by medical researchers during the course of a research endeavor. It is interesting to see that author's citations ranges from Journal articles, questionnaires to pamphlets. As a result, the references were sorted and combined into 9 distinct categories for ease of analysis.

**Table 4. 34 Bibliographic Format of References**

| | Bibliographic format | No of Citations | Percentage (%) |
|---|---|---|---|
| 1 | Journal Articles | 6100 | 87.67 |
| 2 | Books And Book Chapters | 333 | 4.79 |
| 3 | Conference, Seminars Etc | 59 | 0.85 |
| 4 | Web Resources | 52 | 0.75 |
| 5 | Government Publications | 193 | 2.77 |
| 6 | Publication By International Org. (WHO, UNICEF Etc) | 55 | 0.79 |
| 7 | Thesis And Dissertations | 10 | 0.14 |
| 8 | Newspapers And Press Release | 71 | 1.02 |
| 9 | Others | 85 | 1.22 |
| | | 6958 | 100.00 |

Most of the cited references came from articles published in several medical related journals and bulletins. These also include editorials and reviews. Scholarly journals recorded 6100 (87.67%) of citations, followed by Books 333 (4.79%). Wide range of data, reports, records statistics, and standards were cited and these category of sources fall under Government Publications (mostly by the Ministry of Health, Malaysia), 193 (2.77%), and International organizations (WHO, UNICEF etc) 55 (0.79). Sources that falls under "Others" are: questionnaires, survey, software, course notes, atlas, booklets, pamphlets etc.

**Table 4. 35 Reference Sources Used by Year**

| Bibliographic format | 2004 | 2005 | 2006 | 2007 | 2008 | Total | Percentage (%) |
|---|---|---|---|---|---|---|---|
| Journal Articles | 1636 | 1186 | 1048 | 1123 | 1107 | 6100 | 87.67 |
| Books And Book Chapters | 95 | 77 | 65 | 38 | 58 | 333 | 4.79 |
| Conference, Seminars Etc | 17 | 17 | 8 | 4 | 13 | 59 | 0.85 |
| Web Resources | 13 | 14 | 5 | 10 | 10 | 52 | 0.75 |
| Government Publications | 56 | 29 | 43 | 45 | 20 | 193 | 2.77 |
| Publication By International Org. (WHO, UNICEF, Etc) | 22 | 10 | 4 | 10 | 9 | 55 | 0.79 |
| Thesis And Dissertations | 5 | 2 | 1 | 1 | 1 | 10 | 0.14 |
| Newspapers And Press Release | 29 | 10 | 13 | 13 | 6 | 71 | 1.02 |
| Others | 25 | 19 | 13 | 14 | 14 | 85 | 1.22 |
| | 1898 | 1364 | 1200 | 1258 | 1238 | 6958 | 100.00 |

### 4.18.2. Age of References

Examining the age of references encountered in *MJM* tells that medical research process that took place centuries ago may still be very relevant in the laboratories of today. Medical journal articles and books dating back to 18$^{th}$ and early 19$^{th}$ century were cited a couple of times, although most cited references were materials of about 1 to 11 years old (63.87%). Authors continue to cite sources published or produced several decades ago (Table 4.35).

**Table 4. 36 Age of Citations**

| Age of Citation in years | No of Citations | Percentage (%) |
|---|---|---|
| Up to 1 | 73 | 1.05 |
| 2 | 230 | 3.31 |
| 3 | 448 | 6.44 |
| 4 | 502 | 7.21 |
| 5 | 535 | 7.69 |
| 6 | 557 | 8.01 |
| 7 | 463 | 6.65 |
| 8 | 468 | 6.73 |
| 9 | 433 | 6.22 |
| 10 | 365 | 5.25 |
| 11 | 370 | 5.32 |
| 12 | 257 | 3.69 |
| 13 | 249 | 3.58 |
| 14 | 215 | 3.09 |
| 15 | 201 | 2.89 |
| 16 | 177 | 2.54 |
| 17 | 141 | 2.03 |
| 18 | 105 | 1.51 |
| 19 | 117 | 1.68 |
| 20 | 96 | 1.38 |
| 21 | 93 | 1.34 |
| 22 | 91 | 1.31 |
| 23 | 73 | 1.05 |
| 24 | 64 | 0.92 |
| 25 | 65 | 0.93 |
| 26 | 57 | 0.82 |
| 27 | 53 | 0.76 |
| 28 | 55 | 0.79 |
| 29 | 31 | 0.45 |
| 30 | 33 | 0.47 |
| 31 - 40 | 188 | 2.70 |
| 41 - 50 | 55 | 0.79 |
| 51 and above | 45 | 0.65 |
| Undated | 53 | 0.76 |
| **Total** | **6958** | **100.00** |

The study also found the half – life of cited references to be 11 years. This was achieved by plotting the year of citations against cumulative number of citations according to data from (Table 4.36) on a linear graph (Figure 4.3).

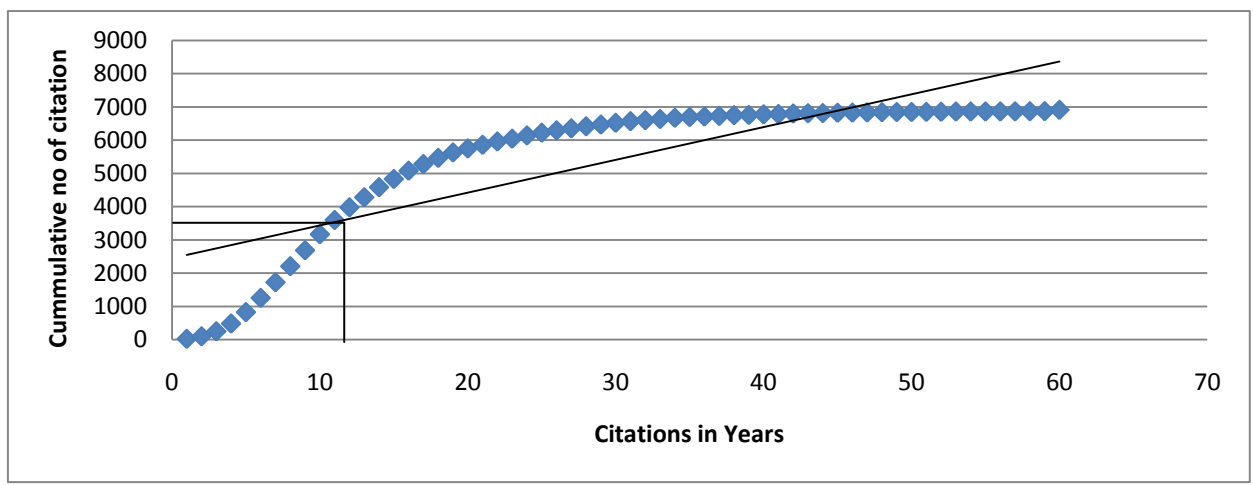

**Figure 4. 3 Half – Life of Cited References**

There are also citations that are less than or about 1 year old prior to citations. Table 4.36 shows a diagonal representation of the year references were produced against the cited year. The Table shows that most references cited were those published between 2002 (471) and 1999 (482). Again, it is prove that sources published more than 10 years continued to be cited by medical researchers.

**Table 4. 37 Publication Year of References against Year of Citation**

| Production year of reference sources | Number of Citations per Year | | | | | Total |
|---|---|---|---|---|---|---|
| | 2008 | | | | | |
| 2008 | 26 | **2007** | | | | **26** |
| 2007 | 65 | 10 | **2006** | | | **75** |
| 2006 | 100 | 46 | 6 | **2005** | | **152** |
| 2005 | 101 | 86 | 32 | 14 | **2004** | **233** |
| 2004 | 116 | 81 | 85 | 44 | 17 | **343** |
| 2003 | 110 | 110 | 85 | 78 | 43 | **426** |
| 2002 | 78 | 110 | 62 | 122 | 99 | **471** |
| 2001 | 72 | 89 | 100 | 108 | 113 | **482** |
| 2000 | 81 | 95 | 80 | 84 | 139 | **479** |
| 1999 | 55 | 90 | 82 | 102 | 153 | **482** |
| 1998 | 68 | 54 | 82 | 108 | 114 | **426** |

| | | | | | |
|---|---|---|---|---|---|
| 1997 | 49 | 69 | 71 | 82 | 111 | **382** |
| 1996 | 34 | 42 | 52 | 76 | 98 | **302** |
| 1995 | 31 | 40 | 42 | 81 | 109 | **303** |
| 1994 | 32 | 27 | 40 | 48 | 100 | **247** |
| 1993 | 38 | 28 | 44 | 63 | 76 | **249** |
| 1992 | 21 | 27 | 38 | 41 | 72 | **199** |
| 1991 | 15 | 31 | 40 | 27 | 72 | **185** |
| 1990 | 17 | 20 | 26 | 25 | 76 | **164** |
| 1989 | 10 | 22 | 18 | 23 | 47 | **120** |
| 1988 | 13 | 15 | 23 | 18 | 40 | **109** |
| 1987 | 14 | 14 | 16 | 23 | 34 | **101** |
| 1986 | 3 | 14 | 16 | 19 | 32 | **84** |
| 1985 | 11 | 13 | 22 | 15 | 36 | **97** |
| 1984 | 6 | 9 | 16 | 16 | 35 | **82** |
| 1983 | 6 | 18 | 9 | 14 | 25 | **72** |
| 1982 | 2 | 5 | 15 | 10 | 27 | **59** |
| 1981 | 4 | 8 | 12 | 12 | 25 | **61** |
| 1980 | 3 | 12 | 12 | 12 | 14 | **53** |
| 1979 | 2 | 7 | 8 | 13 | 22 | **52** |
| 1978 | 5 | 3 | 7 | 16 | 18 | **49** |
| 1977 | 3 | 5 | 5 | 3 | 15 | **31** |
| 1976 | 9 | 3 | 4 | 8 | 11 | **35** |
| 1975 | 8 | 4 | 6 | 3 | 15 | **36** |
| 1974 | 3 | 1 | 4 | 5 | 10 | **23** |
| 1973 | 1 | 4 | 1 | 1 | 7 | **14** |
| 1972 | 0 | 2 | 1 | 4 | 11 | **18** |
| 1971 | 2 | 6 | 5 | 1 | 9 | **23** |
| 1970 | 1 | 1 | 0 | 1 | 4 | **7** |
| 1969 | 2 | 2 | 0 | 6 | 9 | **19** |
| 1968 | 0 | 2 | 3 | 1 | 1 | **7** |
| 1967 | 1 | 3 | 5 | 1 | 7 | **17** |
| 1966 | 1 | 2 | 0 | 2 | 3 | **8** |
| 1965 | 1 | 2 | 2 | 0 | 0 | **5** |
| 1964 | 1 | 1 | 2 | 1 | 4 | **9** |
| 1963 | 0 | 0 | 1 | 2 | 2 | **5** |
| 1962 | 0 | 1 | 0 | 2 | 2 | **5** |
| 1961 | 1 | 1 | 1 | 1 | 5 | **9** |
| 1960 | 0 | 1 | 1 | 1 | 0 | **3** |
| 1959 | 0 | 0 | 0 | 0 | 1 | **1** |
| 1958 | 1 | 0 | 0 | 0 | 3 | **4** |
| 1957 | 1 | 1 | 0 | 1 | 1 | **4** |
| 1956 | 0 | 0 | 1 | 3 | 3 | **7** |
| 1955 | 0 | 1 | 0 | 0 | 0 | **1** |
| 1954 | 0 | 0 | 0 | 1 | 0 | **1** |
| 1953 | 0 | 1 | 0 | 0 | 0 | **1** |
| 1952 | 1 | 0 | 0 | 0 | 2 | **3** |
| 1951 | 0 | 0 | 0 | 0 | 2 | **2** |
| 1950 | 0 | 0 | 0 | 2 | 1 | **3** |
| 1949 and below | 5 | 9 | 9 | 7 | 9 | **39** |
| undated | 8 | 10 | 8 | 13 | 14 | **53** |
| | **1238** | **1258** | **1200** | **1364** | **1898** | **6958** |

Authors also cited books and book chapters and most of the cited reports, data, records statistics, and standards were from ministries and government agencies, especially

"Ministry of Health Malaysia" which was cited 90 times followed by the Department of Statistics, which recorded 32 citations. The most cited international organization is WHO (World Health Organization). The result shows that the age of materials refereed by medical researchers are between 3 - 11 years. Furthermore, results also shows that older publications of over 20 years, and few over 100 years are still referenced, suggesting the longevity of the usefulness of works published in the medical field.

### 4.18.3. Journal Referenced

Journal articles received the largest share (87.67%) of citations in the *Medical Journal of Malaysia* published from 2004 to 2008, and those journals that were cited at least 20 times is presented in Table 4.37. The top 5 most cited are: *Journal of Bone and Joint Surgery* (144 citations), *New England Journal of Medicine* (142) *Lancet* (139), *Clinical Orthopedics and Related Research* (112), and *British Medical Journal* (107). Most medical researchers cite foreign journals and seldom cite regional medical journals. This is similar to the results obtained by Hashimah (1997), when she researched the "citation analysis of the *Medical Journal of Malaysia*". She reported that journals constituted 82.8% of the citations, majority of which were published in the more developed countries.

According to the result revealed by this present study, it can be concluded that, the most cited and most popular regional journals are *Singapore Medical Journal* (38 citations) and *Southeast Asian Journal of Tropical Medicine Public Health* (28 citations). *Singapore Medical Journal* is published by the Singapore Medical Association, while *Asian Journal of Tropical Medicine Public Health* is published by SEAMEO Regional Tropical Medicine and Public Health Network (SEAMEO TROPMED Network) currently hosted at Mahidol University

**Table 4. 38 Journal Referenced by Authors**

| | Journals | Frequency |
|---|---|---|
| 1 | Journal of Bone Joint Surgery | 144 |
| 2 | New England J Med | 142 |
| 3 | Lancet | 139 |
| 4 | Clinical Orthopedic and Related Research | 112 |
| 5 | BMJ | 107 |
| 6 | Diabetes Medicine | 104 |
| 7 | Laryogoscope | 79 |
| 8 | JAMA | 77 |
| 9 | Chest | 65 |
| 10 | Plastic Reconstructive Surgery | 58 |
| 11 | Spine | 53 |
| 12 | Journal Laryngol and Oto | 44 |
| 13 | British J Surg | 41 |
| 14 | Arch Intern Med | 40 |
| 15 | Pediatrics | 40 |
| 16 | Singapore Medical Journal | 38 |
| 17 | Circulation. | 37 |
| 18 | Clinical Infect Dis | 37 |
| 19 | Otoloryngol Head Neck Surg | 33 |
| 20 | Hypertension | 32 |
| 21 | Arch Otolaryngology Head Neck Surgery | 29 |
| 22 | Am J Obstet Gynecol | 28 |
| 23 | Ann Intern Med | 28 |
| 24 | Southeast Asian Journal of Tropical Medicine Public Health | 28 |
| 25 | Transplantation Proceedings | 28 |
| 26 | Radiology | 26 |
| 27 | Cancer | 26 |
| 28 | Journal Hypertens | 25 |
| 29 | Journal Infect Dis | 25 |
| 30 | Stroke | 25 |
| 31 | Am J Med | 24 |
| 32 | Am J Respir Crit Care Med | 24 |
| 33 | Am J Surgical Pathol | 24 |
| 34 | Journal Neurosurgery | 24 |
| 35 | Am J of Epidemiology | 23 |
| 36 | Cancer Res | 23 |
| 37 | Journal Clin Microbiology | 23 |
| 38 | Obstet Gynecol | 23 |
| 39 | Ophthalmology | 23 |
| 40 | Trans R Soc Trop Med Hyg | 23 |
| 41 | Am J Surg | 22 |
| 42 | Arch Dermatol | 22 |
| 43 | Gastoenterology | 22 |
| 44 | Journal Am Acad Dermatol | 21 |
| 45 | Journal Pediatr | 21 |
| 46 | Neurology | 20 |
| 47 | Science | 20 |
| 48 | Journal Clin Epidemiol | 20 |
| 49 | Arthritis Rheum | 20 |
| 50 | British J Psychiatry | 20 |
| 51 | Am J Hypertens | 20 |
| 52 | Ann Otol Rhinol Laryngol | 20 |

### 4.18.4. The Core Journals Referenced

Bradford (1948) published his study on the frequency distribution of papers over journals. He found that "if scientific journals are arranged in order of decreasing productivity on a given subject, they may be divided into a nucleus of journals more particularly devoted to the subject and several groups or zones containing the same number of articles as the nucleus when the numbers of periodicals in the nucleus and the succeeding zones will be as 1: b: b² …" (Glänzel, 2003). Therefore, in order to identify core scholarly journals very relevant to the field, the study applied Bradford's Law of Scattering to the resulting list of cited journals in *MJM* between the years 2004 to 2008.

Three zones were created, each producing approximately one third of the cited references. The list of core journals comprising Zones 1 (Table 4.38) is shown along with the total number of citations each title received.

**Table 4. 39 Distribution of Cited Journals in Zone 1**

| | Journals (Zone 1) | No of citation |
|---|---|---|
| 1 | Journal of Bone Joint Surgery | 144 |
| 2 | New England J Med | 142 |
| 3 | Lancet | 139 |
| 4 | Clinical Orthopedic and Related Research | 112 |
| 5 | BMJ | 107 |
| 6 | Diabetes Medicine | 104 |
| 7 | Laryogoscope | 79 |
| 8 | JAMA | 77 |
| 9 | Chest | 65 |
| 10 | Plastic Reconstructive Surgery | 58 |
| 11 | Spine | 53 |
| 12 | Journal Laryngol and Oto | 44 |
| 13 | British J Surg | 41 |
| 14 | Arch Intern Med | 40 |
| 15 | Pediatrics | 40 |
| 16 | Singapore Medical Journal | 38 |
| 17 | Circulation. | 37 |
| 18 | Clinical Infect Dis | 37 |
| 19 | Otoloryngol Head Neck Surg | 33 |
| 20 | Hypertension | 32 |
| 21 | Arch Otolaryngology Head Neck Surgery | 29 |
| 22 | Am J Obstet Gynecol | 28 |

| 23 | Ann Intern Med | 28 |
| 24 | Southeast Asian Journal of Tropical Medicine Public Health | 28 |
| 25 | Transplantation Proceedings | 28 |
| 26 | Radiology | 26 |
| 27 | Cancer | 26 |
| 28 | Journal Hypertens | 25 |
| 29 | Journal Infect Dis | 25 |
| 30 | Stroke | 25 |
| 31 | Am J Med | 24 |
| 32 | Am J Respir Crit Care Med | 24 |
| 33 | Am J Surgical Pathol | 24 |
| 34 | Journal Neurosurgery | 24 |
| 35 | Am J of Epidemiology | 23 |
| 36 | Cancer Res | 23 |
| 37 | Journal Clin Microbiology | 23 |
| 38 | Obstet Gynecol | 23 |
| 39 | Ophthalmology | 23 |
| 40 | Trans R Soc Trop Med Hyg | 23 |
| 41 | Am J Surg | 22 |
| 42 | Arch Dermatol | 22 |
| 43 | Gastoenterology | 22 |

As illustrated in Table 4.38, 43 journals (1990 citations) are in (Zone 1), 210 (1996 citations) are in (Zone 2), while 1270 journals (1941 citations) are in (Zone 3). The core journal titles which are in the nucleus zone are journals that have 22 or more citations (Table 4.38). This result is in line with Bradford's Law of Scattering, due to the fact that journal titles showed a wide dispersion among a small core, with only about 2.82% of the journals accounting for one-thirds of all the citations. Hence, the proportion of number of journals in the three zones is 43: 210: 1270, which is approximately in the ratio 1: 5: 25. It follows that, our $b \approx 5$ in the general proportion 1: b: b².

**Table 4. 40 Referenced Journal Rank in Descending Order of Citations**

| No of Journals (A) | Cumulative No of Journals (B) | No of Citations | Cumulative No of Citations | Log of (B) |
|---|---|---|---|---|
| 1 | 1 | 144 | 144 | 0.00 |
| 1 | 2 | 142 | 286 | 0.69 |
| 1 | 3 | 139 | 425 | 1.10 |
| 1 | 4 | 112 | 537 | 1.39 |
| 1 | 5 | 107 | 644 | 1.61 |
| 1 | 6 | 104 | 748 | 1.72 |
| 1 | 7 | 79 | 827 | 1.95 |
| 1 | 8 | 77 | 904 | 2.08 |
| 1 | 9 | 65 | 969 | 2.20 |
| 1 | 10 | 58 | 1027 | 2.30 |
| 1 | 11 | 53 | 1080 | 2.40 |
| 1 | 12 | 44 | 1124 | 2.48 |
| 1 | 13 | 41 | 1165 | 2.56 |
| 2 | 15 | 40 | 1205 | 2.71 |
| 1 | 16 | 38 | 1243 | 2.77 |
| 2 | 18 | 37 | 1280 | 2.89 |
| 1 | 19 | 33 | 1313 | 2.94 |
| 1 | 20 | 32 | 1345 | 3.00 |
| 1 | 21 | 29 | 1374 | 3.04 |
| 4 | 25 | 28 | 1402 | 3.22 |
| 2 | 27 | 26 | 1428 | 3.96 |
| 3 | 30 | 25 | 1453 | 3.40 |
| 4 | 34 | 24 | 1477 | 3.53 |
| 6 | 40 | 23 | 1500 | 3.69 |
| 3 | 43 | 22 | 1522 | 3.76 |
| 2 | 45 | 21 | 1543 | 3.81 |
| 7 | 52 | 20 | 1563 | 3.95 |
| 1 | 53 | 19 | 1582 | 3.97 |
| 3 | 56 | 18 | 1600 | 4.03 |
| 4 | 60 | 17 | 1617 | 4.09 |
| 11 | 71 | 16 | 1633 | 4.26 |
| 7 | 78 | 15 | 1648 | 4.36 |
| 6 | 84 | 14 | 1662 | 4.43 |
| 7 | 91 | 13 | 1675 | 4.51 |
| 14 | 105 | 12 | 1687 | 4.65 |
| 10 | 115 | 11 | 1698 | 4.74 |
| 10 | 125 | 10 | 1708 | 4.83 |
| 15 | 140 | 9 | 1717 | 4.94 |
| 21 | 161 | 8 | 1725 | 5.08 |
| 24 | 185 | 7 | 1732 | 5.22 |
| 28 | 213 | 6 | 1738 | 5.36 |
| 40 | 253 | 5 | 1743 | 5.53 |
| 75 | 328 | 4 | 1747 | 5.79 |
| 109 | 437 | 3 | 1750 | 6.08 |
| 228 | 665 | 2 | 1752 | 6.50 |
| 858 | 1523 | 1 | 1753 | 7.33 |

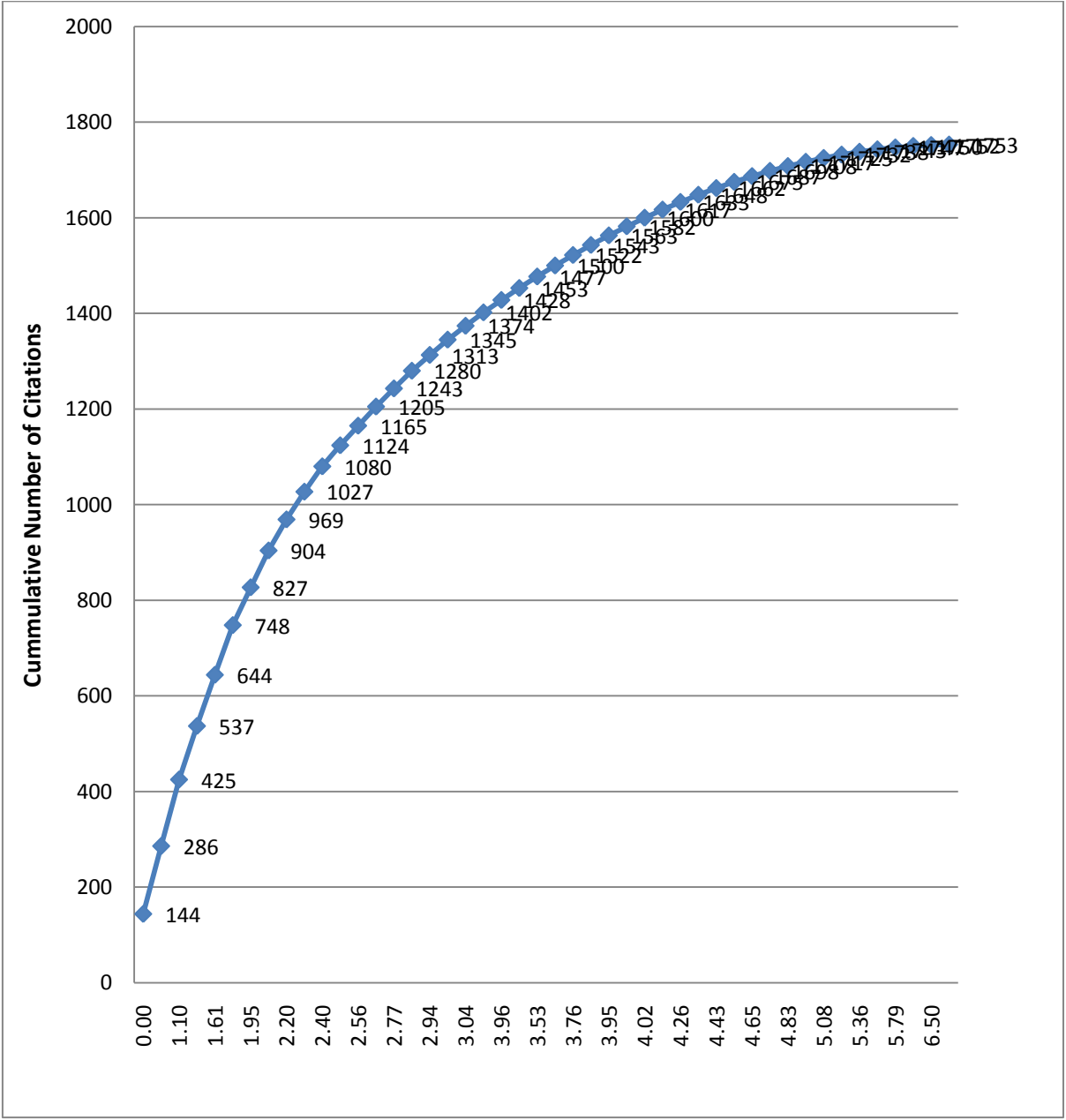

Journals Cumulative (log)

**Figure 4. 4 Frequency of Journal Citations (Bradfords Law)**

### 4.18.5. Journal Self-Citation

Journal self-citation is a situation whereby articles published in a particular journal also refer to the same journal. Table 4.40 shows that 90 *MJM* articles among the total 580 produced practiced journal self-citations.

**Table 4. 41 Journal Self-Citation**

| Year | No of Articles with Self Citations | No of Articles without self-citation | Total No of Articles |
|------|------------------------------------|--------------------------------------|----------------------|
| 2004 | 25 | 114 | 139 |
| 2005 | 18 | 84 | 102 |
| 2006 | 17 | 87 | 104 |
| 2007 | 16 | 84 | 100 |
| 2008 | 14 | 121 | 135 |
| **Total** | **90 (15.5%)** | **490** | **580** |

In Table 4.41, the study observes that *MJM* received 173 citations (2.49%) of the total 6958 citations during the period under study. Thus, this may suggest that journal self-citation is low in *MJM* and the journal has yet to establish itself as an important channel of reference for medical researchers and authors even though it is the oldest medical journal in Malaysia.

**Table 4. 42 Journal Self-Citation and Yearly Percentage**

| Year | Total No of Citations | Journal Self-Citations | Journal Self-Citation (%) |
|------|-----------------------|------------------------|---------------------------|
| 2004 | 1898 | 58 | 3.06 |
| 2005 | 1364 | 31 | 2.27 |
| 2006 | 1200 | 30 | 2.50 |
| 2007 | 1258 | 28 | 2.23 |
| 2008 | 1238 | 26 | 2.10 |
| **Total** | **6958** | **173** | **2.49** |

The highest number of journal self-citation recorded was in year 2004, where 25 articles out of total 139 produced, practiced journal self – citation, and cited *MJM* 58 times. It should also be noted that year 2004 recorded the highest number of article publication (139 articles), and the highest number of citation (1898). In general, the results indicate that, journal self-citations ranges between 2.23% to 3.06% of the total citations and this is quite small.

### 4.18.6. Language of Reference Items

Literature sources published in the English language are mostly referred to by authors publishing in *MJM*, which comprises 99.71% of all the citation, while other languages (Table 4.42) comprises just 0.279% of all cited references.

**Table 4. 43 Language of Reference Items Other than English**

| No | Journal Title | No of times cited | Language |
|----|---------------|-------------------|----------|
| 1 | Changgeng Yi Xue Za Zhi | 1 | Chinese |
| 2 | Chung Hua Liu Hsing Ping Hsueh Tsa Chih | 1 | Chinese |
| 3 | Chung Hua-I-Hsued-Tsa-Chih-Taipei | 1 | Chinese |
| 4 | Yi Xue Ke Xue Za Zhi | 1 | Chinese |
| 5 | Zhonghua Nei Ke Za Zhi | 1 | Chinese |
| 6 | Zhonghua Xue Ye Xue Za Zhi | 1 | Chinese |
| 7 | Zhonghua Yi Xue Za Zhi | 2 | Chinese |
| 8 | Gan To Kagaku Ryoho | 1 | Japanese |
| 9 | Nippon Ketsueki Gakkai Zassho | 1 | Japanese |
| 10 | No Shinkei Geka | 3 | Japanese |
| 11 | Yakugaku Zasshi | 1 | Japanese |
| 12 | Kansenshogaku Zasshi | 1 | Japanese |
| 13 | Nihoh Kyobu Shikkan Gakki Zasshi (abstract) | 1 | Japanese |
| 14 | Kao Hsiung I Hsueh Ko Hsueh Tsa Chih | 1 | Taiwanese |
| 15 | Revue de Chirugie Orthopedique et Raparatrice de la Appareil Moteur | 1 | French |
| 16 | Medicina del Lavoro | 3 | Italian |
| 17 | Cadernos de Saude Publica | 2 | Portuguese |
| 18 | Ginecologiay Obstetricia de Mexico | 1 | Mexican |
| 19 | Deutsch Medicine wochenschr | 1 | Dutch |
| | | 25 | |

The finding here is similar to the findings of (Al-Qallaf, 2003; Chiu et al 2004; Ho and Chiu 2005; Crawley-Low 2006; and Chiu and Ho 2007). They all reported that English language publications dominated the reference sources cited by authors in the field of medical and health sciences.

## 4. 19. Citations Received by *Medical Journal of Malaysia (MJM)*

The citation analysis is the citation count for a journal, an article, a field, or a country's publications. This is the frequency with which papers published in a journal are cited in other papers (Chiu and Ho, 2005). Traditionally, the most commonly used source of bibliometric data is *Thomson ISI Web of Science* and the *Journal Citation Reports* (*JCR*) database, which provide the yearly Journal Impact Factors (JIF) (Harzing and Van der Wal, 2008). Recently, an alternative source of data is being presented by *Google Scholar*, which is a good alternative for journals that are not *ISI*-indexed. It appears to be strongest in the sciences, particularly medicine, and secondarily in the social sciences (Vine, 2006). Šember, et al (2010) analyze the 2007 citation count of articles published by the *Croatian Medical Journal* in 2005-2006 based on data from the *Web of Science*, *Scopus,* and *Google Scholar*. The study observed that the *Web of Science* databases covered the highest-impact scientific journals as the source of citation for the *Croatian Medical Journal*, but that the coverage by *Scopus* and especially of *Google Scholar* was broader and included additional local sources.

Therefore, since *MJM* is not included in the journal citation report, *Google Scholar* is employed, which provides citation information about "*MJM*". The data was retrieved from *Harzing Publish and Perish (http://www.harzing.com/),* which is a bibliometric tool that harvests data from *Google Scholar* to provide structured bibliometrics information, which includes total citations received by a journal and the journal's impact analysis. Table 4.43 shows the total number of citations received by *MJM* between years 2004 and 2008 (as at July 31, 2010).

**Table 4. 44 Citations received by *MJM* Articles Published between Years 2004 to 2008.**

| Publication Year | No of *MJM* articles cited | Year Cited (No of times ) | Total Citations |
|---|---|---|---|
| 2004 | 129 | 2004 (10) | 452 |
| | | 2005(39) | |
| | | 2006(82) | |
| | | 2007(93) | |
| | | 2008(100) | |
| | | 2009(80) | |
| | | 2010(48) | |
| 2005 | 102 | 2005(2) | 268 |
| | | 2006(28) | |
| | | 2007(62) | |
| | | 2008(72) | |
| | | 2009(73) | |
| | | 2010(31) | |
| 2006 | 102 | 2006(7) | 270 |
| | | 2007(65) | |
| | | 2008(61) | |
| | | 2009(93) | |
| | | 2010(94) | |
| 2007 | 56 | 2007(1) | 101 |
| | | 2008(14) | |
| | | 2009(54) | |
| | | 2010(32) | |
| 2008 | 57 | 2008(7) | 73 |
| | | 2009(35) | |
| | | 2010(31) | |
| | **446** | | **1164** |

From Table 4.43, it can be observed that out of the 580 articles published by *MJM* between 2004 and 2008, 76.8% (446) have been cited one time or the other. This implies that *MJM* articles are appealing to local and international researchers as the study noted that *MJM* articles are sources of knowledge in scholarly journals, students' theses and dissertations, books, conference presentations and government publications (Table 4.44).

**Table 4. 45 Types of Documents Citing *MJM* Articles Published between 2004 and 2008**

| | Bibliographic format | Frequency |
|---|---|---|
| 1 | Journal Articles | 1082 |
| 2 | Thesis and Dissertations | 40 |
| 3 | Books and Book Chapters | 22 |
| 4 | Conference Proceedings, Meeting, Seminar etc | 11 |
| 5 | Government Publications, Reports, Statistics etc | 9 |
| | **Total** | **1164** |

As indicated in Table 4.44 *MJM* received citations mainly from journal articles (1082 times), thesis and dissertations (40 times), books and book chapters (22 times), conference proceedings, meetings, seminars (11 times) and publications by government (9 times). These citations were to articles published within the years 2004 to 2008 only.

**Table 4. 46 Top Journals Citing *MJM* Articles Published between Years 2004 to 2008**

| | Journal Title | No of times |
|---|---|---|
| 1 | Malaysian Family Physicians | 19 |
| 2 | *MJM* | 17 |
| 3 | Chinese Journal of Clinical Rehabilitation | 14 |
| 4 | Singapore Medical Journal | 13 |
| 5 | Journal of Bone and Joint Surgery AM | 11 |
| 6 | BMC Med | 10 |
| 7 | Journal of Biomedical Materials Research Part A | 10 |
| 8 | Asian Pacific Journal of Cancer | 8 |
| 9 | Malaysian Journal of Medical Sciences | 8 |
| 10 | Journal of Pediatric Surgery | 7 |
| 11 | International Journal of Pediatric Otorhinolaryngology | 6 |
| 12 | Nature proceedings | 6 |
| 13 | Spine | 6 |
| 14 | World Journal of Gastroenterology | 6 |
| 15 | Bone | 5 |
| 16 | Current Opinion in Otolaryngology and Head and Neck Surgery | 5 |
| 17 | International Journal of Ophthalmology | 5 |
| 18 | Journal of Orthopaedic Surgery | 5 |
| 19 | Malaysian J Pathol | 5 |
| 20 | Medicine | 5 |
| 21 | Otolaryngol Head Neck Surg | 5 |
| 22 | Pediatrics | 5 |
| 23 | Tissue Engineering | 5 |
| 24 | Acimed | 4 |
| 25 | Acta Otorhinolaryngol | 4 |
| 26 | China Orthopedic Surgery | 4 |
| 27 | Chinese Journal of Reparative and Reconstructive Surgery | 4 |
| 28 | Clinics | 4 |
| 29 | Knee Surgery, Sports Traumatology, Arthroscopy | 4 |
| 30 | Malaysian Orthopaedic Journal | 4 |
| 31 | Patient Education and Counseling | 4 |
| 32 | Southeast Asian J Trop Med Public Health | 4 |
| 33 | Virologica Sinica | 4 |
| 34 | Arch Ophthalmol | 3 |
| 35 | Artificial Cells, Blood Substitutes | 3 |
| 36 | ASEAN Journal of Psychiatry | 3 |
| 37 | Asia Pacific Journal of Public Health | 3 |
| 38 | Biomaterials | 3 |
| 39 | Biomedical Materials | 3 |
| 40 | China Tropical Medicine | 3 |
| 41 | Chinese Medical Journal | 3 |
| 42 | Clinical Pediatric Surgery | 3 |
| 43 | Clinical Surgery | 3 |
| 44 | Emerging Infectious Diseases | 3 |
| 45 | European cells and materials | 3 |
| 46 | Fertility and Sterility | 3 |
| 47 | Inter J of Social Psychiatry | 3 |
| 48 | International Journal of Biomedical Engineering | 3 |
| 49 | International Journal of Surgery | 3 |

| 50 | Journal of Ayub Med Coll | 3 |
| 51 | Journal of Laryngology and Otology | 3 |
| 52 | Journal of Materials  science | 3 |
| 53 | Journal of Neurosurgery | 3 |
| 54 | Journal of the University of Malaya Medical Centre | 3 |
| 55 | Jurnal Sains Kesihatan Malaysia | 3 |
| 56 | Laryngoscope | 3 |
| 57 | Marine biotechnology | 3 |
| 58 | Nepal Med Coll J | 3 |
| 59 | Nephrology | 3 |
| 60 | Ophthalmic Research | 3 |
| 61 | Organogenesis | 3 |
| 62 | Pak J Med Sci | 3 |
| 63 | Public health | 3 |
| 64 | Rev. CEFAC | 3 |
| 65 | Revista Colombia Médica | 3 |
| 66 | Saudi medical journal | 3 |
| 67 | Shanghai Jiaotong University | 3 |

The study also observed that *MJM* articles obtained citations from top international medical and health science journals (Table 4.45) *Journal of Bone and Joint Surgery AM* (11 citations), *BMC Med* (10), *Journal of Biomedical Materials Research Part A* (10). The journals with the highest frequency of *MJM* citations are: *Malaysian Family Physicians* (17), *Chinese Journal of Clinical Rehabilitation* (14) *and Singapore Medical Journal* (13).

The result implies that medical researchers in Malaysia are making contributions to the field and the research front would continue to improve and develop due to the stability and consistency observed in the publication productivity of *MJM* and furtherance of international research collaborations.

**Table 4. 47 Countries Affiliations of Authors Citing *MJM* Articles Published in 2004 -2008**

| | Region | Countries | Frequency of citations | Region Total |
|---|---|---|---|---|
| 1 | Africa | | | 23 |
| | | Nigeria | 9 | |
| | | Egypt | 8 | |
| | | South Africa | 3 | |
| | | Uganda | 1 | |
| | | Tunisia | 1 | |
| | | Kenya | 1 | |
| 2 | North America | | | 156 |
| | | United States | 123 | |
| | | Canada | 18 | |
| | | Cuba | 10 | |
| | | Mexico | 5 | |
| 3 | South America | | | 50 |
| | | Brazil | 31 | |
| | | Columbia | 9 | |

| | | | |
|---|---|---|---|
| | | Venezuela | 3 | |
| | | Argentina | 3 | |
| | | Uruguay | 2 | |
| | | Chile | 1 | |
| | | Peru | 1 | |
| **4** | **Europe** | | | **212** |
| | | United Kingdom | 40 | |
| | | Germany | 35 | |
| | | Spain | 27 | |
| | | France | 21 | |
| | | Italy | 17 | |
| | | Greece | 11 | |
| | | Netherlands | 10 | |
| | | Switzerland | 10 | |
| | | Russia | 9 | |
| | | Portugal | 5 | |
| | | Czech republic | 3 | |
| | | Ireland | 3 | |
| | | Poland | 3 | |
| | | Slovakia | 3 | |
| | | Denmark | 2 | |
| | | Serbia | 2 | |
| | | Sweden | 2 | |
| | | Bolivia | 1 | |
| | | Cyprus | 1 | |
| | | Finland | 1 | |
| | | Lithuania | 1 | |
| | | Norway | 1 | |
| | | Romania | 1 | |
| | | Ukraine | 1 | |
| | | Armenia | 1 | |
| | | Austria | 1 | |
| **5** | **South East Asia** | | | **187** |
| | | Malaysia | 171 | |
| | | Singapore | 12 | |
| | | Thailand | 2 | |
| | | Vietnam | 1 | |
| | | Brunei | 1 | |
| **6** | **Australasia** | Australia | 35 | **35** |
| **7** | **South Asia** | | | **61** |
| | | India | 43 | |
| | | Pakistan | 10 | |
| | | Nepal | 5 | |
| | | Bangladesh | 3 | |
| **8** | **East Asia** | | | **258** |
| | | China / Hong Kong | 227 | |
| | | Korea | 14 | |
| | | Japan | 11 | |
| | | Taiwan | 6 | |
| **9** | **Middle East** | | | **66** |
| | | Turkey | 20 | |
| | | Israel | 12 | |
| | | Iran | 12 | |
| | | Saudi | 8 | |
| | | Jordan | 4 | |
| | | Iraq | 2 | |
| | | Lebanon | 2 | |
| | | UAE | 2 | |
| | | Kuwait | 1 | |
| | | Oman | 1 | |
| | | Qatar | 1 | |
| | | Yemen | 1 | |
| | **Total** | | | **1048** |

Authors from seventy-six different countries have cited *MJM* published articles (2004 – 2008) at one time or more. Most of the authors are from China (227), followed by Malaysia (171), United States (123), India (43), United Kingdom (40), Germany (35), Australia (35), Brazil (31), Spain (27), and Turkey (20). In general, *MJM articles* received more citations from East Asia (258 citations), Europe (212), Southeast Asia (187), North America (156), Middle East (66) South Asia (61), South America (50), Australia (35) and Africa (23).

Collaborations that involves 2 or more authors from different countries producing articles citing *MJM* is shown in (Table 4.47).

**Table 4. 48 Countries Collaborations of Papers Citing *MJM* Articles**

| | Collaborating countries | No of citations |
|---|---|---|
| 1 | Japan and India and Brunei and Indonesia and Philippines and Malaysia | 5 |
| 2 | Luxembourg and Germany | 3 |
| 3 | UK and Mal | 3 |
| 4 | UK and US | 3 |
| 5 | US and Germany | 3 |
| 6 | France and US | 2 |
| 7 | Mal and Austria | 2 |
| 8 | Malaysia and Japan | 2 |
| 9 | UK and Netherlands | 2 |
| 10 | US and China | 2 |
| 11 | US and India | 2 |
| 12 | US and Israel | 2 |
| 13 | Argentina and Canada | 1 |
| 14 | Australia and Hong Kong | 1 |
| 15 | Belgium and Korea and china and Thailand and Malaysia and Taiwan and Switzerland | 1 |
| 16 | Bosnia and Herzegovina | 1 |
| 17 | Brazil and Portugal | 1 |
| 18 | Brazil and Portugal and turkey | 1 |
| 19 | Canada and Japan | 1 |
| 20 | Chile and US | 1 |
| 21 | China and Germany | 1 |
| 22 | Denmark and Netherlands and UK | 1 |
| 23 | Egypt and Belgium and Italy and Greece | 1 |
| 24 | Egypt and Saudi | 1 |
| 25 | France and Austria | 1 |
| 26 | France and UK | 1 |
| 27 | Germany and Austria | 1 |
| 28 | Germany and Vietnam and Gabon | 1 |
| 29 | Greece and Egypt | 1 |
| 30 | Greece and US | 1 |
| 31 | India and Nepal | 1 |
| 32 | India and Pakistan | 1 |
| 33 | India and UK | 1 |

| 34 | Iran and Canada | 1 |
| 35 | Italy and Germany | 1 |
| 36 | Italy and UK and Hungary and Belgium | 1 |
| 37 | Mal and Singapore | 1 |
| 38 | Mal and Singapore and Japan and Indonesia and Philippines and India | 1 |
| 39 | Malaysia and Korea | 1 |
| 40 | Netherlands and Switzerland | 1 |
| 41 | Oman and Malaysia | 1 |
| 42 | Peru and US | 1 |
| 43 | Poland and Austria | 1 |
| 44 | Kuwait and Pakistan | 1 |
| 45 | Spain and Costa Rica | 1 |
| 46 | Srilanka and Singapore | 1 |
| 47 | Sweden and Iran and Finland | 1 |
| 48 | Sweden and Norway | 1 |
| 49 | Switzerland and Singapore and US and Laos | 1 |
| 50 | TunISIa and Albania | 1 |
| 51 | UAE and UK | 1 |
| 52 | UK and Bangladesh | 1 |
| 53 | UK and Canada | 1 |
| 54 | UK and INDIA and US | 1 |
| 55 | UK and Iran | 1 |
| 56 | UK and Japan | 1 |
| 57 | UK and Qatar | 1 |
| 58 | UK and Scotland | 1 |
| 59 | UK and Sweden | 1 |
| 60 | US and Australia | 1 |
| 61 | US and Bangladesh | 1 |
| 62 | US and Belgium | 1 |
| 63 | US and Columbia | 1 |
| 64 | US and Iran | 1 |
| 65 | US and Italy | 1 |
| 66 | US and Qatar | 1 |
| 67 | US and Senegal | 1 |
| 68 | US and Spain | 1 |
| 69 | US and Thailand and china and France and Australia and Switzerland | 1 |
| 70 | Wales and Hong Kong | 1 |
| 71 | Algeria and Canada | 1 |
| | **Total** | **90** |

Collaborations that involves United States citing *MJM* articles occurred 20 times, and United Kingdom (16). This result is supported by (Chiu and Ho, 2005) who did a study on homeopathy research and noted that among all countries, the US, the UK, and Germany contributed the most not only in terms of publications but also in citations. Lee et al (2010) examined the effect of author and article characteristics on citation in economic research, and reported that articles written by authors in the USA or UK are roughly twice more cited than articles written by authors outside the USA or the UK. The reason as explained is that authors in the USA and (or) UK are exposed to more competitive environment, where promotion, salary and research grants are typically allocated based on research performance, making it easier to collaborate with colleagues abroad. Chuang et al (2007) examined stroke-related research in Taiwan and discovered that there was an over-reliance on the US as a collaborator, which has proven to be successful, but worried that this may limit future development of research in Taiwan. By these results, it is believed that research articles published in the *Medical Journal of Malaysia (MJM)* meet international standard and that researchers in the country have been able to convince specialist colleagues (or peers) around the world of the significance of their work and to cite.

## 4.20. Journal Impact factor

Journal citation measures are one of the most widely used bibliometric tools. They are used in information retrieval, scientific information, library science and research evaluation, and they are applied at all levels of aggregation (Glanzel, 2003). Publication in a journal with high reputation or high JIF provides greater respect among the peers (Sharma, 2007). The Journal Impact factors are calculated each year by Thomson Scientific for those journals which it index. It is an indicator made available by the Institute of Scientific Information (*ISI*) in its *Journal Citation Report* (*JCR*) which is calculated from *Web of Science (WoS)* databases. Journal impact factor was initially invented in the early 1960s by Gene Garfield, founder of *ISI*. Thus, A journal's JIF for year *n* is defined as the ratio between the number of citations during year *n* of the journal's articles published during years n–1 and n–2 (and only then), and the total number of articles published during these two years (Garfield, 1979 quoted in Merlet et al, ( 2007). Sharma, (2007) quoting Garfield (1999) explained that: A journal's impact factor is based on two elements: the numerator, which is the number of citations in the current year to any items published in a journal in the previous two years, and the denominator, which is the number of substantive articles (source items) published in the same two years. The equation is elaborated in chapter 3 (pg. 55). Hence, the results of the calculations for each year are as follows:

For year 2009

89 = Citations in 2009 to articles published in 2007 and 2008
235 = Number of articles published in 2007 and 2008
*IF = 89/235 = 0.378*

For year 2008

75 = Citations in 2008 to articles published in 2006 and 2007
204 = Number of articles published in 2006 and 2007
*IF = 75/204 = 0.367*

For year 2007:

127 = Citations in 2007 to articles published in 2005 and 2006
206 = Number of articles published in 2005 and 2006
*IF = 127/206 = 0.616*

For year 2006:

110 = Citations in 2006 to articles published between 2004 and 2005
241 = Number of articles published between 2004 and 2005
*IF = 110/241 = 0.456*

A five-year impact factor may be useful to some users and can be calculated by combining the statistical data available from consecutive years (Utap, 2008). Therefore, the impact factor for the five-year period (2004 - 2008) is as follows:

Five year IF:

335 = Citations in 2009 to articles published between 2004 and 2008
580 = Number of articles published between 2004 and 2005
*IF = 335 /580 = 0. 577*

As shown above, the highest *IF* was recorded in year 2007 (0.616) followed by 2006 (0.456), 2009 (0.378) and 2008 (0.367) while for the five years *IF* we have (0.577).

Furthermore, from the Journal Citations Statistics module calculated in *MyAis (Malaysian abstracting and indexing system http://Myais.fsktm.um.edu.my/)*, from eight hundred (800) *MJM* articles indexed, dividing the total cites by the total articles gives (0.64625). Finally, the result above has shown that *Medical Journal of Malaysia (MJM)* has a relatively good IF.

## 4.21. Summary

This bibliometric study has shed light on the publication practice of the *Medical Journal of Malaysia* from year (2004 - 2008). In this chapter, the main research questions have been answered, by presenting the results of the data analysis in accordance with the objectives of this study. The study analyzed a five-year period in the publication process of *Medical Journal of Malaysia* drawing data from the *MyAis (Malaysian abstracting and indexing system http://Myais.fsktm.um.edu.my/)*. Data were analyzed descriptively and results have it that, the number of articles during the five years (2004 - 2008) period of publication is five hundred and eighty (580). The average number produced per year is 116 articles, which indicate a consistent balance in the journal's publication productivity. Productivity pattern of authors conforms slightly to *Lokta's law* with little marginal $c$ value as the frequency of authors publication observed was very close at c = 2.4 when compared with the frequency of authors publication expected.

The most productive authors are senior medical researchers affiliated to Higher Institutions. In terms of collaborations, only 56 articles (9.65%) were single authored. Joint authorship by 3 authors have the highest frequency of articles (23.28%), while the highest number of collaboration was 15 authors with frequency of 1 (0.17%). This suggests that authors in the medical field have a custom of collaborating with their peers. On the average, more than 90% of all the articles were collaboratively produced. The study also observe that 516 (88.96%) articles were contributed by authors affiliated to Malaysia alone compared to 36 (6.20%) contributed by foreign affiliated authors alone. In total, only 64 articles were either authored or co-authored by foreign contributors during the five year period under study. This is relatively small a number. Furthermore, according to results from collaboration pattern, Malaysia has a stronger connection in medical and health science research with Australia, India and the UK.

There are more authors from higher institutions than hospitals. Universiti Kebangsaan Malaysia tops the list of active producers with (157) authors, followed by Universiti Sains Malaysia (133), Hospital Universiti Kebangsaan Malaysia (102) University of Malaya (101) and Universiti Putra Malaysia (96). Among the top collaborators are "Universiti Putra Malaysia with Hospital Kuala Lumpur" (8 times) "Universiti Kebangsaan Malaysia Medical Centre with Ampang Puteri Specialist Hospital Selangor" (8 times). For the keywords that appeared in Journal articles, the word "Diabetes" has the highest frequency of (28) followed by Cancer (13), Endoscopic (13), hypertension (13), Tuberculosis (12) etc. The highest number of keywords found in an article is 10 keywords (1 article), followed by 9 keywords (1 article). The minimum number of words in article title is 2 words (4 articles), while the maximum is 26 words (1 articles), the mode is 9 words (57 articles), while the mean for the number of words in article titles is 10.79. Funding, during the 5 year period is mentioned in only 61 papers (Original research articles). Most of the 61 papers reported receiving financial support from the Ministry of Science, Technology and Innovations as well as Universities own research funds.

Moving forward by examining the references, the average number of items referenced per year is 1391 citations, while the average number of citation per article is 12. The range of average citation per year is from 9 to 12 citations. Most of the cited references came from articles published in several medical related journals (87.67%) of citations, followed by Books (4.79%), which attest to the fact that scholarly journals are the main source of scientific knowledge sharing. Most cited references were materials of about 1 to 11 years old (63.87%), however, older publications of over 20 years, and few over 100 years are still referenced, suggesting the longevity of the usefulness of works published in the medical field. The core journal, which are in the nucleus zone revealed 22 titles. This result is in line with Bradford's Law of Scattering, due to the fact that journal titles showed a wide dispersion among a small core, with only about 2.82% of the journals accounting for one-thirds of all the citations. Through *Harzing Publish and Perish,* which harvested data from *Google scholar,* it was observed that 76.8% of all the articles published have been cited one time or the other. Most of the documents citing *MJM* articles are journal articles (1082 times), followed by thesis and dissertations (40 times), books and book chapters (22 times), conference papers (11 times) and publications by government (9 times). The top journals citing *MJM* articles are: *Malaysian Family Physicians* (17), *Chinese Journal of Clinical Rehabilitation* (14)*, Singapore Medical Journal* (13)*, Journal of Bone and Joint Surgery AM* (11 citations)*, BMC Med* (10)*,* and *Journal of Biomedical Materials Research Part A* (10)*.*

Talking about citations received, *MJM* articles were cited at least once by authors across seventy-six different countries of the world. On top of the list are authors from China (227), followed by Malaysia (171), United States (123), India (43), United Kingdom (40), Germany (35), Australia (35), Brazil (31), Spain (27), and Turkey (20). In addition, *MJM* recorded its highest *IF* in year 2007 (0.616) followed by 2006 (0.456), 2009 (0.378) and 2008 (0.367) while for the five years we have the *IF* to be (0.577). Hence, the results of this study indicate that research activities in Malaysian field of medicine is attracting recognitions from not only the Asian continent and neighboring countries, but the world at large. The study hereby submits that sustained efforts would be required to further achieve a reputable status in the field, and this would be attained through increase in research funding and partnership.

# CHAPTER 5

# CONCLUSION

## 5.1. Introduction

The main objective of the study is to examine the publication productivity pattern of articles published in the *Medical Journal of Malaysia (MJM)* between a 5-year period (2004 - 2008). This was achieved by applying bibliometric methodologies to study the publication productivity, authorship productivity, co-authorship pattern, degree of collaboration, types of collaborations, distribution of contributing countries, types of affiliations, distribution of keywords in articles, numbers of words in titles, the agencies and other organizations funding medical research, distribution of citations according to bibliographic format, age of citations, half-life of citation, the core journals by bradford's law and the total number of citations received by *Medical Journal of Malaysia (MJM)*.

Hence, as regards to the goal of the study, this section discusses the results of the study, and subsequently conclude with recommendations for future bibliometric studies.

## 5.2.  Findings and Discussions

### 5.2.1. Article Productivity of *Medical Journal of Malaysia* from Year 2004 – 2008

The number one aim of the study is to find the publication productivity of *MJM* between the periods of five years (2004 - 2008). As regards to this, the study observed that five hundred and eighty (580) articles were produced, with an average of 116 articles per year. Hashimah (1997) conducted "a citation analysis of *MJM* from 1991 - 1995" and observed that a total of four hundred and fifteen (415) articles were produced during the five year period, with an average of 83 articles per year. Comparing Hashimah's result (average of 83 articles per year) with the findings of this present study (average of 116 articles per year), it shows that *MJM* have recorded 39.7% increase in article productivity.

This is an improvement from previous years and indicates a consistent balance in publication productivity.

Author's productivity pattern in *MJM* from year 2004 - 2008 conforms slightly to *Lokta's law* with little marginal c value. The frequency of authors publications observed and the frequency of authors expected when c = 2.4 is very close. It is conceived that the journal would be able to maintain a strong position in term of publication productivity for many years.

### 5.2.2. Authors Productivity Pattern and the Productive Authors

The finding shows that two thousand one hundred and seventy seven (2177) names were recorded. Among these, 568 authors were the highest number in year 2008 and the least number of authors recorded was 352 authors in year 2007. However, since an author can publish more than once during a period, we crosscheck the names to remove duplicates and mistakes that may arise as a result of punctuation errors. As regards to that, one thousand four hundred and thirty five (1435) unique names were returned. Among these, Ruszymah B.H.I, who is affiliated to Universiti Kebangsaan Malaysia is the most productive of all, with 19 papers, followed by Aminuddin B.S (15), Gendeh B.S (15), Chua, K.H (14), Chua, K.B (12), Philip, R (12), Prepageran, N (12), Halim A.S (11), Kwan, M.K. (10). Only one (Philip, R) among these top authors is associated to an affiliation other than Higher Institution. This suggests that the most prolific researchers are authors affiliated to higher institution, this may be because the allocation of research funds to medical researchers in institutions of higher learning are easily mobilized through the four research universities in Malaysia.

However, Hashimah (1997) employed a different approach and listed the four most outstanding authors to be "Ministry of Health Malaysia, World Health Organization, MJM-Editorial and Department of Statistics Malaysia". In Hashimah's study, articles and contributions authored by these top non-academic institutions and MJM-Editorial members, that are published in *MJM* at that period (1991 - 1995), were not assigned to individual researchers (either single or joint authored), but rather considered authored by the parent-bodies (Ministry of Health Malaysia, World Health Organization, MJM-Editorial and Department of Statistics Malaysia).

In general, the result of Hashimah's study attested to the findings of this present study, that in Malaysia, the Ministry of Health, WHO, Department of Statistics, and Hospitals are very active contributors to medical and health science research in Malaysia.

### 5.2.3. Co-Authorship Pattern

It is part of the objective of the study to find out about the co-authorship practice of authors. It was revealed that co-authorship is the rule rather than the exception. 580 articles were produced during the period under study and only 56 articles (9.65%) were single authored. Joint authorship by 3 authors have the highest frequency of 135 articles (23.28%), the highest number of collaboration was 15 authors with frequency of 1 (0.17%). This suggests that authors in the medical field have a custom of collaborating with their peers (Macías-Chapula and Mijangos-Nolasco, 2002; Nwagwu, 2007), the reason as explained by (Gauthier, 1998), that in biomedical science, problems are commonly interdisciplinary in nature and thus it is especially crucial to foster collaborative behavior. Wiles et al, (2010) supporting this idea and noted that as companies have amalgamated into transnational conglomerates in order to compete, it is likely that there is greater collaboration within and across research institutions. On the average, more than 90% of all the articles were collaboratively produced. The finding is also similar to that of (Macías-Chapula and Mijangos-Nolasco, 2002; Al-Qallaf, 2003; and Hazmir, 2008) who observed that there is a trend of multiple authorship in medical and health science journals.

### 5.2.4. Degree of Collaboration

Finding the rate of collaboration of authors is an important objective of this study. This was achieved by analyzing the number of single and multi-authored articles per year, and checking with author's country and institutional affiliations. In all, Malaysian authors collaborated with their foreign counterpart 28 times, while foreign authors collaborated with each other 5 times. In total, 65 articles were either authored or co-authored by foreign contributors. Although contributions from foreign countries were small, however, results obtained from year 2008 foreign contributions suggest that the contribution would continue to grow. Furthermore, the study adopted the formula of (Subramanyam, 1983 cited in Utap 2008) to calculate the proportion of co-authored publications in *MJM*. The result shows that the degree of collaboration in *MJM* between years 2004 – 2008 is 0.9. It ranges from 0.88 to 0.9 indicating that collaboration between authors in the field of medical sciences is a very common practice.

### (a)     Collaborations between Different Countries

Scientific collaboration, above all international co-operation has unquestionably a positive effect on visibility and citation impact (Glänzel, 2008). During the period under study, there are contributions from 16 different countries, making up 8 different region of the world, Southeast Asia (Singapore and Indonesia) leads with 20.11% (36 authors) of the foreign total, followed by South Asia (India and Pakistan) with 17.88% (32 authors), Europe (Netherlands, United Kingdom, France, and Ireland) with 16.76% (30 authors), Middle-Eastern countries (Iran, Saudi-Arabia, and Yemen,) with 10.06% (18 authors).

Authors from Turkey alone produced 13.97% (25 authors), Japan 8.38% (15 authors), and Australia 7.82% (14 authors), while North America (USA and Canada) produced 5.03% (9 authors) of the total foreign contributors. Among these countries, collaboration between Australia and Malaysia is the highest with 7 articles, next is India and Malaysia (5 times), UK and Malaysia (4 times), Singapore and Malaysia (3 times), USA and Malaysia (2 times), Japan and Malaysia (2 times).   From this, it is assumed that, Malaysia has a strong connection in medical and health science research with Australia, India, UK, Singapore, USA and Japan, more than any other country. This is similar to the result obtained by Hazmir (2008), who observed that Malaysian active collaborating countries in "biomedicine and health sciences research as reflected through the ISI database from 1990 – 2005" are: United Kingdom, followed closely by the USA, Japan, Australia, Singapore, and China. This result partially supported that of Gupta et al (2002), who investigated Indian collaboration with south-East-Asian countries in Science and Technology and discovered the strongest collaboration was with Malaysia especially in chemistry, while the second strongest collaboration was with Singapore.

In addition, countries from Asia have the highest number of contribution. It is thus, suggested that more needs to be done to attract foreign contributions from different regions around the world. Bibliometric studies (Omotayo, 2004; Glänzel, 2008; Davarpana and Behrouzfar, 2008; and Zainab, 2009), have attested to the fact that collaborations is a key to garner more research funds, visibility and strengthening the position of a country in the scientific community. Uzun (2004) examines the patterns of foreign authorship of articles, and international composition of journal editorial boards in five leading journals in the field of information science, and scientometrics. It covers an American journal and four European journals. The findings revealed that the number of foreign countries contributing in all journals have increased rapidly since 1996 which could be explained by the percentage of foreign members on the editorial boards of the journals. This also suggests that a formidable and balanced editorial board also have a positive effect in the publication process of a quality highly rated scientific journal. This is part of the criteria for a journal to be deem international in nature, if the journal exhibit high contributions from international authors, as well as articles published in a journal are referenced in published articles (Zainab, 2008). Thus, more international collaboration is expected to lead to more output due to the sharing of ideas and workloads (Chuang, et al 2007).

### (b) Collaboration between Foreign Countries

It is also an important aspect of the study to know if authors affiliated to foreign countries make contributions to *MJM* and how much, if they do. The study revealed that 36 articles were contributed solely by foreign authors. Singapore and India produced 8 articles each, Turkey alone with 5 articles, Iran (4), Indonesia (2), Saudi Arabia (2), United States (2), while, Yemen, Australia, Ireland, United Kingdom and Japan produced 1 each. The results also show that, Malaysian authors collaborated with their foreign counterparts to produce 28 articles. Malaysian authors collaborated with Australians to produce 7 articles, with India to produce 5 articles, with United Kingdom to produce 4 articles, with Singapore to produce 3 articles, with the United States and Japan producing 2 articles respectively, and with Pakistan, Canada, and Saudi Arabia to produce one each. Thus, during the five year period under study, we found no collaborations with African countries and also weak collaboration between Europe and America, but observe that there are numbers of *MJM* articles cited couple of times by authors from these regions.

The study hereby suggests that, perhaps opportunities for further or renew collaboration with these regions should be the focus in future research endeavor and strong cooperation should be encouraged. Furthermore, collaboration from different affiliations of the same country was also observed. This type of collaboration from foreign countries produced 10 articles during the period under study. From Turkey there are 4 articles, Singapore 3 articles as well as 1 article respectively from India, Indonesia, and the United Kingdom.

### 5.2.5. Types of Affiliations

Authors publishing in *MJM* are from different countries around the world and are associated to different affiliations which were grouped into 8. One hundred and seventy three (173) unique affiliation types were observed and 35.26 % of this were hospitals, followed by higher institutions (34.10%), ministries and government agencies (14.45%), medical centers (6.94%), clinics (4.62%), private organizations (4.05%), and international organizations make up 0.58% of the total. Most of the authors are affiliated to institution of higher learning (60.91%) followed by hospitals (27.74%). Although hospitals are ranked the highest number in types of unique affiliations, there are more authors from institutions of higher learning than hospitals. This is because authors affiliated to higher institutions are concentrated to few top institutions whereas those affiliated to hospitals spread across many (university, college, state and private hospitals). This results corroborates Hazmir's (2008) study, who observed that Malaysian research publication in the field of biomedical and health Sciences (BHM) was highly active among a few dominant institutions especially among the public institutions of higher learning (IHL) such as UKM (Universiti Kerbangsan Malaysia), USM (Universiti Sains Malaysia), UM (University of Malaya), and UPM (Universiti Putra Malaysia). He further explained that the four public universities acquired the status of research intensive university under the 9th Malaysia Plan (9MP), and that these institutions are dominant and seen to be playing a greater decisive role in the field of BHM research in Malaysia. It is assumed by Hazmir (2008) that these four top public higher Institutions would continue to dominate the research front in the future.

There are 3 authors affiliated to the World Health Organization (WHO) and there are no foreign authors affiliated to University Medical Centers. When collaborations were examined between different affiliations, findings of this study have revealed that medical practitioners in hospitals and medical centers are active authors in Malaysia.

This is also in par with the result obtained by Hashimah (1997), who observed that government institutions like the Ministry of Health Malaysia and Department of Statistics Malaysia top the list of affiliations recorded. Hashimah also noted that, Kuala Lumpur General Hospital recorded the highest frequency of occurrence among hospitals, while the World Health Organization (WHO) stood out among contributing international organizations.

### 5.2.6. Distribution of Keywords in Articles

The main objective of examining the keywords in article is to determine the focus area of research during the period under examination. 10 keywords were counted in an article, followed by 9 keywords in another. Most of the articles (188 articles) have 3 keywords each; another 131 articles have 4 keywords each and the least number of keywords is 1 which was identified in 17 articles. The active research areas were Diabetes counted in 28 articles, followed by Cancer, Endoscopic, and Hypertension in 13 articles each, and Tuberculosis in 12 articles.

### 5.2.7. The Agencies and Organizations Funding Medical Research

During the 5-year period of publication examined in this study, it was observed that only 61 papers (Original research articles) received financial support. The study only focused on the number of funded papers against the total number of "Original research articles" published yearly, since other papers that belongs to the category of "continuing medical education (CME), case reports, short communications correspondence and brief abstracts "are not original research and may not have attracted or in need of sponsorship or financial aid". It also observed that most of the financial aids were awarded by the top public higher institutions in the country with the Malaysian government playing a key role.

There are 2 papers sponsored by the Iranian government, 1 by Japan and 2 by private organizations. The results suggest that some other types of national and international research funds should be solicited by medical researchers and practitioners.

### 5.2.8. Distribution of Citations

Another important aspect of the study is to examine the type of reference sources used by researchers in their articles. The study revealed that authors make use of a variety of resources and wide range of data, reports, records, and statistics published by government, associations and organizations. However, most of the cited references came from scholarly journals, which also includes editorials and reviews. Scholarly journals recorded 6100 (87.67%) of citations, followed by books 333 (4.79%). This result may suggest that changes have occurred over time in medical schoarlarly communication, when compared with the result of Hashimah (1997). Hashimah has observed that journal articles and books make up to 82.8% and 9.7% respectively of the sources used by authors in *MJM*. This reveals a 5% decrease in the usage of books, and 4.87% increase in the usage of journal articles, which may be due to the recent popularity and accessibility of online databases that provides easy and faster access to journal articles.  This finding is also in line with Crawley and Low (2006), who reported that researchers publishing in the *American Journal of Veterinary Research* mostly cited journals (88.8%), followed by books. The same result was obtained by Hashimah (1997) Al-Qallaf (2003) Chiu and Ho (2007) and Hazmir (2008)

In addition, wide range of data, reports, records statistics, and standards, were cited, and these category of sources fall under government publications (mostly by the Ministry of Health, Malaysia) 193 (2.77%), and international organizations (WHO, UNICEF etc) 55 (0.79). Moreover, the average number of citations per year is 1391 citations, while the average number of citation per article is approximately 12. This is very close with the result obtained by Hashimah (1997) who observed that the average citations per article in *MJM* from 1991 – 1995 is 11.4.

In this study, the range of average citation per article is from 9 to 14 citations, most citations are within the range of 0 - 10 (325 articles) (56.0%), followed by 11 – 20 (156 articles) (26.90%). The largest range is between 80 – 90 citations (1 article) (0.17%) and 71 – 80 citations (2 articles) (0.34%).

### 5.2.9. Age of Citations

The important question to be answered here is how long will a research article continue to be used after it has been published? Findings resulting from the analysis of the "Age of Citation" of reference sources enable libraries to make decisions on their collections "usage and time". It also helps and guides library policy makers on acquisition and weeding process as regards to obsolescence of a publication. Findings show that medical researchers still use sources that are several years old; sources dating back to 40 – 50 years. Although 63.87% of the sources are about 11 years old, this is in contrast with a field like computer science as observed by Utap (2008), where currency of information is very important. Utap noted that articles more than 6 years old might be less relevant in the field of computer science. This is also in concordance with Glanzel and Schoepflin (1995), who suggested that the idea of ageing and obsolescence seems to be specific to the field rather than to the individual journal, and that obsolescence of the social science journals in the study is slower than for the medical and chemistry journals. Also, Omotayo (2004)'s study on content analysis of *Ife psychologia* shows that most of the papers cited were published within the last 11–20 years (31.5%), and only 6.8% of the citations were between 0 and 2 years. Tsay (1998) examined library journal use and citation age in medical science, using the Library of Veterans General Hospital, Taipei, Taiwan. Tsay found that about 80% of uses are attributed to journals less than ten years old, while these journals contribute about 70% of total citations. While analyzing the citation patterns of researchers publishing in *The American Journal of Veterinary Research (AJVR)*, Crawley-Low (2006) revealed that more than one-half (65%) of the cited journal titles were published within the previous ten years. Gomez (2003) presented the result of a bibliometric study to manage a journal collection in an astronomical library and stated that most of the journals cited are less than 10 years old. The results infer that in the medical field currency of items is not an important criteria for an item to be cited as older authoritative articles and books continue to be cited even if they were more than 10 years old.

### 5.2.10. Core Journals Referenced by Articles

This present study applied *Bradford's Law of Scattering* (Bradford 1948) to the resulting list of cited journals in *MJM* between years 2004 to 2008. As a result, three zones were created, each producing approximately one third of the cited references. The core journal titles which are in the nucleus zone are journals that have 22 or more citations. 43 journals (1990 citations) are in Zone 1, 210 (1996 citations) are in Zone 2, while 1270 journals (1941 citations) are in Zone 3. This result is in line with Bradford's Law of Scattering, because journal titles showed a wide dispersion among a small core. Only about 2.82% of the journals accounted for one-thirds of all the citations. The proportion of number of journals in the three zones is 43: 210: 1270 is approximately in the ratio 1: 5: 25. Therefore, it follows that, our $b \approx 5$ *in* the general proportion 1: b: b².

Hence, this result suggests that a core collection of medical and health science related journals would include 43 journals from zones 1. Medical Libraries may also include medical journals from Zone 2 and 3.

The top 5 most cited journals are: *Journal of Bone and Joint Surgery* (144 citations), *New England Journal of Medicine* (142) *Lancet* (139), *Clinical Orthopedics and Related Research* (112), and *British Medical Journal* (107). It can be observed that the top five most cited journals are from USA and UK who are the world leaders in Health Science research. In addition, these journals are one of the oldest, widely read, cited medical journal in the world with high impact factors. In the citation study of Hashimah (1997) on *MJM*, the top 5 most cited journals revealed are: *Lancet, MJM, New England Journal of Medicine, BMJ,* and *JAMA*. Bloom, et al (2007) acknowledges that the five leading general medical journals are: *The Lancet*, *British Medical Journal (BMJ)*, *Journal of American Medical Association (JAMA)*, *New England Journal of Medicine (NEJM)*, and *Annals of Internal Medicine (AIM)*. All these 5 journals acknowledged by (Bloom et al, 2007) are top 10 in our list of most cited journals except for *Annals of Internal Medicine (AIM)* which is in top 23 most cited in this present study and top 16 in Hashimah (1997)'s study. Levi, (2009) also noted that the two leading clinical journals are *New England Journal of Medicine (NEJM)* and *The Lancet*. The study by Chiu and Ho (2005) on bibliometric analysis of all homeopathy-related publications in *Science Citation Index (SCI)* revealed that, the most frequently cited article was published in the *Journal of American Medical Association (JAMA)* which is the second highest impact factor journal in the category of General and Internal Medicine. In the work of (Macías-Chapula and Mijangos-Nolasco, 2002), it was reported that *"Lancet"* was among the five top most cited journal by AIDS researchers in Sub-Saharan Africa.

The result shows that Malaysian medical researcher's sources of choice have not really changed for the past 18 years or so, when result of this study is compared with Hashimah's, and that these sources of choice are similar to other world medical researchers. *Journal of Bone and Joint Surgery* is a peer reviewed medical journal in the field of orthopedic surgery, first published in 1887. *The Lancet* is a weekly peer-reviewed general medical journal founded in 1823. *Clinical Orthopaedics and Related Research (CORR)* is an international peer-reviewed medical journal, established in 1953 by the Association of Bone and Joint Surgeons. *British Medical Journal* is a partially open access medical journal, which was founded in 1840. While the *New England Journal of Medicine* is an English-language peer-reviewed medical journal published by the Massachusetts Medical Society, founded in 1812. These peer review process remain the most important means of communicating science today. Researchers depend on them to learn about the latest advances in their fields and to report their own findings. In principle, this system enables science to move forward on the collective confidence of previously published works (Neylon and Wu 2009).

Furthermore, it has been revealed that authors very rarely cites regional journals, and the most cited and most popular regional journals are *Singapore Medical Journal* (38 citations) and *Southeast Asian Journal of Tropical Medicine Public Health* (28 citations). This result partially confirmed that of Hashimah (1997), who listed *Singapore Medical Journal* as the top most cited regional journal in *MJM* articles published from 1991 – 1995. *Singapore Medical Journal* is published by the Singapore Medical Association, while *Asian Journal of Tropical Medicine Public Health* is published by SEAMEO Regional Tropical Medicine and Public Health Network (SEAMEO TROPMED Network) currently hosted at Mahidol University.

### 5.2.11.  Journal Self-Citation

It has been noted that self-citations can make up a significant portion of the citations a journal gives and receives each year.  The study also find out how often *MJM* authors refers to articles from *MJM*. As regards, *MJM* received 173 citations (2.49%) of the total citations, during this period. The highest number of Journal self-citation recorded was in year 2004, where 25 articles out of total 139 produced cited *MJM* 58 times. It should also be noted that year 2004 recorded the highest number of article publication (139 articles), and the highest number of citation (1898). Overall, the finding shows that the practice of Journal Self-Citation is minimal in *MJM* with just 2.49% of the total citations.

### 5.2.12. Citations Received by *Medical Journal of Malaysia* (*MJM*)

Citation counts though are not actually an indicator of the quality of a paper (Chiu and Ho, 2005), is an excellent measure of influence, impact and visibility (Neylon and Wu 2009). Citing is an established way for authors to declare their sources of information and politely recognize someone's intellectual property. The importance of an article, author or journal can be estimated through the number of citations each acquires (Kovaˇcic´and Miˇsak, 2004). *Medical Journal of Malaysia* is the first medical journal in the country and it's the medium frequently employed by medical researchers in the country to promote the dissemination of their ideas. The journal has also kept pace in attracting contributions by and large from foreign authors even though it is still very small in number.

In essence, it is essential to find the contribution and relevancy of *MJM* to the wider community. If articles published were been used and cited by other journals in the field, this citation would demonstrate the importance of the journal. The total number of citations received by *Medical Journal of Malaysia (MJM)* from 2004 - 2008 was obtained in *Google scholar* through "*Harzing publish and perish*". *Google Scholar* provides a more comprehensive coverage of citations than *ISI*, including citations in books, conference and working papers and non-*ISI* indexed journals. Harzing and Van der Wal (2008) further explained that the use of *Google Scholar (GS)* generally results in more comprehensive citation coverage —which particularly benefits academics publishing in sources that are not (well) covered or not covered at all in *ISI*. *Google Scholar* is a good source of information about citations and does it well if the articles in the journals are accessible electronically and the citing articles are available on the web in its full text format (Zainab, 2006).

Thus, the results of the findings from *Google Scholar (GS)* showed that 76.8% of all the articles published between the years 2004 to 2008 have been cited one time or the other, that most of the documents citing *MJM* articles are journal articles (1082 times), followed by thesis and dissertations (40 times), books and book chapters (22 times), Conference proceedings, meetings etc (11 times) and publications by government (9 times).

The top journals citing *MJM* articles are: *Malaysian Family Physicians* (17), *Chinese Journal of Clinical Rehabilitation* (14)*, Singapore Medical Journal* (13)*, Journal of Bone and Joint Surgery AM* (11 citations)*, BMC Med* (10)*,* and *Journal of Biomedical Materials Research Part A* (10)*. MJM articles* received more citations from East – Asia (258 citations,) Europe (212), Southeast Asia (187), North America (156), Middle – East (66), South Asia (61), South America (50), Australia (35) and Africa (23). Findings further showed that *MJM* articles were cited at least once by authors across seventy six (76) different countries of the world. Topping the list of citers are authors from China (227 times), followed by Malaysia (171), United States (123), India (43), United Kingdom (40), Germany (35), Australia (35), Brazil (31), Spain (27), and Turkey (20). In addition, *MJM* recorded its highest *IF* in year 2007 (0.616) followed by 2006 (0.456), 2009 (0.378) and 2008 (0.367), while the five year *IF* is 0.577. Hence, the result of this study implies that medical researchers in Malaysia are making contributions to the field and the medical research front is expected to grow.

## 5.3. Recommendation

The findings of this study suggest that bibliometric results must be used with care and caution because of a number of factors that limit the study. Hence

a)  The study suggests that further research should pay more attention on examining the content of article published in *MJM*, by analyzing the abstract, length of the articles appendixes, illustrations, and research methodology. In addition, gender distribution of authors, distribution of members of the editorial board and reviewers may be examined. Some of these analyses could have been possible in this study, if *MyAis* have included the full text of all the articles and improved the database to contain those elements.

b)  The study suggest that future research objectives should be looking at the proportion of the papers devoted to topics of more local interest against that of international or regional appeal.

c)  The study also suggests that future bibliometric research should check the understanding and influence of open access publications and self-archiving on trends and developments in the medical field.

## 5.4. Conclusion

The bibliometric studies on *MJM* have highlighted the following:

a)  *MJM* has proven to be an important channel for communicating research amongst Malaysian medical researchers and practitioners. Over the years, including the 5-year span under study, *MJM* has managed to sustain and improve its publication with an impressive production of sometimes over 100 articles per year. This is indeed an impressive feat and it is undoubtedly the mouthpiece of the Malaysian Medical Association. This is clearly indicated by the high contribution of articles by Malaysian researchers and practitioners in both the institutions of higher learning, the hospitals, research centers and clinics.

b)  The editorial members of *MJM* are mainly Malaysians from various medical related fields and institutions and this is reflected by the mainly Malaysian contributions to the journals. The journal published very few foreign contributions. This indicates that *MJM* is still very much a Malaysian-based journal reporting on issues and findings closely related to Malaysia. Even though it is being covered by *Index Medicus,* it is not indexed by the *ISI Science Citation Index*, which prefers journals with a more international appeal. This is an area where *MJM* can be improved by extending the composition of its editorial members to include foreign experts and to apply a more rigorous reviewing practice. *MJM* has longevity, has an Asian approach but needs adopt a more regional outlook to encourage more foreign contributions.

c)  Even though *MJM* is a Malaysian journal, publishing mainly Malaysian papers, it is encouraging to discover that the articles published by *MJM* are being cited. This is clearly indicated by the current study, using citation data provided by *Google Scholar*. Articles in *MJM* are being cited by journal articles in a variety of journals both mainstream and local. *MJM* is making contribution to medical literature as reflected by the citations it receives in each of the 5 years under study. As a result of the citation information, the journal's impact factor can be derived even though it is neither being covered by the *ISI* databases nor listed in the Journal Citation Report (JCR). *MJM* has managed to make itself visible through its coverage by *MyAis (Malaysian Abstracting and Indexing System).* As *MyAis'*s content is indexed by *Google Scholar*, *MJM* is therefore automatically covered by *Google scholar*. As a result of this visibility, their articles began to be assessable, picked, and used by global medical researchers, which eventually lead to citation. This also indicates the viability of using *Google Scholar* instead of total dependence on the very selective *ISI* databases for citation and impact factor information.

# Appendixes

Appendix 1: Ranked List of Authors with Numbers of Contributions and their Affiliation

| No of Articles | Author's Names | Affiliations |
|---|---|---|
| 19 | Ruszymah B.H.I. | University Kebangsaan Malaysia |
| 15 | Aminuddin B.S. | Universiti Kebangsaan Malaysia |
| 15 | Gendeh, B.S. | University Kerbangsan Malaysia |
| 14 | Chua, K.H. | Universiti Kebangsaan Malaysia |
| 12 | Philip, R | Hospital Ipoh |
| 12 | Chua, K.B. | Ministry of Health |
| 12 | Prepageran, N. | University of Malaya |
| 11 | Halim A.S. | Universiti Sains Malaysia |
| 10 | Kwan, M.K. | University of Malaya |
| 9 | Sukumar, N. | University Kebangsaan Malaysia |
| 8 | Sopyan I. | International Islamic University Malaysia |
| 8 | Sherina M.S | Universiti Putra Malaysia |
| 8 | Abdullah J.M | Universiti Sains Malaysia |
| 8 | Zulmi W. | Universiti Sains Malaysia. |
| 7 | Gurdeep, S. | Hospital Ipoh |
| 7 | Raymond, A.A. | Hospital University Kebangsaan Malaysia |
| 7 | Loh, K.Y., | International Medical University |
| 7 | Teng, C.L. | International Medical University |
| 7 | Kumarasamy,V | Ministry of Health |
| 7 | Tan, G.C. | Universiti Kebangsaan Malaysia |
| 7 | Rampal, L. | Universiti Putra Malaysia |
| 6 | Harvinder, S. | Hospital Ipoh |
| 6 | Hamidon B.B. | Hospital Universiti Kebangsaan Malaysia |
| 6 | Sivalingam, N | International Medical University Malaysia |
| 6 | Chan, K.Y. | Universiti Kebangsaan Malaysia |
| 6 | Saim L | Universiti Kebangsaan Malaysia |
| 6 | Zulfiqar M.A | Universiti Kebangsaan Malaysia |
| 6 | Shashinder, S | University Malaya |
| 6 | Gopala, K.G. | University of Malaya |
| 5 | Mallina, S | Hospital Ipoh |
| 5 | Rosalind, S | Hospital Ipoh |
| 5 | Leong, C.F. | Hospital Universiti Kebangsaan Malaysia |
| 5 | Teoh, C.M | Hospital University Kebangsaan Malaysia |
| 5 | Yeap, J.S | International Medical University |
| 5 | Loh, L.C. | International Medical University Malaysia |
| 5 | Norizah I | Kementerian Kesihatan |
| 5 | Chan, S.C | Royal College of Medicine Perak |
| 5 | Khalid B.A.K | Universiti Kebangsaan Malaysia |

| 5 | Hamzaini A.H. | Universiti Kebangsaan Malaysia. |
|---|---|---|
| 5 | Reddy, S.C | Universiti Putra Malaysia |
| 5 | Subha, S.T. | Universiti Putra Malaysia |
| 5 | Biswal, B.M. | Universiti Sains Malaysia |
| 5 | Mafauzy M | Universiti Sains Malaysia |
| 5 | Naing L | Universiti Sains Malaysia |
| 5 | Faisham W.I | Universiti Sains Malaysia. |
| 5 | Kuljit, S | University Malaya |
| 5 | Saw, A | University Malaya Medical Centre |
| 5 | Choon, S.K. | University of Malaya |
| 4 | Fazir M | Hospital Kuala Lumpur |
| 4 | Saffari H.M | Hospital Kuala Lumpur |
| 4 | Chew, Y.K | Hospital Pakar Sultanah Fatimah |
| 4 | Ng, S.P | Hospital UNiversiti Kebangsaan Malaysia |
| 4 | Ramzisham A.R.M | Hospital Universiti Kebangsaan Malaysia |
| 4 | Rohana A.G | Hospital Universiti Kebangsaan Malaysia |
| 4 | Boo, N.Y | Hospital Universiti Kebangsaan Malaysia. |
| 4 | How, S.H. | International Islamic University Malaysia |
| 4 | Koh, C.N | Klinik Kesihatan Seremban |
| 4 | Sulaiman A. | Malaysian Armed Forces Health Services DivISIon. |
| 4 | Shashikiran, U | Melaka Manipal Medical College |
| 4 | Sudha, V | Melaka Manipal Medical College |
| 4 | Zuridah H | Ministry of health |
| 4 | Khairuddin H | Terendak Camp. Armed Forces Health Training Institute |
| 4 | Mohd Zin B | Terendak Camp. Malaysian Armed Forces |
| 4 | Rozali A | Terendak Camp. Malaysian Armed Forces |
| 4 | Abdul Rahman H | Universiti Kebangsaan Malaysia |
| 4 | Cheong, S K | Universiti Kebangsaan Malaysia |
| 4 | Fadilah S.A.W | Universiti Kebangsaan Malaysia |
| 4 | Hazim M.Y.S. | Universiti Kebangsaan Malaysia |
| 4 | Ng, A.M.H | Universiti Kebangsaan Malaysia |
| 4 | Primuharsa Putra S.H.A | Universiti Kebangsaan Malaysia |
| 4 | Tan, A.E | Universiti Kebangsaan Malaysia |
| 4 | Thambi Dorai, C.R | Universiti Kebangsaan Malaysia |
| 4 | Nazri J. | Universiti Kebangsaan Malaysia Hospital |
| 4 | Zulkifli Z. | Universiti Kebangsaan Malaysia Hospital |
| 4 | Lokman B.S. | Universiti Kebangsaan Malaysia Medical Centre |
| 4 | Samsudin O.C | Universiti Kebangsaan Malaysia Medical Centre. |
| 4 | Baharudin A | Universiti Sains Malaysia |
| 4 | Imran Y | Universiti Sains Malaysia |
| 4 | Winn, T | Universiti Sains Malaysia |
| 4 | Ramesh, S | Universiti Tenaga Nasional |
| 4 | Raman, R | University Malaya Medical Centre |

| 4 | Tang, I.P | University of Malaya |
|---|---|---|
| 3 | Abd Halim M | Armed Forces Hospital Lumut |
| 3 | Bairy, K.L | Asian Institute of Medicine |
| 3 | Jegan, T | Hosital Universiti Kebangsaan Malaysia |
| 3 | Selladurai, B.M | Hosital Universiti Kebangsaan Malaysia |
| 3 | Manjit, S | Hospital Ipoh |
| 3 | Azmi A. | Hospital Kuala Lumpur |
| 3 | Latif A.Z | Hospital Kuala Lumpur |
| 3 | Valalarmathi, S. | Hospital Kuala Lumpur |
| 3 | Zainal, A.A | Hospital Kuala Lumpur |
| 3 | Hooi, L.S | Hospital Pakar Sultanah Fatimah |
| 3 | Brito-Mutunayagam, S | Hospital Pakar Sultanah Fatimah, |
| 3 | Megat Shiraz M.A.R | Hospital Universiti Kebangsaan Malaysia |
| 3 | Razman J | Hospital Universiti Kebangsaan Malaysia |
| 3 | Tan, H.J | Hospital Universiti Kebangsaan Malaysia |
| 3 | Aminudin C.A. | International Islamic University Malaysia |
| 3 | Mel, M | International Islamic University Malaysia |
| 3 | Abu Bakar S | International Medical University |
| 3 | Kareem B.A | International Medical University |
| 3 | Chem, Y.K | Kementerian Kesihatan |
| 3 | Anuradha, S | Maulana Azad Medical College, India. |
| 3 | Jayaprakash, B | Melaka Manipal Medical College |
| 3 | Phang, K.S. | National University of Malaysia |
| 3 | Ong, C.K | Penang Hospital |
| 3 | Tan, W.C | Penang Hospital. |
| 3 | Wong, J.S | Sarawak General Hospital. |
| 3 | Wan Hazmy C.H | Seremban Hospital |
| 3 | Mazlyzam A.L | UKM Medical Centre. |
| 3 | Asma A | Universiti Kebangsaan Malaysia |
| 3 | Fauzi, A.R. | Universiti Kebangsaan Malaysia |
| 3 | Jamil M.A. | Universiti Kebangsaan Malaysia |
| 3 | Jeevanan, J | Universiti Kebangsaan Malaysia |
| 3 | Khairani O. | Universiti Kebangsaan Malaysia |
| 3 | Kong, N.C.T | Universiti Kebangsaan Malaysia |
| 3 | Makpol S | Universiti Kebangsaan Malaysia |
| 3 | Sani A | Universiti Kebangsaan Malaysia |
| 3 | Tan, K.K | Universiti Kebangsaan Malaysia |
| 3 | Yong, S.C. | Universiti Kebangsaan Malaysia |
| 3 | Sritharan, S | Universiti Kebangsaan Malaysia Hospital |
| 3 | Lee, D.J.K | Universiti Putra Malaysia |
| 3 | Sabariah A.R | Universiti Putra Malaysia |
| 3 | Ramasamy, R | Universiti Putra Malaysia. |
| 3 | Dorai, A.A | Universiti Sains Malaysia |

| 3 | George, J | Universiti Sains Malaysia |
|---|---|---|
| 3 | Iskandar Z.A | Universiti Sains Malaysia |
| 3 | Rusli B.N | Universiti Sains Malaysia |
| 3 | Sallehudin A.Y | Universiti Sains Malaysia |
| 3 | Samsuddin A.R. | Universiti Sains Malaysia |
| 3 | Shahid H | Universiti Sains Malaysia |
| 3 | Sulaiman A.R | Universiti Sains Malaysia |
| 3 | Tengku M.A | Universiti Sains Malaysia |
| 3 | Zakaria A | Universiti Sains Malaysia |
| 3 | Mutum, S.S | Universiti Sains Malaysia. |
| 3 | Penafort, R | University Malaya Medical Centre |
| 3 | Sara Ahmad T | University Malaya Medical Centre |
| 3 | Sithasanan, N | University Malaya Medical Centre |
| 3 | Shailendra S. | University Malaya Medical Centre. |
| 3 | Hamdi M. | University of Malaya |
| 3 | Kamarul T | University of Malaya |
| 3 | Khoo, E.M | University of Malaya |
| 3 | Liam, C.K | University of Malaya |
| 3 | Sangkar, J.V | University of Malaya |
| 3 | Tajunisah I | University of Malaya |
| 3 | Gan, G.G | University of Malaya. |
| 3 | Chen, H.C | University Putra Malaysia |
| 3 | Sobri M | UPM-HKL.Universiti Putra Malaysia |
| 2 | Ruhaini I. | Gombak District Health Office. Disease Control Unit. |
| 2 | Kodama, S | Harvard University, USA. Brigham and Women's Hospital |
| 2 | Kojima, K. | Harvard University, USA. Brigham and Women's Hospital |
| 2 | Vacanti, A.C. | Harvard University, USA. Brigham and Women's Hospital |
| 2 | M*ISI*ran K | Hosital Universiti Kebangsaan Malaysia |
| 2 | Nurliza I | Hospital Alor Setar |
| 2 | Chong, A.W | Hospital Ipoh |
| 2 | Chooi, Y.S | Hospital Kuala Lumpur |
| 2 | Fauziah K | Hospital Kuala Lumpur |
| 2 | Gangaram, H.B | Hospital Kuala Lumpur |
| 2 | Liu, W.J. | Hospital Kuala Lumpur |
| 2 | Muhd Borhan T.A | Hospital Kuala Lumpur |
| 2 | Suryati M.Y | Hospital Kuala Lumpur |
| 2 | Eng, J.B | Hospital Lam Wah Ee |
| 2 | Noorizan Y | Hospital Pakar Sultanah Fatimah, |
| 2 | Khir A | Hospital Pakar Sultanah Fatimah, Muar |
| 2 | Sreetharan, S.S | Hospital Queen Elizabeth |
| 2 | Ong, S.G | Hospital Selayang |
| 2 | Thayaparan, T | Hospital Seremban |
| 2 | Vijayasingham, P. | Hospital Seremban |
| 2 | Balan, S | **Hospital Sultanah Aminah** |
| 2 | Chow, Y.W | **Hospital Sultanah Aminah** |
| 2 | Khoo, J.J. | **Hospital Sultanah Aminah** |
| 2 | Thiyagar, N | **Hospital Sungai Petani** |
| 2 | Ang, Y.M. | **Hospital Tengku Ampuan Afzan** |
| 2 | Kuan, Y.C | **Hospital Tengku Ampuan Afzan** |
| 2 | Sapari S. | **Hospital Tengku Ampuan Afzan,** |
| 2 | Ho, C.C.K | **Hospital Universiti Kebangsaan Malaysia** |
| 2 | Hyzan M.Y. | **Hospital Universiti Kebangsaan Malaysia** |
| 2 | Jasmi A.Y | **Hospital Universiti Kebangsaan Malaysia** |
| 2 | Joanna O.S.M | **Hospital Universiti Kebangsaan Malaysia** |
| 2 | Lukman M.R | **Hospital Universiti Kebangsaan Malaysia** |
| 2 | Norlaila M | **Hospital Universiti Kebangsaan Malaysia** |
| 2 | Salina H. | **Hospital Universiti Kebangsaan Malaysia** |
| 2 | Teh, H.S | **Hospital Universiti Kebangsaan Malaysia** |
| 2 | Zaleha O | **Hospital Universiti Kebangsaan Malaysia** |
| 2 | Zamrin D.M | **Hospital Universiti Kebangsaan Malaysia** |
| 2 | Naicker, A.S. | **Hospital Universiti Kebangsaan Malaysia** |
| 2 | Bastion, M.L.C | **Hospital Universiti Kebangsaan Malaysia.** |
| 2 | Jamil M.A.Y | **Hospital University Kebangsaan Malaysia** |
| 2 | Shaharin S | **Hospital University Kebangsaan Malaysia** |
| 2 | Kenali M.S. | **Hospital University Kebangsaan Malaysia** |
| 2 | Norlijah O | **Institute for Medical Research Malaysia** |
| 2 | Ahmad R | **International Islamic University Malaysia** |
| 2 | Fauzi A.R. | **International Islamic University Malaysia** |
| 2 | Jamal P | **International islamic University Malaysia** |
| 2 | Nazri M.Y | **International Islamic University Malaysia** |
| 2 | Ng, T.H. | **International Islamic University Malaysia** |
| 2 | Noriah M.N | **International Islamic University Malaysia** |
| 2 | Wan Ishlah L | **International Islamic University Malaysia** |
| 2 | Yusof A | **International Islamic University Malaysia** |
| 2 | Sharaf I. | **International Islamic University Malaysia.** |
| 2 | Azizi A. | **International Islamic University of Malaysia** |
| 2 | Rafidah H.M | **International Islamic University of Malaysia** |
| 2 | Kathiravan, C | **International Medical University** |
| 2 | Lukman H | **International Medical University** |
| 2 | Yadav, H | **International Medical University** |
| 2 | Yeap, R | **International Medical University** |
| 2 | Achike, F.I. | **International Medical University** |
| 2 | Vijayasingham, P | **International Medical University** |
| 2 | Beevi, Z | **International Medical University Malaysia** |
| 2 | Hebbar, S | **Kasturba Medical College** |
| 2 | Ismail Y | **Kedah Medical Centre** |

| 2 | Asmah Hani A.W | Kementerian Kesihatan |
|---|---|---|
| 2 | Fouda, M.A | King Khalid University Hospital |
| 2 | Malabu, U.H. | King Khalid University Hospital |
| 2 | Mariam A.M | Klinik Kesihatan in Negri Sembilan |
| 2 | Mastura I | Klinik Kesihatan in Negri Sembilan |
| 2 | Narayanan, S | Klinik Kesihatan in Negri Sembilan |
| 2 | Norsiah A | Klinik Kesihatan in Negri Sembilan |
| 2 | Sabariah I. | Klinik Kesihatan in Negri Sembilan |
| 2 | Sharifah I | Klinik Kesihatan in Negri Sembilan |
| 2 | Siti Rokiah K | Klinik Kesihatan in Negri Sembilan |
| 2 | Ariza A. | Kuala Lumpur Hospital. |
| 2 | Aziah A.M | Kuala Lumpur Hospital. Clinical Research Centre. |
| 2 | Rugayah B | Kuala Lumpur Hospital. Clinical Research Centre. |
| 2 | Koay, T.K | Kudat Health Office, Sabah |
| 2 | Mariam M | Makmal Kesihatan Awam Kebangsaan |
| 2 | Sonkusare, S. | Melaka Manipal Medical College |
| 2 | Adinegara L.A | Melaka Manipal Medical College |
| 2 | Yalcin, A | Mersin University, Turkey |
| 2 | Lim, T.O | Ministry of Health |
| 2 | Tee, G.H | Ministry of Health, Malaysia |
| 2 | Helvaci , M.R. | Mustafa Kemal University |
| 2 | Kaya, H | Mustafa Kemal University |
| 2 | Premnath, N | Penang General Hospital. |
| 2 | Lo Kang, S.C | Penang Hospital |
| 2 | Abdul Razak M | Penang Hospital. |
| 2 | Hooi, L.N. | Public Specialist Centre, Penang. |
| 2 | Harjit, K | Putrajaya Hospital. |
| 2 | Chua, C.N | Sarawak General Hospital |
| 2 | Venugopalan, B | Selangor State Health Dept. Disease Control Unit. |
| 2 | Siow, Y.C | Selayang Hospital |
| 2 | Parwathi, A. | Seremban Hospital |
| 2 | Azizi, F | Shaheed Beheshti University of Medical Sciences, Iran |
| 2 | Loh, Y.C. William | Southport District General Hospital |
| 2 | Sathappan, S | Subang Jaya Medical Centre |
| 2 | Teh, A | Subang Jaya Medical Centre |
| 2 | Pany, A | Sunway Medical Centre |
| 2 | Sathananthar, K.S | Tengku Ampuan Afzan |
| 2 | Abdul Rahim N | Universiti Kebangsaan Malaysia |
| 2 | Al-Joudi, F.S | Universiti Kebangsaan Malaysia |
| 2 | Aljunid, S M | Universiti Kebangsaan Malaysia |
| 2 | Anwar A. | Universiti Kebangsaan Malaysia |
| 2 | Edariah A.B | Universiti Kebangsaan Malaysia |
| 2 | Goh, B.S. | Universiti Kebangsaan Malaysia |

| 2 | Halim A.G | Universiti Kebangsaan Malaysia |
|---|---|---|
| 2 | Haritharan, T. | Universiti Kebangsaan Malaysia |
| 2 | Ilina I | Universiti Kebangsaan Malaysia |
| 2 | Marina M.B. | Universiti Kebangsaan Malaysia |
| 2 | Mohd Manzor N.F | Universiti Kebangsaan Malaysia |
| 2 | Mohd Yusof Y.A | Universiti Kebangsaan Malaysia |
| 2 | Norazlina M. | Universiti Kebangsaan Malaysia |
| 2 | Norazmi M.K | Universiti Kebangsaan Malaysia |
| 2 | Norella K | Universiti Kebangsaan Malaysia |
| 2 | Norlia A | Universiti Kebangsaan Malaysia |
| 2 | Nurismah M.I | Universiti Kebangsaan Malaysia |
| 2 | Nurshaireen A. | Universiti Kebangsaan Malaysia |
| 2 | Rizal A.M., | Universiti Kebangsaan Malaysia |
| 2 | Rohana J | Universiti Kebangsaan Malaysia |
| 2 | Tan, G.H. | Universiti Kebangsaan Malaysia |
| 2 | Yeap, J.K. | Universiti Kebangsaan Malaysia |
| 2 | Soehardi Z. | Universiti Kebangsaan Malaysia |
| 2 | Neo, E.N | Universiti Kebangsaan Malaysia Hospital |
| 2 | Ismail S | Universiti Kebangsaan Malaysia Hospital. |
| 2 | Alfaqeh H | Universiti Kebangsaan Malaysia Medical Centre |
| 2 | Mazita A | Universiti Kebangsaan Malaysia Medical Centre |
| 2 | Wan Ngah W.Z | Universiti Kebangsaan Malaysia Medical Centre |
| 2 | Ghafar N.A | Universiti Kebangsaan Malaysia. |
| 2 | Kanaheswari, Y | Universiti Kebangsaan Malaysia. |
| 2 | Simat S.F | Universiti Kebangsaan Malaysia. |
| 2 | Alhady M | Universiti Malaysia Sarawak (UNIMAS) |
| 2 | Pan K.L | Universiti Malaysia Sarawak. |
| 2 | Rasit A.H | Universiti Malaysia Sarawak. |
| 2 | Bahaman A.R. | Universiti Putra Malaysia |
| 2 | Gul, Y.A | Universiti Putra Malaysia |
| 2 | Lim, T.A | Universiti Putra Malaysia |
| 2 | Nathan, S. | Universiti Putra Malaysia |
| 2 | Ng, W.M | Universiti Putra Malaysia |
| 2 | Nordin N | Universiti Putra Malaysia |
| 2 | Shiran M.S. | Universiti Putra Malaysia |
| 2 | Subramaniam, S | Universiti Putra Malaysia |
| 2 | Hejar, A.R | Universiti Putra Malaysia. |
| 2 | Mustaqim A | Universiti Putra Malaysia. |
| 2 | Othman F | Universiti Putra Malaysia. |
| 2 | Rahmat A | Universiti Putra Malaysia. |
| 2 | Saidi M. | Universiti Putra Malaysia. |
| 2 | Al-Joudi, F.S | Universiti Sains Malaysia |
| 2 | Azidah A.K | Universiti Sains Malaysia |

| 2 | Azizah Y | Universiti Sains Malaysia |
|---|---|---|
| 2 | Iskandar M.A | Universiti Sains Malaysia |
| 2 | Lee, N.N.A | Universiti Sains Malaysia |
| 2 | Othman, N. | Universiti Sains Malaysia |
| 2 | Saiful Azli M.N | Universiti Sains Malaysia |
| 2 | Saini R | Universiti Sains Malaysia |
| 2 | Sayuthi, S | Universiti Sains Malaysia |
| 2 | Shafie M.A | Universiti Sains Malaysia |
| 2 | Wan Aasim W.A | Universiti Sains Malaysia |
| 2 | Zainuddin N | Universiti Sains Malaysia |
| 2 | Zamzuri I | Universiti Sains Malaysia |
| 2 | Watihayati M.S | Universiti Sains Malaysia. |
| 2 | Zabidi-Hussin A.M.H | Universiti Sains Malaysia. |
| 2 | Zilfalil B.A | Universiti Sains Malaysia. |
| 2 | Karuthan, C | Universiti Teknologi MARA |
| 2 | Yusoff K. | Universiti Teknologi MARA |
| 2 | Froemming, G.A | Universiti Teknologi MARA. |
| 2 | Aw, K.L | Universiti Tenaga Nasional |
| 2 | Yeo, W.H | Universiti Tenaga Nasional |
| 2 | Chan, Y.F | University Malaya |
| 2 | Rahmat O | University Malaya |
| 2 | Chan, C.Y.W | University Malaya Medical Centre |
| 2 | Chan, K.Y | University Malaya Medical Centre |
| 2 | Chuah, K.H. | University Malaya Medical Centre |
| 2 | Jalaludin M.A | University Malaya Medical Centre |
| 2 | Kihne, M | University Malaya Medical Centre |
| 2 | Lee, W.S | University Malaya Medical Centre |
| 2 | Merican A.M. | University Malaya Medical Centre |
| 2 | Pang, Y.K | University Malaya Medical Centre |
| 2 | Ramanujam, T.M | University Malaya Medical Centre |
| 2 | Wong, E.L.W | University Malaya Medical Centre |
| 2 | Low, W.Y | University Malaya. |
| 2 | Amalourde, A | University of Malaya |
| 2 | Chan, L.L | University of Malaya |
| 2 | Eow, G.I | University of Malaya |
| 2 | Looi, L.M. | University of Malaya |
| 2 | Naveed, N. | University of Malaya |
| 2 | Ng, C.J | University of Malaya |
| 2 | Noran N.H | University of Malaya |
| 2 | Peh, S.C | University of Malaya |
| 2 | Quek, K.F | University of Malaya |
| 2 | Rabia K | University of Malaya |
| 2 | Shyamala, P. | University of Malaya |

| 2 | Vinayaga, P | University of Malaya |
|---|---|---|
| 2 | Sara T.A | University of Malaya Medical Center. |
| 2 | Ariffin H | University of Malaya Medical Centre |
| 2 | Chong, L.A | University of Malaya Medical Centre |
| 2 | Kamarulzaman A. | University of Malaya Medical Centre |
| 2 | Saravanan, S. | University of Malaya Medical Centre |
| 2 | Farizah H. | University of Malaya. |
| 2 | Hon, S.K. | University Putra Malaysia |
| 2 | Vidyadaran, S | University Putra Malaysia |
| 2 | Yunus, A Gul | University Putra Malaysia |
| 2 | Sharifah Salmah S.H | University Putra Malaysia |
| 2 | Adlina S. | University Technology Mara |
| 1 | Nurani, L.H | Ahmad Dahlan University |
| 1 | Baba, K. | AIMST University Malaysia. |
| 1 | Xavier, R | AIMST University Malaysia. |
| 1 | Loh, S.Y | Alexandra Hospital, Singapore. |
| 1 | Wilfred, C.G. Peh | Alexandra Hospital, Singapore. |
| 1 | Goriman Khan, M.A.K | Ar-Raudhah Biotech. Farm, Sdn. Bhd. |
| 1 | Muthu, K | Arunamary Medical Centre, Klang |
| 1 | Kota, R. | Asian Institute of Medicine |
| 1 | Rathinam, X | Asian Institute of Medicine |
| 1 | Shalini S | Asian Institute of Medicine |
| 1 | Sivagnanam, G | Asian Institute of Medicine |
| 1 | Vidyadaran, M.K., | Asian Institute of Medicine |
| 1 | Al-Jawad, M. | Asian Institute of Medicine, Science and Technology, Malaysia |
| 1 | Ganaraja, B | Asian Institute of Medicine, Science and Technology, Malaysia |
| 1 | Indira, B | Asian Institute of Medicine, Science and Technology, Malaysia |
| 1 | Narayan, K.A | Asian Institute of Medicine, Science and Technology, Malaysia |
| 1 | Rashid A.K | Asian Institute of Medicine, Science and Technology, Malaysia |
| 1 | Hyatt, Alex D | Australian Animal Health Laboratory (CSIRO) |
| 1 | Noitie, L. | Beaufort Health Office. |
| 1 | Tan, E | Beaufort Hospital |
| 1 | Khebir, B.V | Cawangan Penyakit Bawaan Vektor Negeri Johor |
| 1 | Cropp, Bruce C | Center for Diseases Control and Prevention, USA |
| 1 | Andrew, J.S. Tan | Changi General Hospital, Singapore |
| 1 | Gerald, J.S. Tan | Changi General Hospital, Singapore |
| 1 | Verhoeven, W.J., | Changi General Hospital, Singapore |
| 1 | CRP Kuala Kangsar. | Community Residency Programme Kuala Kangsar Group |
| 1 | Ergun, U.G.O | Cukurova University, Turkey |
| 1 | Seydaoglu, G | Cukurova University, Turkey |
| 1 | Abdul Rashid A.R | Cyberjaya University College of Medical Sciences |
| 1 | Azizi B.H.O | Damansara Specialist Hospital |
| 1 | Idrus R.M | Damansara Specialist Hospital. |
| 1 | Gan, E.C. | Damansara Specialists Hospital. |
| 1 | Zahari Z. | Department of Survey amd Mapping Malaysia |
| 1 | Adam M.A | Dept. of Health, Johor |
| 1 | Daud A.R | Dept. of Health, Johor |
| 1 | Shahrom C.M.D | Dept. of Health, Johor |
| 1 | Jorawar, S | District Tuberculosis Centre and Hospital |
| 1 | Kajal, N.C | District Tuberculosis Centre and Hospital |
| 1 | Sianturi, G.P | Dr. Kariadi General Hospital, Indonesia. |
| 1 | Widiastuti-Samekto, M | Dr. Kariadi General Hospital, Indonesia. |
| 1 | Magetsari, R. | Dr. Sardjito General Hospital, Indonesia |
| 1 | Dewo, P | Dr. Sardjito General Hospital, Indonesia/University of Groningen (The Netherlands) |
| 1 | Sahni, V | Drexel University College of Medicine (USA) |
| 1 | Algin, M.C | Dumlupinar University, Turkey |
| 1 | Hasan, B.U | Erciyes University, Turkey. |
| 1 | Huseyin, K | Erciyes University, Turkey. |
| 1 | Ibrahim, K.H. | Erciyes University, Turkey. |
| 1 | Kazim, U. | Erciyes University, Turkey. |
| 1 | Selim, K | Erciyes University, Turkey. |
| 1 | Bolhuis, R.L.H. | Esramus Medical Center, Netherlands |
| 1 | Nooter, K | Esramus Medical Center, Netherlands |
| 1 | Oostrum, R.G | Esramus Medical Center, Netherlands |
| 1 | Ng, Y.S | Fatimah Hospital, Ipoh |
| 1 | Abu, T.A | Gadjah Mada University, Indonesia |
| 1 | Ibnu, G.G | Gadjah Mada University, Indonesia |
| 1 | Mae, S.H.W | Gadjah Mada University, Indonesia |
| 1 | Sofia, M | Gadjah Mada University, Indonesia |
| 1 | Subagus, W | Gadjah Mada University, Indonesia |
| 1 | Mustofa | Gadjah Mada University, Indonesia |
| 1 | Sholikhah, E.N | Gadjah Mada University, Indonesia |
| 1 | Wijayawati, M.A. | Gadjah Mada University, Indonesia |
| 1 | Koley, S | Guru Nanak Dev University |
| 1 | Sandhu, J.S. | Guru Nanak Dev University |
| 1 | Sharma, G | Guru Nanak Dev University |
| 1 | Badaruddoza, | Guru Nanak Dev University, India. |
| 1 | Kaur, A. | Guru Nanak Dev University, India. |
| 1 | Sidhu, S., | Guru Nanak Dev University, India. |
| 1 | Westerman, K | Harvard University, USA. Brigham and Women's Hospital |
| 1 | Oztuzun, S. | Haydarpasa Numune Education and Research Hospital, Turkey. |
| 1 | Doi, M. | Hosital Universiti Kebangsaan Malaysia |
| 1 | Sakina M.S | Hosital Universiti Kebangsaan Malaysia |
| 1 | Soon, Y.Y. | Hospital Batu Gajah, Perak |
| 1 | Emilia, S.H | Hospital Batu Pahat |

| 1 | Moret, J | Hospital de la Foundation Rothschild, Paris |
|---|---|---|
| 1 | Pany, A | Hospital de la Foundation Rothschild, Paris |
| 1 | Sobri A. | Hospital de la Foundation Rothschild, Paris |
| 1 | Amanjit, K | Hospital Ipoh |
| 1 | Anil, S. | Hospital Ipoh |
| 1 | Balachandran, A. | Hospital Ipoh |
| 1 | Chandran, A. | Hospital Ipoh |
| 1 | Go, K.W | Hospital Ipoh |
| 1 | Kalyani, S | Hospital Ipoh |
| 1 | Mann, G.S | Hospital Ipoh |
| 1 | Pathma, L | Hospital Ipoh |
| 1 | Sabaria M.N | Hospital Ipoh |
| 1 | Teo, S.M | Hospital Ipoh |
| 1 | Ismail A.I | Hospital Ipoh.Perak. |
| 1 | Abdullah A.A | Hospital Kuala Lumpur |
| 1 | Adnan J.S | Hospital Kuala Lumpur |
| 1 | Azizan N.Z | Hospital Kuala Lumpur |
| 1 | Bavanandan, S. | Hospital Kuala Lumpur |
| 1 | Chye, P.C | Hospital Kuala Lumpur |
| 1 | Eusni R.T | Hospital Kuala Lumpur |
| 1 | Hayati A.N | Hospital Kuala Lumpur |
| 1 | Hussein S.H | Hospital Kuala Lumpur |
| 1 | Jeyaledchumy, M | Hospital Kuala Lumpur |
| 1 | Johari A. | Hospital Kuala Lumpur |
| 1 | Kamarul A | Hospital Kuala Lumpur |
| 1 | Kan, C.H. | Hospital Kuala Lumpur |
| 1 | Khoo, T.B | Hospital Kuala Lumpur |
| 1 | Lee, F.S | Hospital Kuala Lumpur |
| 1 | Lee, S.K. | Hospital Kuala Lumpur |
| 1 | Mangalam, S | Hospital Kuala Lumpur |
| 1 | Mansar A.H | Hospital Kuala Lumpur |
| 1 | Mazri Y.M | Hospital Kuala Lumpur |
| 1 | Merican, J.S | Hospital Kuala Lumpur |
| 1 | Morad Z. | Hospital Kuala Lumpur |
| 1 | Naresh, G | Hospital Kuala Lumpur |
| 1 | Ng, C.M. | Hospital Kuala Lumpur |
| 1 | Normayah K | Hospital Kuala Lumpur |
| 1 | Nyanaveelan, M | Hospital Kuala Lumpur |
| 1 | Ozlan I.M. Kamil | Hospital Kuala Lumpur |
| 1 | Rozaini R | Hospital Kuala Lumpur |
| 1 | Salina A.A | Hospital Kuala Lumpur |
| 1 | Samat N.A | Hospital Kuala Lumpur |
| 1 | Santhi, D.P | Hospital Kuala Lumpur |
| 1 | Saw, C.B | Hospital Kuala Lumpur |
| 1 | Shenaz Banu, S.K | Hospital Kuala Lumpur |
| 1 | Sidik Che Kob | Hospital Kuala Lumpur |
| 1 | Suib I. | Hospital Kuala Lumpur |
| 1 | Sundram, M. | Hospital Kuala lumpur |
| 1 | Suraiya H | Hospital Kuala Lumpur |
| 1 | Zahari N | Hospital Kuala Lumpur |
| 1 | Zainab S. | Hospital Kuala Lumpur |
| 1 | Zaki, M. | Hospital Kuala Lumpur |
| 1 | Abdullah A.H.R. | Hospital Kuala Lumpur. |
| 1 | Ismail Z. | Hospital Kuala Lumpur. |
| 1 | Norzila M.Z. | Hospital Kuala Lumpur. |
| 1 | Rusanida A | Hospital Kuala Lumpur. |
| 1 | Bharathi, V | Hospital Kuala Lumpur. Universiti Putra Malaysia Clinical Campus |
| 1 | Azhar M.Z | Hospital Kuala Lumpur. Universiti Putra Malaysia. |
| 1 | Sachchithanantham | Hospital Melaka |
| 1 | Taye, G.A.W.C | Hospital Melaka |
| 1 | Abdul R. | Hospital Melaka Medical College |
| 1 | Hisa M.K., | Hospital Melaka. |
| 1 | Sa'iah A | Hospital Orang Asli |
| 1 | Chow, H.L | Hospital Pakar Sultanah Fatimah |
| 1 | Leong, C.L | Hospital Pakar Sultanah Fatimah |
| 1 | Mujait K | Hospital Pakar Sultanah Fatimah |
| 1 | Shamsuddin A.A. | Hospital Pakar Sultanah Fatimah |
| 1 | Rozina G | Hospital Penang |
| 1 | Bhupinder, S | Hospital Pulau Pinang |
| 1 | Ismail O | Hospital Pulau Pinang |
| 1 | Subathra, S. | Hospital Pulau Pinang |
| 1 | Gooi, B.H. | Hospital Pulau Pinang. |
| 1 | Chern, P.M | Hospital Putrajaya |
| 1 | Yusniza M.Y | Hospital Putrajaya |
| 1 | Hussien Z | Hospital Putrajaya. |
| 1 | Baskaran, S | Hospital Queen Elizabeth |
| 1 | Nahulan, T | Hospital Queen Elizabeth |
| 1 | Jayaram, M | Hospital Queen Elizabeth, Kota kinabalu |
| 1 | Azam A | Hospital Selayang |
| 1 | Cardosa, M S | Hospital Selayang |
| 1 | Hamid A. | Hospital Selayang |
| 1 | Mariam I. | Hospital Selayang |
| 1 | Nor Fariza N | Hospital Selayang |
| 1 | Norrashidah A.W | Hospital Selayang |
| 1 | Pagalavan, L | Hospital Selayang |
| **1** | **Poh, E.P.** | **Hospital Selayang** |
| **1** | **Rajasingam, R** | **Hospital Selayang** |
| **1** | **Rohan, M.J** | **Hospital Selayang** |
| **1** | **Yudisthra, G** | **Hospital Serdang, Selangor** |
| **1** | **Harwant, S.** | **Hospital Seremban** |
| **1** | **Khamizar W.** | **Hospital Seremban** |
| **1** | **Ramesh, G** | **Hospital Seremban** |
| **1** | **Shahrul H** | **Hospital Seremban** |
| **1** | **Suryani M.Y** | **Hospital Seremban** |
| **1** | **Wong, S.L.** | **Hospital Seremban** |
| **1** | **Rasan, M.I** | **Hospital Sultan Haji Ahmad Shah** |
| **1** | **Ahmad N.** | **Hospital Sultanah Aminah** |
| **1** | **Aida Z** | **Hospital Sultanah Aminah** |
| **1** | **Bala Sundaram, M.** | **Hospital Sultanah Aminah** |
| **1** | **Hooi L S,** | **Hospital Sultanah Aminah** |
| **1** | **Liam, K.I.** | **Hospital Sultanah Aminah** |
| **1** | **Lim, B.B.** | **Hospital Sultanah Aminah** |
| **1** | **Singaraveloo, M.** | **Hospital Sultanah Aminah** |
| **1** | **Tan, C.C** | **Hospital Sultanah Aminah** |
| **1** | **Tan,C.C** | **Hospital Sultanah Aminah** |
| **1** | **Yoong, H.F** | **Hospital Sultanah Aminah** |
| **1** | **Zanariah, Y** | **Hospital Sultanah Aminah** |
| **1** | **Abbas A.A** | **Hospital Sultanah Aminah (Johor Bahru).** |
| **1** | **Kamari Z.H** | **Hospital Sultanah Aminah (Johor Bahru).** |
| **1** | **Loh, S.S.** | **Hospital Sultanah Aminah, Johor Bahru.** |
| **1** | **Prakash, K** | **Hospital Sultanah Aminah, Johor Bahru.** |
| **1** | **Hor, J.Y.** | **Hospital Sultanah Bahiyah** |
| **1** | **Lee, Christopher K.C.** | **Hospital Sungai Buloh.** |
| **1** | **Choo, C.M** | **Hospital Sungai Petani** |
| **1** | **See, C.K** | **Hospital Sungai Petani** |
| **1** | **Chin P.S** | **Hospital Taiping, Perak** |
| **1** | **Khoo, A.P.C** | **Hospital Taiping, Perak** |
| **1** | **Cheong, B.M.K.** | **Hospital Teluk Intan** |
| **1** | **Norra H** | **Hospital Tengku Ampuan Afzan** |
| **1** | **Ramachandram, K** | **Hospital Tengku Ampuan Afzan** |
| **1** | **Chin, S.P** | **Hospital Tengku Ampuan Afzan,** |
| **1** | **Anna Liza R** | **Hospital Tengku Ampuan Afzan, Kuantan** |
| **1** | **Ghazali I** | **Hospital Tengku Ampuan Afzan, Kuantan** |
| **1** | **Fahmi A.M** | **Hospital Tengku Ampuan Afzan.** |
| **1** | **Srinovianti, N** | **Hospital Tengku Ampuan Afzan.** |
| **1** | **Cheong, S.M** | **Hospital Tengku Ampuan Rahimah, Klang** |
| **1** | **Mohd Farouk A** | **Hospital Tengku Ampuan Rahimah, Klang** |
| **1** | **Ooi, H.L.** | **Hospital Tengku Ampuan Rahimah, Klang** |

| 1 | Yogeswary, S. | Hospital Tengku Ampuan Rahimah, Klang |
|---|---|---|
| 1 | Satwi S | Hospital Tunku Ampuan Afzan. |
| 1 | Zuridah H | Hospital Umum Sarawak. |
| 1 | Ahmad Zakuan K | Hospital Universiti Kebangsaan Malaysia |
| 1 | Badrulhisham B | Hospital Universiti Kebangsaan Malaysia |
| 1 | Hamizah R | Hospital Universiti Kebangsaan Malaysia |
| 1 | Hayati A.R. | Hospital Universiti Kebangsaan Malaysia |
| 1 | Jaafar M.Z. | Hospital Universiti Kebangsaan Malaysia |
| 1 | Johann K.F | Hospital Universiti Kebangsaan Malaysia |
| 1 | Joseph, J.P. | Hospital Universiti Kebangsaan Malaysia |
| 1 | Marjmin O | Hospital Universiti Kebangsaan Malaysia |
| 1 | Marliza M.Y | Hospital Universiti Kebangsaan Malaysia |
| 1 | Masbah O | Hospital Universiti Kebangsaan Malaysia |
| 1 | Mohd Nizlan M.N | Hospital Universiti Kebangsaan Malaysia |
| 1 | Muraly, S | Hospital Universiti Kebangsaan Malaysia |
| 1 | Ngoo, K.S. | Hospital Universiti Kebangsaan Malaysia |
| 1 | Noorddin Y | Hospital Universiti Kebangsaan Malaysia |
| 1 | Noorfaizan S. | Hospital Universiti Kebangsaan Malaysia |
| 1 | Nor Azmi M | Hospital Universiti Kebangsaan Malaysia |
| 1 | Norleza A.N | Hospital Universiti Kebangsaan Malaysia |
| 1 | Ong, T.Z | Hospital Universiti Kebangsaan Malaysia |
| 1 | Praveen, S. | Hospital Universiti Kebangsaan Malaysia |
| 1 | Qureshi, M.A | Hospital Universiti Kebangsaan Malaysia |
| 1 | Rabani R | Hospital Universiti Kebangsaan Malaysia |
| 1 | Raha A.R | Hospital Universiti Kebangsaan Malaysia |
| 1 | Rashid M.O | Hospital Universiti Kebangsaan Malaysia |
| 1 | Roohi, S.A | Hospital Universiti Kebangsaan Malaysia |
| 1 | Rozaidi S.H.W | Hospital Universiti Kebangsaan Malaysia |
| 1 | Rozilaila R | Hospital Universiti Kebangsaan Malaysia |
| 1 | Salmi A | Hospital Universiti Kebangsaan Malaysia |
| 1 | Salwati S | Hospital UNiversiti Kebangsaan Malaysia |
| 1 | Sangar, P | Hospital Universiti Kebangsaan Malaysia |
| 1 | Sharifa Ezat W.P | Hospital Universiti Kebangsaan Malaysia |
| 1 | Shukri J. | Hospital Universiti Kebangsaan Malaysia |
| 1 | Siti Aishah M.A | Hospital Universiti Kebangsaan Malaysia |
| 1 | Suhail A | Hospital Universiti Kebangsaan Malaysia |
| 1 | Surianti S | Hospital Universiti Kebangsaan Malaysia |
| 1 | Talal A.R | Hospital Universiti Kebangsaan Malaysia |
| 1 | Zahiah M. | Hospital Universiti Kebangsaan Malaysia |
| 1 | Zarina A.L | Hospital UNiversiti Kebangsaan Malaysia |
| 1 | Zulkarnaen A.N. | Hospital Universiti Kebangsaan Malaysia |
| 1 | Adeeb S.M.S.J. | Hospital Universiti Kebangsaan Malaysia. |
| 1 | Jong, Y.H | Hospital Universiti Kebangsaan Malaysia. |
| 1 | Kok, H.S | Hospital Universiti Kebangsaan Malaysia. |
| 1 | Leong, K.K | Hospital Universiti Kebangsaan Malaysia. |
| 1 | Mallina S | Hospital Universiti Kebangsaan Malaysia. |
| 1 | Muhaya, M | Hospital Universiti Kebangsaan Malaysia. |
| 1 | Shukur M.H. | Hospital Universiti Kebangsaan Malaysia. |
| 1 | Rahman, A J A | Hospital Universiti Kerbangsan Malaysia |
| 1 | Cik Fareha A. | Hospital Universiti Sains Malaysia |
| 1 | Lim, C.K | Hospital Universiti Sains Malaysia |
| 1 | Azizi A.B | Hospital University Kebangsaan Malaysia |
| 1 | Fadzilah I. | Hospital University Kebangsaan Malaysia |
| 1 | Loo, C.Y | Hospital University Kebangsaan Malaysia |
| 1 | Maizaton A.A | Hospital University Kebangsaan Malaysia |
| 1 | Marlyn, M | Hospital University Kebangsaan Malaysia |
| 1 | Mohammad A.R | Hospital University Kebangsaan Malaysia |
| 1 | Muizatul W.M.N | Hospital University Kebangsaan Malaysia |
| 1 | Nasir Z.M. | Hospital University Kebangsaan Malaysia |
| 1 | Satpal, S. | Hospital University Kebangsaan Malaysia |
| 1 | Somasundaram, S. | Hospital University Kebangsaan Malaysia |
| 1 | Zurin A.A.R | Hospital University Kebangsaan Malaysia |
| 1 | Azhar A.A | Hospital University Kebangsaan Malaysia. |
| 1 | Mujahid S.H. | Hospital University Kebangsaan Malaysia. |
| 1 | Chua, C.B | Hospital University of Malaya |
| 1 | Razack, A.H | Hospital University of Malaya |
| 1 | Nazihah M., | Hospital USM Kubang Kerian |
| 1 | Murugan, S | Hulu Langat District Health Office. Disease Control Unit |
| 1 | Meftahuddin T., | Hulu Langat District Health Office. Disease Control Unit. |
| 1 | Nik Rubiah N.A.R. | Hulu Langat District Health Office. Disease Control Unit. |
| 1 | Arif M. | Hyderabad Medical Complex, Pakistan |
| 1 | Siang, T.K | IMU Clinical School, Seremban |
| 1 | Amal M.D | Inno Bio Diagnostics Sdn Bhd, Malaysia. |
| 1 | Caroline, J.S.Y. | Inno Bio Diagnostics Sdn Bhd, Malaysia. |
| 1 | Rafidah A.Z | Inno Bio Diagnostics Sdn Bhd, Malaysia. |
| 1 | Mohd Nazlee K. | Inno Bio Ventures Sdn Bhd |
| 1 | Alwi M. | Institut Jantung Negara |
| 1 | Awang Y. | Institut Jantung Negara |
| 1 | Leman H. | Institut Jantung Negara |
| 1 | Yakub A | Institut Jantung Negara |
| 1 | Azman A.B | Institute for Health Systems Research |
| 1 | Low, L.L | Institute for Health Systems Research |
| 1 | Sararaks, S. | Institute for Health Systems Research |
| 1 | Kuldip, K | Institute for Medical Research |
| 1 | Rahimah A. | Institute for Medical Research |
| 1 | Zubaidah Z. | Institute for Medical Research |
| 1 | Lye, M.S. | **Institute for Medical Research Malaysia** |
| 1 | Rohani M.Y. | **Institute for Medical Research Malaysia** |
| 1 | Sherina M.S | **Institute of Army Health.** |
| 1 | Azizah R. | **Institute of Medical Research Malaysia** |
| 1 | Wan Nazaimoon W.M | **Institute of Medical Research Malaysia** |
| 1 | Murad S. | **Institute of Medical Research Malaysia.** |
| 1 | Lai, M.W. | **Institute of Paediatrics** |
| 1 | Ahmad Nor Y. | **International Islamic University Malaysia** |
| 1 | Alam M.Z | **International islamic University Malaysia** |
| 1 | Fairus M | **International Islamic University Malaysia** |
| 1 | Hafiz A | **International Islamic University Malaysia** |
| 1 | Ishlah W | **International Islamic University Malaysia** |
| 1 | Khalid K.A | **International Islamic University Malaysia** |
| 1 | Muyibi S.A | **International islamic University Malaysia** |
| 1 | Natasha A.N. | **International Islamic University Malaysia** |
| 1 | Raihana M.F | **International Islamic University Malaysia** |
| 1 | Rosli A | **International Islamic University Malaysia** |
| 1 | Suhaimi H | **International Islamic University Malaysia** |
| 1 | Suhani F | **International islamic University Malaysia** |
| 1 | Syarif W.M | **International islamic University Malaysia** |
| 1 | Toibah A.R | **International Islamic University Malaysia** |
| 1 | Zainab K | **International Islamic University Malaysia** |
| 1 | Lokman S. | **International Islamic University Malaysia.** |
| 1 | Mardziah C.M | **International Islamic University Malaysia.** |
| 1 | Rathor, M.Y. | **International Islamic University Malaysia.** |
| 1 | Shah A. | **International Islamic University Malaysia.** |
| 1 | Yusof M. | **International Islamic University Malaysia.** |
| 1 | Zamzuri Z | **International Islamic University Malaysia.** |
| 1 | Ishlah L | **International Islamic University of Malaysia** |
| 1 | Ang, T.H | **International Medical University** |
| 1 | Chelliah, A. | **International Medical University** |
| 1 | Codati, A. | **International Medical University** |
| 1 | Ding, C.H | **International Medical University** |
| 1 | Jamil, M. | **International Medical University** |
| 1 | Kandasami, P., | **International Medical University** |
| 1 | Khuzaiah R. | **International Medical University** |
| 1 | Leong K.C. | **International Medical University** |
| 1 | Lim, B.K | **International Medical University** |
| 1 | Lim, V.K.E | **International Medical University** |
| 1 | Loh, Li Cher | **International Medical University** |
| 1 | Mohd Ali, A. | **International Medical University** |
| 1 | Mohd Noor Z | **International Medical University** |
| 1 | Tan, K.L. | **International Medical University** |

| 1 | Teh, P.N | International Medical University |
|---|----------|----------------------------------|
| 1 | Wong, Pei Se | International Medical University |
| 1 | Zailinawati A.H | International Medical University |
| 1 | Chong, K.H | International Medical University (Kuala Lumpur) |
| 1 | Abdul Samah S.Z | International Medical University Malaysia |
| 1 | Isa M.N | International Medical University Malaysia |
| 1 | Mohamadou G. | International Medical University Malaysia |
| 1 | Ogle, J. | International Medical University Malaysia |
| 1 | Wong, G.L.S | International Medical University Malaysia |
| 1 | Zainudin A. | International Medical University Malaysia |
| 1 | Bong, Y.C | International Medical University, Kuala Lumpur |
| 1 | Judson, J.P. | International Medical University, Kuala Lumpur |
| 1 | Nadarajah, V.D | International Medical University, Kuala Lumpur |
| 1 | Subramaniam, K. | International Medical University, Kuala Lumpur |
| 1 | Au Yeung, P.S | International Medical University, Seremban |
| 1 | Lim, J.W | International Medical University, Seremban |
| 1 | Phua, K.L. | International Medical University, Seremban |
| 1 | Vergis, M. | International Medical University, Seremban |
| 1 | Halim A.Y | International University Malaysia |
| 1 | Shukrimi A | International University Malaysia |
| 1 | Amar H.S.S | Ipoh Hospital |
| 1 | Jai Mohan | Ipoh Hospital |
| 1 | Kelishadi, R. | Isfahan University of Medical Sciences, Iran. |
| 1 | Roohafza, H.R. | Isfahan University of Medical Sciences, Iran. |
| 1 | Sadeghi, M. | Isfahan University of Medical Sciences, Iran. |
| 1 | Lim, B.K. Aaron | Island Hospital (Penang). |
| 1 | Junaidi I | Jabatan Kesihatan Negeri Perak |
| 1 | Abdul Samad B.H. | Johor State Health Department |
| 1 | Baba MD, N | Johor State Health Department |
| 1 | Rajasekaran, G. | Johor State Health Department |
| 1 | Suhaili M.R | Johor State Health Department |
| 1 | Baizura J. | Kajang Hospital |
| 1 | Sachidananda, A. | Kasturba Medical College and Dr. TMA Pai Hospital, India, |
| 1 | Anindya, C. | Kasturba Medical College and Hospital |
| 1 | Bhanu, P | Kasturba Medical College and Hospital |
| 1 | Mohan, M. | Kasturba Medical College and Hospital |
| 1 | Rao, P.L.N.G. | Kasturba Medical College and Hospital |
| 1 | Vijay, K. | Kasturba Medical College and Hospital |
| 1 | Dipak, R.N. | Kasturba Medical College and Hospital, India |
| 1 | Hazarika, P. | Kasturba Medical College and Hospital, India |
| 1 | Kailesh, P | Kasturba Medical College and Hospital, India |
| 1 | Parul, P. | Kasturba Medical College and Hospital, India |
| 1 | Abhishek, M | Kasturba Medical College, India |
| 1 | Annappa, K | Kasturba Medical College, India |
| 1 | Mukhyaprana, M.P. | Kasturba Medical College, India |
| 1 | Samjhana, K | Katurba Medical College |
| 1 | Mimi O | Kelana Jaya Health Clinic |
| 1 | Prathapa, S. | Kementerian Kesihatan |
| 1 | Faridah M.N | Kementerian Kesihatan Malaysia |
| 1 | Juliana R | Kementerian Kesihatan Malaysia |
| 1 | Bhimji, S. | King Fahd Medical Centre, Saudi Arabia |
| 1 | Lim, K.B. | Klang Hospital. |
| 1 | Chua, W.T. | Klinik Chua (Sitiawan). |
| 1 | Kamil M.A | Klinik Kamil Arif, Arau |
| 1 | Nor Asiah H | Klinik Kesihatan in Negri Sembilan |
| 1 | Siti Zubaidah M.A | Klinik Kesihatan in Negri Sembilan |
| 1 | Rahimah N. | Klinik Kesihatan Tanah Puteh, Sarawak. |
| 1 | John, J | Klinik Pergigian Hospital Pasir Mas, Kelantan |
| 1 | Matsuo, M. | Kobe University. |
| 1 | Nishio, H | Kobe University. |
| 1 | Sutomo, R. | Kobe University. |
| 1 | Imran A.K | Kota Bharu General Hospital |
| 1 | Azmi S | Kuala Lumpur Hospital |
| 1 | Hussien I. | Kuala Lumpur Hospital |
| 1 | Zubaidah A.W | Kuala Lumpur Hospital |
| 1 | Norhaya M.R | Kuala Terengganu Hospital |
| 1 | Fung, Y.K. | Kuching General Hospital |
| 1 | Tan, C | Kuching General Hospital |
| 1 | Koh, K.H | Kuching Sarawak General Hospital |
| 1 | Tan, Clare H.H. | Kuching Sarawak General Hospital |
| 1 | Inatomi, T | Kyoto Prefectural University of Medicine, Japan. |
| 1 | Kinoshita, S., | Kyoto Prefectural University of Medicine, Japan. |
| 1 | Koizumi, N | Kyoto Prefectural University of Medicine, Japan. |
| 1 | Nakamura, T | Kyoto Prefectural University of Medicine, Japan. |
| 1 | Sotozono, C. | Kyoto Prefectural University of Medicine, Japan. |
| 1 | Malini, A | Madras University, India |
| 1 | Rekha, K. | Madras University, India |
| 1 | Lim, B.A. | Mahkota Medical Centre, Melaka |
| 1 | Selvanesan, S | Makmal Kesihatan Awam Kebangsaan |
| 1 | Au, L.F | Malaysian Nuclear Agency |
| 1 | Besar I | Malaysian Nuclear Agency |
| 1 | Fazliana M.S | Malaysian Nuclear Agency |
| 1 | Mazleha M | Malaysian Nuclear Agency |
| 1 | Mustaffa R | Malaysian Nuclear Agency |
| 1 | Norimah Y | Malaysian Nuclear Agency |
| 1 | Shafii K | Malaysian Nuclear Agency |
| 1 | Md Top G | Malaysian Palm Oil Board |
| 1 | Lamont, A.C | Mater Children's Hospital, Australia. |
| 1 | Parker, A | Mater Children's Hospital, Australia. |
| 1 | Sahrir S. | Mater Children's Hospital, Australia. |
| 1 | Agarwal, S.K | Maulana Azad Medical College and Associated Hospitals, India |
| 1 | Kaur, R | Maulana Azad Medical College and Associated Hospitals, India |
| 1 | Prakash, A | Maulana Azad Medical College and Associated Hospitals, India |
| 1 | Singh, N.P | Maulana Azad Medical College and Associated Hospitals, India |
| 1 | Gupta, N | Maulana Azad Medical College and Associated Lok Nayak Hospital, G.B. Pant Hospital and Guru Nanak Eye Hospital (India). |
| 1 | Kar, P | Maulana Azad Medical College and Associated Lok Nayak Hospital, G.B. Pant Hospital and Guru Nanak Eye Hospital (India). |
| 1 | Tatke, M | Maulana Azad Medical College and Associated Lok Nayak Hospital, G.B. Pant Hospital and Guru Nanak Eye Hospital (India). |
| 1 | Chin, C.K | Medical Specialist Centre |
| 1 | Ho | Megah Medical Specialist Group, Petaling Jaya |
| 1 | Arokiasamy, J.T | Melaka Manipal Medical College |
| 1 | Razzak M.S | Melaka Manipal Medical College |
| 1 | Chew, S.M | Ministry of Health |
| 1 | Faridah A | Ministry of Health |
| 1 | Haniza S. | Ministry of Health |
| 1 | Lai, L.S | Ministry of Health |
| 1 | Lim, C.M | Ministry of Health |
| 1 | Rohaizat Y | Ministry of Health |
| 1 | Sameerah S.A.R | Ministry of Health |
| 1 | Sarojini, S | Ministry of Health |
| 1 | Merican M.I. | Ministry of Health (Malaysia). |
| 1 | Ang, K.T. | Ministry of Health Malaysia |
| 1 | Marzuki I | Ministry of Health Malaysia |
| 1 | Narimah A. | Ministry of Health Malaysia |
| 1 | Nor Khamisah A | Ministry of Health Malaysia |
| 1 | Sinniah, M. | Ministry of Health Malaysia |
| 1 | Wang, H.B | Ministry of health Malaysia |
| 1 | Gurpreet, K. | Ministry of Health. |
| 1 | Wan Mansor H. | Ministry of Health. Communicable Disease Control Div*ISI*on |
| 1 | Sha'ari B.N | Ministry of Health. Disease Control Div*ISI*on. AIDS/STD Section. |
| 1 | Nirmal, S | Ministry of Health. Vectorborne Disease Control. |
| 1 | Wijesinha, S | Monash University, Australia |
| 1 | Piterman, L | Monash University, Australia. |
| 1 | Chew, Stephen T.H | Mount Elizabeth Medical Centre, Singapore |
| 1 | Ozer, C | Mustafa Kemal University |
| 1 | Ebisawa, K. | Nagoya University (Japan) |
| 1 | Kamei, Y | Nagoya University (Japan) |
| 1 | Kato, R | Nagoya University (Japan) |
| 1 | Narita, Y | **Nagoya University (Japan)** |
| 1 | Okada, M | **Nagoya University (Japan)** |
| 1 | Kagami, H. | **Nagoya University (Japan) and University of Tokyo (Japan)** |
| 1 | Ueda, M. | **Nagoya University (Japan) and University of Tokyo (Japan)** |
| 1 | Mustaffa B.E | **National Diabetes Institute, Kuala Lumpur** |
| 1 | Siti Norlasiah I. | **National Population and Family Development Board. Genetic Laboratory** |
| 1 | Lai, P.S. | **National University Hospital. National University of Singapore** |
| 1 | Aziz N.A | **National University of Malaysia** |
| 1 | Chin, G.L | **National University of Malaysia** |
| 1 | Tong, S.F. | **National University of Malaysia** |
| 1 | Ting, F. | **Normah Medical Specialist Centre** |
| 1 | Murli N.I. | **Penang Adventist Hospital** |
| 1 | Lim, Y.H | **Penang General Hospital** |
| 1 | Yeap, B.H | **Penang General Hospital.** |
| 1 | Hussain, I.H.M.I | **Penang Hospital** |
| 1 | Leong, K.N. | **Penang Hospital** |
| 1 | Thomas, G.S.T., | **Penang Hospital** |
| 1 | Eow, G.B | **Penang Hospital.** |
| 1 | Goh, A.S | **Penang Hospital.** |
| 1 | Lim, S.L | **Penang Hospital.** |
| 1 | Looi, I. | **Penang Hospital.** |
| 1 | Norlia A.M. | **Penang Hospital.** |
| 1 | Ong, E.E | **Penang Hospital.** |
| 1 | Lee, T.W. | **Perak College of Medicine** |
| 1 | Harbinder, K | **Perak College of Medicine.** |
| 1 | Ho, J.J | **Perak College of Medicine.** |
| 1 | Khoo, Y.L | **Perak College of Medicine.** |
| 1 | Ow Yeang, Y.L | **Perak College of Medicine.** |
| 1 | Sethuraman, K. | **Perak College of Medicine.** |
| 1 | Teoh, L.C. | **Perak College of Medicine.** |
| 1 | Siva Achana, K | **Perak Royal College of Medicine** |
| 1 | Chan, G.C | **Poliklinik Komuniti Peringgit** |
| 1 | Tey, A | **Princess Alexandra Eye Pavilion, UK** |
| 1 | Maimunah A.H. | **Public Health Institute Malaysia** |
| 1 | Fatimah O. | **Putrajaya Hospital.** |
| 1 | Hisham A.N. | **Putrajaya Hospital.** |
| 1 | Yun, S.I | **Putrajaya Hospital.** |
| 1 | Boay, A.G.I | **Queen Elizabeth Hospital** |
| 1 | Chidambaram, S | **Queen Elizabeth Hospital** |
| 1 | Chuah, J.A | **Queen Elizabeth Hospital** |
| 1 | Kumar, V.M. | **Queen Elizabeth Hospital** |
| 1 | Rahimah M.S | **Queen Elizabeth Hospital** |

| 1 | Ramu, P. | Queen Elizabeth Hospital |
|---|---|---|
| 1 | Menon, J. | Queen Elizabeth Hospital, Kota Kinabalu |
| 1 | Harries, R. | Queen Elizabeth Hospital, South Australia |
| 1 | Ang, G.S. | Raigmore Hospital (United Kingdom). |
| 1 | Ng, W.S | Raigmore Hospital (United Kingdom). |
| 1 | Subrayan, V. | Raigmore Hospital (United Kingdom). |
| 1 | Nooriah S | Royal College of Medicine |
| 1 | Mokhtar N | Royal College of Medicine Perak |
| 1 | Colman, P.G. | Royal Melbourne Hospital, Australia |
| 1 | Tress, B | Royal Melbourne Hospital, Australia |
| 1 | Muhd Yusof A.B | Rumah Sakit Angkatan Tentera, |
| 1 | Al-Mansoob, M.A.K | Sana'a University, Yemen |
| 1 | Al-Mazzah, M.M. | Sana'a University, Yemen |
| 1 | Chan, G. | Sarawak General Hospital |
| 1 | Chang, L.K | Sarawak General Hospital |
| 1 | Khairul Faizi A | Sarawak General Hospital |
| 1 | Lee, P.I. | Sarawak General Hospital |
| 1 | Liew, N.S | Sarawak General Hospital |
| 1 | Loi, H.D.K | Sarawak General Hospital |
| 1 | Noor Zairul M. | Sarawak General Hospital |
| 1 | Norzalina E. | Sarawak General Hospital |
| 1 | Parhr, A.S | Sarawak General Hospital |
| 1 | Rozario, F | Sarawak General Hospital |
| 1 | Sim, K.H. | Sarawak general Hospital |
| 1 | Soh, H.L | Sarawak General Hospital |
| 1 | Subramaniam, S.K | Sarawak General Hospital |
| 1 | Tan, F. | Sarawak General Hospital |
| 1 | Tan, S.Z. | Sarawak General Hospital |
| 1 | Wong, C.C | Sarawak General Hospital |
| 1 | Wong, S.H | Sarawak General Hospital |
| 1 | Zabri K | Sarawak General Hospital, Kuching |
| 1 | Chan, L.G | Sarawak General Hospital. |
| 1 | M. Ameenudeen S.A.K. | Sarawak General Hospital. |
| 1 | Ng, H.P | Sarawak General Hospital. |
| 1 | Jamal A | Selangor Medical Centre. |
| 1 | Prema, R | Selangor State Health Dept. Disease Control Unit. |
| 1 | Lim, A.K.E | Selayang Hospital |
| 1 | Lim, K.S | Selayang Hospital |
| 1 | Rashdeen F.M.N | Selayang Hospital |
| 1 | V. Ulagantheran | Selayang Hospital |
| 1 | Pathmanathan, V. | Selayang Hospital. |
| 1 | Yap, C.M | Sentosa Medical Centre |
| 1 | Mohd Safdar N.A. | Seremban General Hospital. |
| 1 | Siow, K.Y | Seremban General Hospital. |
| 1 | Gan, W.H | Seremban Hospital |
| 1 | Yusuf W.S. | Seremban Hospital |
| 1 | Chiu, C.K | Seremban Hospital. |
| 1 | Singh, H | Seremban Hospital. |
| 1 | Ghanbarian, A | Shaheed Beheshti University of Medical Sciences, Iran |
| 1 | Hadaegh, F. | Shaheed Beheshti University of Medical Sciences, Iran |
| 1 | Harati, H. | Shaheed Beheshti University of Medical Sciences, Iran |
| 1 | Rezaei-Ghaleh, N | Shaheed Beheshti University of Medical Sciences, Iran |
| 1 | Salehi, P | Shaheed Beheshti University of Medical Sciences, Iran |
| 1 | Zabetian, A | Shaheed Beheshti University of Medical Sciences, Iran |
| 1 | Lenton, E.A. | Sheffield Fertility Centre, United Kingdom. |
| 1 | Chai, S.C. | Sibu Hospital. |
| 1 | Wong, C.W | Sibu Hospital. |
| 1 | Chow, W.C | Singapore General Hospital |
| 1 | Gue, C.S. | Singapore General Hospital |
| 1 | Kwek, A.B.E | Singapore General Hospital |
| 1 | Lui, H.F. | Singapore General Hospital |
| 1 | Luman, W | Singapore General Hospital |
| 1 | Ng, H.S. | Singapore General Hospital |
| 1 | Ong, W.C | Singapore General Hospital |
| 1 | Tan, E.K | Singapore General Hospital |
| 1 | Tan, H.M | Singapore General Hospital |
| 1 | Vathsala, A | Singapore General Hospital |
| 1 | Woo, K.T. | Singapore General Hospital |
| 1 | Yap, C.K | Singapore General Hospital |
| 1 | Yew, B.S. | Singapore General Hospital |
| 1 | Chiang, Gilbert S.C | Singapore General Hospital. |
| 1 | Soon, J.L | Singapore Health Services |
| 1 | Peh, W.C.G | Singapore Medical Journal |
| 1 | Teng, W.D. | SIRIM Berhad. Ceramics Technology Group |
| 1 | Marican, A.M | Southport District General Hospital |
| 1 | Ayob A | State Epidemiological Unit, Perak |
| 1 | Selviedran, N. | State Epidemiological Unit, Perak |
| 1 | Rosnah I | State Health Department |
| 1 | Bina Rai, S | State Health Department, Penang |
| 1 | Flora Ong. | State Health Department, Sarawak |
| 1 | Anita S. | State Health Dept |
| 1 | Rahimah M.A | State Health Dept. of Selangor. |
| 1 | Lock, L.T | State University of New York |
| 1 | Tzanakakis, E.S | State University of New York |
| 1 | Choong, S.N | StemLife Berhad, Malaysia. |
| 1 | Goh, E.H. | StemLife Berhad, Malaysia. |
| 1 | Kamalan, A | **StemLife Berhad, Malaysia.** |
| 1 | Lee, M.J | **StemLife Berhad, Malaysia.** |
| 1 | Ng, Y.K | **StemLife Berhad, Malaysia.** |
| 1 | Saraswathy, S. | **StemLife Berhad, Malaysia.** |
| 1 | Wong, Y.T | **StemLife Berhad, Malaysia.** |
| 1 | Yan, S.V | **StemLife Berhad, Malaysia.** |
| 1 | Ng, S.C | **Subang Jaya Medical Centre** |
| 1 | Pathmanathan, R | **Subang Jaya Medical Centre** |
| 1 | Santahapan, S. | **Subang Jaya Medical Centre** |
| 1 | Pee, S. | **Sultanah Aminah Hospital** |
| 1 | Thevarajah, B | **Sultanah Aminah Hospital** |
| 1 | Yap, Y.C. | **Sultanah Aminah Hospital** |
| 1 | Loh, C.S | **Sunway Medical Centre** |
| 1 | Mohammadi, G | **Tabriz University of Medical Science, Iran** |
| 1 | Naderpour, M. | **Tabriz University of Medical Science, Iran** |
| 1 | Sayyah Meli, M.R | **Tabriz University of Medical Science, Iran** |
| 1 | Pang, K.P | **Tan Tock Seng Hospital** |
| 1 | Siow, J.K. | **Tan Tock Seng Hospital** |
| 1 | Arumainathan, U.D | **Tawakal Hospital** |
| 1 | Norie A | **Tengku Ampuan Afzan** |
| 1 | Shaharudin M.H | **Tengku Ampuan Afzan** |
| 1 | Vasantha, T. | **Timur Laut, Penang** |
| 1 | Mohd Yusuf S. | **Tuanku Jaafar Hospital.** |
| 1 | Raman, S. | **Tuanku Jaafar Hospital.** |
| 1 | Low, K.C | **UKM Medical Centre** |
| 1 | Mohd Adha P.R | **UKM Medical Centre.** |
| 1 | Ali M | **Universiti Darul Iman** |
| 1 | Ahmad Murad Z | **Universiti Islam Antarabangsa. Kulliyyah of Medicine** |
| 1 | Alik R.Z | **Universiti Islam Antarabangsa. Kulliyyah of Medicine** |
| 1 | Abd Halim A.R | **Universiti Kebangsaan Malaysia** |
| 1 | Abdullah A | **Universiti Kebangsaan Malaysia** |
| 1 | Abdullah S.A. | **Universiti Kebangsaan Malaysia** |
| 1 | Aini A.A | **Universiti Kebangsaan Malaysia** |
| 1 | Amaramalar, S.N. | **Universiti Kebangsaan Malaysia** |
| 1 | Aminuddin A | **Universiti Kebangsaan Malaysia** |
| 1 | Angela Ng, M.H. | **Universiti Kebangsaan Malaysia** |
| 1 | Anuar A | **Universiti Kebangsaan Malaysia** |
| 1 | Ariffin A. K | **Universiti Kebangsaan Malaysia** |
| 1 | Ashwaq, A.M | **Universiti Kebangsaan Malaysia** |
| 1 | Azmi M.T. | **Universiti Kebangsaan Malaysia** |
| 1 | Bahariah K. | **Universiti Kebangsaan Malaysia** |
| 1 | Chan, A.C. | **Universiti Kebangsaan Malaysia** |
| 1 | Chua, C.W | **Universiti Kebangsaan Malaysia** |

| 1 | Chua, M.K. | Universiti Kebangsaan Malaysia |
|---|---|---|
| 1 | Dhachayani, S | Universiti Kebangsaan Malaysia |
| 1 | Farah Wahida I. | Universiti Kebangsaan Malaysia |
| 1 | Faridah H.A | Universiti Kebangsaan Malaysia |
| 1 | Fauzi M.A. | Universiti Kebangsaan Malaysia |
| 1 | Fazilah A.H. | Universiti Kebangsaan Malaysia |
| 1 | Ganusegaram, T. | Universiti Kebangsaan Malaysia |
| 1 | Hapizah M.N | Universiti Kebangsaan Malaysia |
| 1 | Hesham R | Universiti Kebangsaan Malaysia |
| 1 | Hesham, M.S. | Universiti Kebangsaan Malaysia |
| 1 | Ibrahim S. | Universiti Kebangsaan Malaysia |
| 1 | Idris, M.N. | Universiti Kebangsaan Malaysia |
| 1 | Ima Nirwana S | Universiti Kebangsaan Malaysia |
| 1 | Islah M | Universiti Kebangsaan Malaysia |
| 1 | Jeyabalan, N. | Universiti Kebangsaan Malaysia |
| 1 | Kamaruddin N.A | Universiti Kebangsaan Malaysia |
| 1 | Khalid Y | Universiti Kebangsaan Malaysia |
| 1 | Liew, W.F. | Universiti Kebangsaan Malaysia |
| 1 | Lo, H.L | Universiti Kebangsaan Malaysia |
| 1 | Long, C.W | Universiti Kebangsaan Malaysia |
| 1 | Mahfudz Z. | Universiti Kebangsaan Malaysia |
| 1 | Maniam, T. | Universiti Kebangsaan Malaysia |
| 1 | Mazlina S. | Universiti Kebangsaan Malaysia |
| 1 | Md Isa S.H. | Universiti Kebangsaan Malaysia |
| 1 | Mohamad A.R | Universiti Kebangsaan Malaysia |
| 1 | Mukari, S.Z.M | Universiti Kebangsaan Malaysia |
| 1 | Naqiyah I. | Universiti Kebangsaan Malaysia |
| 1 | Ng, Y.W | Universiti Kebangsaan Malaysia |
| 1 | Nik Azlan N.M | Universiti Kebangsaan Malaysia |
| 1 | Noor Hassim I. | Universiti Kebangsaan Malaysia |
| 1 | Nor Azmi K | Universiti Kebangsaan Malaysia |
| 1 | Nor Hayati S | Universiti Kebangsaan Malaysia |
| 1 | Nor Zaidah A.H | Universiti Kebangsaan Malaysia |
| 1 | Norhayati M | Universiti Kebangsaan Malaysia |
| 1 | Norimah K. | Universiti Kebangsaan Malaysia |
| 1 | Norlaili T. | Universiti Kebangsaan Malaysia |
| 1 | Norlela S | Universiti Kebangsaan Malaysia |
| 1 | Normah C.D. | Universiti Kebangsaan Malaysia |
| 1 | Noryati M. | Universiti Kebangsaan Malaysia |
| 1 | Omar M.H | Universiti Kebangsaan Malaysia |
| 1 | Ong, F.B | Universiti Kebangsaan Malaysia |
| 1 | Ong, L.C | Universiti Kebangsaan Malaysia |
| 1 | Onhmar H.T.W.E | Universiti Kebangsaan Malaysia |
| 1 | Reddy, J.P. | Universiti Kebangsaan Malaysia |
| 1 | Rica, M.A.I | Universiti Kebangsaan Malaysia |
| 1 | Rosdinom R. | Universiti Kebangsaan Malaysia |
| 1 | Roslan Abdul A.R | Universiti Kebangsaan Malaysia |
| 1 | Roszalina R. | Universiti Kebangsaan Malaysia |
| 1 | Rozita M | Universiti Kebangsaan Malaysia |
| 1 | Rozman Z. | Universiti Kebangsaan Malaysia |
| 1 | Ruzanna Z.Z. | Universiti Kebangsaan Malaysia |
| 1 | Sabarul Afian M | Universiti Kebangsaan Malaysia |
| 1 | Sagap, I. | Universiti Kebangsaan Malaysia |
| 1 | Samiah Yasmin A.K. | Universiti Kebangsaan Malaysia |
| 1 | Shaharom M.H | Universiti Kebangsaan Malaysia |
| 1 | Shahila T | Universiti Kebangsaan Malaysia |
| 1 | Shahizon A.M.M | Universiti Kebangsaan Malaysia |
| 1 | Shamsul B.S. | Universiti Kebangsaan Malaysia |
| 1 | Sharifah Teh N.S.M.N. | Universiti Kebangsaan Malaysia |
| 1 | Suganthi, C | Universiti Kebangsaan Malaysia |
| 1 | Suraya A | Universiti Kebangsaan Malaysia |
| 1 | Syarif J | Universiti Kebangsaan Malaysia |
| 1 | Tajunisah M.E | Universiti Kebangsaan Malaysia |
| 1 | Tan, S.M.K. | Universiti Kebangsaan Malaysia |
| 1 | Vikneswaran, T | Universiti Kebangsaan Malaysia |
| 1 | Vincent, E.S. Tan, | Universiti Kebangsaan Malaysia |
| 1 | Wong, C.Y. | Universiti Kebangsaan Malaysia |
| 1 | Wong, M. | Universiti Kebangsaan Malaysia |
| 1 | Wong, Ming | Universiti Kebangsaan Malaysia |
| 1 | Yazid A.G.M | Universiti Kebangsaan Malaysia |
| 1 | Yeo, C.K | Universiti Kebangsaan Malaysia |
| 1 | Zainul M.R | Universiti Kebangsaan Malaysia |
| 1 | Zainul Rashid M.R | Universiti Kebangsaan Malaysia |
| 1 | Zaiton S. | Universiti Kebangsaan Malaysia |
| 1 | Zakinah Y., | Universiti Kebangsaan Malaysia |
| 1 | Zaleha A.M | Universiti Kebangsaan Malaysia |
| 1 | Zamzarina M.A | Universiti Kebangsaan Malaysia |
| 1 | Zawawi, M.F | Universiti Kebangsaan Malaysia |
| 1 | Mohd Heikal M.Y | Universiti Kebangsaan Malaysia and Hospital Universiti Kebangsaan Malaysia |
| 1 | Lee, B.C. | Universiti Kebangsaan Malaysia Hospital |
| 1 | Chai, Ping | Universiti Kebangsaan Malaysia Hospital. |
| 1 | Haron A | Universiti Kebangsaan Malaysia Hospital. |
| 1 | im, Y.T | Universiti Kebangsaan Malaysia Hospital. |
| 1 | Omar A.R. | Universiti Kebangsaan Malaysia Hospital. |
| 1 | Tan, H.C | Universiti Kebangsaan Malaysia Hospital. |
| 1 | Azida Z.N. | Universiti Kebangsaan Malaysia Medical Centre |
| 1 | Chau, K.H | Universiti Kebangsaan Malaysia Medical Centre |
| 1 | Hairul Nizam M.H | Universiti Kebangsaan Malaysia Medical Centre |
| 1 | Hamzah J.C | Universiti Kebangsaan Malaysia Medical Centre |
| 1 | Ibnubaidah M.A | Universiti Kebangsaan Malaysia Medical Centre |
| 1 | Ishak M.F | Universiti Kebangsaan Malaysia Medical Centre |
| 1 | Norhamdan M.Y | Universiti Kebangsaan Malaysia Medical Centre |
| 1 | Zainudin A | Universiti Kebangsaan Malaysia Medical Centre |
| 1 | Hamid A.A | Universiti Kebangsaan Malaysia Medical Centre |
| 1 | Munirah S. | Universiti Kebangsaan Malaysia Medical Centre and UKM Medical Centre. |
| 1 | Wan Kamarul Zaman W.S | Universiti Kebangsaan Malaysia Medical Centre. |
| 1 | Hasni H. | Universiti Kebangsaan Malaysia. |
| 1 | Meah F.A | Universiti Kebangsaan Malaysia. |
| 1 | Rampal, K.G | Universiti Kebangsaan Malaysia. |
| 1 | Sharifah N.A., | Universiti Kebangsaan Malaysia. |
| 1 | Wong, S.W | Universiti Kebangsaan Malaysia. |
| 1 | Goh, K.S.K | Universiti Malaya Medical Centre |
| 1 | Naicker, M. | Universiti Malaya Medical Centre |
| 1 | Ong, T.A | Universiti Malaya Medical Centre |
| 1 | Yuen, H.L | Universiti Malaya Medical Centre |
| 1 | Lua, P.L | Universiti Malaysia Sabah |
| 1 | Choo, K.E | Universiti Malaysia Sarawak |
| 1 | Ngo, C.T | Universiti Malaysia Sarawak |
| 1 | Ong, G.B. | Universiti Malaysia Sarawak |
| 1 | Tan, A.K. | Universiti Malaysia Sarawak |
| 1 | Finlay, I.G | Universiti of Wales |
| 1 | Salek, M.S | Universiti of Wales |
| 1 | Al-Edrus, S.A | Universiti Putra Malaysia |
| 1 | Arfah Hanim M. | Universiti Putra Malaysia |
| 1 | Awang Hazmi A.J | Universiti Putra Malaysia |
| 1 | Chan, C.L. | Universiti Putra Malaysia |
| 1 | Chong, Y.M | Universiti Putra Malaysia |
| 1 | Ferdaos N | Universiti Putra Malaysia |
| 1 | Fuad A | Universiti Putra Malaysia |
| 1 | George, E | Universiti Putra Malaysia |
| 1 | Goh, Y.M. | Universiti Putra Malaysia |
| 1 | Govindaraju, R | Universiti Putra Malaysia |
| 1 | Hairuzah I. | Universiti Putra Malaysia |
| 1 | Hazilawati H | Universiti Putra Malaysia |
| 1 | Heng, F.S | Universiti Putra Malaysia |
| 1 | Jalila A | Universiti Putra Malaysia |
| 1 | Jammal, A.B.E | Universiti Putra Malaysia |

| 1 | Khor, G.L. | Universiti Putra Malaysia |
|---|---|---|
| 1 | Kulanthayan, K.C.M. | Universiti Putra Malaysia |
| 1 | Kumar, S. Mahesh | Universiti Putra Malaysia |
| 1 | Law, T.H | Universiti Putra Malaysia |
| 1 | Lee, L.M. | Universiti Putra Malaysia |
| 1 | Loh, J.W | Universiti Putra Malaysia |
| 1 | Maizaton A.A. | Universiti Putra Malaysia |
| 1 | Merlin, A | Universiti Putra Malaysia |
| 1 | Mirnalini K | Universiti Putra Malaysia |
| 1 | Mohd Azlan A | Universiti Putra Malaysia |
| 1 | Mohd Azmi M.L | Universiti Putra Malaysia |
| 1 | Mujahir H | Universiti Putra Malaysia |
| 1 | Mutalib A.R | Universiti Putra Malaysia |
| 1 | Nasaruddin A.A. | Universiti Putra Malaysia |
| 1 | Noordin M.M | Universiti Putra Malaysia |
| 1 | Nor Afiah M.Z | Universiti Putra Malaysia |
| 1 | Nor Mariah A | Universiti Putra Malaysia |
| 1 | Nordiyana M | Universiti Putra Malaysia |
| 1 | Nurulaini O | Universiti Putra Malaysia |
| 1 | Othman N | Universiti Putra Malaysia |
| 1 | Raja B Hisham | Universiti Putra Malaysia |
| 1 | Raman, R | Universiti Putra Malaysia |
| 1 | Ramesh, K.N | Universiti Putra Malaysia |
| 1 | Sarmadi, V.H. | Universiti Putra Malaysia |
| 1 | Sekawi Z | Universiti Putra Malaysia |
| 1 | Seow, H.F | Universiti Putra Malaysia |
| 1 | Suhba, S.T | Universiti Putra Malaysia |
| 1 | Tan, B.L. | Universiti Putra Malaysia |
| 1 | Tan, P.O | Universiti Putra Malaysia |
| 1 | Teh, P.C. | Universiti Putra Malaysia |
| 1 | Thomas, N | Universiti Putra Malaysia |
| 1 | Thong, P.L | Universiti Putra Malaysia |
| 1 | Tong, C.K | Universiti Putra Malaysia |
| 1 | Yip, C.W | Universiti Putra Malaysia |
| 1 | Yushak A.W | Universiti Putra Malaysia |
| 1 | Zalilah M.S | Universiti Putra Malaysia |
| 1 | Zuki A.B.Z | Universiti Putra Malaysia |
| 1 | Chong, F.B. | Universiti Putra Malaysia. |
| 1 | Kaneson, N. | Universiti Putra Malaysia. |
| 1 | Loqman M.Y. | Universiti Putra Malaysia. |
| 1 | Masrudin S.S. | Universiti Putra Malaysia. |
| 1 | Ooi, Y.Y. | Universiti Putra Malaysia. |
| 1 | Rosnan H. | Universiti Putra Malaysia. |
| 1 | Zailina H. | Universiti Putra Malaysia. |
| 1 | Noor Airini I. | Universiti Putra Malaysia/ Hospital Kuala Lumpur. |
| 1 | Poh, K.S | Universiti Putra Malaysia/ Hospital Kuala Lumpur. |
| 1 | Abd Rahman I.G | Universiti Sains Malaysia |
| 1 | Abdul Kareem M.M | Universiti Sains Malaysia |
| 1 | Alias N.A. | Universiti Sains Malaysia |
| 1 | Al-Salihi K.A.M | Universiti Sains Malaysia |
| 1 | Ariff A.R.M | Universiti Sains Malaysia |
| 1 | Aziah B,D | Universiti Sains Malaysia |
| 1 | Azmi A.S | Universiti Sains Malaysia |
| 1 | Azril A | Universiti Sains Malaysia |
| 1 | Chew, K.S | Universiti Sains Malaysia |
| 1 | Choy, W.P | Universiti Sains Malaysia |
| 1 | Dinsuhaimi S | Universiti Sains Malaysia |
| 1 | Effat O | Universiti Sains Malaysia |
| 1 | Eid, M | Universiti Sains Malaysia |
| 1 | Faridah A.R. | Universiti Sains Malaysia |
| 1 | Ghani A.R.I | Universiti Sains Malaysia |
| 1 | Ghani N.B | Universiti Sains Malaysia |
| 1 | Ghazali M.M. | Universiti Sains Malaysia |
| 1 | Gurjeet, K | Universiti Sains Malaysia |
| 1 | Harun Nor Rahid S.A | Universiti Sains Malaysia |
| 1 | Hisamuddin N.A.R | Universiti Sains Malaysia |
| 1 | Huda B.Z | Universiti Sains Malaysia |
| 1 | Idzwan Z.M | Universiti Sains Malaysia |
| 1 | Imran A.G.A. | Universiti Sains Malaysia |
| 1 | Jaafar H. | Universiti Sains Malaysia |
| 1 | Jafri A.M. | Universiti Sains Malaysia |
| 1 | Jain George, P | Universiti Sains Malaysia |
| 1 | Jihan, W.S | Universiti Sains Malaysia |
| 1 | Juwita S | Universiti Sains Malaysia |
| 1 | Kamaruddin J | Universiti Sains Malaysia |
| 1 | Kantha, R | Universiti Sains Malaysia |
| 1 | Kiflie A | Universiti Sains Malaysia |
| 1 | Ling, J.M. | Universiti Sains Malaysia |
| 1 | Ling, S.S.N. | Universiti Sains Malaysia |
| 1 | Madhavan, M | Universiti Sains Malaysia |
| 1 | Magosso, E. | Universiti Sains Malaysia |
| 1 | Mani, S.A. | Universiti Sains Malaysia |
| 1 | Mat W. | Universiti Sains Malaysia |
| 1 | Md Salzihan M.S | Universiti Sains Malaysia |
| 1 | Mohd Khairi M.D | Universiti Sains Malaysia |
| 1 | Mohd Nizam I | Universiti Sains Malaysia |
| 1 | Mohd Yusoff A.A. | Universiti Sains Malaysia |
| 1 | Muhammad N.A. | Universiti Sains Malaysia |
| 1 | Naing N.N | Universiti Sains Malaysia |
| 1 | Nazarah M.Y | Universiti Sains Malaysia |
| 1 | Nazri S.M | Universiti Sains Malaysia |
| 1 | Ng, B.H. | Universiti Sains Malaysia |
| 1 | Nizam A | Universiti Sains Malaysia |
| 1 | Nordin S. | Universiti Sains Malaysia |
| 1 | Normastura A.R. | Universiti Sains Malaysia |
| 1 | Prakash, R.G | Universiti Sains Malaysia |
| 1 | Quah, B.S | Universiti Sains Malaysia |
| 1 | Rosediani M | Universiti Sains Malaysia |
| 1 | Rusli J | Universiti Sains Malaysia |
| 1 | Rusli N | Universiti Sains Malaysia |
| 1 | Salmah W.M | Universiti Sains Malaysia |
| 1 | Samarendra, S.M., | Universiti Sains Malaysia |
| 1 | Santhanam, J. | Universiti Sains Malaysia |
| 1 | Sarina S | Universiti Sains Malaysia |
| 1 | Sasidaran, R. | Universiti Sains Malaysia |
| 1 | Sayuti, R.M. | Universiti Sains Malaysia |
| 1 | Selasawati H.G. | Universiti Sains Malaysia |
| 1 | Shahidan Y | Universiti Sains Malaysia |
| 1 | Shaiful B.I | Universiti Sains Malaysia |
| 1 | Shaza A.M | Universiti Sains Malaysia |
| 1 | Shen, T.H | Universiti Sains Malaysia |
| 1 | Socorro Pieter, M | Universiti Sains Malaysia |
| 1 | Sulaiman A.S | Universiti Sains Malaysia |
| 1 | Sulaiman W.A. | Universiti Sains Malaysia |
| 1 | Sulong S | Universiti Sains Malaysia |
| 1 | Tang, T.H. | Universiti Sains Malaysia |
| 1 | Tengku Muzaffar T.S | Universiti Sains Malaysia |
| 1 | Tharakan, J | Universiti Sains Malaysia |
| 1 | Ur-Rahman, N. | Universiti Sains Malaysia |
| 1 | Van Rostenberghe, H. | Universiti Sains Malaysia |
| 1 | Wong, J.W | Universiti Sains Malaysia |
| 1 | Yuen, K.H. | Universiti Sains Malaysia |
| 1 | Zahir W.M | Universiti Sains Malaysia |
| 1 | Zainal M. | Universiti Sains Malaysia |
| 1 | Ankathil, R | Universiti Sains Malaysia Health Campus |
| 1 | Hoh, B.P | Universiti Sains Malaysia Health Campus |
| 1 | Mardziah M.D.S. | Universiti Sains Malaysia Health Campus |
| 1 | Marini M | Universiti Sains Malaysia Health Campus |
| 1 | Salmi A.A | Universiti Sains Malaysia Health Campus |

| 1 | Zahri M.K., | Universiti Sains Malaysia Health Campus |
|---|---|---|
| 1 | Baba A.A | Universiti Sains Malaysia. |
| 1 | Ja'afar R | Universiti Sains Malaysia. |
| 1 | Long, G | Universiti Sains Malaysia. |
| 1 | Mat Saim A.H | Universiti Sains Malaysia. |
| 1 | Mohamed Izham M.I | Universiti Sains Malaysia. |
| 1 | Ngai, S. | Universiti Sains Malaysia. |
| 1 | Rozainah M.Y | Universiti Sains Malaysia. |
| 1 | Syed Azhar S.S | Universiti Sains Malaysia. |
| 1 | Yusof M.I. | Universiti Sains Malaysia. |
| 1 | Abdul Gaffar R | Universiti Teknologi Malaysia |
| 1 | Abdul Majid F.A | Universiti Teknologi Malaysia |
| 1 | Sarmidi M.R | Universiti Teknologi Malaysia |
| 1 | Ishak A.K | Universiti Teknologi Mara |
| 1 | Mohd Ariff F | Universiti Teknologi MARA |
| 1 | Narimah A.H.H | Universiti Teknologi MARA |
| 1 | Nuraliza A.S | Universiti Teknologi MARA |
| 1 | Ramli M | Universiti Teknologi MARA |
| 1 | Soe-Soe-Aye | Universiti Teknologi MARA |
| 1 | Suthahar, A | Universiti Teknologi MARA |
| 1 | Abdul Rahman S | Universiti Teknologi MARA. |
| 1 | Mohamad Nawawi H | Universiti Teknologi MARA. |
| 1 | Mohamed Said M.S. | Universiti Teknologi MARA. |
| 1 | Salin N | Universiti Teknologi MARA. |
| 1 | Al-Shaham, A.A.H | Universiti Teknology Mara |
| 1 | Tan, C.Y. | Universiti Tenaga Nasional |
| 1 | Lim, V.W.L | University College Dublin, Ireland |
| 1 | Staines, A | University College Dublin, Ireland |
| 1 | Ang, H.H | University College Sedaya International, Kuala Lumpur. School of Pharmacy |
| 1 | Bahari M.B. | University College Sedaya International, Kuala Lumpur. School of Pharmacy |
| 1 | Saw, J.T. | University College Sedaya International, Kuala Lumpur. School of Pharmacy |
| 1 | Kalyani, A., | University Kebangsaan Malaysia |
| 1 | Mazzre, M. | University Kebangsaan Malaysia |
| 1 | Nur Hidayah H | University Kebangsaan Malaysia |
| 1 | Shalimar A | University Kebangsaan Malaysia |
| 1 | Nawawi H | University Kebangsaan Malaysia. |
| 1 | Sapiah S | University Kebangsaan Malaysia. |
| 1 | Usha, D.A | University Malaya |
| 1 | Jaais F. | University Malaya Medical Center |
| 1 | Ahmad Zubaidi A.L. | University Malaya Medical Centre |
| 1 | Azhar M.M. | University Malaya Medical Centre |
| 1 | Cheah, P.L | **University Malaya Medical Centre** |
| 1 | Chooi, W.K | **University Malaya Medical Centre** |
| 1 | Hari Chandran, T. | **University Malaya Medical Centre** |
| 1 | Izzuddin Poo W | **University Malaya Medical Centre** |
| 1 | Laim, C.K | **University Malaya Medical Centre** |
| 1 | Leow, C.H. | **University Malaya Medical Centre** |
| 1 | Lim, H.H | **University Malaya Medical Centre** |
| 1 | Ling, H.T | **University Malaya Medical Centre** |
| 1 | Makundala, V. | **University Malaya Medical Centre** |
| 1 | Mansor M | **University Malaya Medical Centre** |
| 1 | Mohamad J.A | **University Malaya Medical Centre** |
| 1 | Naidu, R.R | **University Malaya Medical Centre** |
| 1 | Nam, H.Y | **University Malaya Medical Centre** |
| 1 | Rajen, G. | **University Malaya Medical Centre** |
| 1 | Sham E.H | **University Malaya Medical Centre** |
| 1 | Siva Kumar, A | **University Malaya Medical Centre** |
| 1 | Soo Hoo, T.S | **University Malaya Medical Centre** |
| 1 | Tai, C.C | **University Malaya Medical Centre** |
| 1 | Wang, C.Y. | **University Malaya Medical Centre** |
| 1 | Zal, A.R. | **University Malaya Medical Centre** |
| 1 | Leong, B.D.K. | **University Malaya Medical Centre.** |
| 1 | Ayu M | **University Malaya.** |
| 1 | Bastiaan, P.K. | **University Medical Center Groningen, University of Groningen, Netherlands.** |
| 1 | Henk, J.B | **University Medical Center Groningen, University of Groningen, Netherlands.** |
| 1 | Henny, C.M. | **University Medical Center Groningen, University of Groningen, Netherlands.** |
| 1 | Roel, K | **University Medical Center Groningen, University of Groningen, Netherlands.** |
| 1 | Busscher, H.J. | **University Medical Centre Groningen** |
| 1 | Sharma, P.K | **University Medical Centre Groningen** |
| 1 | Timmer, M. | **University Medical Centre Groningen** |
| 1 | van der Houwen, E.B | **University Medical Centre Groningen** |
| 1 | van der Tas, H.F | **University Medical Centre Groningen** |
| 1 | van Horn, J.R | **University Medical Centre Groningen** |
| 1 | Verkerke, G.J | **University Medical Centre Groningen/University of Twente (The Netherlands)** |
| 1 | Akman, Y | **University of Abant Izzet Baysal, Turkey** |
| 1 | Alper, M. | **University of Abant Izzet Baysal, Turkey** |
| 1 | Arbak, P. | **University of Abant Izzet Baysal, Turkey** |
| 1 | Balbay, O | **University of Abant Izzet Baysal, Turkey** |
| 1 | Cam, K. | **University of Abant Izzet Baysal, Turkey** |
| 1 | Li, M | **University of Edinburgh, UK** |
| 1 | Mason, J.O. | **University of Edinburgh, UK** |
| 1 | Nuryastuti, T | University of Groningen, Netherlands, and Gadjah Mada University, Indonesia |
| 1 | Abd Rahim S. | University of Malaya |
| 1 | Adchalingam, K. | University of Malaya |
| 1 | Asma H. | University of Malaya |
| 1 | Azhana N.H | University of Malaya |
| 1 | Bee, P.C | University of Malaya |
| 1 | Chan, S.P | University of Malaya |
| 1 | Chang, N.L.W | University of Malaya |
| 1 | Chia, Y.C | University of Malaya |
| 1 | Chua, K.T | University of Malaya |
| 1 | Fong, M.Y | University of Malaya |
| 1 | Goh, K.L | University of Malaya |
| 1 | Goh, K.Y | University of Malaya |
| 1 | Haris A.R | University of Malaya |
| 1 | Indran, M. | University of Malaya |
| 1 | Karina R | University of Malaya |
| 1 | Kim, L.H | University of Malaya |
| 1 | Kong, W.H. | University of Malaya |
| 1 | Kuppusamy, U.R. | University of Malaya |
| 1 | Lang, C.C | University of Malaya |
| 1 | Lean, Q.Y | University of Malaya |
| 1 | Lee, M.L. | University of Malaya |
| 1 | Loganathan, A | University of Malaya |
| 1 | Loh, S.Y | University of Malaya |
| 1 | Mimiwati Z. | University of Malaya |
| 1 | Nabilah H | University of Malaya |
| 1 | Ng, C.L.L | University of Malaya |
| 1 | Noor S.M | University of Malaya |
| 1 | Phil, M | University of Malaya |
| 1 | Phipps, M.E. | University of Malaya |
| 1 | Phuah, S.J. | University of Malaya |
| 1 | Raveenthiran, R | University of Malaya |
| 1 | Rokiah P. | University of Malaya |
| 1 | Roslani A.C | University of Malaya |
| 1 | Sam, I.C. | University of Malaya |
| 1 | Saub R | University of Malaya |
| 1 | Savithri, D.P | University of Malaya |
| 1 | Selvaratnam, L | University of Malaya |
| 1 | Sengupta, S. | University of Malaya |
| 1 | Shamsuddin N | University of Malaya |
| 1 | Sureshan, S. | University of Malaya |
| 1 | Tan, J.A.M.A | University of Malaya |
| 1 | Tan, K.L. | University of Malaya |
| 1 | Tan, Y.M | University of Malaya |
| 1 | Teo, S.C | University of Malaya |
| 1 | Thaneemalai, J | University of Malaya |
| 1 | Velayutham, P | University of Malaya |
| 1 | Wan Ahmad W.A. | University of Malaya |
| 1 | Wong, Elsie M.H. | University of Malaya |
| 1 | Wong, J.H.D | University of Malaya |
| 1 | Wong, Y.L | University of Malaya |
| 1 | Yap, L.Y. | University of Malaya |
| 1 | Zaini M. | University of Malaya |
| 1 | Zakiah M.A. | University of Malaya |
| 1 | Cheok, C.Y | University of Malaya Medical Center. |
| 1 | Chai, P.F | University of Malaya Medical Centre |
| 1 | Chan, P.W.K | University of Malaya Medical Centre |
| 1 | Chua, Y.P. | University of Malaya Medical Centre |
| 1 | Deepak, A.S | University of Malaya Medical Centre |
| 1 | Gan, C.S. | University of Malaya Medical Centre |
| 1 | Harun F | University of Malaya Medical Centre |
| 1 | Josephine, P. | University of Malaya Medical Centre |
| 1 | Krshnan, H | University of Malaya Medical Centre |
| 1 | Saw, L.B. | University of Malaya Medical Centre |
| 1 | Tay, P.Y.S | University of Malaya Medical Centre |
| 1 | Thong, C.L. | University of Malaya Medical Centre |
| 1 | Vivek, A.S | University of Malaya Medical Centre |
| 1 | Yong, C.K | University of Malaya Medical Centre |
| 1 | Abd. Rashid M | University of Malaya. |
| 1 | Chai, C.C | University of Malaya. |
| 1 | Dymna, V.K | University of Malaya. |
| 1 | Elmuntser, A. | University of Malaya. |
| 1 | Gilbert W | University of Malaya. |
| 1 | Hazreen A.M | University of Malaya. |
| 1 | Leong, K.M | University of Malaya. |
| 1 | Myint Myint, S | University of Malaya. |
| 1 | Ng, K.H. | University of Malaya. |
| 1 | Noor Huzaimah H | University of Malaya. |
| 1 | Seri Diana M.A | University of Malaya. |
| 1 | Shekhar, K. | University of Malaya. |
| 1 | Sia, S.F | University of Malaya. |
| 1 | Sivakumar, K. | University of Malaya. |
| 1 | Sri Rahayu S | University of Malaya. |
| 1 | Vijayananthan, A. | University of Malaya. |
| 1 | Waran, V | University of Malaya. |

| 1 | Wong, B | University of Malaysia Sarawak |
|---|---------|-------------------------------|
| 1 | Minas, I.H | University of Melbourne, Australia |
| 1 | Hoque, M.E | University of Nottingham Malaysia Campus |
| 1 | Zainal N.H | University of Nottingham Malaysia Campus |
| 1 | Navin, I.D | University of Queensland |
| 1 | Wahab N.A. | University of Sciences of Malaysia |
| 1 | Locker, D. | University of Toronto, Canada |
| 1 | Jabar, M.F | University Putra Malaysia |
| 1 | Manohar, A. | University Putra Malaysia |
| 1 | Md Akim A | University Putra Malaysia |
| 1 | Megat Ahmad M.M.H | University Putra Malaysia |
| 1 | Mo'min N | University Putra Malaysia |
| 1 | Ooi, S.S | University Putra Malaysia |
| 1 | Radin Umar R.S | University Putra Malaysia |
| 1 | Wong, S.V | University Putra Malaysia |
| 1 | Mohd Sham K. | University Putra Malaysia. |
| 1 | Nooraudah A.R. | University Putra Malaysia. |
| 1 | Nor Shamdudin M | University Putra Malaysia. |
| 1 | Nor Azman M.Z | University Sains Malaysia |
| 1 | Urban D'Souza | University Sains Malaysia |
| 1 | Fadhilah, J. | University Technology MARA |
| 1 | Hanafiah, H | Western General Hospital, UK |
| 1 | Ngui, N.K.T | Westmead Hospital, Australia. |
| 1 | Barr, I.G. | WHO Collaborating Centre for Reference and Research on Influenza, Melbourne and Monash University Australia |
| 1 | Hampson, A.W | WHO Collaborating Centre for Reference and Research on Influenza, Melbourne and Monash University Australia |
| 1 | Akbayram, S | Yuzuncu Yil University |
| 1 | Atas, B. | Yuzuncu Yil University |
| 1 | Caksen, M.D | Yuzuncu Yil University |
| 1 | Kirimi, E | Yuzuncu Yil University |
| 1 | Oner, A.F | Yuzuncu Yil University |
| 1 | Tuncer, O. | Yuzuncu Yil University |
| 1 | Huseyin, C | Yuzuncu Yil University, Turkey |
| 1 | Azian A.A | unknown |
| 1 | Geeta, S | unknown |

Appendix 2: Names of Affiliations with Numbers of Contributors

| Affiliation | No of Authors | Country |
|---|---|---|
| UniversitI Kebangsaan Malaysia | 157 | Malaysia |
| Universiti Sains Malaysia. | 133 | Malaysia |
| Hospital Universiti Kebangsaan Malaysia | 102 | Malaysia |
| University of Malaya | 101 | Malaysia |
| Universiti Putra Malaysia | 96 | Malaysia |
| Hospital Kuala Lumpur | 64 | Malaysia |
| University of Malaya Medical Centre | 59 | Malaysia |
| Ministry of Health Malaysia | 47 | Malaysia |
| International Medical University, Seremban | 46 | Malaysia |
| International Islamic University of Malaysia | 39 | Malaysia |
| Sarawak General Hospital. | 23 | Malaysia |
| Penang Hospital. | 22 | Malaysia |
| Universiti Technology Mara | 20 | Malaysia |
| Hospital Ipoh.Perak. | 18 | Malaysia |
| Hospital Sultanah Aminah, Johor Bahru. | 18 | Malaysia |
| Hospital Selayang | 17 | Malaysia |
| Hospital Seremban | 16 | Malaysia |
| Singapore General Hospital | 16 | Singapore |
| Hospital Tunku Ampuan Afzan. | 15 | Malaysia |
| Katurba Medical College | 15 | India |
| Asian Institute of Medicine, Science and Technology, Malaysia | 11 | Malaysia |
| University Medical Centre Groningen | 11 | Netherlands |
| Queen Elizabeth Hospital, Kota Kinabalu | 11 | Malaysia |
| Hospital Pakar Sultanah Fatimah | 9 | Malaysia |
| Institute for Medical Research Malaysia | 9 | Malaysia |
| Klinik Kesihatan in Negri Sembilan | 9 | Malaysia |
| Gadjah Mada University, Indonesia | 8 | Indonesia |
| Maulana Azad Medical College and Associated Lok Nayak Hospital, G.B. Pant Hospital and Guru Nanak Eye Hospital (India). | 8 | India |
| Perak Royal College of Medicine | 8 | Malaysia |
| StemLife Berhad, Malaysia. | 8 | Malaysia |
| Malaysian Nuclear Agency | 7 | Malaysia |
| Melaka Manipal Medical College | 7 | Malaysia |
| Nagoya University (Japan) and University of Tokyo (Japan) | 7 | Japan |
| Shaheed Beheshti University of Medical Sciences, Iran | 7 | Iran |
| Universiti Malaysia Sarawak. | 7 | Malaysia |
| Yuzuncu Yil University, Turkey | 7 | Turkey |
| Putrajaya Hospital. | 7 | Malaysia |
| Guru Nanak Dev University, India. | 6 | India |
| Erciyes University, Turkey. | 5 | Turkey |
| Hospital Melaka | 5 | Malaysia |
| Kyoto Prefectural University of Medicine, Japan. | 5 | Japan |
| Subang Jaya Medical Centre | 5 | Malaysia |
| University of Abant Izzet Baysal, Turkey | 5 | Turkey |
| Harvard University, USA. Brigham and Women's Hospital | 4 | USA |
| Institut Jantung Negara | 4 | Malaysia |
| Kuching General Hospital | 4 | Malaysia |
| Malaysian Armed Forces | 4 | Malaysia |
| National University of Malaysia | 4 | Malaysia |
| Universiti Tenaga Nasional | 4 | Malaysia |
| Changi General Hospital, Singapore | 3 | Singapore |
| Damansara Specialist Hospital | 3 | Malaysia |
| Esramus Medical Center, Netherlands | 3 | Neitherlands |
| Hospital de la Foundation Rothschild, Paris | 3 | France |
| Hospital Sungai Petani | 3 | Malaysia |
| Hospital Universiti Sains Malaysia | 3 | Malaysia |
| Hulu Langat District Health Office. Disease Control Unit | 3 | Malaysia |
| Inno Bio Diagnostics Sdn Bhd, Malaysia. | 3 | Malaysia |
| Institute for Health Systems Research | 3 | Malaysia |
| Isfahan University of Medical Sciences, Iran. | 3 | Iran |
| Kobe University. | 3 | Japan |
| Mater Children's Hospital, Australia. | 3 | Australia |
| Mustafa Kemal University | 3 | Turkey |
| Raigmore Hospital (United Kingdom). | 3 | UK |
| Royal College of Medicine Perak | 3 | Malaysia |
| Sultanah Aminah Hospital | 3 | Malaysia |
| Tabriz University of Medical Science, Iran | 3 | Iran |
| Tengku Ampuan Afzan | 3 | Malaysia |
| University College Sedaya International, Kuala Lumpur. School of Pharmacy | 3 | Malaysia |
| AIMST University Malaysia. | 2 | Malaysia |
| Alexandra Hospital, Singapore. | 2 | Singapore |
| Beaufort Health Office. | 2 | Malaysia |
| Cukurova University, Turkey | 2 | Turkey |
| District Tuberculosis Centre and Hospital | 2 | Malaysia |
| Dr. Kariadi General Hospital, Indonesia. | 2 | Indonesia |
| Dr. Sardjito General Hospital, Indonesia | 2 | Indonesia |
| Hospital Taiping, Perak | 2 | Malaysia |
| Hospital University of Malaya | 2 | Malaysia |
| International University Malaysia | 2 | Malaysia |
| Ipoh Hospital | 2 | Malaysia |
| King Khalid University Hospital | 2 | Saudi |
| **Madras University, India** | **2** | **India** |
| **Makmal Kesihatan Awam Kebangsaan** | **2** | **Malaysia** |
| **Monash University, Australia** | **2** | **Austalia** |
| **Royal Melbourne Hospital, Australia** | **2** | **Australia** |
| **Sana'a University, Yemen** | **2** | **Yemen** |
| **Sibu Hospital.** | **2** | **Malaysia** |
| **Southport District General Hospital** | **2** | **Malaysia** |
| **State University of New York** | **2** | **USA** |
| **Sunway Medical Centre** | **2** | **Malaysia** |
| **Tan Tock Seng Hospital** | **2** | **Malaysia** |
| **Tuanku Jaafar Hospital.** | **2** | **Malaysia** |
| **University kerbangsan malaysia medical center** | **2** | **Malaysia** |
| **Universiti of Wales** | **2** | **UK** |
| **University College Dublin, Ireland** | **2** | **Ireland** |
| **University of Edinburgh, UK** | **2** | **UK** |
| **University of Nottingham Malaysia Campus** | **2** | **Malaysia** |
| **WHO Collaborating Centre for Reference and Research on Influenza, Melbourne and Monash University Australia** | **2** | **Australia** |
| **Ahmad Dahlan University** | **1** | **Indonesia** |
| **Armed Forces Hospital Lumut** | **1** | **Malaysia** |
| **Ar-Raudhah Biotech. Farm, Sdn. Bhd.** | **1** | **Malaysia** |
| **Arunamary Medical Centre, Klang** | **1** | **Malaysia** |
| **Australian Animal Health Laboratory (CSIRO)** | **1** | **Austalia** |
| **Cawangan Penyakit Bawaan Vektor Negeri Johor** | **1** | **Malaysia** |
| **Center for Diseases Control and Prevention, USA** | **1** | **USA** |
| **Community Residency Programme Kuala Kangsar Group** | **1** | **Malaysia** |
| **Cyberjaya University College of Medical Sciences** | **1** | **Malaysia** |
| **Department of Survey amd Mapping Malaysia** | **1** | **Malaysia** |
| **Drexel University College of Medicine (USA)** | **1** | **USA** |
| **Dumlupinar University, Turkey** | **1** | **Turkey** |
| **Fatimah Hospital, Ipoh** | **1** | **Malaysia** |
| **Haydarpasa Numune Education and Research Hospital, Turkey.** | **1** | **Turkey** |
| **Hospital Alor Setar** | **1** | **Malaysia** |
| **Hospital Batu Gajah, Perak** | **1** | **Malaysia** |
| **Hospital Batu Pahat** | **1** | **Malaysia** |
| **Hospital Lam Wah Ee** | **1** | **Malaysia** |
| **Hospital Orang Asli** | **1** | **Malaysia** |
| **Hospital Serdang, Selangor** | **1** | **Malaysia** |
| **Hospital Sultan Haji Ahmad Shah** | **1** | **Malaysia** |
| **Hospital Sultanah Bahiyah** | **1** | **Malaysia** |
| **Hospital Sungai Buloh.** | **1** | **Malaysia** |
| **Hospital Teluk Intan** | **1** | **Malaysia** |
| **Hospital Umum Sarawak.** | **1** | **Malaysia** |
| Hyderabad Medical Complex, Pakistan | 1 | Pakistan |
| IMU Clinical School, Seremban | 1 | Malaysia |
| Inno Bio Ventures Sdn Bhd | 1 | Malaysia |
| Institute of Army Health. | 1 | Malaysia |
| Institute of Paediatrics | 1 | Malaysia |
| Island Hospital (Penang). | 1 | Malaysia |
| Jabatan Kesihatan Negeri Perak | 1 | Malaysia |
| Kajang Hospital | 1 | Malaysia |
| Kedah Medical Centre | 1 | Malaysia |
| Kelana Jaya Health Clinic | 1 | Malaysia |
| King Fahd Medical Centre, Saudi Arabia | 1 | Saudi Arabia |
| Klang Hospital. | 1 | Malaysia |
| Klinik Chua (Sitiawan). | 1 | Malaysia |
| Klinik Kamil Arif, Arau | 1 | Malaysia |
| Klinik Kesihatan Seremban | 1 | Malaysia |
| Klinik Kesihatan Tanah Puteh, Sarawak. | 1 | Malaysia |
| Klinik Pergigian Hospital Pasir Mas, Kelantan | 1 | Malaysia |
| Kota Bharu General Hospital | 1 | Malaysia |
| Kuala Terengganu Hospital | 1 | Malaysia |
| Kudat Health Office, Sabah | 1 | Malaysia |
| Mahkota Medical Centre, Melaka | 1 | Malaysia |
| Malaysian Palm Oil Board | 1 | Malaysia |
| Medical Specialist Centre | 1 | Malaysia |
| Megah Medical Specialist Group, Petaling Jaya | 1 | Malaysia |
| Mersin University Turkey | 1 | Turkey |
| Mount Elizabeth Medical Centre, Singapore | 1 | Singapore |
| National Diabetes Institute, Kuala Lumpur | 1 | Malaysia |
| National Population and Family Development Board. Genetic Laboratory | 1 | Malaysia |
| National University Hospital. National University of Singapore | 1 | Singapore |
| Normah Medical Specialist Centre | 1 | Malaysia |
| Poliklinik Komuniti Peringgit | 1 | Malaysia |
| Princess Alexandra Eye Pavilion, UK | 1 | UK |
| Public Health Institute Malaysia | 1 | Malaysia |
| Public Specialist Centre, Penang. | 1 | Malaysia |
| Queen Elizabeth Hospital, South Australia | 1 | Australia |
| Rumah Sakit Angkatan Tentera, | 1 | Malaysia |
| Selangor Medical Centre. | 1 | Malaysia |
| Sentosa Medical Centre | 1 | Malaysia |
| Sheffield Fertility Centre, United Kingdom. | 1 | UK |
| SIRIM Berhad. Ceramics Technology Group | 1 | Malaysia |
| Tawakal Hospital | 1 | Malaysia |
| Timur Laut, Penang | 1 | Malaysia |
| **Universiti Darul Iman** | **1** | **Malaysia** |
| **Universiti Malaysia Sabah** | **1** | **Malaysia** |
| **University of Malaysia Sarawak** | **1** | **Malaysia** |
| **University of Melbourne, Australia** | **1** | **Australia** |
| **University of Queensland** | **1** | **Australia** |
| **University of Toronto, Canada** | **1** | **Canada** |
| **University of Twente (The Netherlands)** | **1** | **Netherlands** |
| **Western General Hospital, UK** | **1** | **UK** |
| **Westmead Hospital, Australia.** | **1** | **Australia** |

**Appendix 3. Ranked List of Article Keywords**

| Keywords | Frequency |
|---|---|
| Diabetes | 28 |
| Cancers | 13 |
| Endoscopic | 13 |
| hypertension | 13 |
| Tuberculous | 12 |
| Bone | 10 |
| Children | 10 |
| Elderly | 10 |
| Primary care | 10 |
| Clinical trials | 8 |
| Medical Students | 8 |
| Pregnancy | 8 |
| Risk-factors | 8 |
| Urinary | 8 |
| Blood pressure | 7 |
| Chronic | 7 |
| HIV/AIDS | 7 |
| Knee | 7 |
| Nasal | 7 |
| Neonatal | 7 |
| Outbreaks | 7 |
| Prevention | 7 |
| Stem cells | 7 |
| Surgery | 7 |
| Attitude | 6 |
| Congenital | 6 |
| Hospital | 6 |
| Knowledge | 6 |
| Abdominal | 5 |
| Adolescence | 5 |
| Aged | 5 |
| Benign | 5 |
| Body Mass Index | 5 |
| Dengue | 5 |
| Depression | 5 |
| diseases | 5 |
| Epidemiology | 5 |
| Hepatitis | 5 |
| **Pituitary** | **5** |
| **Posterior** | **5** |
| **Prognosis** | **5** |
| **Survival** | **5** |
| **Adults** | **4** |
| **Asthma** | **4** |
| **cataract** | **4** |
| **Cerebral** | **4** |
| **Cervical** | **4** |
| **Community** | **4** |
| **Dyslipidaemia** | **4** |
| **Emergency** | **4** |
| **Ethnicity** | **4** |
| **Foreign body** | **4** |
| **glaucoma** | **4** |
| **Mortality** | **4** |
| **Paediatric** | **4** |
| **Pancreas** | **4** |
| **Pulmonary** | **4** |
| **Thrombocytopaenia** | **4** |
| **Validity** | **4** |
| **anaesthesia** | **3** |
| **Angiotensin** | **3** |
| **Aspiration** | **3** |
| **Audit** | **3** |
| **carcinoma** | **3** |
| **Chikungunya virus** | **3** |
| **Compliance** | **3** |
| **Day care** | **3** |
| **Epistaxis** | **3** |
| **Fracture** | **3** |
| **General practice** | **3** |
| **Guidelines** | **3** |
| **Health care workers** | **3** |
| **Ischaemia** | **3** |
| **Laparoscopic** | **3** |
| **Laryngeal Mask Airway** | **3** |
| **Lupus** | **3** |
| **Malignant** | **3** |
| **Management** | **3** |
| **Maxillary sinus** | **3** |
| **Nephrotic syndrome** | **3** |
| **Oral Health** | **3** |
| Orthopaedic | 3 |
| Outcome | 3 |
| Overweight | 3 |
| Pain | 3 |
| Placenta | 3 |
| Pseudoaneurysm | 3 |
| Radiation | 3 |
| Rubella | 3 |
| Selangor | 3 |
| Severe Acute Respiratory Syndrome | 3 |
| Shoulder dislocation | 3 |
| Tracheostomy | 3 |
| Treatment | 3 |
| Acute cerebellar ataxia | 2 |
| Acute Disseminated Encephalomyelitis | 2 |
| Adrenalectomy | 2 |
| allergy | 2 |
| Anagesia | 2 |
| Anatomical | 2 |
| Anterior shoulder dislocation | 2 |
| Antibiotics | 2 |
| Aortic arch aneurysm | 2 |
| Appropriateness | 2 |
| Arterial Compliance | 2 |
| Articular cartilage | 2 |
| Artificial conduit | 2 |
| Autologous bone graft | 2 |
| Autosomal dominant | 2 |
| Bcl-2 | 2 |
| Botulinum toxin | 2 |
| Brain tumors | 2 |
| Breast | 2 |
| Bronchoscopy | 2 |
| cardiovacular | 2 |
| Case control study | 2 |
| Ceftriaxone | 2 |
| Cell culture | 2 |
| Cerebellar | 2 |
| chest | 2 |
| Child Disability | 2 |
| Chondrocytes | 2 |
| Chondrosarcoma | 2 |
| Clitoris | 2 |
| Internal consistency | 2 |
| Internal fixation | 2 |
| Intraocular pressure | 2 |
| Inverted papilloma | 2 |
| Kimura's Disease | 2 |
| Laser | 2 |
| Latissimus dorsi | 2 |
| Limb salvage | 2 |
| Malaria | 2 |
| Mastectomy | 2 |
| Mastoid | 2 |
| Medical therapy | 2 |
| Meliodosis | 2 |
| Meningioma | 2 |
| Mental Health | 2 |
| mesenchymal | 2 |
| Microalbuminuria | 2 |
| Middle ear | 2 |
| Mother | 2 |
| Mutation | 2 |
| Myringoplasty | 2 |
| Needle stick injury, | 2 |
| Neuroimaging | 2 |
| Obesity | 2 |
| Occupational divers | 2 |
| Orang Asli | 2 |
| osteomyelitis | 2 |
| Paranasal sinuses | 2 |
| Parapharyngeal space | 2 |
| Peripheral arterial disease | 2 |
| Pleomorphic adenoma | 2 |
| Pneumocephalus | 2 |
| Pneumomediastinum | 2 |
| Pre-Clinical Students | 2 |
| Pre-Operative Assessment | 2 |
| Primary Hyperparathyroidism | 2 |
| Proteinuria | 2 |
| Pyrexia | 2 |
| Quality assurance | 2 |
| Quality of Life | 2 |
| Questionnaire study | 2 |
| Radiotherapy | 2 |
| Recurrence | 2 |
| **Acne** | **1** |
| **Acoustic Neuroma** | **1** |
| **Acquired choanal atresia** | **1** |
| **Acquired subglottic stenosis** | **1** |
| **ACTH syndrome** | **1** |
| **Actinomyces infection** | **1** |
| **Acute Coronary Syndromes** | **1** |
| **Acute flaccid paralysis** | **1** |
| **Acute gastroenteritis** | **1** |
| **Acute ischaemic stroke** | **1** |
| **Acute renal failure** | **1** |
| **Acute respiratory distress syndrome** | **1** |
| **Adenoidectomy** | **1** |
| **Adhesive capsulitis** | **1** |
| **Adhesive small bowel obstruction** | **1** |
| **Admission** | **1** |
| **Advanced cancer** | **1** |
| **Aeroallergens** | **1** |
| **agro-based liquid wastes** | **1** |
| **Albuminuria** | **1** |
| **ALLHAT** | **1** |
| **Allografts** | **1** |
| **Alor Gajah** | **1** |
| **Amniotic fluids** | **1** |
| **amputation** | **1** |
| **Amyloidosis** | **1** |
| **Anaplastic large cell lymphoma** | **1** |
| **Anastomosis leak** | **1** |
| **ANCA** | **1** |
| **Ancient schwannoma** | **1** |
| **Angiogram** | **1** |
| **Angiolymphoid hyperplasic** | **1** |
| **Ankle brachial pressure** | **1** |
| **Anorectal anomaly** | **1** |
| **Anterior capsule staining** | **1** |
| **Anterior craniotomy** | **1** |
| **Anterior resection** | **1** |
| **Ante-thoracic skin-tube neo-oesophagus** | **1** |
| **Anti aging therapy** | **1** |
| **Antiemetics** | **1** |
| **Antifungal therapy** | **1** |
| **Antioxidant enzymes** | **1** |
| **Antituberculosis** | **1** |
| Antrochonal Polyp | 1 |
| Apical Vertebral Rotation | 1 |
| Aplastic anemia | 1 |
| Apoptosis | 1 |
| Aqueous extract | 1 |
| Arteriovenous malformation | 1 |
| Aseptic meningitis | 1 |
| Aseptic Non-Union | 1 |
| Asians | 1 |
| Assisted Reproduction | 1 |
| Associated-Factors | 1 |
| Atherosclerotic plaque | 1 |
| Atopy | 1 |
| Augmented | 1 |
| auricular cartilage | 1 |
| Autoantibodies | 1 |
| Autofluroscence | 1 |
| Autografts | 1 |
| Awareness | 1 |
| Base of Tongue | 1 |
| Basedow's Paraplegia | 1 |
| Beaufort | 1 |
| Behcet's disease | 1 |
| below-the knee-amputation | 1 |
| Beta-HCG | 1 |
| Bilateral corneal perforations | 1 |
| Bilateral epidural extension | 1 |
| bilayered corneal construct | 1 |
| Bioavailability | 1 |
| Bioequivalence | 1 |
| Biofilm | 1 |
| biomechanical testing | 1 |
| Bladder catheterisation | 1 |
| Blindness | 1 |
| Blood glucose | 1 |
| blood safety | 1 |
| Blood transfusion | 1 |
| Blood-borne diseases | 1 |
| Bloodstream Infection | 1 |
| Blunt chest trauma | 1 |
| Body Composition Components | 1 |
| Bogota bag | 1 |
| Bowel Care Programme | 1 |
| **Bowel Dysfunction** | **1** |
| **Brachial plexus injuries** | **1** |
| **Breathing disorder during sleep** | **1** |
| **Bronchiolitis obliterans organ*ISI*ng pneumonia** | **1** |
| **Bronchogenic** | **1** |
| **Bronchopulmonary dysplasia** | **1** |
| **Budesonide/Formoterol** | **1** |
| **Burkitt lymphoma** | **1** |
| **Burns** | **1** |
| **Cadmium** | **1** |
| **Caecal metastasis** | **1** |
| **Caecum** | **1** |
| **Calcaneum** | **1** |
| **calculi** | **1** |
| **Cancellation** | **1** |
| **Carbamazepine intoxication** | **1** |
| **Carcinoid tumour** | **1** |
| **Cardiac** | **1** |
| **Cardiopulmonary Resuscitation** | **1** |
| **Carpometacarpal joint** | **1** |
| **Case-Mix** | **1** |
| **Cashew leaf extracts** | **1** |
| **Catamenial** | **1** |
| **Catecholamine** | **1** |
| **Caucasian Nose** | **1** |
| **Cavernous arigioma** | **1** |
| **CD15** | **1** |
| **CD30** | **1** |
| **CD30 positive** | **1** |
| **CD45-CD106⁺ phenotype** | **1** |
| **CD99 positive** | **1** |
| **Centella Asiotica** | **1** |
| **Central Venous Catheter** | **1** |
| **Central/East African genotype** | **1** |
| **Ceramics** | **1** |
| **cerebellopontine angle (CPA)** | **1** |
| **Cerebrospinal Fluid** | **1** |
| **cerebrovascular disease** | **1** |
| **Cervicoparotid** | **1** |
| **Challenges** | **1** |
| **Changing Trend** | **1** |
| **Chemical Classification** | **1** |
| **Chemoport** | **1** |
| **Differentiation** | **1** |
| **Diffuse cutaneous systemic sclerosis** | **1** |
| **Diffuse Large B-Cell Lymphoma** | **1** |
| **Disability** | **1** |
| **Disabled children** | **1** |
| **Disclosure** | **1** |
| **Disinfection** | **1** |
| **Dislocation** | **1** |
| **Diving Accidents** | **1** |
| **DNA typing** | **1** |
| **DNET** | **1** |
| **Doctor-patient reiationship** | **1** |
| **Down syndrome** | **1** |
| **Dropped Nucleus** | **1** |
| **Drug Cost Prescribing Pattern** | **1** |
| **Drug Use** | **1** |
| **Duchenne Muscular Dystrophy (DMD)** | **1** |
| **Duodenal perforation** | **1** |
| **Dynamic traction** | **1** |
| **Dysfunctional Uterine Bleeding** | **1** |
| **Dystrophin Gene** | **1** |
| **E. longifolia** | **1** |
| **Ear mould impression** | **1** |
| **Ear surgery** | **1** |
| **Early complications** | **1** |
| **Early results** | **1** |
| **ears** | **1** |
| **Echovirus 11** | **1** |
| **Ectopic** | **1** |
| **Effectiveness** | **1** |
| **Efficacy** | **1** |
| **Electroencephalography** | **1** |
| **Electrogustometry** | **1** |
| **Electropherotype** | **1** |
| **ELISA** | **1** |
| **Embryonic stem cells** | **1** |
| **Emotional incontinence** | **1** |
| **Emotionalism** | **1** |
| **Empathy** | **1** |
| **Emphysematous pyelonephritis** | **1** |
| **Empirical treatment** | **1** |
| **Empty follicle syndrome** | **1** |
| **Empyema** | **1** |
| **Encephalitis** | **1** |
| **End stage renal failure** | **1** |
| **Endobronchialnodules** | **1** |
| **Endophthalmitis** | **1** |
| **Endothelial cells** | **1** |
| **Endotracheal tube** | **1** |
| **Endovascular** | **1** |
| **Engraftment** | **1** |
| **epithelial transplantation** | **1** |
| **ERCP** | **1** |
| **Esophageal perforation** | **1** |
| **Esophageal varices** | **1** |
| **Euthanasia** | **1** |
| **Evaluation tools** | **1** |
| **Excision** | **1** |
| **Exercise stress test** | **1** |
| **Exposure** | **1** |
| **External ear canal** | **1** |
| **External Fixation** | **1** |
| **Facial hemangioma** | **1** |
| **Facial Nerve Palsy** | **1** |
| **Factory workers** | **1** |
| **Familial aggregation** | **1** |
| **Familial hemophagocytic lymphohistiocytosis** | **1** |
| **Fatal** | **1** |
| **Feasibility** | **1** |
| **Female rats** | **1** |
| **fever** | **1** |
| **fibroblast** | **1** |
| **Ficus Deltoidea var. agustifolia** | **1** |
| **fillet flap** | **1** |
| **Fine needle aspiration cytology** | **1** |
| **Finite element analysis** | **1** |
| **Fish bone** | **1** |
| **Flexible Bronchoscopy** | **1** |
| **Flexor tendon** | **1** |
| **Fludrocortisone** | **1** |
| **Fluoroscopic** | **1** |
| **Focal segmental glomerulosclerosis** | **1** |
| **Folic acid** | **1** |
| **Food Allergens** | **1** |
| **Formalin dab** | **1** |
| **Frontal sinus** | **1** |
| Frozen section | 1 |
| Frozen shoulder syndrome | 1 |
| Fructosamine | 1 |
| Functional dyspepsia | 1 |
| Functional impairment, | 1 |
| Funnel technique | 1 |
| Future Physicians | 1 |
| Gastric carcinoma | 1 |
| Gastric outlet obstruction | 1 |
| Gastrointestinal Bleeding | 1 |
| Gene Interaction | 1 |
| General medical clinic | 1 |
| Generic Drugs Prescribing Rate | 1 |
| Genotoxic | 1 |
| Geriatric dentistry | 1 |
| Gestational trophoblastic disease | 1 |
| Giant aneurysm | 1 |
| Giant cell tumour of the bone and pulmonary metastases | 1 |
| Giant serpentine aneurysm | 1 |
| Girls | 1 |
| GJB2 | 1 |
| Glial gene expression | 1 |
| Glioma | 1 |
| Globulin compensation index | 1 |
| Glomerulonephritis | 1 |
| Glomus tympanicum | 1 |
| Glycaemic Excursion | 1 |
| Glycated Haemoglobin | 1 |
| Glyclosylated hemoglobin A | 1 |
| Gonococcal conjunctivitis | 1 |
| Government | 1 |
| Grafts | 1 |
| Graves' disease | 1 |
| Graves' Opthalmopathy | 1 |
| Groshong catheter | 1 |
| Group A rotavirus | 1 |
| Growing endoprosthesis | 1 |
| Growth factor | 1 |
| Growth hormone | 1 |
| Guillain-Barre syndrome | 1 |
| GVHD | 1 |
| H.pylori | 1 |
| Habits and Attitudes | 1 |
| Habitual eye rubbing | 1 |
| Haemangioma | 1 |
| Haematoma | 1 |
| Hair cells in ears | 1 |
| Hand rehabilitation | 1 |
| Hand, Foot and Mouth Disease | 1 |
| Hashimoto's thyroiditis | 1 |
| HCG | 1 |
| Health | 1 |
| Health planning | 1 |
| Health policy | 1 |
| Health-seeking behaviour | 1 |
| Hearing aids | 1 |
| Hearing screening | 1 |
| Heavy metals | 1 |
| Heel defect | 1 |
| Heimlich valve | 1 |
| Helicobacter pylori | 1 |
| Hemangiopericytoma | 1 |
| Hemongioblastoma | 1 |
| Hemophagocytic syndrome | 1 |
| Hemophiliacs | 1 |
| Hemopneumothorax | 1 |
| Hepcitotoxicity | 1 |
| Herbal Medicine | 1 |
| Heterotopic pancreas | 1 |
| High grade mucoepidermoid carcinoma | 1 |
| High-flux dialyzer membranes | 1 |
| Highly active antiretroviral therapy | 1 |
| Hill-Sachs lesion | 1 |
| Hip | 1 |
| Hip dislocation | 1 |
| Hip transplant | 1 |
| Histological recurrence | 1 |
| Histoplasmosis | 1 |
| HMW-CK | 1 |
| Hoemophagocytic syndrome | 1 |
| Hoemostosis | 1 |
| home Confinement | 1 |
| Homer's Syndrome | 1 |
| Honey | 1 |
| Hormone cement therapy | 1 |
| Horner's syndrome | 1 |
| Hospital Universiti Kebangsaan | 1 |
| Housekeeping genes | 1 |
| HPV | 1 |
| Human bone marrow | 1 |
| Human cells | 1 |
| Human immunodeficiency virus | 1 |
| Humanistic | 1 |
| Humeral head fracture-dislocation | 1 |
| Humeral head migration | 1 |
| HUSM | 1 |
| Hybrid Capture II | 1 |
| Hydatidiform mole | 1 |
| Hydrothermal process | 1 |
| Hydroxyapatite (HA, mechanical properties | 1 |
| Hydroxyapatite(HA) powder | 1 |
| Hyperbaric oxygen therapy | 1 |
| Hyperglycaemic Range | 1 |
| Hypernasality | 1 |
| Hyperprolactinaemia | 1 |
| Hypocalcaemia | 1 |
| Hyponasality | 1 |
| Hypophysectomy | 1 |
| Hypothyroidism | 1 |
| IADL | 1 |
| Iatrogenic | 1 |
| Iatrogenic perforation | 1 |
| Idiopathic peripheral neuropathy | 1 |
| Idiopathic Scoliosis | 1 |
| Iliac fossa | 1 |
| Ilizarov | 1 |
| Image analyzer | 1 |
| Immune properties | 1 |
| Immunocompromised patient | 1 |
| Immunohistochemistry | 1 |
| Immunomarker | 1 |
| Immunosuppressive | 1 |
| Impalpable breast lesions | 1 |
| In vitro antiplasmodial activity | 1 |
| Inadequate tuberculosis therapy | 1 |
| Inappropriate utilization | 1 |
| Incisional Hernia | 1 |
| Incontinence | 1 |
| Indication | 1 |
| Inert gas | 1 |
| Infant Birth Weight | 1 |
| Infected closed fracture | 1 |
| Infection control | 1 |
| Infectious arthritis | 1 |
| infectious diseases | 1 |
| Inferior vena caval thrombosis | 1 |
| Inflammatoty bowel disease | 1 |
| Influenza A | 1 |
| Information Leaflet | 1 |
| In-patient Hospital Phototherapy | 1 |
| Insect resistance | 1 |
| Instep Island flap | 1 |
| Insulin dependant diabetes mellitus | 1 |
| Insulinoma | 1 |
| Integrated pest management | 1 |
| Interlocking nailing | 1 |
| Intermittent | 1 |
| Intermittent claudication | 1 |
| Intermittent respiratory obstruction | 1 |
| Internal pudendal artery flap | 1 |
| Intestinal Parasitic Infections | 1 |
| Intestinal tuberculosis | 1 |
| Intra-articular injection | 1 |
| Intracerebral trauma | 1 |
| Intraclass | 1 |
| Intraclass correlation coefficient | 1 |
| Intracranial Aneurysm | 1 |
| Intracytoplasmic Sperm Injection | 1 |
| Intraocular lens | 1 |
| Intraoperative consultation | 1 |
| Intravenous drug user | 1 |
| Intravenous Immunoglobulin | 1 |
| Intussusception | 1 |
| Invasive ductal carcinoma | 1 |
| In-vitro Fertilisation | 1 |
| Iron | 1 |
| Iron Deficiency Anaemia | 1 |
| Issues | 1 |
| Jack | 1 |
| JCQ | 1 |
| Jejunal serosal patch | 1 |
| Jejunum | 1 |
| **Job Dissatisfaction** | **1** |
| **Job Strain Model** | **1** |
| **Job-related depression** | **1** |
| **KAP study** | **1** |
| **Karasek's Job Content Questionnaire** | **1** |
| **Kelantan** | **1** |
| **Kencing Manis** | **1** |
| **Kidney** | **1** |
| **KKM Hospitals** | **1** |
| **Klang Valley** | **1** |
| **Knotting, Breakage** | **1** |
| **Kota Bharu** | **1** |
| **Kuala Kangsar district** | **1** |
| **Kudat** | **1** |
| **Laboratory technicians** | **1** |
| **Lacrimal stents** | **1** |
| **Lactation** | **1** |
| **Lactic acidosis** | **1** |
| **Laryngeal Infections** | **1** |
| **Laryngeal Tube** | **1** |
| **Laryngectomy** | **1** |
| **Laser Tonsillotomy** | **1** |
| **Late-onset congenital adrenal hyperplasia (CAH)** | **1** |
| **Lateral sinus thrombosis(LST)** | **1** |
| **Latex agglutination** | **1** |
| **Learning resources** | **1** |
| **Lecturers** | **1** |
| **Legionella** | **1** |
| **Lens dislocation** | **1** |
| **Leptospirosis** | **1** |
| **Leucopenia** | **1** |
| **Lifestyle** | **1** |
| **Likelihood Ratio** | **1** |
| **limb numbness** | **1** |
| **Limited cutaneous systemic sclerosis** | **1** |
| **Linear Regression** | **1** |
| **Linezolid** | **1** |
| **Lingual thyroid** | **1** |
| **Lipid peroxidation** | **1** |
| **Lipids** | **1** |
| **Lipoprotein(a)** | **1** |
| **Lisfranc fracture** | **1** |
| **Long bone metastasis** | **1** |
| **Loss of heterozygosity** | **1** |
| **Low Dose Spinal Anaesthesia** | **1** |
| **Lung cavities** | **1** |
| **Lung Inflamatory** | **1** |
| **Luteal phase** | **1** |
| **Lymphadenopathy** | **1** |
| **Lymphoma** | **1** |
| **M. Pneumoniae** | **1** |
| **Magnesium ions** | **1** |
| **Magnetic resonance arteriography** | **1** |
| **Magnetic resonance imaging** | **1** |
| **Major Amputation** | **1** |
| **Malabsorption syndrome** | **1** |
| **Male** | **1** |
| **Malnutrition** | **1** |
| **Mammography** | **1** |
| **Mangenese** | **1** |
| **Manual In-Line Neck Stabilization** | **1** |
| **Massive epistaxis** | **1** |
| **Massive, Embolization** | **1** |
| **Maternal weight** | **1** |
| **Maxillary Antrum** | **1** |
| **Maxillectomy** | **1** |
| **Mechanical properties** | **1** |
| **Meconium ileus obstruction** | **1** |
| **Median Blood Glucose** | **1** |
| **Mediastinal abscess** | **1** |
| **Medical audit** | **1** |
| **Membranous glomerulonephritis** | **1** |
| **Mercury** | **1** |
| **Mesh Repair** | **1** |
| **Mesocolic hernia** | **1** |
| **Metacarpophalangeal joint** | **1** |
| **Metal on metal articulation** | **1** |
| **metallic bone plates** | **1** |
| **Metered-dose inhaler** | **1** |
| **Methods of suicide** | **1** |
| **Methyl methacrylate** | **1** |
| **Methylene blue** | **1** |
| **MIC** | **1** |
| **Micronutrient Deficiency** | **1** |
| **Microsatelite instability** | **1** |
| **Microsatellite markers** | **1** |
| Middle cerebral artery aneurysm | 1 |
| Mid-term result | 1 |
| Migraine | 1 |
| Migrant worker | 1 |
| Military Diving | 1 |
| Mineralocorticoid deficiency | 1 |
| MMQOL-CSF | 1 |
| Mobile bearing | 1 |
| Mobile home phototherapy | 1 |
| Modified traction | 1 |
| Mold | 1 |
| Mondini's dysplasia, | 1 |
| Morning sickness | 1 |
| Motorcycle road crash | 1 |
| MRI | 1 |
| MSC cell cuture | 1 |
| Mucinous odenoccircinomct | 1 |
| Mucocele | 1 |
| Mucociliary action | 1 |
| Multicultural | 1 |
| Multiple choroiditis | 1 |
| Multiple Endcrine Neoplasia | 1 |
| Multisegmented Hook-Rod System | 1 |
| Muscle Infections | 1 |
| Muscle Strength Measurement | 1 |
| Mycobacterium infection | 1 |
| Mycophenolate mofetil | 1 |
| Myeloclysplasia | 1 |
| Myeloid leukemia | 1 |
| Myocardial infarct | 1 |
| Myocarditis | 1 |
| Nasopharynx | 1 |
| National Medicines Use Survey | 1 |
| Nausea and vomiting of pregnancy | 1 |
| Neck Abscess | 1 |
| Necrot*ISI*ng Fasciitis | 1 |
| Negeri Sembilan | 1 |
| Neisseria meningitidis | 1 |
| Nephrectomy | 1 |
| Nerve defect | 1 |
| Nerve graft | 1 |
| Nerve injuries | 1 |
| Nerve regeneration | 1 |
| **Neural lube defect** | **1** |
| **Neurogenetics** | **1** |
| **Neurological examination** | **1** |
| **Neuromuscular complications of hyperthyroidism** | **1** |
| **Neuronal** | **1** |
| **Neuronal Apoptosis Inhibitory Protein (NAIP) gene** | **1** |
| **Neutropenia** | **1** |
| **Newborn** | **1** |
| **Nipah virus** | **1** |
| **Nitrous oxide pollution** | **1** |
| **Nocardiosis** | **1** |
| **Noce** | **1** |
| **Nocturia** | **1** |
| **Nocturnal polyuria** | **1** |
| **Nodular fasciitis** | **1** |
| **Nodules** | **1** |
| **Noise induced hearing loss** | **1** |
| **Non randomized** | **1** |
| **Non Response (NR)** | **1** |
| **Non Shockable Rhythms** | **1** |
| **Non ulcer dyspepsia** | **1** |
| **Non Von-Hippel Lindau** | **1** |
| **Non-Hodgkin's** | **1** |
| **Non-invasive** | **1** |
| **Non-union** | **1** |
| **Normal Weight** | **1** |
| **North East Malaysia** | **1** |
| **Nucleoside analogue** | **1** |
| **Nutritional intervention** | **1** |
| **Nutritional Status** | **1** |
| **Obstructed paraesophageal hernia** | **1** |
| **Occupational exposure** | **1** |
| **Ocular tuberculosis** | **1** |
| **Oesophageal Cancer** | **1** |
| **Oesophageal varices** | **1** |
| **Open label** | **1** |
| **Open septorhinoplasty** | **1** |
| **Open-access** | **1** |
| **Operating Rooms** | **1** |
| **Operative procedure** | **1** |
| **Operative technique.** | **1** |
| **Opthamology** | **1** |
| **Optic disc angioma** | **1** |
| Orchitis | 1 |
| Organic farming. | 1 |
| Oriental Nose | 1 |
| Orientals | 1 |
| Ossifying fibromyxoid tumor | 1 |
| Osteosarcoma | 1 |
| Osteosarcoma and Limb salvage surgery | 1 |
| Osteosynthesis Plate | 1 |
| Otolaringology | 1 |
| Otosclerosis | 1 |
| Out of Hospital Cardiac Arrest | 1 |
| Ovarian Stimulation | 1 |
| Oxidative stress | 1 |
| P16 gene | 1 |
| P63 | 1 |
| Painless swelling | 1 |
| Palliative care | 1 |
| Palm oil mill effluent | 1 |
| Paraganglioma | 1 |
| Parapharyngeal | 1 |
| Parasite | 1 |
| Parathyroid hormone | 1 |
| Parathyroidectomy | 1 |
| Parotid Mass | 1 |
| Parotidectomy | 1 |
| Pars Plana Vitrectomy | 1 |
| Partial Response (PR) | 1 |
| Passive mobilization | 1 |
| Passive smoking | 1 |
| Patella | 1 |
| Patellar thickness | 1 |
| Pathological crying | 1 |
| Pathological fracture | 1 |
| Patients with Pulmonary Tuberculosis | 1 |
| Pattern and Perceptions | 1 |
| PBSCT Chronic myeloid leukemia | 1 |
| PCR | 1 |
| Peak Force | 1 |
| Pedicle Screw System and Fulcrum Bending Index | 1 |
| Penicillin | 1 |
| Peninsular Malaysia | 1 |
| Pentoxifylline | 1 |
| Peptic ulcer bleeding | 1 |
| Percutaneous transpedicular | 1 |
| Perioncil fistula | 1 |
| Perioperative Pregnancy-Related Deaths | 1 |
| Peripheral neuropathy | 1 |
| Persistent diarrhoea | 1 |
| Pesticides | 1 |
| Phacoemulsification | 1 |
| Phaeohyphomycosis | 1 |
| pharyngeal airway | 1 |
| Pharyngeal-cervical-brachial variant | 1 |
| Phenotypic variation | 1 |
| Pheochromocytoma | 1 |
| physicochemical properties | 1 |
| Physiotherapy | 1 |
| Plasma Glucose | 1 |
| Pleuritis | 1 |
| Pneumatic Reduction | 1 |
| Poediatric | 1 |
| Poisoning | 1 |
| Pollen | 1 |
| Pollutants | 1 |
| Polycystic kidney disease | 1 |
| Polypharmacy | 1 |
| porous calcium phosphate | 1 |
| Portal hypertension | 1 |
| positive pressure ventilation | 1 |
| Positive regards | 1 |
| Positivity rate | 1 |
| Post exposure prophylaxis | 1 |
| Post operative adhesions | 1 |
| Post radiotherapy | 1 |
| Post-intubation | 1 |
| Postnatal Depression | 1 |
| Postoperative outcome | 1 |
| Post-Partum Haemorrhage | 1 |
| postpartum mothers | 1 |
| Post-traumatic pseudoaneurysm | 1 |
| Practice Care | 1 |
| Preauricular sinus | 1 |
| Preconception care | 1 |
| Prednisolone | 1 |
| Pre-Eclampsia | 1 |
| Premature ovarian failure | 1 |
| Prenatal diagnosis | 1 |
| Prescribing | 1 |
| Prescribing, Primary Care | 1 |
| Prescription Drug | 1 |
| Presentation | 1 |
| Pressure Management Inventory (PMI) | 1 |
| Preterm Delivery | 1 |
| Primary ovarian lymphoma | 1 |
| Primitive neuroectodermal tumour (PNET)/medulloblastoma | 1 |
| Private | 1 |
| Problem-based learning | 1 |
| Progesterone | 1 |
| progressive neurological | 1 |
| Prolactinoma | 1 |
| Prolonged fever | 1 |
| Prophylactic Antibiotic | 1 |
| Prospective survey | 1 |
| Prostatic carcinoma | 1 |
| Protein C Deficiency | 1 |
| Protein-calorie Malnutrition | 1 |
| Psoriasis | 1 |
| Psychiatry | 1 |
| Psychological stress | 1 |
| PTEN | 1 |
| Public hospital | 1 |
| Pudendal flap | 1 |
| Punjabi Females | 1 |
| PVNS | 1 |
| Pyoderma Gangrenosum | 1 |
| Pyridoxine | 1 |
| QOL | 1 |
| QT dispersion | 1 |
| QT interval | 1 |
| QuickClot | 1 |
| Radical exc*ISI*on | 1 |
| Radical nephrectomy with inferior vena caval thrombectomy | 1 |
| Radiological assessment | 1 |
| Radiological imaging | 1 |
| Radiosurgery | 1 |
| Radiotherapy and Oncology | 1 |
| Ramadan | 1 |
| Rape myths | 1 |
| Rare | 1 |
| Rare Disorder | 1 |
| Rash | 1 |
| Raynaud's | 1 |
| Reamed Interlocking Nail | 1 |
| Rebleeding | 1 |
| Recompression chamber | 1 |
| Recompression Treatment | 1 |
| Reconstruction | 1 |
| Recreational Diving | 1 |
| rectal formalin | 1 |
| Rectovaginal fistula | 1 |
| Rectus abdominis myocutaneous flap, | 1 |
| Recurrent dislocation | 1 |
| Recurrent episodes | 1 |
| Recurrent glomerular diseases | 1 |
| Recurrent postventricular septal defect | 1 |
| Recurrent spontaneous abortion | 1 |
| regenerative medicine | 1 |
| Registry | 1 |
| Renal angiomyolipoma | 1 |
| Renal autotransplantation | 1 |
| Renal cell carcinoma | 1 |
| Replantation | 1 |
| Rescue Medication | 1 |
| Research | 1 |
| Resection | 1 |
| Resection-arthrodesis | 1 |
| Resources consumed | 1 |
| Respiratory airways | 1 |
| Respiratory obstruction | 1 |
| Respiratory Problem in Abdominoplasty | 1 |
| Restructuring | 1 |
| Resurfacing | 1 |
| Retained surgical gauze | 1 |
| Retinoblastoma | 1 |
| Retroperitoneal vascular access | 1 |
| Retroperitoneum | 1 |
| Retropharyngeal abscess | 1 |
| Retropharyngeal mass | 1 |
| Return of Spontaneous Circulation | 1 |
| Revascularisation | 1 |
| Review | 1 |
| Rheumatoid nodules | 1 |
| Rhinitis | 1 |
| Rhino - orbito - cerebral mucormycosis | 1 |
| Rhinoplasty | 1 |
| Rhinorrhoea | 1 |
| Rhinosinusitis | 1 |
| Rhinosporidiosis | 1 |
| River water faecal contamination | 1 |
| RNA-PAGE | 1 |
| Rocker bed | 1 |
| Rotating platform | 1 |
| Rural | 1 |
| Rural and Urban | 1 |
| Rural community | 1 |
| Sabah | 1 |
| Safety measures | 1 |
| Salmonella | 1 |
| Schwann Cells | 1 |
| Schwannomas | 1 |
| Sciatica | 1 |
| Seasonal Variation | 1 |
| Secondary Care | 1 |
| Secondary Hyperparathyroidism | 1 |
| Secondary School Students | 1 |
| Secondary Schools | 1 |
| Sedation | 1 |
| Selective Serotonin Reuptake Inhibitors | 1 |
| Self Esteem | 1 |
| Self-limiting | 1 |
| Self-monitoring of blood glucose | 1 |
| Self-Rated Health in Pregnancy | 1 |
| Sepsis | 1 |
| Septic Shock With Multi-Organ Dysfunction | 1 |
| Seremban | 1 |
| Seroprevalence | 1 |
| Serum globular proteins | 1 |
| Services | 1 |
| Severe open fracture of the tibia | 1 |
| Sexual assault | 1 |
| Sexual attitudes | 1 |
| Sexual health | 1 |
| Sexual knowledge | 1 |
| SF-36 | 1 |
| Shaken Baby Syndrome | 1 |
| Sheep bone marrow | 1 |
| Shift workers | 1 |
| Shockable Rhythms | 1 |
| Shock-wave | 1 |
| Shortened | 1 |
| Shoulder arthrodesis | 1 |
| Single-stranded | 1 |
| Sinonasal | 1 |
| Sinonasal malignant melanocytic melanoma | 1 |
| Sinonasal undifferentiated carcinoma | 1 |
| Skin grafting | 1 |
| Skin tissue engineering | 1 |
| Skin-sparing mastectomy | 1 |
| Skull fracture | 1 |
| Sleeve fracture | 1 |
| Small cell variant | 1 |
| Smear positive | 1 |
| Smear-negative | 1 |
| Social Health | 1 |
| Socio-Cultural Practices | 1 |
| Sociodemography | 1 |
| Soft tissue tumour | 1 |
| Solid pseudopapillary tumor | 1 |
| Solitary | 1 |
| Solitary fibrous tumour | 1 |
| Source of information | 1 |
| Spastic Children's Association Malaysia | 1 |
| Specimen radiograph | 1 |
| Speech Disorders | 1 |
| Spielberger State-Trait Anxiety Inventory | 1 |
| Spinal biopsy | 1 |
| Spinal fusion | 1 |
| Spinal Muscular Atrophy (SMA) | 1 |
| Spine injuries | 1 |
| Spleen | 1 |
| Splenic abscess | 1 |
| Splenic rupture | 1 |
| Spontaneous pneumothorax | 1 |
| Spontaneous ventilation | 1 |
| Sports | 1 |
| Standard of care | 1 |
| Standard Operating Procedure | 1 |
| Staphylococcus aureus | 1 |
| Stomach | 1 |
| Stone | 1 |
| Strangulation of bowel | 1 |
| Stress-induced premature senescence (SIPS) model | 1 |
| Stroke | 1 |
| Subarachnoid Haemorrhage | 1 |
| Subclavian artery pseudoaneurysm | 1 |
| Subclinical Cerebral infarcts | 1 |
| Subconjunctival hemorrhage | 1 |
| Subcutaneous fat | 1 |
| Subdural haemorrhage | 1 |
| Subretinal haemorrhage | 1 |
| Substance Abuse | 1 |
| Subtemporal approach | 1 |
| SUDEP | 1 |
| Suicide | 1 |
| Sulphated Glycosaminoglycans | 1 |
| Supratentorial | 1 |
| Surgical audit | 1 |
| Surgical debridement | 1 |
| Surgical exc*ISI*on | 1 |
| Surgical Outcome | 1 |
| Surgical Technique | 1 |
| Survivin expression | 1 |
| Sustained Release | 1 |
| swellings | 1 |
| Swimming | 1 |
| Symbicort | 1 |
| Sympathetic Nervous System | 1 |
| Synchronous tumour | 1 |
| Synovitis, Systemic Lupus Erythematosus | 1 |
| Syphilis | 1 |
| Systemic lupus erythematosus | 1 |
| Systemic sclerosis | 1 |
| Tako Tsubo cardiomyopathy | 1 |
| Tarsal tunnel syndrome | 1 |
| Taste Threshold | 1 |
| TB/HIV co-infection | 1 |
| Telomerase activity | 1 |
| Temporal bone resection | 1 |
| Teratogenic | 1 |
| Tertiary Hyperparathyroidism | 1 |
| Testosterone | 1 |
| Tetanus | 1 |
| Thalassaemia | 1 |
| Therapeutic effects | 1 |
| Therapeutic in pigmentation problem | 1 |
| Thermal injuries | 1 |
| Thigh reconstruction | 1 |
| Third ventricular tumor | 1 |
| Thoracic endometriosis | 1 |
| Thoracic pedicle screw | 1 |
| Throat | 1 |
| Thumb | 1 |
| Thyroid autoantibodies | 1 |
| Thyrotoxic neuropathy | 1 |
| Tlgs | 1 |
| Tolerability | 1 |
| Tongue | 1 |
| Tonometric success | 1 |
| Tonsillar hyperplasia | 1 |
| Tonsillectomy | 1 |
| Total femur endoprosthesis | 1 |
| Total hip arthroplasty | 1 |
| Total joint replacement | 1 |
| Total knee arthroplasty | 1 |
| Tourniquet test | 1 |
| Trabeculectomy | 1 |
| Tracheal mucociliary | 1 |
| Tracheal neoplasm | 1 |
| Tracheobronchial stenting | 1 |
| Tracheo-oesophageal speech | 1 |
| Traditional medicine | 1 |
| Traffic personnel | 1 |
| TRAM flap | 1 |
| Transbronchial needle aspiration | 1 |
| Transcervical | 1 |
| Transcolumellar | 1 |
| Transient otoacoustic emission | 1 |
| Translabyrinthine approach | 1 |
| Translocation | 1 |
| Transmandibular | 1 |
| Transoral | 1 |
| Transpec | 1 |
| Transseptal transphenoid surgery | 1 |
| Transsphenoidal | 1 |
| Trans-vaginal Ultrasound | 1 |
| Transverse colon | 1 |
| Tricalcium phosphate | 1 |
| Triceps | 1 |
| Trochanteric fracture | 1 |
| Tropical pyomyositis | 1 |
| Troponin T | 1 |
| Tuboovarian abscess | 1 |
| Tumorigenesis | 1 |
| Turn-around-time | 1 |
| Twin pregnancy | 1 |
| Type II | 1 |
| UKM | 1 |
| Ultrasound | 1 |
| Umbilical necrosis | 1 |
| Uncomplicated hyperbilirubinaemia | 1 |
| Undergraduate | 1 |
| Undergraduate medical curriculum | 1 |
| Undergraduate Medical Students | 1 |
| Underutilization | 1 |
| Underwater-seal | 1 |
| Underweight | 1 |
| University hospital | 1 |
| University students | 1 |
| Unrelated cord blood transplantation | 1 |
| Unsafe abortion | 1 |
| Unstable intra-articular fracture of the distal radius | 1 |
| Unusual clinical features | 1 |
| Unwanted pregnancy | 1 |
| Upper Airway Obstruction | 1 |
| Upper cervical instability | 1 |
| Upper extremity | 1 |
| Upper gastrointestinal endoscopy | 1 |
| Upper Respiratory Tract Infection | 1 |
| Use of complementary medicine | 1 |
| User-guided request form | 1 |
| USM | 1 |
| V371 | 1 |
| Vacuum extraction | 1 |
| Vagal nerve stimulation | 1 |
| Vancomycin-resistant enterococci (VRE) | 1 |
| Variceal bleeding | 1 |
| Varicose Veins | 1 |
| Varied clinical presentation | 1 |
| Vascular malformation | 1 |
| Vascular tumors | 1 |
| Vascularized fibula graft | 1 |
| Venepuncture | 1 |
| Vertebra hemangioma | 1 |
| Very low birthweight infants | 1 |
| Vesical calculus | 1 |
| Vickers hardness | 1 |
| Virkon S | 1 |
| Vitamin A Deficiency | 1 |
| VNTR | 1 |
| Vocal cord immobility | 1 |
| Vocal Cord Palsy | 1 |
| Voice prosthesis | 1 |
| Voice-Master | 1 |
| Volvulus | 1 |
| Vomiting | 1 |
| Von Hippel Lindau Disease | 1 |
| Washing Machine Injury | 1 |
| Water- bourne diseases | 1 |
| Wellness programme | 1 |
| WHO declaration of Health for all | 1 |
| Work stress | 1 |
| Wound Dressing | 1 |
| X-Linked Recessive | 1 |
| Yemen | 1 |
| Yield | 1 |
| Zidovudine | 1 |

## Appendix 4: Journal Title Rank with Frequency of Citations

| | |
|---|---|
| **Journal Of Bone & Joint Surgery** | **144** |
| **N England J Med** | **142** |
| **Lancet** | **139** |
| **Clinical Orthopedic And Related Research** | **112** |
| **Bmj** | **107** |
| **Diabetes Medicine** | **104** |
| **Laryogoscope** | **79** |
| **Jama** | **77** |
| **Chest** | **65** |
| **Plastic Reconstructive Surgery** | **58** |
| **Spine** | **53** |
| **Journal Laryngol And Oto** | **44** |
| **British J Surg** | **41** |
| **Arch Intern Med** | **40** |
| **Pediatrics** | **40** |
| **Singapore Medical Journal** | **38** |
| **Circulation.** | **37** |
| **Clinical Infect Dis** | **37** |
| **Otoloryngol Head Neck Surg** | **33** |
| **Hypertension** | **32** |
| **Arch Otolaryngology Head Neck Surgery** | **29** |
| **Am J Obstet Gynecol** | **28** |
| **Ann Intern Med** | **28** |
| **Southeast Asian Journal Of Tropical Medicine Public Health** | **28** |
| **Transplantation Proceedings** | **28** |
| **Cancer** | **26** |
| **Radiology** | **26** |
| **Journal Hypertens** | **25** |
| **Journal Infect Dis** | **25** |
| **Stroke** | **25** |
| **Am J Med** | **24** |
| **Am J Respir Crit Care Med** | **24** |
| **Am J Surgical Pathol** | **24** |
| **Journal Neurosurgery** | **24** |
| **Am J Of Epidemiology** | **23** |
| **Cancer Res** | **23** |
| **Journal Clin Microbiology** | **23** |
| **Obstet Gynecol** | **23** |
| **Ophthalmology** | **23** |
| **Trans R Soc Trop Med Hyg** | **23** |
| **Am J Surg** | **22** |
| **Arch Dermatol** | **22** |
| **Gastoenterology** | **22** |
| **Journal Am Acad Dermatol** | **21** |
| **Journal Pediatr** | **21** |
| **Am J Hypertens** | **20** |
| **Ann Otol Rhinol Laryngol** | **20** |
| **Arthritis Rheum** | **20** |
| **British J Psychiatry** | **20** |
| **Journal Clin Epidemiol** | **20** |
| **Neurology** | **20** |
| **Science** | **20** |
| **Crit Care Med** | **19** |
| **Am J Sports Med** | **18** |
| **Arch Dis Child** | **18** |
| **Journal Urol** | **18** |
| **Gastrointest Endosc** | **17** |
| **Int J Pediatr Otorhinolaryngol** | **17** |
| **Journal Hand Surg** | **17** |
| **Quality Of Life Research** | **17** |
| **Acta Psychiatr Scand** | **16** |
| **Am J Clin Nutr** | **16** |
| **Am J Gastroenterol** | **16** |
| **Am J Trop Med Hyg** | **16** |
| **Arch Surgery** | **16** |
| **British J Cancer** | **16** |
| **Diabetes Respiratory Clinical Practice** | **16** |
| **Diabetologica** | **16** |
| **Journal Neurol Neurosurg Psychiatry** | **16** |
| **Neurosurgery** | **16** |
| **Thorax** | **16** |
| **Am J Opthalmol** | **15** |
| **Am J Otoloryngol** | **15** |
| **Ann Acad Med Singapore** | **15** |
| **Blood** | **15** |
| **Cochrane Database Syst Rev** | **15** |
| **Journal Clin Endocrinol Metab** | **15** |
| **Medical J Aust** | **15** |
| **Acta Orthop Scandinavia** | **14** |
| **Ann Plast Surg** | **14** |
| **Ann Surg** | **14** |
| **Journal Hum Hypertens** | **14** |
| **Journal Trauma** | **14** |
| **Respir Med** | **14** |
| **Am J Cardiol** | **13** |
| **Am J Kidney Dis** | **13** |
| **British J Of Haematology** | **13** |
| **Hum Reprod** | **13** |
| **Injury** | **13** |
| **Journal Am Coll Cardiol** | **13** |
| **Journal Pediatric Surgery** | **13** |
| **Am Fam Physician** | **12** |
| **Am Rev Respir Dis** | **12** |
| **Anesthesia And Analgesia** | **12** |
| **Arch Ophthalmol** | **12** |
| **Asia Pacific Journal Public Health** | **12** |
| **British J Anaesth** | **12** |
| **Gut** | **12** |
| **Journal Mater Sci** | **12** |
| **Mayo Clinic Proceedings** | **12** |
| **Medical Care** | **12** |
| **Medical Education** | **12** |
| **Otolaryngologic Clinics Of North America** | **12** |
| **Postgradl.Med** | **12** |
| **Urology** | **12** |
| **Am J Public Health** | **11** |
| **Anaesthesia** | **11** |
| **Asia Pacific J Pharmacol** | **11** |
| **British J Ophthalmol** | **11** |
| **Cell.** | **11** |
| **Diabetes** | **11** |
| **Eur J Cancer** | **11** |
| **Journal Pediatr Gastroenterol Nutr** | **11** |
| **Journal Pediatr Orthop** | **11** |
| **Sleep** | **11** |
| **Ann Pharmacother** | **10** |
| **Asia Pacific Journal Of Clinical Nutrition** | **10** |
| **Clinical Geriatr Med** | **10** |
| **Contraception** | **10** |
| **Eur Heart J** | **10** |
| **Fertil Steril** | **10** |
| **Journal Rheumatol** | **10** |
| **Journal Vasc Surg** | **10** |
| **Nature** | **10** |
| **Orthop Clin N A** | **10** |
| **Acta Neurochir(Wien)** | **9** |
| **Am J Rhinology** | **9** |
| **Anticancer Res** | **9** |
| **Arthroscopy** | **9** |
| **Australian Family Physician** | **9** |
| **Blood Pressure Monitoring** | **9** |
| **British J Obstet Gynaecol** | **9** |
| **Clinical Endocrinol** | **9** |
| **Emerg Infect Dis** | **9** |
| **Head And Neck** | **9** |
| **Journal Epid And Comm Health** | **9** |
| **Journal Int Med** | **9** |
| **Kidney Int** | **9** |
| **Saudi Medical Journal** | **9** |
| **Stem Cells** | **9** |
| **Am Heart J** | **8** |
| **Am J Roentgenol** | **8** |
| **Arch Neurol** | **8** |
| **Archives Of General Psychiatry** | **8** |
| **British J Dermatol** | **8** |
| **Cochrane Library** | **8** |
| **Eur Respir J** | **8** |
| **Int J Epidemiolo** | **8** |
| **Int J Of Occupational Medicine And Environmental Health** | **8** |
| **Int J Tuberc Lung Dis** | **8** |
| **Journal Allergy Clin Immunol** | **8** |
| **Journal Clin Pathol** | **8** |
| **Journal Paediatr Child Health** | **8** |
| **Journal Spinal Dis** | **8** |
| **Malaysian Journal Of Medical Sciences** | **8** |
| **Nephrol Dial Transplant** | **8** |
| **Neuroradiology** | **8** |
| **Pediatr Infect Dis** | **8** |
| **Proc Natl Acad Sci Usa** | **8** |
| **Q J Med** | **8** |
| **Surg Gynecol Obstet** | **8** |
| **Acta Otorrinolaringol Esp** | **7** |
| **Adolescence** | **7** |
| **Allergy** | **7** |
| **Am J Clin Pathology** | **7** |
| **Am J Pathol** | **7** |
| **Ann Rheum Dis** | **7** |
| **Arch Pathol Lab Med** | **7** |
| **Atherosclerosis** | **7** |
| **Bju Int** | **7** |
| **Clinical Otolaryngology** | **7** |
| **Clinical Sci** | **7** |
| **Dis Colon Rectum** | **7** |
| **Family Physician** | **7** |
| **Family Practice** | **7** |
| **Heart** | **7** |
| **Journal Am Geriatr Soc** | **7** |
| **Journal Bone Miner Res** | **7** |
| **Journal Reconstr Microsurgery** | **7** |
| **Journal Trop Ped Environ Child Hlth** | **7** |
| **Medicine** | **7** |
| **N Z Med J** | **7** |
| **Occup Environ Med** | **7** |
| **Scand J Gastroenterol** | **7** |
| **World J Surg** | **7** |
| **Am J Of Emergency Medicine** | **6** |
| **Anesthesiology** | **6** |
| **Ann Neurol** | **6** |
| **Ann Thor Surg** | **6** |
| **Antimicrobial Agents And Chemotherapy** | **6** |
| **Arch Gynecol Obstet** | **6** |
| **British J Gen Pract** | **6** |
| **British J Rheum** | **6** |
| **British J Uro** | **6** |
| **Can J Anaesth** | **6** |
| **Clinical Chem** | **6** |
| **Clinical Microbiol** | **6** |
| **Clinical Nephro** | **6** |
| **Cmaj** | **6** |
| **Eur J Gastroenterol Hepatol** | **6** |
| **Hum Biol** | **6** |
| **Journal Am Diet Assoc** | **6** |
| **Journal Cataract Refrac Surg** | **6** |
| **Journal Family Practice** | **6** |
| **Journal Natl Cancer Inst** | **6** |
| **Journal Neurol Sc** | **6** |
| **Journal Shoulder Elbow Surg** | **6** |
| **Journal Thorac Cardiovasc Surg** | **6** |
| **Pain** | **6** |
| **Pathology** | **6** |
| **South Med J** | **6** |
| **Stat Med** | **6** |
| **Transfusion** | **6** |
| **Acta Derm Venereol** | **5** |
| **Acta Obstet Gynaecol Scand** | **5** |
| **Am J Psychiatry** | **5** |
| **Anz J Surg** | **5** |
| **Arterioscler Thromb Vasc Biol** | **5** |
| **Asia Pacific Family Medicine** | **5** |
| **Auris, Nasus, Larynx** | **5** |
| **Aust N Z J Surg** | **5** |
| **Bone Marrow Transplant** | **5** |
| **Brain** | **5** |
| **Clinical Exp Allergy** | **5** |
| **Clinical Obstet Gynaecol** | **5** |
| **Clinical Radiol** | **5** |
| **Dermatologica** | **5** |
| **Developmental Medicine And Child Neurology** | **5** |
| **Endoscopy** | **5** |
| **Eur J Clin Nutrn** | **5** |
| **Eur J Epidemiol** | **5** |
| **Eur J Obstet Gynecol Reprod Biol** | **5** |
| **Foot Ankle Lnt** | **5** |
| **Gerontology** | **5** |
| **Hepatology** | **5** |
| **Hum Pathol** | **5** |
| **Indian J Med Res** | **5** |
| **Int J Clin Phamacol Ther Toxicol** | **5** |
| **Int J Dermatol** | **5** |
| **Intensive Care Med** | **5** |
| **Journal Child Neurol** | **5** |
| **Journal Clin Oncol** | **5** |
| **Journal Consulting And Clinical Psychology** | **5** |
| **Journal Emerg Med** | **5** |
| **Journal Invest Derm** | **5** |
| **Journal Obstet And Gynaecol** | **5** |
| **Journal Orthop Trauma** | **5** |
| **Journal Otolaryngology And Otology** | **5** |
| **Journal Surg Oncol** | **5** |
| **Journal The American Academy Of Child And Adolesce, Psychiatry** | **5** |
| **Journal Youth And Adolescence** | **5** |
| **Malaysian Journal Of Pathology** | **5** |
| **Medical Journal Of Australia** | **5** |
| **Am J Nephrol** | **4** |
| **Am J Neuroradiol** | **4** |
| **Ann Emerg Med** | **4** |
| **Ann Epidemiol** | **4** |
| **Ann R Coil Surg Engl** | **4** |
| **Ann Saudi Med** | **4** |
| **Ann Trop Paediatr** | **4** |
| **Arch Orthop Trauma Surg** | **4** |
| **Arch Pediatr Adolesc Med** | **4** |
| **Arch Phys Med Rehab** | **4** |
| **Arch Virol** | **4** |
| **Aust N Z J Public Health** | **4** |
| **Biomaterials** | **4** |
| **British J Of Anaesthesia** | **4** |
| **Burns** | **4** |
| **Can Med Assoc J** | **4** |
| **Clinical Cancer Res** | **4** |
| **Clinical Diabetes** | **4** |
| **Clinical Paediatr** | **4** |
| **Clinical Pharmacol And Therapeutics** | **4** |
| **Dent Clin North Am** | **4** |
| **Dev Med Child Neurol.** | **4** |
| **Diabetes Obes Metab** | **4** |
| **East Afr Med J** | **4** |
| **Endocrinol** | **4** |
| **Eur J Vasc Endovasc Surg** | **4** |
| **Eur Radiol** | **4** |
| **Exp Cell Res** | **4** |
| **Eye** | **4** |
| **Gen Dent** | **4** |
| **Geriatrics** | **4** |
| **Hum Mol Genet** | **4** |
| **Infect Immun** | **4** |
| **Int J Colorectal Dis** | **4** |
| **Journal Abnormal Psychology** | **4** |
| **Journal Am Coll Surg** | **4** |
| **Journal Am Soc Nephrol** | **4** |
| **Journal Asean Fed Endo Soc** | **4** |
| **Journal Clin Hypertens** | **4** |
| **Journal Clin Invest** | **4** |
| **Journal Clin Neurosci** | **4** |
| **Journal Craniofac Surg** | **4** |
| **Journal Gastroenterol** | **4** |
| **Journal Hepatol** | **4** |
| **Journal Hosp Infec** | **4** |
| **Journal Hyg** | **4** |
| **Journal Korean Med Sci** | **4** |
| **Journal Med Assoc Thai** | **4** |
| **Journal Med Ed** | **4** |
| **Journal Med Genet** | **4** |
| **Journal Med Virol** | **4** |
| **Journal Neuoro Oncol** | **4** |
| **Journal Neuropathol Exp Neurol** | **4** |
| **Journal Nutr** | **4** |
| **Journal Pathol** | **4** |
| **Journal Perinatol** | **4** |
| **Microsurgery** | **4** |
| **National Medical Journal Of Lndia** | **4** |
| **Neurol India** | **4** |
| **Oncogene** | **4** |
| **Orthopaedics** | **4** |
| **Palliat Med** | **4** |
| **Placenta** | **4** |
| **Resuscitation** | **4** |
| **S Af Med J** | **4** |
| **Social Sciences And Medicine** | **4** |
| **Spec Care Dentist** | **4** |
| **Surg Neurol** | **4** |
| **Surgery** | **4** |
| **Surgical Clinics Of North America** | **4** |
| **Tissue Eng** | **4** |
| **Trop Geog Med** | **4** |
| **Trop Med Int Health** | **4** |
| **Tubercle** | **4** |
| **Ultrasound Obstet Gynecol** | **4** |
| **Acad Med** | **3** |
| **Accident Analysis And Prevention** | **3** |
| **Acta Anaesthesiologica Scandinavia** | **3** |
| **Acta Cardiol** | **3** |
| **Acta Chir Scand** | **3** |
| **Acta Cytol** | **3** |
| **Acta Diabetol** | **3** |
| **Acta Ophthalmol** | **3** |
| **Acta Trop** | **3** |
| **Acts Paediatr Jpn** | **3** |
| **Addiction** | **3** |
| **Adv Ren Replace Ther** | **3** |
| **Adv Virus Res** | **3** |
| **Aesthetic Plast Surg** | **3** |
| **Age And Ageing** | **3** |
| **Aids** | **3** |
| **Ajnr Am J Neuradiol** | **3** |
| **Am J Hum Biol** | **3** |
| **Am J Hum Genet** | **3** |
| **Am J Med Genet** | **3** |
| **Am Psychol** | **3** |
| **Anaesth Intens Care** | **3** |
| **Angiology** | **3** |
| **Ann Nutr Metab** | **3** |
| **Ann Oncol** | **3** |
| **Arch Gen Psychiatry** | **3** |
| **Arq Neuropsiquiatr** | **3** |
| **Aust N Z J Ophthalmol** | **3** |
| **Brain Dev** | **3** |
| **Braz J Med Biol Res** | **3** |
| **British J Neurosurg** | **3** |
| **British J Nutr** | **3** |
| **Canadian Journal Of Public Health** | **3** |
| **Caries Research** | **3** |
| **Clao J** | **3** |
| **Clinical Chim Acta** | **3** |
| **Clinical Exp Hypertens** | **3** |
| **Community Dent Oral Epidemiol** | **3** |
| **Cornea** | **3** |
| **Curr Med Res Opin** | **3** |
| **Current Opinion Infect Dis** | **3** |
| **Dan Med Bull** | **3** |
| **Dermatol Clin** | **3** |
| **Dermatol Surg** | **3** |
| **Dermatol Therapy** | **3** |
| **Diabetes Metab** | **3** |
| **Disabil Rehabil** | **3** |
| **eJournal of the International AIDS Society** | **3** |
| **eMedicine Journal** | **3** |
| **Environmental Health Perspectives** | **3** |
| **Eur J Clin Microbiol Infect Dis** | **3** |
| **Eur J Haematol** | **3** |
| **Eur J Surg** | **3** |
| **Health Psychology** | **3** |
| **Hum Genet** | **3** |
| **Indian J Tuberc** | **3** |
| **Indonesian Journal of Pharmacy.** | **3** |
| **Infect Control Hosp Epidemiol** | **3** |
| **Int J Antimicrob Agents** | **3** |
| **Int J Haematol** | **3** |
| **Int J of Oncology.** | **3** |
| **Int J of Oral and Maxillofacial Surgery** | **3** |
| **Int J Urol** | **3** |
| **Intern Med** | **3** |
| **International Ophthalmol** | **3** |
| **International Orthop** | **3** |
| **Irish Med J** | **3** |
| **Journal Aerosol Med** | **3** |
| **Journal Am Acad Orthop Surg** | **3** |
| **Journal American Medical Association** | **3** |
| **Journal Antimicrob Chemother** | **3** |
| **Journal Asthma** | **3** |
| **Journal Bronchology** | **3** |
| **Journal Cancer Educ** | **3** |
| **Journal Cardiovasc Risk** | **3** |
| **Journal Chemother** | **3** |
| **Journal Clin Ultrasound** | **3** |
| **Journal Indian Med Assoc** | **3** |
| **Journal Orthop Sci** | **3** |
| **Journal Psychiatric Research** | **3** |
| **Journal R Coll Edinb** | **3** |
| **Journal School Health** | **3** |
| **Journal the ASEAN Federation of Endocrine Societies** | **3** |
| **Journal Trop Med and Hygiene** | **3** |
| **Journal Virol** | **3** |
| **Jpn J Ophthalmol** | **3** |
| **Korean J Gastroenterol** | **3** |
| **Malaysia Patent** | **3** |
| **Maturitas** | **3** |
| **Medical Teacher** | **3** |
| **Medicina del Lavoro** | **3** |
| **Nephrology** | **3** |
| **Nephron** | **3** |
| **Neuroreport** | **3** |
| **Neurourol Urodyn** | **3** |
| **No Shinkei Geka** | **3** |
| **Oncology** | **3** |
| **Pediatr Clin N Am** | **3** |
| **Pediatr Radiol** | **3** |
| **Pediatr Res** | **3** |
| **Phys Ther** | **3** |
| **Prim Care Clin Office Pract** | **3** |
| **Psycho Med** | **3** |
| **Radiographics** | **3** |
| **Radiologic Clinics of North America** | **3** |
| **Rev Neurol** | **3** |
| **Scand J Infect** | **3** |
| **Spinal cord** | **3** |
| **Yemen Med J** | **3** |
| **AJR** | **2** |
| **Alcohol and Alcoholism** | **2** |
| **Am J Anat** | **2** |
| **Am J Chin Med** | **2** |
| **Am J Dis Chil** | **2** |
| **Am J Hematol** | **2** |
| **Am J Hosp Pharm** | **2** |
| **Am J Infect Control** | **2** |
| **Am J on Mental retardation** | **2** |
| **Am J Perinatol** | **2** |
| **Am J Phys Anthropol** | **2** |
| **Ann Acad** | **2** |
| **Ann Allergy** | **2** |
| **Ann Allergy Asthma Immunol** | **2** |
| **Ann Biomed Res Edu** | **2** |
| **Ann Diagn Pathol** | **2** |
| **Ann Fam Med** | **2** |
| **Ann N Y Aca Sci** | **2** |
| **Ann Rev Biochem** | **2** |
| **Ann Rev Med** | **2** |
| **Ann Surg Oncol** | **2** |
| **Ann Thorac Cardiovasc Surg** | **2** |
| **Arch Fam Med** | **2** |
| **Arch Neurol Pshych** | **2** |
| **Arch of Phys Med and Rehab** | **2** |
| **Arch Orthop.** | **2** |
| **Archives of Pathology and Laboratory Medicine** | **2** |
| **Asian Journal of Surgery** | **2** |
| **Asian Med J** | **2** |
| **Asian Pacific Journal of Allergy and Immunology** | **2** |
| **Aust New Zealand J Psychiat** | **2** |
| **Aust Paediatri J** | **2** |
| **Bailliere's Clinical Anaesthesiology** | **2** |
| **Best Pract Res Clin Obstet Gynaecol** | **2** |
| **Biochemical and Biophysical Research Communication** | **2** |
| **BJOG** | **2** |
| **Blackwell Science** | **2** |
| **BMC Fam Pract** | **2** |
| **Bone** | **2** |
| **Braz J Infect Dis** | **2** |
| **British Heart J** | **2** |
| **British J Clin Pharmacol** | **2** |
| **British J Oral Maxillofac Surg** | **2** |
| **British J Radiol** | **2** |
| **Bull Exp Biol Med** | **2** |
| **Cadernos de Saude publica** | **2** |
| **Can J Cardiol** | **2** |
| **Can J Psychiatry** | **2** |
| **Card Electophysiology Rev** | **2** |
| **Ceram, Inter** | **2** |
| **Chem Senses** | **2** |
| **Chinese Med J** | **2** |
| **Circ** | **2** |
| **Cleft Palate-Craniofac J** | **2** |
| **Clinical Biomechanics** | **2** |
| **Clinical Diag Lab Immunol** | **2** |
| **Clinical J Sports Med** | **2** |
| **Clinical Liver Dis** | **2** |
| **Clinical Oncol** | **2** |
| **Clinical Updates** | **2** |
| **Clinics in Dermatology** | **2** |
| **Contemporary Surgery** | **2** |
| **Control Clinical Trials** | **2** |
| **Corrosion Science** | **2** |
| **Curr Opin Neurol** | **2** |
| **Current opinion in oncol** | **2** |
| **Current Opinion in Paediatrics** | **2** |
| **Current Opinion in Psychiatry** | **2** |
| **Current opinion Nephrol. Hypertens** | **2** |
| **Current opinion Obstet Gynecol** | **2** |
| **Current opinion Ophthalmol** | **2** |
| **Current opinion Otolaryngol Head Neck Surg** | **2** |
| **Drugs** | **2** |
| **Ear and Hearing** | **2** |
| **Ear Nose Throat J** | **2** |
| **Edin Med J** | **2** |
| **Emerg Med J** | **2** |
| **Endocrine Practice** | **2** |
| **Epilepsia** | **2** |
| **Eur J Appl Physiol** | **2** |
| **Eur J Intern Med** | **2** |
| **Eur J Neurol** | **2** |
| **Eur J Paediatr** | **2** |
| **Eur J Pain** | **2** |
| **Exp Hematol** | **2** |
| **FEBS Lett** | **2** |
| **Folia Phoniatr Logop** | **2** |
| **Food Chemistry** | **2** |
| **Gen Hosp Psychiatry** | **2** |
| **Gene Therapy** | **2** |
| **Genet Med** | **2** |
| **Gynecol Endocrinol** | **2** |
| **Haematologica** | **2** |
| **Headache** | **2** |
| **Hernia** | **2** |
| **Histopathology** | **2** |
| **Human and Experimental Toxicology** | **2** |
| **Immunology and Allergy clinics of North America** | **2** |
| **Indian J Pediatr** | **2** |
| **Indian Journal of Otolaryngology and Head and Neck Surgery** | **2** |
| **Infection** | **2** |
| **Int Angiol** | **2** |
| **Int Arch Occup Enviro Health** | **2** |
| **Int Heart J** | **2** |
| **Int J Cancer** | **2** |
| **Int J Nurs Stud** | **2** |
| **Int J Obes** | **2** |
| **Int J of Clin Practice** | **2** |
| **Int J Radiat Oncol Biol Phys** | **2** |
| **Int J STD and AIDS** | **2** |
| **International Medical Journal** | **2** |
| **International Quarterly of Community Health Education** | **2** |
| **Invest Ophthalmol Vis Sci** | **2** |
| **Ital Heart J** | **2** |
| **JACC** | **2** |
| **JAFES** | **2** |
| **Journal Acquir Immune Defic Syndr Hum Retrovirol** | **2** |
| **Journal Adv Nurs** | **2** |
| **Journal Affective Disorders** | **2** |
| **Journal Aging and Health** | **2** |
| **Journal Am Board Fam Pract** | **2** |
| **Journal Assoc Physicians India** | **2** |
| **Journal Auton Neev System** | **2** |
| **Journal Biol Chem** | **2** |
| **Journal Biomedical Material Research** | **2** |
| **Journal Biosci** | **2** |
| **Journal Burn Care Rehabil** | **2** |
| **Journal Card Surg** | **2** |
| **Journal Cardiovasc Pharmacol** | **2** |
| **Journal Clin Gastroenterol** | **2** |
| **Journal Clinical Anaesthesia** | **2** |
| **Journal Clinical and Child Psychology** | **2** |
| **Journal computer assisted tomography** | **2** |
| **Journal Dent Edu** | **2** |
| **Journal Derm** | **2** |
| **Journal Dev Behav Pediatr** | **2** |
| **Journal Endocrinol Invest** | **2** |
| **Journal Ethnopharmacology** | **2** |
| **Journal Eval Clin Pract** | **2** |
| **Journal Exp Med** | **2** |
| **Journal Gen Intern Med** | **2** |
| **Journal Gen Virol** | **2** |
| **Journal Health Psychology** | **2** |
| **Journal Health Sci** | **2** |
| **Journal Heart valve Dis** | **2** |
| **Journal Immunol** | **2** |
| **Journal Microbiol Immunol Infect** | **2** |
| **Journal Neuroradiol** | **2** |
| **Journal Neurotrauma** | **2** |
| **Journal Occup Environ Med** | **2** |
| **Journal Occup Health** | **2** |
| **Journal Occup Health Psychol** | **2** |
| **Journal Ophthalmol** | **2** |
| **Journal Oral Surg** | **2** |
| **Journal Orthop Res** | **2** |
| **Journal Paediatr Surg** | **2** |
| **Journal Paediatric and Child Health** | **2** |
| **Journal Prosthet Dent** | **2** |
| **Journal Public Health Medicine** | **2** |
| **Journal R Soc Med** | **2** |
| **Journal Reprod Med** | **2** |
| **Journal Respir Crit Care Med** | **2** |
| **Journal Traditional Medicine** | **2** |
| **Journal Vasc Interv Radiol** | **2** |
| **Jpn J Clin Oncol** | **2** |
| **Jpn J Med Sci Biol** | **2** |
| **Lab Invest** | **2** |
| **Leukemia** | **2** |
| **Lupus** | **2** |
| **Malaysian Journal of Nutrition** | **2** |
| **Malaysian Journal of Obstetrics and Gynaecology** | **2** |
| **Malaysian Journal of Public Health Medicine** | **2** |
| **Manage Review** | **2** |
| **Mater. Lett** | **2** |
| **Materials Science and Engineering C** | **2** |
| **Medical  Vet Entomol** | **2** |
| **Microbes Infect** | **2** |
| **Mod Pathol** | **2** |
| **Mol Cell Biol** | **2** |
| **Molecular Biology of the Cell** | **2** |
| **Morbidity And Mortality Weekly Report (MMWR) October** | **2** |
| **Nat Biotechnol** | **2** |
| **Neurochirurgie** | **2** |
| **Neurol J southeast Asia** | **2** |
| **Neurological Research** | **2** |
| **Neurologist** | **2** |
| **Neuroscience** | **2** |
| **Nurse Pract** | **2** |
| **Nutrition Research** | **2** |
| **Oncol Nurs Forum** | **2** |
| **Operative Techniques In Otolaryngology-Head And Neck Surgery** | **2** |
| **Otol Rhinol Laryngol** | **2** |
| **Paediatr Perinat Epidemiol** | **2** |
| **Parasitology** | **2** |
| **Ped Infect Dis J** | **2** |
| **Pediatr Allergy Immunol** | **2** |
| **Pediatr Neurol** | **2** |
| **Pharm World Science** | **2** |
| **Pharmacotherapy** | **2** |
| **Prey Med** | **2** |
| **Prob Gen Surg** | **2** |
| **Psychiatric Clinics of North America** | **2** |
| **Psychosom Med** | **2** |
| **Reviews in Med Microbiol** | **2** |
| **Rheumatology** | **2** |
| **Rhinology** | **2** |
| **Scand J Rheumatol** | **2** |
| **Scand J Work Environ Health** | **2** |
| **Sex Transm Dis** | **2** |
| **Sur Laparosc. Endosc** | **2** |
| **Surg, Gynecol Obstet** | **2** |
| **Surv Ophthalmol** | **2** |
| **Swiss Med Wkly** | **2** |
| **Tex Heart Inst J** | **2** |
| **Thorac Cardiovasc Surg** | **2** |
| **Tobacco Control** | **2** |
| **Toxicological Sciences** | **2** |
| **Tuber Lung Dis** | **2** |
| **Tubercle and Lung Disease** | **2** |
| **Tumor Biol** | **2** |
| **Ultrasound Med Biol** | **2** |
| **UNAIDS** | **2** |
| **Violence and Victims** | **2** |
| **Virchows Arch** | **2** |
| **Virus Res** | **2** |
| **West J Med** | **2** |
| **World J Gastroenterol** | **2** |
| **Zhonghua Yi Xue Za Zhi** | **2** |
| **Acad Emerg Med** | **1** |
| **Accident Emergency Nursing** | **1** |
| **ACI international** | **1** |
| **Acta Anaesth. Belg** | **1** |
| **Acta Biochem Ateneo Parmense** | **1** |
| **Acta Cient Venez** | **1** |
| **Acta Endocrinologica** | **1** |
| **Acta haematol** | **1** |
| **Acta Histochem Cytochem** | **1** |
| **Acta Orthop Traumatol Turc** | **1** |
| **Acta Radiol** | **1** |
| **Acta Scandinavia** | **1** |
| **Addictive Behaviors** | **1** |
| **Admin Sci Quarter** | **1** |
| **Adv Endocrinol Metab** | **1** |
| **Adv Metab Dis** | **1** |
| **Adv Surg** | **1** |
| **Advanced Drug Delivery Reviews** | **1** |
| **AIDS Care** | **1** |
| **AIDS Education and Prevention** | **1** |
| **AJNR** | **1** |
| **AJR AM J Roentgenol** | **1** |
| **Aktuelle Radiol** | **1** |
| **Aliment Pharmacol Ther** | **1** |
| **Allergic et Immunologic** | **1** |
| **Am Acad Orthop Surg** | **1** |
| **Am and Deaf** | **1** |
| **Am Clin Nutr** | **1** |
| **Am Coil Cardiol** | **1** |
| **Am Epidemiol** | **1** |
| **Am Intern Med** | **1** |
| **Am J of Medical Sciences** | **1** |
| **Am J of Physidal Medicine and Rehabilitation** | **1** |
| **Am J of Respiratory and Critical Care Medicine** | **1** |
| **Am J Clin Oncol** | **1** |
| **Am J Crit Care** | **1** |
| **Am J Ger Psych** | **1** |
| **Am J Health Promotion** | **1** |
| **Am J Health-Syst Pharm** | **1** |
| **Am J Hyg** | **1** |
| **Am J of Cardiology** | **1** |
| **Am J of Health Education** | **1** |
| **Am J of Managed Care** | **1** |
| **Am J of Roent** | **1** |
| **Am J of Sociology** | **1** |
| **Am J Orthod Dentofacial Orthop** | **1** |
| **Am J Paediatr Haematol Docol** | **1** |
| **Am J Phys Med Rehabil** | **1** |
| **Am J Physiol Endocrinol Metab** | **1** |
| **Am J Plast Surg** | **1** |
| **Am J Pres Med** | **1** |
| **Am J Prev Med** | **1** |
| **Am J Roentology** | **1** |
| **Am Jof Human Genet** | **1** |
| **Am Respir Crit Care Med.** | **1** |
| **Am Rev Nutr** | **1** |
| **American Academy of Ophthalmology** | **1** |
| **American Cancer Society** | **1** |
| **American Sociological Review** | **1** |
| **Amersham Biosci.** | **1** |
| **An Med Interna** | **1** |
| **Analytical Biochemistry** | **1** |
| **Anat Embryol** | **1** |
| **Anesthesiol Clin North America** | **1** |
| **Ann Agricultural and Environmental Medicine.** | **1** |
| **Ann Behav Med** | **1** |
| **Ann Biochem** | **1** |
| **Ann Biomed Eng** | **1** |
| **Ann Burns and Fire Disasters** | **1** |
| **Ann Chir Gynaecol** | **1** |
| **Ann Clin Biochem** | **1** |
| **Ann Clin Microbiol Antimicrob** | **1** |
| **Ann Coil Surg** | **1** |
| **Ann de Biologie Clinique** | **1** |
| **Ann Econ Soc Measure** | **1** |
| **Ann Hematol** | **1** |
| **Ann Medit Burns Club** | **1** |
| **Ann Microscopy** | **1** |
| **Ann Occupational Hygiene** | **1** |
| **Ann Ophthalmol** | **1** |
| **Ann Rev Pub Health** | **1** |
| **Ann the New York Academy of Sciences** | **1** |
| **Antivir Ther** | **1** |
| **Aplan Journal of Rheumatology** | **1** |
| **Apoptosis** | **1** |
| **Appl Environ Microbiol** | **1** |
| **Appl Immunohistochem Mol Morpho** | **1** |
| **Applied and Environmental Microbiology** | **1** |
| **Apuzzo MLJ** | **1** |
| **Arc Path Lab Med** | **1** |
| **Arch In Med** | **1** |
| **Arch Iranian Med** | **1** |
| **Arch of Paed and Adolescent Med** | **1** |
| **Arch Ohrenh** | **1** |
| **Arch Womens Ment Health** | **1** |
| **Archive of Oncology** | **1** |
| **Archives of Diseases of Childhood** | **1** |
| **Archives of Pediatric and Adolescent Medicine** | **1** |
| **Ark Med Soc** | **1** |
| **Arm J Kidney Dis** | **1** |
| **Arq Bras Oftalmol** | **1** |
| **Arthritis Res Ther** | **1** |
| **Artif Organs** | **1** |
| **Asean Journal of Anaesthesiology** | **1** |
| **Assess** | **1** |
| **Asthma and Immunology** | **1** |
| **Aten Primaria** | **1** |
| **Audiology** | **1** |
| **Audit Commission of Local Authorities and National Health Services in England and Wales** | **1** |
| **Aust Clin Rev** | **1** |
| **Aust Crit Care** | **1** |
| **Aust Health Review** | **1** |
| **Aust J Derm** | **1** |
| **Aust J Ophthal** | **1** |
| **Aust J Wound Management** | **1** |
| **Aust N Z J Obstet Gynaecol** | **1** |
| **Aust NZ Med** | **1** |
| **Australas Radiol** | **1** |
| **Australasian Psychiatry** | **1** |
| **Australian Dental Journal** | **1** |
| **Australian Doctor** | **1** |
| **Baillieres Best Pract Res Clin Haematol** | **1** |
| **Bangladesh Med Res Counc Bull** | **1** |
| **Best Pract Res Clin Gastroenterol** | **1** |
| **Best Pract Res Clin Gynaecol** | **1** |
| **Best Practice and Research Clin Haematol** | **1** |
| **Best Practice and Research Clinical Anaesthesiology** | **1** |
| **Bethseda Maryland** | **1** |
| **Biochem Biophys Res Commun** | **1** |
| **Biochem. J.** | **1** |
| **Biokhimiia** | **1** |
| **Biol. Cell.** | **1** |
| **Bio-Med Mat. Eng** | **1** |
| **Biomedica** | **1** |
| **Biomedical Sciences Instrumentation** | **1** |
| **Bioproc. Eng.** | **1** |
| **Biotechniques** | **1** |
| **BJA** | **1** |
| **Blond Press Monit** | **1** |
| **Blood Reviews** | **1** |
| **BMC Cancer** | **1** |
| **BMC Complementary and Alternative Medicine** | **1** |
| **BMC Endocrine Disorders** | **1** |
| **BMC Health Services Research** | **1** |
| **BMC Infect Dis** | **1** |
| **BMC Pediatr** | **1** |
| **Bone joint Surg** | **1** |
| **Br J Oral Maxillofac Surg** | **1** |
| **Brain Inj** | **1** |
| **Brain Pathol** | **1** |
| **Bratisl Lek Listy** | **1** |
| **Breast Cancer Res Treat** | **1** |
| **Brief** | **1** |
| **British Dent J** | **1** |
| **British J Audiol** | **1** |
| **British J Clin Pract** | **1** |
| **British J Dis Chest** | **1** |
| **British J Hosp Med** | **1** |
| **British j of Health Psychology** | **1** |
| **British j of Medical psychology** | **1** |
| **British Med Bull** | **1** |
| **British National Formulary** | **1** |
| **Bronchus** | **1** |
| **Bull. Mater. Sci.** | **1** |
| **Bulletin. The Royal College of Surgeons of England Bulletin** | **1** |
| **CA** | **1** |
| **CA Cancer J Clin** | **1** |
| **Can Fam Physician** | **1** |
| **Can J Clin Pharmacol** | **1** |
| **Can J Microbiol** | **1** |
| **Can J Neurol Sc** | **1** |
| **Can J Public Health** | **1** |
| **Can J Urol** | **1** |
| **Can Respir J** | **1** |
| **Canadian Breast Cancer Initiative** | **1** |
| **Canadian Journal of Neurological Sciences** | **1** |
| **Canadian journal of Otolaryngology** | **1** |
| **Cardiology** | **1** |
| **Cardiovasc Med** | **1** |
| **Catheter Cardiovas Interv** | **1** |
| **CDA J** | **1** |
| **Cellular and Molecular Life Sciences** | **1** |
| **Cent Afr J Med** | **1** |
| **Cephalalgia** | **1** |
| **Cesk Slov Oftalmol** | **1** |
| **Ceylon Med J** | **1** |
| **Changgeng Yi Xue Za Zhi** | **1** |
| **Chemosphere** | **1** |
| **Child Psychiatry and Human Development** | **1** |
| **Chin MED Assoc** | **1** |
| **Chinese J Cancer Res** | **1** |
| **Chirurgie** | **1** |
| **Chronobiol Int** | **1** |
| **Circ J** | **1** |
| **Circ Res** | **1** |
| **Cleft Palate Journal** | **1** |
| **Cleft Palate Speech** | **1** |
| **Clev Clin J Med** | **1** |
| **Clevelan Clinic Foundation Experience Laryngoscope** | **1** |
| **Clin Pulm Med** | **1** |
| **Clin Radiol** | **1** |
| **Clin Sci** | **1** |
| **Clinical** | **1** |
| **Clinical and experimental hypertension** | **1** |
| **Clinical Auton Res** | **1** |
| **Clinical Biochem** | **1** |
| **Clinical Cardiol** | **1** |
| **Clinical Chest Med** | **1** |
| **Clinical Diag Virol** | **1** |
| **Clinical Electroencephalogr** | **1** |
| **Clinical Exp Obstet Gynecol** | **1** |
| **Clinical Expt Derm** | **1** |
| **Clinical Genetics** | **1** |
| **Clinical Immuno Immunopathol** | **1** |
| **Clinical Immunol** | **1** |
| **Clinical Invest Med** | **1** |
| **Clinical J Pain** | **1** |
| **Clinical Lab Med** | **1** |
| **Clinical Lymphoma** | **1** |
| **Clinical Med.** | **1** |
| **Clinical Modma** | **1** |
| **Clinical Neurosurg** | **1** |
| **Clinical North Am** | **1** |
| **Clinical Nuclear Medicine** | **1** |
| **Clinical Nurse Spec** | **1** |
| **Clinical Plast Surg** | **1** |
| **Clinical Rheumatol** | **1** |
| **Clinical Ther** | **1** |
| **Clinical Toxicol** | **1** |
| **Clinical Transplant** | **1** |
| **Clinics in Geriatric Medicine** | **1** |
| **Cloning and Stem Cells** | **1** |
| **CME Bulletin** | **1** |
| **Colorectal Dis** | **1** |
| **Commun Dis Intell** | **1** |
| **Companion to Psychiatric studies** | **1** |
| **Complement Ther Clin Pract** | **1** |
| **Composites Science and Technology** | **1** |
| **Cont Lens Anterior Eye** | **1** |
| **Contemporary Neurosurgery** | **1** |
| **Contr Nephrol** | **1** |
| **Contrib Nephrol** | **1** |
| **CRC Press** | **1** |
| **Crit Rev Food Sci Nutr** | **1** |
| **Critical Care** | **1** |
| **Critical Care and Resuscitation** | **1** |
| **Critical Review Oral Biology Med** | **1** |
| **Critical Reviews in Immunology** | **1** |
| **Critical Reviews in Oncology/Hematology** | **1** |
| **Curr Control Trials Cardiovasc Med** | **1** |
| **Curr Hypertens Rep** | **1** |
| **Curr Med Chem** | **1** |
| **Curr Opin Lipidol** | **1** |
| **Curr Stem Cell Res Ther** | **1** |
| **Curr. Probl. Diagn. Radiol** | **1** |
| **Current Anaesthesia and Critical Care** | **1** |
| **Current opinion Anaesth** | **1** |
| **Current opinion Cardiol** | **1** |
| **Current opinion Cut Care** | **1** |
| **Current opinion Hematol** | **1** |
| **Current Opinion in Microbiology** | **1** |
| **Current opinion in Otolaryngol Head Neck Surg** | **1** |
| **Current opinion Orthop** | **1** |
| **Current opinion Rheumatology** | **1** |
| **Current opinion Urol** | **1** |
| **Current Probl Surg** | **1** |
| **Current Topics Microbiol Immunol** | **1** |
| **Cutis** | **1** |
| **CVD prevention** | **1** |
| **Dermatology Surgery** | **1** |
| **Dev Dialogue** | **1** |
| **Dev Pharmacol** | **1** |
| **Developmental Dynamics** | **1** |
| **Diabetes mellitus, hypertension and their associated factors** | **1** |
| **Diabetes Print Care** | **1** |
| **Diagn Cytopathol** | **1** |
| **Diagn Interv Radiol** | **1** |
| **Diagn Microbiol Infect Dis** | **1** |
| **Dig Dis Sci** | **1** |
| **Dig Liver Dis** | **1** |
| **Digestion** | **1** |
| **Dis Ear** | **1** |
| **Dis Esophagus** | **1** |
| **Dis Manag** | **1** |
| **Dis Mon** | **1** |
| **Dis. Colon Rectum** | **1** |
| **Dordrecht** | **1** |
| **Drug Discovery Today** | **1** |
| **Drug Invest** | **1** |
| **Drug Metabolism and Drug Interactions** | **1** |
| **Drug Resistance Updates** | **1** |
| **Ear Hear** | **1** |
| **Ear. Preg: Bio. and Med** | **1** |
| **Early Human Dev** | **1** |
| **Eastern Mediterranean Health Journal** | **1** |
| **EBMT** | **1** |
| **Eksp Klin Gastroenterol** | **1** |
| **Electronic Journal of Biotechnology** | **1** |
| **EMBO J** | **1** |
| **Emergency orthopaedics and trauma** | **1** |
| **Encephale** | **1** |
| **Endocr Rev** | **1** |
| **Eng. Mat** | **1** |
| **ENT Journal** | **1** |
| **Environ Health Prev Med** | **1** |
| **Environmental International** | **1** |
| **Epidemiol Community Health** | **1** |
| **Epidemiol Infect** | **1** |
| **Epidemiol Rev** | **1** |
| **Epidemiologic Reviews** | **1** |
| **Epidemiology** | **1** |
| **Eur Cytokine Netw** | **1** |
| **Eur J Anaesthesiol** | **1** |
| **Eur J Biochem** | **1** |
| **Eur J Clin Invest** | **1** |
| **Eur J Eur Med** | **1** |
| **Eur J Gen Pract** | **1** |
| **Eur J Hum Genet** | **1** |
| **Eur J Med Res** | **1** |
| **Eur J Nucl Med** | **1** |
| **Eur J Oral Sci** | **1** |
| **Eur J Respir Dis** | **1** |
| **Eur J Surg Oncol** | **1** |
| **Eur Urol** | **1** |
| **Eur. Ceram. Soc** | **1** |
| **Eur. J. Clin. Pharmacol** | **1** |
| **Eur. J. Inorg. Chem** | **1** |
| **Eur. J. Pharmacol** | **1** |
| **Europ. Cells Mats** | **1** |
| **European J Anaesth** | **1** |
| **European Radiology** | **1** |
| **Exp Clin Transplant** | **1** |
| **Exp Dermatol** | **1** |
| **Exp. Gerontol** | **1** |
| **exp. Hemato** | **1** |
| **Experimental Neurology** | **1** |
| **expression. Oncogene** | **1** |
| **Eye** | **1** |
| **Eye and Contact Lens** | **1** |
| **Eye, Ear, Nose, Throat Mon** | **1** |
| **Facial Plast Surg Clin North Am** | **1** |
| **Fakta Kesihatan Johor** | **1** |
| **Family Matters** | **1** |
| **Family Planning Perspectives** | **1** |
| **FEMS Iminunol Med Microbiol** | **1** |
| **FEMS Microbiol Rev** | **1** |
| **Fertil Control Rev** | **1** |
| **Field's Virology** | **1** |
| **First Asean Medical Conference** | **1** |
| **Fortschr Neuro Psychiatr** | **1** |
| **Fox Chase Cancer Centre Scientific Report** | **1** |
| **FRIM** | **1** |
| **Front Biosci** | **1** |
| **Fund Clin Pharmacol** | **1** |
| **Fur Neurol** | **1** |
| **Gait Posture** | **1** |
| **Gan To Kagaku Ryoho** | **1** |
| **Gastrointest Endosc Clin N Am** | **1** |
| **Gene** | **1** |
| **General Pharmac** | **1** |
| **Genes Chrom Cancer** | **1** |
| **Genes Devel** | **1** |
| **Genet Epidmiol** | **1** |
| **Geneva** | **1** |
| **Genome** | **1** |
| **Geriatr Dent Update** | **1** |
| **Ginecologiay Obstetricia de Mexico** | **1** |
| **Graefes Arch Clin Exp Ophthalmol** | **1** |
| **Gynecol Obstet Invest** | **1** |
| **Gynecol Oncol** | **1** |
| **Hand Surg** | **1** |
| **Healing. Clin. Orthop** | **1** |
| **Health and Quality of Life Outcomes** | **1** |
| **Health (Millwood)** | **1** |
| **Health Care Management Science** | **1** |
| **Health Care Women Int** | **1** |
| **Health Communication Australia** | **1** |
| **Health Facts** | **1** |
| **Health Promotion International** | **1** |
| **Health Serv Res** | **1** |
| **Health technology Assessment** | **1** |
| **HepatogastroenterOl** | **1** |
| **Hepatogastroenterology** | **1** |
| **HKMJ** | **1** |
| **HNO Bilateral recurrent laryngeal nerve palsy. HNO** | **1** |
| **Hong Kong Med J** | **1** |
| **Hosp J** | **1** |
| **Hum Hered.** | **1** |
| **Hum Immunol** | **1** |
| **Hum Mutat** | **1** |
| **Human Factors** | **1** |
| **Human Kinetics** | **1** |
| **Human Mutation** | **1** |
| **Human Pathol** | **1** |
| **Hypertension in pregnancy** | **1** |
| **I Cancer Educ** | **1** |
| **IIUM Engineering Journal** | **1** |
| **Immun Infekt** | **1** |
| **Immunol Cell Biol** | **1** |
| **In Vivo** | **1** |
| **Ind Biologist** | **1** |
| **Ind J Hum Genet** | **1** |
| **Ind J Physiol Pharmacol** | **1** |
| **Ind J Radiol Imag** | **1** |
| **Ind Practitioner** | **1** |
| **Indian Gastroenderol** | **1** |
| **Indian J Med Microbiol** | **1** |
| **Indian J Ophthalmol** | **1** |
| **Indian J Pathol Microbiol** | **1** |
| **Indian J Psychiatry** | **1** |
| **Indian Journal Med** | **1** |
| **Indian Journal of Malariology** | **1** |
| **Industrial Health** | **1** |
| **Infect Control** | **1** |
| **Inorganica Chimica Acta.** | **1** |
| **Instr Course Lect** | **1** |
| **Int Anesthesiol Clin** | **1** |
| **Int J Addict** | **1** |
| **Int J Cardiol** | **1** |
| **Int J Clin Lab Res** | **1** |
| **Int J Gynecol Obstet** | **1** |
| **Int J Health Serv** | **1** |
| **Int J Obes Relat Metab Disord** | **1** |
| **Int J Obstet Anesth** | **1** |
| **Int J of Aging and Human Development** | **1** |
| **Int J of Health Services** | **1** |
| **Int J Paediatr Dent** | **1** |
| **Int J Psychiatry Med** | **1** |
| **Int J Qual Health Care** | **1** |
| **Int J Soc Psychiatry** | **1** |
| **Int J Sports Med** | **1** |
| **Int Journal Radiation Oncology Biology Physiology** | **1** |
| **Int Rev Pyschiatry** | **1** |
| **Int Surg** | **1** |
| **Inter J Obstet Gynecol** | **1** |
| **Internat Sociol Assoc** | **1** |
| **International Archives of Occupational and Environmental Health** | **1** |
| **International Medical Research Journal** | **1** |
| **International Review of Vitamin Research** | **1** |
| **International Society for Trophoblastic Disease** | **1** |
| **International Statistical Rev** | **1** |
| **Internatl J Nurs Stud** | **1** |
| **Intervirology** | **1** |
| **Invest Radiol** | **1** |
| **IRSC Med Sci** | **1** |
| **Isr J Med Sci** | **1** |
| **Isr Med Assoc J** | **1** |
| **ISTA** | **1** |
| **J of mol** | **1** |
| **J of Mov** | **1** |
| **Jap. J. Cl. Ophtalmol** | **1** |
| **JARO** | **1** |
| **Journal Accid Emerg Med** | **1** |
| **Journal Acous Scoc. Am** | **1** |
| **Journal Adoles Res** | **1** |
| **Journal Adolescent Health** | **1** |
| **Journal Agricultural Safety and Health** | **1** |
| **Journal Agricultural and Food Chemistry** | **1** |
| **Journal AM Coll Nutr** | **1** |
| **Journal Am Med Women's Assoc** | **1** |
| **Journal Am Nutraceitical Assoc** | **1** |
| **Journal Am Soc Surg Hand** | **1** |
| **Journal Appl Physiol** | **1** |
| **Journal Arthroplasty** | **1** |
| **Journal Assist Reprod Genet** | **1** |
| **Journal Asst. Reprod. Genet** | **1** |
| **Journal Atheroscler Res** | **1** |
| **Journal Bacteriol** | **1** |
| **Journal Behav Med** | **1** |
| **Journal Biomechanics.** | **1** |
| **Journal Biomol Tech** | **1** |
| **Journal Bioorganis and Medicinal Chemistry Letters** | **1** |
| **Journal Biotechnol** | **1** |
| **Journal Bone Miner Metab** | **1** |
| **Journal Burn Care Rehab** | **1** |
| **Journal California Dent** | **1** |
| **Journal Canadian Dental Association** | **1** |
| **Journal Cardiopulm Rehabil** | **1** |
| **Journal Cardiovasc Electrophysiol** | **1** |
| **Journal Cell Science** | **1** |
| **Journal Ceram. Soc. Jpn.** | **1** |
| **Journal Cereb Blood Flow Metab** | **1** |
| **Journal Chronic Dis** | **1** |
| **Journal Clin Pharm Ther** | **1** |
| **Journal Clin Pharmacol** | **1** |
| **Journal Clin Psychiatry** | **1** |
| **Journal Clin Psychol Med Settings** | **1** |
| **Journal Clin Rheu** | **1** |
| **Journal Clin Virol** | **1** |
| **Journal Cognitive Psychotherapy** | **1** |
| **Journal Common Health** | **1** |
| **Journal Community Health** | **1** |
| **Journal Cont. Educ Health Prof.** | **1** |
| **Journal Counsel Develop** | **1** |
| **Journal Counselling Psych** | **1** |
| **Journal Cross-Cultural Gerontology** | **1** |
| **Journal Cross-Cultural Psychol** | **1** |
| **Journal Dent Res** | **1** |
| **Journal Diabetes and its Complications** | **1** |
| **Journal Digit Imaging** | **1** |
| **Journal Divorce** | **1** |
| **Journal Econometrics** | **1** |
| **Journal Emerg Nurs** | **1** |
| **Journal Endourol** | **1** |
| **Journal Environmental Quality** | **1** |
| **Journal Euro. Ceram. Soc.** | **1** |
| **Journal Europ Acad Dermatol Venerol** | **1** |
| **Journal Family Psychology** | **1** |
| **Journal Fed Child Health** | **1** |
| **Journal Forensic Sci** | **1** |
| **Journal Form Med Ass** | **1** |
| **Journal Gastroenterology and Hepatology** | **1** |
| **Journal Hand Ther** | **1** |
| **Journal Heart Lung Transplant** | **1** |
| **Journal Hong Kong College Radiol** | **1** |
| **Journal Hum Ecol** | **1** |
| **Journal Hum Genet** | **1** |
| **Journal Humanistic Psych** | **1** |
| **Journal Hyg Epidemiol Microbiol Immunol** | **1** |
| **Journal Ind Anthrop Soc** | **1** |
| **Journal Indian Academy Clin Med** | **1** |
| **Journal Korean Acad Rehabil Med** | **1** |
| **Journal Korean Neurol Assoc** | **1** |
| **Journal Lab Clin Med** | **1** |
| **Journal Laparoendosc Adv Surg Tech** | **1** |
| **Journal Malaysia Society of Health** | **1** |
| **Journal Matern Fetal Med** | **1** |
| **Journal Matern Fetal Medicine** | **1** |
| **Journal Max-Fac Surg** | **1** |
| **Journal Med Sci** | **1** |
| **Journal Medical Imaging** | **1** |
| **Journal Mental Hlth.** | **1** |
| **Journal Midwifery Womens Health** | **1** |
| **Journal Mol Biol** | **1** |
| **Journal Mol Cell Cardiol.** | **1** |
| **Journal Mol Med** | **1** |
| **Journal Mol. Endocrinol** | **1** |
| **Journal National Medical Association** | **1** |
| **Journal Neurocytol** | **1** |
| **Journal Neurol** | **1** |
| **Journal Neurol Rehab** | **1** |
| **Journal Neuropsychiatry Clin Neurosci** | **1** |
| **Journal Neurosci Methods** | **1** |
| **Journal New Ment Dis** | **1** |
| **Journal Nippon Med Sch** | **1** |
| **Journal Nucl Med** | **1** |
| **Journal Nurs Manag** | **1** |
| **Journal Occup. Org. Psychol.** | **1** |
| **Journal of Diab** | **1** |
| **Journal of Gast** | **1** |
| **Journal of Orthop** | **1** |
| **Journal Oral Maxillofac Surg** | **1** |
| **Journal Oral Tissue Engineering.** | **1** |
| **Journal Ortho Sports Phys Ther** | **1** |
| **Journal Orthop. Surg** | **1** |
| **Journal Orthopaedic and Sports Physical Therapy** | **1** |
| **Journal Paediatr Orthop** | **1** |
| **Journal Paediatric Nursing** | **1** |
| **Journal Pain Palliat Care Pharmacother** | **1** |
| **Journal Pain Symptom Manage** | **1** |
| **Journal Pak Med Assoc** | **1** |
| **Journal Palliative Med** | **1** |
| **Journal Pediatr Adolesc Gynecol** | **1** |
| **Journal Pediatr Endocrinol Metab** | **1** |
| **Journal Pers Soc Psychol** | **1** |
| **Journal Pharmacol Exp Ther** | **1** |
| **Journal Pharmacology** | **1** |
| **Journal Plast Reconstr Aesthet Surg** | **1** |
| **Journal Plast Reconstr Surg** | **1** |
| **Journal Polynes Soc** | **1** |
| **Journal Postgraduate Med** | **1** |
| **Journal Psy Obstet Gynecol** | **1** |
| **Journal Psychosoc Nurs Ment Health Serv** | **1** |
| **Journal Psychosom Res** | **1** |
| **Journal Public Health** | **1** |
| **Journal Public Health Dent** | **1** |
| **Journal Public Health Manag Prac** | **1** |
| **Journal Qual Clin Pract** | **1** |
| **Journal R Coil Physicians Lond** | **1** |
| **Journal R. Coll. Surg.** | **1** |
| **Journal Research** | **1** |
| **Journal Roy Soc Med** | **1** |
| **Journal Singapore Paed Soc** | **1** |
| **Journal Soc Gynaecol Investig** | **1** |
| **Journal Soc Psychol** | **1** |
| **Journal Social Science Medicine** | **1** |
| **Journal Solid State Chem** | **1** |
| **Journal Sound and Vibration** | **1** |
| **Journal Surg** | **1** |
| **Journal the American Geriatrics Society** | **1** |
| **Journal the American Medical Association** | **1** |
| **Journal the American Society of Nephrology** | **1** |
| **Journal the Irish Dental Association** | **1** |
| **Journal the Malaysian Society of Health** | **1** |
| **Journal Toxicol** | **1** |
| **Journal Toxicology and Environmental Health** | **1** |
| **Journal Tropical Pediatrics** | **1** |
| **Jpn J Cancer Res** | **1** |
| **Jpn. J Nephrol** | **1** |
| **Jur of ind va** | **1** |
| **Kansenshogaku Zasshi** | **1** |
| **Kaohsiung Journal of Medical Sciences** | **1** |
| **Key Eng Mater** | **1** |
| **Kobe J. Med. Sci.** | **1** |
| **Korean J Infect Dis** | **1** |
| **Langenbecks Arch Chir Suppl Kongressbd** | **1** |
| **Laser Med Surg** | **1** |
| **Law Hum Behav** | **1** |
| **Leuk Lyphoma** | **1** |
| **Life** | **1** |
| **lnt J Obes Relat Metab Disord** | **1** |
| **lnt Orthop** | **1** |
| **Lung Cancer** | **1** |
| **Malaysia J Psychiatry** | **1** |
| **Malaysian Journal** | **1** |
| **Malaysian Journal of Biochemistry and Molecular Biology** | **1** |
| **Maryland State Medical Journal** | **1** |
| **Mater. Chem. Phys** | **1** |
| **Mater. Res** | **1** |
| **Materials and Design** | **1** |
| **MD Comput** | **1** |
| **Medical  Clin Barc** | **1** |
| **Medical  Clin North Am** | **1** |
| **Medical  Dig** | **1** |
| **Medical  Laser Appl** | **1** |
| **Medical  Monit** | **1** |
| **Medical  Virol** | **1** |
| **Medicine (Baltimore)** | **1** |
| **Medicine of the Americas** | **1** |
| **Medika** | **1** |
| **Medycyna Pracy** | **1** |
| **Mental Health** | **1** |
| **Mental Retard and Dev Dis Res Rev** | **1** |
| **Mickroskopie** | **1** |
| **Micosurgery** | **1** |
| **Microbiol Immunol** | **1** |
| **Microbiology** | **1** |
| **Mil Med** | **1** |
| **Minerva Anestesiol** | **1** |
| **Minerva Ginecol** | **1** |
| **MMWR Morb Mortal Wkly Rep** | **1** |
| **MNWR. Legionellosis** | **1** |
| **Mol' Cell Endocrinol** | **1** |
| **Mol Diagn** | **1** |
| **Mol Microbiol.** | **1** |
| **Mol Pharmacol** | **1** |
| **Molecular Brain Research** | **1** |
| **Molecular Therapy** | **1** |
| **Monte Carlo, Monaco** | **1** |
| **Mov Disorder** | **1** |
| **Mutation Res** | **1** |
| **Mycoses** | **1** |
| **Nat Clin Prac Gastroenterol Hepatol** | **1** |
| **Nat Genet** | **1** |
| **Nat Med** | **1** |
| **Nat. Rev. Drug Discov** | **1** |
| **National Science Foundation. US** | **1** |
| **Nature Clin Prac Nephrol** | **1** |
| **Nature Reviews: Drug Discovery.** | **1** |
| **Nature Structural Biology** | **1** |
| **Naturwissenschaften** | **1** |
| **NCI Monographs** | **1** |
| **Neonatal Netw** | **1** |
| **Nephron (Switzerland)** | **1** |
| **Nephron Clin Pract** | **1** |
| **Neuroendocrinol** | **1** |
| **Neuroimaging Clinics of North America** | **1** |
| **Neurol Clin** | **1** |
| **Neurol Med Chir** | **1** |
| **Neurol Neurochir Pol** | **1** |
| **Neuromuscul Disord** | **1** |
| **Neuro-oncol** | **1** |
| **Neuropsychiatry Clin Neurosci** | **1** |
| **Neuroscientist** | **1** |
| **Nippon Rinsho** | **1** |
| **NMCJ** | **1** |
| **Nucl Acids Res** | **1** |
| **Nucleic Acids Res** | **1** |
| **Nurs Res** | **1** |
| **Nurs Stand** | **1** |
| **Nursing Standard** | **1** |
| **Nutr Cancer** | **1** |
| **Nutr Health** | **1** |
| **Nutr Metab Cardiovasc Dis** | **1** |
| **Nutr Rev** | **1** |
| **Nutrition** | **1** |
| **O Med O Surg O Pathol O Radiology Endodontics** | **1** |
| **Obes Metab** | **1** |
| **Obesity Reviews** | **1** |
| **Obstet Gynecol Clin Nort Am** | **1** |
| **Obstet Gynecol Surv** | **1** |
| **Oncol Rep** | **1** |
| **Opthalmic surgery** | **1** |
| **Osteoarthritis Cartilage** | **1** |
| **Otol Neutrotol** | **1** |
| **Otoneurotol** | **1** |
| **Outcome in Pediatric Tracheotomy** | **1** |
| **Pacing Clin Electrophysiol** | **1** |
| **Paediatr Neurosci** | **1** |
| **Pam Pract** | **1** |
| **Pan American Journal of Public Health** | **1** |
| **Parasitology Today** | **1** |
| **Pathol International** | **1** |
| **Pathol Res Pract** | **1** |
| **Pediatr Ann** | **1** |
| **Pediatr Cardiol** | **1** |
| **Pediatr Nurs** | **1** |
| **Pediatr Pathol** | **1** |
| **Pediatr Surg Int** | **1** |
| **Pediatric Infect Dis J** | **1** |
| **Pediatric Nursing** | **1** |
| **Peripheral Neuropathy.** | **1** |
| **Peritoneal Dialysis International** | **1** |
| **pers. comm** | **1** |
| **Pharm Res** | **1** |
| **Pharmacol Biochem Behav** | **1** |
| **Pharmacol Rev** | **1** |
| **Pharmacy International** | **1** |
| **Phil J Infect Dis** | **1** |
| **Philadelphia** | **1** |
| **Phlebologie** | **1** |
| **Physician Sports Med** | **1** |
| **Physiol. Rev** | **1** |
| **Phytochem** | **1** |
| **Plant Mol Biol Reporter** | **1** |
| **Practical Diabetes Int** | **1** |
| **Press Med** | **1** |
| **Preventive Med** | **1** |
| **Principles and Practices of Clinical Virology** | **1** |
| **Principles of Orthopaedic Practice** | **1** |
| **Proc R Soc Lond B Biol Sci** | **1** |
| **Proc R Soc Med** | **1** |
| **Proc. Natl Acad.** | **1** |
| **Prof Nurse** | **1** |
| **Prog Cardiovasc Dis** | **1** |
| **Prophylaxis MMWR** | **1** |
| **Prosthet . Orthop Int** | **1** |
| **Prov Med** | **1** |
| **Psi Chi Journal Winter** | **1** |
| **Psychiatr Serv** | **1** |
| **Psychiatric Bulletin** | **1** |
| **Psychiatric Services** | **1** |
| **Psychiatrica Scandinavica** | **1** |
| **Psychiatry and Clinical Neurosciences** | **1** |
| **Psychological Medicine** | **1** |
| **Psychologists Press** | **1** |
| **Psychometrika** | **1** |
| **Psycho-Oncol** | **1** |
| **Psychother Psychosom** | **1** |
| **Psychotherapy Research** | **1** |
| **Psycological Assessment** | **1** |
| **Q Rev Biol** | **1** |
| **Radiol Med (Torino)** | **1** |
| **Radiology Clinics of North America** | **1** |
| **Radiother. Oncol.** | **1** |
| **Reconstr. Surg** | **1** |
| **Reprod Toxicol** | **1** |
| **Reprod. Biomed. Onl** | **1** |
| **Res Exp Med (Berl)** | **1** |
| **Res Virol** | **1** |
| **Research in Nursing and Health** | **1** |
| **Rev Clin Exp Hematol** | **1** |
| **Rev Infect Dis** | **1** |
| **Rev Med Interns** | **1** |
| **Rev Rhum** | **1** |
| **Rev Saude Publica** | **1** |
| **Rev scin tech Off int Epiz** | **1** |
| **Rev. Esp. Cardiol** | **1** |
| **Rev. Otoneuroophthalmol** | **1** |
| **Review of endoprosthetic reconstruction in the limb sparing surgery** | **1** |
| **Revista Brasileira de Psiquiatria** | **1** |
| **Revista de Biologica Tropical** | **1** |
| **Rheumatol Int** | **1** |
| **Rheumatology and Rehabilitation** | **1** |
| **S Afr J Surg** | **1** |
| **Saccine** | **1** |
| **Sains Malaysiana** | **1** |
| **Salud Publica Mex** | **1** |
| **Saunders** | **1** |
| **Scan J Clin Lab Invest** | **1** |
| **Scand J Dent Res** | **1** |
| **Scand J Plast Reconstr Hand Surg.** | **1** |
| **Scand J Prim Health Care** | **1** |
| **Scandinavian Audiology** | **1** |
| **Scandinavian Journal of Behaviour Therapy** | **1** |
| **Scandinavian Journal of Work, Environment and Health** | **1** |
| **Schizophrenia Bull** | **1** |
| **Sci Technol Adv Mater** | **1** |
| **ScienceAsia.** | **1** |
| **Scientific America** | **1** |
| **Sem Pediatr Infect Dis** | **1** |
| **Semin Reprod Med** | **1** |
| **Semin Surg Oncol** | **1** |
| **Semin. Cardiothorac. Vasc. Anesth** | **1** |
| **Senin Arthritis Rheum** | **1** |
| **Sensorineural Hearing Loss Pediatric** | **1** |
| **Shock** | **1** |
| **Shoulder** | **1** |
| **Sing J Obstet Gynaecol** | **1** |
| **Skel. Radiol.** | **1** |
| **Skeletal Radiology** | **1** |
| **Sleep Med Rev** | **1** |
| **SOC PREV MED** | **1** |
| **South Asian J of Preventive Cardiology** | **1** |
| **Soz Praventivmed** | **1** |
| **Steroids** | **1** |
| **Stress and Health** | **1** |
| **Studies on mucoceles of the maxillary sinuses Rhinology** | **1** |
| **Surg** | **1** |
| **Surg Engl** | **1** |
| **Surg Radiol Anat** | **1** |
| **Surg. Edin** | **1** |
| **Surgical Neurology** | **1** |
| **Surgical Oncology** | **1** |
| **TAMA** | **1** |
| **Tani Gir*ISI*m Radyol** | **1** |
| **Thai StemLife** | **1** |
| **Thromb. Vasc. Biol.** | **1** |
| **Thyroid** | **1** |
| **Thyroidology** | **1** |
| **Topics in Emerg Med** | **1** |
| **Toxicology Letters** | **1** |
| **Trans Ophthalmol. Soc. UK** | **1** |
| **Trans Orthop Res Soc.** | **1** |
| **Trans Soc Pathol Jpn** | **1** |
| **Transactions of the Royal Society of Tropical Medicine and Hygiene** | **1** |
| **Transcultural Psychiatry** | **1** |
| **Travel Med Infect Dis** | **1** |
| **Treat Respir Med.** | **1** |
| **Trends Biomater. Artif. Organs.** | **1** |
| **Trends in Molecular Medicine** | **1** |
| **Trends of Genetics** | **1** |
| **Trop. Med. Parasitol** | **1** |
| **Tropical Doctor** | **1** |
| **Tumors of the central nervous system; Atlas of tumour pathology : AFIP** | **1** |
| **Turk J Gastroenterol** | **1** |
| **Turk J Haematol** | **1** |
| **Turkish Journal of Pediatrics** | **1** |
| **Unfallchirurg** | **1** |
| **Upper airway tumor, Primary tumor.** | **1** |
| **Vaccine** | **1** |
| **Vasc Surg** | **1** |
| **Virchows Arch A Pathol Anat Histopathol** | **1** |
| **Virology** | **1** |
| **Vitro Cellular and Developmental Biology** | **1** |
| **Vopr Pitan** | **1** |
| **WAO Research and Advocacy** | **1** |
| **Water Qual. Int** | **1** |
| **Water Research** | **1** |
| **Wien Klin Wonchenschr** | **1** |
| **Women at risk for postpartum** | **1** |
| **Women Health** | **1** |
| **Wound Repair Regeneration** | **1** |
| **Yakugaku Zasshi** | **1** |
| **Yi Xue Ke Xue Za Zhi** | **1** |
| **Zentrabl Chir** | **1** |
| **Zhonghua Nei Ke Za Zhi** | **1** |
| **Zhonghua Xue Ye Xue Za Zhi** | **1** |